ADVANCES IN

# SECURITY IN COMPUTING AND COMMUNICATIONS

Edited by **Jaydip Sen**

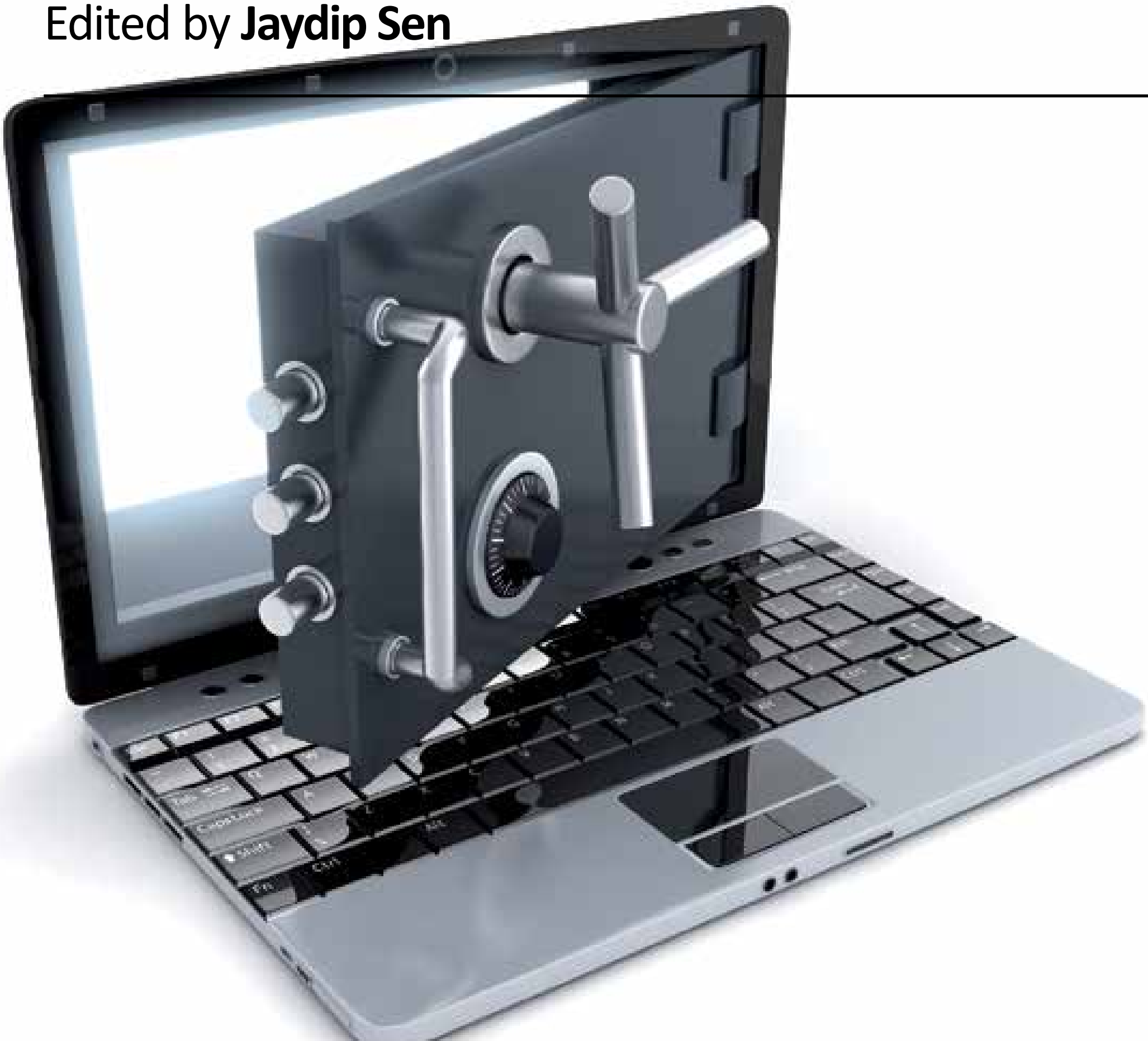

# ADVANCES IN SECURITY IN COMPUTING AND COMMUNICATIONS

Edited by **Jaydip Sen**

**Contributors**

Javier Franco-Contreras, Gouenou Coatrieux, Nilay K Sangani, Haroot Zarger, Faouzi Jaidi, Bob Duncan, Alfred Bratterud, Andreas Happe, Chin-Feng Lin, Che-Wei Liu, Walid Elgeanidi, Muftah Fraifer, Thomas Newe, Eoin OConnell, Avijit Mathur, Ruolin Zhang, Eric Filiol

# Contents

# Preface

The field of cryptography as a separate discipline of computer science and communications engineering announced its arrival onto the world stage in the early 1990s with a full promise to secure the Internet. At that point of time, many envisioned cryptography as a great technological equalizer that could put the weakest privacy-seeking individual on the same platform as the greatest national intelligence agencies with powerful resources at their command. Some political strategists forecasted that the power of cryptography would bring about the downfall of nations when governments would no longer be able to snoop on people in the cyberspace, while others looked forward to it as a fantastic tool for the drug dealers, terrorists, and child pornographers, who would be able to communicate in perfect secrecy. Some proponents also imagined cryptography as a technology that would enable global commerce in this new online world.

Even 25 years later, none of these expectations are met in reality today. Despite the phenomenal advances in cryptographic algorithms, the Internet's national borders are more apparent than ever. The ability to detect and eavesdrop on criminal communications has more to do with politics and manual interventions by human rather than by automatic detection or prevention by mathematics of cryptographic protocols. Individuals still don't stand a chance against powerful and well-funded government agencies by seeking protection under the shield of cryptography. And the rise of global commerce had more to do with open economic policies of the nations than on the prevalence of cryptographic protocols and standards.

While it is true that cryptography has failed to provide its users the real security it promised, the reasons for this failure have less to do with cryptography as a mathematical science. Rather, poor implementation of cryptographic protocols and algorithms has been the major source of problems. Although to a large extent we have been successful in developing cryptographic systems, what we have been less effective at is to convert the mathematical promise and ideas of cryptographic security into a secure working system in practice.

Another aspect of cryptography that is responsible for its failure in real world is that there are too many myths about it. There is no dearth of engineers who consider cryptography as a sort of magic wand that they can wave over their hardware or software in order to achieve the security level promised by the cryptographic algorithms. Far too many users impose their full faith on the word "encrypted" in the products they use and live under the false impression of magical security in their operations. Reviewers have also no exceptions, comparing algorithms and protocols on the basis of key lengths and then falsely believing that products using longer key lengths are more secure.

The literature of cryptography has also served no good in spreading the myths about cryptography. Numerous propositions have been made for increasing the key length of a partic-ular protocol to enhance its mythical security level without any concrete specification and guidelines about how to generate the keys. Sophisticated and complex cryptographic protocols have been designed without adequate considerations about the business and social and computing constraints under which those protocols would have to work. Too much effort has been spent in promoting cryptography as a pure mathematical ideal working in an isolated magic box, untarnished by and oblivious of any real-world constraints and realities. But it's exactly those real-world constraints and realities that make the difference between the promise of cryptographic magic and the reality of digital security.

While the Advanced Encryption Standard (AES) is being embedded into more and more devices and there are some interesting developments in the area of public key cryptography, many implementation challenges confront the security researchers and engineers today. Side channels, poorly designed APIs, and protocol failures continue to break systems. Pervasive computing also has opened up new challenges. As computers and communications become embedded invisibly everywhere in the era of the Internet of Things (IoT), the problems that used to only affect the traditional computers have cropped up in all other devices including smartphones, tablets, refrigerators, air-conditions, televisions, and other household gadgets and devices. Today, security also interacts with safety in applications from cars through utilities to electronic healthcare. Hence, it has become imperative for security engineers and practitioners to understand not only the technicalities of cryptographic algorithms and operating systems but also the economics and human factors of the applications as well. With the advent of ubiquitous computing and the Internet of Things, the issue of security and privacy in computing and communications is no longer a problem challenging some computer scientists and system engineers. Computer forensics is increasingly becoming an important and multidisciplinary subject with many of the crimes today being committed using servers, laptop computers, smartphones, and other specialized handheld digital devices. It is becoming mandatory for lawyers, accountants, managers, bankers, and other professionals whose day-to-day job may not involve technicalities of computer engineering to have working-level awareness of system and communication security, access control, and other privacy-related issues in their computing systems so as to effectively perform their tasks. Exponential growth in the number of users of social networking applications like Facebook, Twitter, Quora, etc. and online services provided by companies like Google and Amazon has changed the world too. Ensuring robust authentication and providing data privacy in massively parallel and distributed systems have posed significant challenges to the security engineers and scientists. Fixing bugs in online applications has become a critical issue to handle as an increasingly large numbers of sensitive applications are launched in the web and smartphones. Securing an operating system and an application software is not enough in today's connected world. What is needed is a complete security analysis of the entire computing system including its online and mobile applications. In other words, we are witnessing a rapidly changing world of extremely fast-evolving techno-socio-economic systems without having much knowledge about how the evolution is being driven and who is in control. The one incident of recent past that has brought about most significant changes in the security industry by altering our perceptions and priorities in design and operations of our systems is the tragic event of September 2001. Since then, terrorism is no longer being just considered as a risk. It is now being treated as a proactive perception of risk and the subsequent manipulation and mitigation, if not elimination of that risk. This has resulted in security being an amalgamation of technology, psychology, politics, and economics. In this current context, security engineers must contribute to political and policy debates so that inappropriate or inadequate reactions to acts of terrorism do not lead to major wastage of precious resources or unforced policy errors.

However, one must not forget that the fundamental problems in security are not new. What have changed over the years are the exponential growth in the number of connected devices, evolution of networks with data communication speed close to terabit/s in the near field, massive increase in the volume of data communication, and availability of high-performance hardware and massively parallel architecture for computing and intelligent software. Before the mid-1980s, mainframe and minicomputers dominated the market, and computer security problems and solutions were phrased in terms of securing files or processes on a single system. With the rise of networking and with the advent of the Internet, the paradigm changed, and security problems started to be defined in terms of not only securing the individual machines but also defending networks of computing against possible attacks. The issues are even more complex today, in the era of Web 2.0 and the Internet of Things, and the problems of security and communication have undergone yet another paradigm shift. However, the fundamental approaches toward developing secure systems have not changed. As an example, let us consider Saltzer and Schroeder's principles to secure design, which was conceived of as far back as in 1975. They focused on three important criteria in designing secure systems: simplicity, confinement, and understanding. However, as security mechanisms become too complex, attackers can evade or bypass the designed security measures in the systems. The argument that the principles are old, and somehow outdated, is not tenable when in reality the systems are vulnerable not due to bad design principles but simply because they are nonsecure systems. A mistake that is committed most often by security specialists is not making a comprehensive analysis of the systems to be secured before making a choice about which security mechanism to deploy. In many occasions, the security mechanism chosen turn out to be either incompatible with or inadequate for handling the complexities of the system. This, however, does not vitiate the ideas, algorithms, and the protocols of the security mechanisms. While the same old security mechanisms even with appropriate extensions and enhancements may not be strong enough to secure multiplicity of complex systems today, the underlying principles will continue to live on for the next-generation systems and indeed for the next era of computing and communications.

The purpose of the book is to discuss and critically analyze some of the important challenges in security and privacy enforcement in real-world computing and communication systems. For effectively mitigating those challenges, the book presents a collection of theoretical and practical research work done by some of the experts in the world in the field of cryptography and security in computing and communications.

The organization of this book is as follows: there are two parts in the book containing eight chapters in total. There are six chapters in Part I, which mainly focus on various challenges in several aspects of security issues in computing. Part II on the other hand, contains two chapters dealing with security issues in wireless communications and signal processing.

I am sure that the book will be very useful for researchers, engineers, graduate and doctoral students, and faculty members of graduate schools and universities, who work in the broad area of cryptography and security in networks and communications. However, since it is not a basic tutorial, the subject matters in the book do not deal with any introductory information on the field of cryptography and network security. The chapters in the book present in-depth cryptography and security-related theories and some of the latest updates in a par-ticular research area that might be useful to advanced readers and researchers in indentifying their research directions and formulating problems for further scientific investigation. It is assumed that the readers have knowledge on mathematical theories of cryptography and security algorithms and protocols.

I express my sincere thanks to all the authors who have contributed their valuable work in this volume. Without their contributions, this project could not have been successfully completed. The authors have been extremely cooperative during the submission, editing, and publication process of the book. I would like to express my special thanks to Ms. Romina Rovan, Publishing Process Manager of InTechOpen Publisher, Croatia, for her support, encouragement, patience, and cooperation during the period of the publication of the volume. My sincere thanks also go to Ms. Ana Pantar, Senior Commissioning Editor of InTechOpen Publisher, Croatia, for reposing faith on me and delegating me with the critical responsibility of editorship of such a prestigious academic volume. I would be failing in my duty if I don't acknowledge the motivation and encouragement that I received from my faculty colleagues in Calcutta Business School and Praxis Business School, Kolkata, India. Prof. Tamal Datta Chaudhuri, Prof. Sanjib Biswas, and Prof. Indranil Ghosh of Calcutta Business School deserve special mention for being my wonderful academic colleagues and for being the sources of motivation for me always. Last but not the least, I would like to thank my mother Ms. Krishna Sen, my wife Ms. Nalanda Sen, and my daughter Ms. Ritabrata Sen, for being my pillars of strength and the major sources of inspiration always.

**Professor Jaydip Sen**
Department of Analytics and Information Technology
Praxis Business School, Kolkata
India

Section 1

# Computing Security

Chapter 1

# Proactive Detection of Unknown Binary Executable Malware

Eric Filiol

Additional information is available at the end of the chapter

**Abstract**

To detect unknown malware, heuristic methods or more generally statistical approaches are the most promising research trends nowadays, but their computing and detection performances are generally not compatible with what users do accept. Hence, most commercial AV products still heavily rely on signature-based detection (opcodes, control flow graph, and so on). This implies that frequent and prior updates must be performed. May their analysis techniques be fully static of dynamic (using sandboxing or virtual machines), commercial AVs do not capture what defines malware compared to benign files: their intrinsic actions. In this chapter, we focus on binary executables and we describe how to effectively synthetize these actions and what are the differences between malware and nonmalicious files. We extract and analyze two tables that are present in executable files: the import address table (IAT) and export address table (EAT). These tables summarize the different interactions of the executable with the operating system. We show how this information can be used in supervised learning to provide effective detection algorithms, which have proven to be very accurate and proactive with respect to unknown malware detection.

## 1. Introduction

To detect unknown malware (or at least malware that are unknown in the antivirus database), heuristic methods or more generally statistical approaches are the most promising research trends nowadays. However, innovative detection algorithms cannot be included in antivirus software due to performance requirements. Among them, we generally face a relatively high false-positive rate, a significant analysis time for a given sample or have memory limit constraints. Having a too high false-positive rate may be a critical issue regarding executable files which are essential for the operating system kernel, for instance. Reducing the risk of false-positive detection by limiting the scope of efficient heuristic methods is still possible but it does not constitute a realistic solution.

Most of commercial AV products rely on signature-based detection or equivalent techniques. They all use the same scheme to detect malware while dealing with the above-mentioned limitations. The classification about malware signatures by antivirus company can be the following:

- **Object file header attribute** in this case, the header of a portable executable is used to detect whether the file is a malware or not, using combination of the different parts of the file structure. Despite the fact that packers may be used, their identification is relatively straightforward. A similar technique has been proposed in [28] by hashing object file feature. The key advantage of this technique lies on the fact that the result is efficient. Malware belonging to the same family (and written by the same programmer) are easy to detect. On the other hand, if the malware has some modifications while compiled or linked, due to compiler options, the header information may change.

- **Byte level approaches** There are three main possibilities about the byte level:

  - File hashing: the concept is to obtain a hash of whole or part of the malware. This a very common techniques which is quite systematically implemented in antivirus software, especially because it is easy to implement and it does not require a lot of computing resources with respect to the detection process. However, the major drawback comes from the fact that any modification of the binary code will result in a totally different hash value.

  - Character String signatures: a static character string present in the binary code of all the malware of the same family is used to detect the complete family. Griffin, Schneider, Hu and Chiueh [14] had proposed a way to automatically extract strings signatures from a set of malware.

  - Code normalization: the most common approach consist in rewriting some parts of the code using optimization techniques [1]. Junk code, dead code, and one-branch tests are removed while expressions with algebraic identities are simplified. The final code is a normal form that can be easily compared to other malware codes under the same form.

- **Instructions distributions**: the detection here is based on the distribution of the binary executable opcodes [2, 10]. A statistical scheme can be created and used to detect a whole family. Another way is to use N-gram analysis using the method given by McBoost [22].

- **Basic blocks**: the main technique with basic blocks deals with the description of the number of insertions, deletions, and substitutions to mutate a string into another one [3, 12]. To classify a malware from that, it is disassembled statistically and all its basic blocks are extracted. They are then compared to other malware blocks in order to get the smallest differences from one block to another.

- **API calls**: this technique consists into disassembling a full malware to extract the API call sequence. This sequence is compared to that of other malware. The SAVE system [26] is using this method.

Even when heuristics are supposedly used, they do not capture and synthetize enough information to be able to detect unknown malware accurately and proactively. This implies that frequent and prior updates must be performed. May their analysis techniques be fully static of dynamic (using sandboxing or virtual machines), commercial AVs do not capture what defines malware compared to benign files: their intrinsic actions.

In this chapter, we describe how effectively synthetize the essential differences (behaviors, structure, internal primitives) between benign files (or goodware) and malware. Aside a few features about the MZ-PE file header [7], we extract and analyze two tables that are present in executable files: the import address table (IAT) and export address table (EAT). These tables summarize the different interactions of the executable with the operating system. We show how this information, once it has been extracted, can be used in supervised learning (Sections 2 and 3) to provide an effective detection algorithm which has proven to be very accurate and proactive with respect to unknown malware detection.

As a main result, we achieve a very high detection rate with a low false-positive rate while our database has not been updated since 2014. All the techniques presented in this chapter have been implemented in the French antivirus project called DAVFI and presented in Section 4.

Because most of the malware are targeting Windows systems, our techniques are mostly designed for this operating system family. However, our approach has been similarly extended and applied to UNIX systems in the same way (up to the technical differences between ELF executables and MZ-PE executables). Even if we implemented our algorithms to be able to detect UNIX malware specifically as well, without loss of generality we will not present them in this chapter since it would be redundant with what has been made for Windows.

The chapter is organized as follows. Section 2 explains which information to extract from the binary code IAT/EAT and how to use it to capture the essential differences between malware and benign files with respect to their intrinsic behaviors. From that, a very efficient and accurate detection algorithm is designed. To improve further the description of binary executable behaviors, we consider the correlation of order 2 or of order 3 between the different functions involved in the IAT. By considering generic combinatorial structure, we derived a second detection algorithm in Section 3. In Section 4, we present the practical implementation of the algorithms of Sections 2 and 3 in the French antivirus project denoted DAVFI. We conclude in Section 5 and explore the possible evolutions for the results presented in this chapter.

## 2. Heuristic and proactive IAT/EAT detection

### 2.1. Technical background: import address table (IAT) and export address table (EAT)

#### *2.1.1. Introduction to IAT and EAT*

Any executable file contains a lot of information in the MZ-PE header [21] but some information can be considered more relevant than the others. Tables like import address table (IAT) and export address table (EAT) are, in our case, enough to describe what a program should do or is supposed to do. The IAT is a list of functions required from the operating system by the program. Technically there are two possibilities of importing functions on Windows. The first one is made explicitly through the IAT during the loading phase of the process before running it, and or during the running phase with the use of the *LoadLibrary* and *GetProcAddress* functions [19, 20]. The second possibility is used by a lot of malware to hide their real functionalities by loading them without referencing them in their IAT. Nonetheless, the functions are used to load libraries and to retrieve functions during runtime and therefore constitute some unavoidable points of passage which can be referenced. In most of the cases, malware, or packers have enough significant IAT to be detectable.

All executable files need an IAT. Without IAT—if this one is empty—it would mean that the targeted program would have no interaction with the operation system. In other words, it is not able to display text or any information at screen, it is not able to access any file on the system and it cannot allocate any segment of memory. Except consuming CPU time — with no result exploitable — it is not supposed to do anything else. Such useless program can be considered as suspicious (since it is suspicious to launch useless programs) or as malware in the most common case. If executable files need IAT, dynamic linked library (Dll) can also provide an EAT. This table describes which functions are exported by a Dll (and which are importable by an executable). Dll generally contains IAT and EAT — except for specific libraries which only export functions or objects. An executable can contain both an IAT and an EAT (the kernel of Windows `ntoskrnl.exe` is a good example). The use of EAT and IAT is a good combination to discriminate most of the libraries since the export and import is quite unique.

However, there are some limits to this system. One lies on the fact that this system only uses and trusts function, executable or library names. If a malware is designed to change every name of function to unknown ones, the system will not be able to give any reliable information any more. In addition, samples which imitate IAT and EAT from real benign files are able to bypass this type of test. Of course, it is a true limit of our model but, surprisingly, in most operational case, such a situation is not common. Most of the packers which are used on malware provides reliable IAT and EAT based on the executable file packed or on the packer itself (which helps to discriminate which packer is used). This observation is extensible to setup programs which are sort of packers.

#### *2.1.2. IAT and EAT extraction*

Before we can extract the IAT and EAT, it is necessary to find whether they are present or not. For this purpose it is necessary to analyze the entries of each table in the DataDirectory array of the IMAGE_OPTIONAL_HEADER (or IMAGE_OPTIONAL_HEADER64 in x64) structure. These entries (whose type is IMAGE_DATA_DIRECTORY) are DataDirectory[IMAGE_DIRECTORY_ENTRY_EXPORT], and DataDirectory[IMAGE_DIRECTORY_ENTRY_IMPORT].

For the IAT and EAT to be present, it is necessary that the VirtualAddress and Size fields in the associated structures are nonzero.

Upon confirmation of the presence of an IAT, it must then be read. Each DLL is stored as a structure of type IMAGE_IMPORT_DESCRIPTOR. From this structure we extract the *Name* field first. It contains the name of the DLL, then the *OriginalFirstThunk* field containing the address where is stored the primary function, the other being stored in sequence. Each function is stored in a structure of type IMAGE_THUNK_DATA, in which the field *AddressOfData* (whose type is IMAGE_IMPORT_BY_NAME) contains:

- the *hint* value (or Hint field). This 16-bit value is an index to the loader that can be the ordinal of the imported function [24],
- and the function name, if present (*Name* field), i.e., if the function has not been imported by ordinal (see further in Section 2.1.3). In the case of imports by ordinal only, it is the *Ordinal* field of IMAGE_THUNK_DATA that contains the ordinal of the function (if the most significant bit is equal to 1 then it means that the least significant 16 bits are the ordinal of the function [16]).

After getting the name of the function, a pair *dll_name/function_name* (*function_name* is the name of the function or its ordinal otherwise) is formed and stored, and the next function is played until all the functions of the DLL are read, and so on for each imported DLL. On output, a set of pairs *dll_name/function_name* is obtained, which will go through a formatting phase (see Section 2.1.4).

The format of the EAT, although also representing a DLL and all of its functions, is different from that of the IAT. All of the EAT is contained in a structure of IMAGE_EXPORT_-DIRECTORY type. From this structure are obtained the name of the DLL (which may be different in the case of renaming) using the *Name* field, the number of functions contained in the EAT (*NumberOfFunctions* field) and the number of named functions among them (since some functions can be exported by ordinal only) (*NumberOfNames* field).

Then we recover the functions and their name/ordinal. For the named functions, we just have to read two arrays in parallel, whose addresses are *AddressOfNames* and *AddressOfNameOrdinals*: at equal index, one contains the name of a function, and the other, its ordinal. For nonnamed functions, we must then retain all ordinals of named functions and then recover in the table with address *AddressOfFunctions* — which is indexed according to the ordinals of the functions it contains — all the functions whose ordinal has not been retained. After obtaining the set of functions/ordinals, in a similar way to that for the IAT, a set of pairs *dll_name/function_name* is built and then formatted (Section 2.1.4).

### *2.1.3. Miscellanous data*

Let us now detail a few technical points that are interesting to understand IAT and EAT in depth. Microsoft's documentation [18] explains how to export functions by ordinal in a DLL: ordinals inside a DLL MUST be from 1 to *N*, where *N* is the number of functions exported by the DLL. This is interesting and leads us to think that maybe some malicious files do not respect this rule. To go further, it is likely that this also applies to the hint of functions, although no documentation about it could be found. However, the analysis of a few Windows system DLL export tables like *kernel32.dll* and *user32.dll* shows that they comply to this rule. After conducting tests on malicious files and benign files, it turns out that only one "healthy" file (*sptd.sys*, a driver from alcohol120%) does not follow this rule, while a number of malicious files do the same.

### *2.1.4. Generation of IAT and EAT vectors*

After getting all the *dll_name/function_name* pairs from a file, two vectors are created (one for the IAT and one for the EAT). These vectors will the base object for our detection algorithm. In order to generate those vectors, we must build a database containing all the known pairs. A unique ID is associated to each unique pair. This database is populated by a base set of files with a known classification (malicious or benign). The population process is the following:

1. EAT and IAT pairs are extracted from files.
2. For each pair, a unique ID is constructed. This ID is a 64-bit number with the 20 most significant bits representing the DLL and the remaining 44 bits representing the function.
3. For the DLL ID: if the DLL is known, its ID is used. In the other case, a new ID is used, corresponding to the number of currently known DLLs (the first is 0).
4. The function ID follows the same process with known functions.

This population process is only executed manually whenever we would update the database; it is not run during file analysis. The two vectors are created according to this database. For each pairs, its ID is recovered from the database. If it does not exist in the database, the pair is discarded. All the 64-bit numbers are then sorted and stored in a file.

## 2.2. The detection algorithm

In this section we are now presenting our supervised detection algorithm which works on the vectors built with the data extracted and presented in the previous section. Usually [17, 25, 29] the database of known samples (training sets) must be built before writing the detection algorithm, as far as supervised algorithms are concerned. Such a procedure is led by the knowledge and the learning of what to detect (malware) and what not to detect (benign files). So the training set contains two subsets summarizing the essence of what malware and benign files really are.

### *2.2.1. How to build the algorithm*

Our solution is quite different. Indeed, if we know beforehand which data to use to perform detection, we did not know how to build the database to make it reliable and accurate enough for our algorithm. Which data to select among a set of millions of malware samples and of benign files, in order to get a representative picture of what a malware is (or is not) for the algorithm, is a complex problem in itself. Our approach has privileged the operational point of view. We have designed the algorithm as formal as possible and we have applied it on sets of malware and on a set of benign files to allow it to learn by itself, building the database after the creation of the algorithm. In other words, the algorithm is designed to use a minimal database of malware and of benign files at the beginning and this one is able to perform minimal detection helping to develop the database with samples undetected to improve results. We thus consider an iterative learning process, somehow similar to boosting procedure [15, 29].

Such an approach privileges experimental results and design of algorithms to detect unknown malware. Indeed, the algorithm uses subsets of malware samples which are the most representative of their families. Derivatives and parts of known malware (or variants) can be recognized since they have been learned previously. "Unknown" malware uses most of the time old fashion technologies, with the same base behaviors, and hence our algorithm is able to detect a lot of them with such a design and approach. For sake of clarity, the description of our algorithm starts with building the detection databases (training sets). To help the reader, we suppose in this part that we (already) have a known detection algorithm which is presented right after in the chapter (refer to Section 2.2.3).

### 2.2.2. *Building the detection database (training set)*

The heuristic algorithm we have designed uses a database of knowledge to help it to make decisions. Of course, algorithm databases are built with the two different types of files it is supposed to process and decide on: benign files and malware. The use of a combination of samples from malware and benign files gives the best results since they are suitably chosen. The way the database is built is the key step of our heuristic algorithm, since it affects directly the results we obtained. However, we must stress on the fact that we would obtain the same results for different malware/benign files subsets, as long as those sets are representative enough of their respective family. Somehow, this step can be seen as a probabilistic algorithm.

From a simple observation, more than the number of samples we could set in the database, the diversity of samples helps better to get the widest possible spectrum of detection. Smaller and more diverse the database is, faster and better are the results obtained. Indeed, if the database is too big, searching inside will be too much time-consuming, thus resulting in the impossibility to use it in real time. Only the most representative malware of a family must be included in the database (and similarly for the benign files).

First, we need a detection function which is the one used by our algorithm. At the beginning, the database used by this function is composed only with a small set of malware arbitrarily selected (denoted $M$) to be representative of the family we want to include. Such a detection function can be defined as follows. From any sample $S$ we want to analyze, we have a prior detection function $D_M$ which is of the form

$$D_M(S) = \begin{cases} 0 & \text{if } S \text{ is a non malicious} \\ 1 & \text{if } S \text{ is a malware} \end{cases} \quad (1)$$

It is not required that function $D_M$ exhibits huge and optimal detection performances. So a known and initial malware (respectively benign file) sample set is enough to initiate the process. To expand the databases (malware and benign files), Algorithm 1 is used. This approach is more or less similar to boosting methods such as Ada-Boost [11, 15].

---

**Algorithm 1** Database creation algorithm (training set)

---

**Require:** A set of files $S_f$ to analyze (which has $n$ files) and a maximal error detection rate $\epsilon$.

**Ensure:** Database files $S_d$ (malware) and $S_{ud}$ (benign files).

**while** $\frac{|S_f|}{n} < \epsilon$ **do**

**for** $\{s\} \in S_f$ **do**

**if** $D_M(s) == 1$ **then**

$S_d \leftarrow \{s\}$

**else**

$S_{ud} \leftarrow \{s\}$

**end if**

**end for**

$M = M \cup S_{ud}$

**if** $|S_d| == 0$ **then**

break

**end if**

**end while**

---

Algorithm 1 also enables to control the error detection rate $\varepsilon$ for a given malware family (with $\epsilon \in [0, 1] \subset \mathbb{R}$). Indeed, if $\varepsilon$ is chosen too small, the algorithm can include all the files from $S_f$. Of course, the representativeness of files in $S_f$ is a key point to use the algorithm. Working with several different samples of the same family is, most of the time, the best approach. Another possibility of control is to use the rate of detected files such as $\frac{|S_d|}{S_{ud}} < \epsilon$ with $\epsilon$ close to zero.

The building of database is performed family per family (of malware). It is possible to make it faster mixing multiple relevant samples from different families in one set. For example, to build the benign file database, one can choose files among those coming from `C:\ Windows`. In fact, the initial choice of incoming files defines the relevance and the diversity of the database. Starting from a small set of these files, we launch Algorithm 1 on the remaining files until we have got enough file detected by the database created on the fly.

One key advantage of this principle lies in the fact that we can increase the size of the database in the future without prior knowledge of a malware family. At the first time we created the database, if the diversity of malware families was enough good, it is possible to include new samples of malware without knowing its type/family. In fact, malware share strong IAT and EAT correspondences and similarities with many other families, in most of the cases. It means that malware can be detected by the database previously built even if we never included any sample from its family. In other words, we can use this property to increase the size of the database by adding undetected malware coming from different families into the current database. Taking a file defined as malware (which could be given by any trusted source or by a prior manual analysis), if this one is not detected by our algorithm, we can include it in our database in order to improve the detection of its family. It is a simple way to improve the accuracy of the detection.

### *2.2.3. The detection procedure: the K-nn algorithm*

Once the structural analysis is achieved and the database (training sets) has been built, then the detection tests occur by using the IAT and EAT vectors which have previously generated. This is the second part our module is in charge of, and which aims at deciding the nature of a file.

Detection tests are split into two sets: the IAT comparison test and the EAT comparison test. The principle of those tests is: the unknown file's IAT (or EAT) is compared to each element of the base of benign files and to each element of the base of malicious files. The $k = 2p + 1$ files that are closest to the unknown file are kept with their respective label (malware or benign file). A decision is then made based on these $k$ files to decide which label to give to the file under analysis. This test thus uses the method of k-Nearest Neighbors [15, 29], which has been modified for the occasion. In both cases, the input consists of the $k$ closest training examples in the feature space.

#### *2.2.3.1. Vector format limits*

While this format allows an optimized storage of the IAT/EAT, it faces several constraints that limit its use. The first constraint is a space constraint, which actually is not an intractable problem. Our encoding limits to $2^{20}$ possible DLLs and to $2^{44}$ functions per DLL. Today, this is more than enough, but we must keep in mind that this limit exists, and could be a problem in a (very far) future.

The second constraint lies in the fact that our vectors do not have a fixed length. It is a problem if we want to use standard distance functions, like the Euclidean distance. We could have used a similar vector format in which each possible couple was given a 0 or 1 number depending on whether it was present in the file or not. But the length would have been around 106 (about the current size of the database) instead of around 103 (for large files) with the current format. It would have a bad impact on the performances of real-time analysis, and hence it would have increased the time of analysis by too a high factor. In order to optimize the computation time, all the vectors in the bases and generated during analysis are sorted.

#### *2.2.3.2. The similarity measure*

In order to determine the nearest neighbors, we need a function to compare two IAT/EAT vectors of different sizes. The format prevents the use of standard distances (because to use a standard distance, the IAT/EAT vectors should have the same size, i.e., always the same number of imported/exported functions in each file, which is quite never the case). It was therefore necessary to find a function fulfilling this role and to apply it our format. Let us adopt a few notations:

- An IAT/EAT vector of size $n$ is written as $\sigma = \sigma_1\sigma_2...\sigma_n$ where $\sigma_i \in \{0, 1\}^{64}$ (64-bit integers). The set of such vectors is denoted $\Sigma_U$.

- The inverse indicator function $I : E,F \rightarrow \{0,1\}$ is defined such that $\forall x \in E, I_F(x) = 0$ if $x \in F$ and 1 otherwise.
- If $v$ is an IAT/EAT vector, $E_v = \{\sigma_i\}$ (this notation describes the fact that vectors are implemented as lists of 64-bit integers).

The function we use to compute the degree of similarity between IAT or EAT vectors is then defined by:

$$\forall a \in \Sigma_U, \forall b \in \Sigma_U,\ f(a,b) = \frac{1}{|a| + |b|}\left(\sum_{i=1}^{|a|} I_{E_b}(a_i) + \sum_{j=1}^{|b|} I_{E_a}(b_j)\right) \tag{2}$$

It is easy to prove that this function satisfies the separation, the symmetry and the coincidence axioms as any similarity measure has to.

*2.2.3.3. The decision algorithm*

The detection algorithm to decide the nature of a file (malware or benign) is given by Algorithm 2. It is composed of two parts in order first to reflect the importance of similarity optimally and second to eliminate some neighbors who are there only due to the lack of data.

The first part consists in filtering the set of neighbors that the k-NN algorithm returns to refine the best decision based on the neighbors that are really close. For this purpose, a threshold is set (50% for now) and only neighbors with a higher degree of similarity (i.e., that the function f returns a value less than 0.5) are kept. Then classical decision is applied to this new set: the file is considered closer to the base with the most representative among the neighbors.

The second part is used in the case when an equal number of representatives in each base, is returned (situation of indecision). All the neighbors are again considered, and again the file is considered closer to the base with the most representatives among the neighbors. If $k$ is odd, it helps to avoid indecision (majority decision rule). It was therefore decided that all $k$ are used odd in order not to fall in the case of indecision.

**Algorithm 2** Algorithm used to classify a file

**Require:** A vector $X$ representing a file to analyze, a malware vector base $B_M$ and a benign vector base $B_B$.

**Ensure:** A Boolean value indicating whether the file is malicious.

$i \leftarrow 0$

**for** $\{b\} \in B_B$ **do**

$d$ = f(X, b)

**if** $d == 0$ **then**

Return(false)

**Else**

*neighbors*[*i*] += (*d, brnidn*)

i++

**end if**

**end for**

**for** $\{m\} \in B_M$ **do**

$d$ = f(X, m)

**if** $d == 0$ **then**

Return(false)

**else**

*neighbors*[*i*] += (*d, malicious*)

i++

**end if**

**end for**

**if** MaxNeighbors(neighbors) == malicious **then**

Return(true)

**else**

Return(false)

**end if**

---

### 2.3. Detection and performances results

In order to test and to tune up our algorithm, we have defined many tests. On the one hand, we have tested the modification of the number of neighbors' parameter in the *k*-nn algorithm. This test is made in order to observe for how many neighbors the test is the most efficient. Then, on the other hand, we performed tests on databases to measure results of the algorithm. Of course, the detection algorithm is used with the most efficient number of neighbors obtained in the first test.

Increasing the number of neighbors by more than 9 does not change the results significantly. In fact, keeping the number of neighbors as minimal as possible is a better choice since it has an impact on the response time of the algorithm — a key point when we used it in real-time detection conditions. The results about this test are displayed in **Figure 1**. For the final test, we have put the algorithm to the proof with two sample sets. One is composed of 10,000 malware (extracted from different families and unknown from our databases) and one composed of legitimate files composed of executable files extracted from a clean Microsoft Windows operating system (around 131,000 files). The results are given in **Table 1** These results show that the algorithm is quite efficient to detect similarities between different executable files. Nonetheless, it is not enough to use it for detection in real time only since the rate of false-positive detection is too high to be acceptable. To prevent such a case, our algorithm in module 5.2 is chained with other techniques (see Section 4). This is the most efficient approach since we succeeded in making the residual false-positive rate tends toward 0.

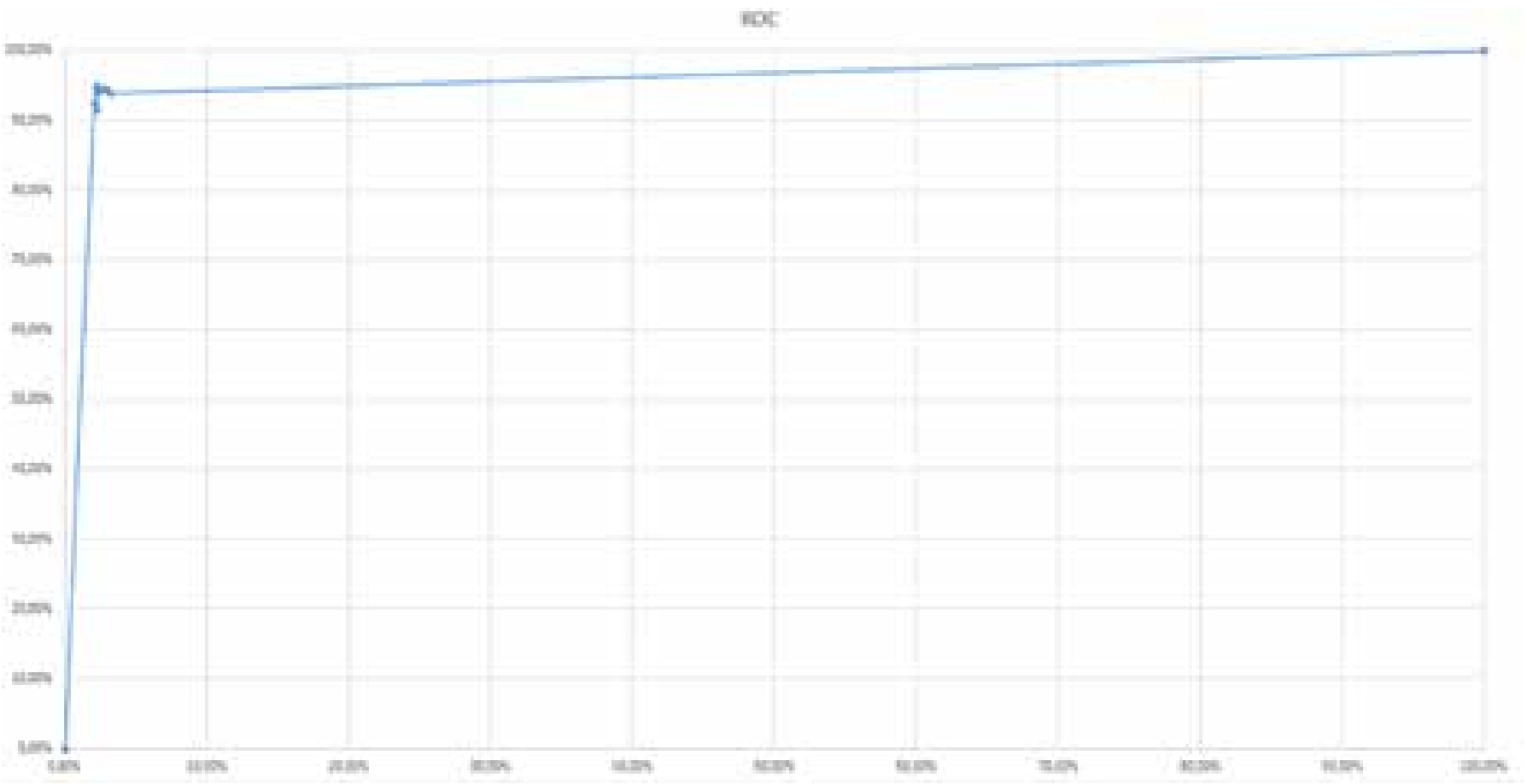

**Figure 1.** ROC summarizing the detection algorithm performances.

| | Malware set | Benign files set |
|---|---|---|
| Detected as malware | 95.028% | 4.972% |
| Detected as benign file | 2.053% | 97.947% |

**Table 1.** Algorithm performances results.

## 3. Combinatorial detection of malware by IAT discrimination

As we did in the previous section, we now consider a mix between the object file header and the API call. We are orienting our research toward the Import Address Table (IAT) and especially the correlation between IAT functions that are used either by malware or by benign files or used by both.

For this purpose, we use supervised learning techniques. The training models aims at building vectors that capture the combined use of specific IAT functions. We have observed that the subsets of specific functions significantly differ depending on the nature of the executable file — malware and benign files. Then, the testing phase enables to detect codes, even when unknown, with a very good true positive rate while keeping the false-positive rate very low.

### 3.1. The IAT functions correlation model

To build our model (our training sets), we have to extract the specific IAT functions and to build specific vectors that describe their combined use by malware, benign files and blacklisted IAT functions. We thus build two vector sets, one set which models malware, the second the benign files. The (unique) blacklist vector set describes specific IAT functions which must be considered as used systematically by malware only (see further).

Each of our vectors is implemented as a multiprecision integer by using GMP [13]. Each bit of this integer represents the presence or absence of a predefined (specific) function in its Import Address Table. This implementation approach allows to perform vectorized computation with simple bitwise logical operators. The predefined functions are derived from the extraction of all the Export Address Table from the dynamic-linked library in the operating system. For example, **Table 2** summarized the occurrences of the predefined functions in both malware and legitimate files.

The vector for the malware is: 001 0011 ⇒ 19, the vector for the benign file 1 is: 101 0101 ⇒ 85 and the vector for the benign file 2 is: 111 0100 ⇒ 116. In this way, we can easily and quickly detect which function from which dynamic-linked library is used by malware or benign files. The Dll name:function name indices are arbitrarily ordered, provided the chosen ordering remains the same for all vectors.

#### 3.1.1. *Creation of initial vectors*

The different sets containing the vectors we generate are essential components in our detection engine. We used a fresh install of Windows 7 professional with all update at January 1st, 2015. Three vector sets are created: for benign files, malware and for the blacklisted functions. In order to obtain a list of all functions, we extracted all of them in each dynamic link library which was present in the Windows system. We obtained a total of 76,669 functions in 1568 dynamic link libraries.

| Dll name | Function name | Malware | Benign file 1 | Benign file 2 |
|---|---|---|---|---|
| $dll_i$ | F1 | 1 | 1 | 0 |
| | F2 | 1 | 0 | 0 |
| | F3 | 0 | 1 | 1 |
| $dll_j$ | F1 | 0 | 0 | 0 |
| | F2 | 1 | 1 | 1 |
| | F3 | 0 | 0 | 1 |
| $dll_k$ | F1 | 0 | 1 | 1 |

**Table 2.** Vectors creation table (drawn from [8]).

#### 3.1.1.1. *Malware vector set*

The vectors are created by extracting the import address table from a set of 3567 malware. This set covers 95% of the different families for the two last years. After analysis and cleaning steps (especially for discarding duplicated vectors), we have obtained more than 1381 vectors. Let us remind that many malware use packers to delay the analysis or to make it less straightforward. Whenever a benign file packs its code a packer that is generally used by malware, we then decide it as malicious.

#### 3.1.1.2. *Goodware vectors*

Goodware vectors are created from the executable files on a clean installation of Windows 7. We have obtained a set of 985 vectors.

#### 3.1.1.3. *Blacklist vector*

This blacklist vector set is created by considering all undocumented functions contained by the Microsoft dynamic-link library on a native Windows 7 professional, as well as a few functions used by malware only. Development standards now make nowadays compulsory not to use undocumented functions (may them be Windows functions or not). As a consequence, it is a key point to keep in mind that there is no real reason for a legit program to rely on or to use undocumented functions from the Microsoft dynamic-link library. Those functions can become deprecated at any time without explanations from Microsoft. As a consequence, any legit program does not have to use them. In order to add some other functions, we also take the PeStudio blacklist [23] into account. The blacklist vector references around 47,040 blacklisted functions.

### 3.1.2. *Correlation between functions and function subsets*

In order to improve the detection scheme presented in Section 2, we decided to use the correlations between functions. Indeed, program behaviors can be described by a set of functions, which are generally indexed by time (in order words, the order according to which functions are called, matters). We thus intend to use the information describing the simultaneous occurrence of subsets of functions. Since a few years, compilers do no longer preserve the time ordering of functions in the IAT. To retrieve this information either we have to reverse the binaries and analyze the code or to perform a dynamical analysis from execution traces. Hence, subsets can be considered in place of vectors (ordered subsets). To model this, we are going to use all subsets of size 2 (pairs) or of size 3 (triplets). In other words, we intend to capture more closely the behaviors by considering the call of any two (resp. three) possible functions.

From the initial vectors of size $n$ we then build pair-vectors or triplet-vectors.

Pair-vectors have size of $\binom{n}{2}$ while triplet-vectors have a size of $\binom{n}{3}$. For an easy implementation, we will keep on representing these vectors as GMP integers.

All possible function pairs and function triplets are ordered according to some arbitrary ordering, for example, $(1;2), (1;3), \ldots(1;n), (2;3), \ldots, (n-1;n)$. For example, when considering data given in **Table 2**, we produce the data given in **Table 3** (due to lack of space, we show only pairs that are effectively present in the binary code of at least one of the files). For ease of writing we call *binomial sets* the function subsets of size 2 and *trinomial sets* the function subsets of size 3. We thus produce three new vectors sets.

#### 3.1.2.1. Binomial set vectors

From the previous initial vectors produced in Section 3.1.1, we generate binomial set vectors for both benign files and malware. For each vector and for any function binomial set, we check whether this set is present (the corresponding vector bit is set to 1) in the executable or not (the bit is set to 0). If again we consider the result of **Table 3** drawn from [8], the malware file is defined by the following binomial sets: (1;2), (1;5) and (2;5). Then the resulting binomial set vector is 000000000000100001001 where the binomial (1;2) is the least significant bit and the binomial ($n$ – 1, $n$) is the most significant bit. Goodware are then similarly defined by the two followings vectors: 010000101000000101010 and 111000111000000000000. There is only one pair, (1;5), in common. **Table 4** summarizes the number of subsets for each category.

| Bit 1 | Bit 2 | Malware | Benign file 1 | Benign file 2 |
|---|---|---|---|---|
| 1 | 2 | 1 | 0 | 0 |
| 1 | 3 | 0 | 1 | 0 |
| 1 | 5 | 1 | 1 | 0 |
| 1 | 7 | 0 | 1 | 0 |
| 2 | 5 | 1 | 0 | 0 |
| 3 | 5 | 0 | 1 | 1 |
| 3 | 6 | 0 | 0 | 1 |
| 3 | 7 | 0 | 1 | 1 |
| 5 | 6 | 0 | 0 | 1 |
| 5 | 7 | 0 | 1 | 1 |
| 6 | 7 | 0 | 0 | 1 |

**Table 3.** IAT function pairs (example drawn from [8]).

| | Count |
|---|---|
| Goodware | 1,753,640 |
| Malware | 2,036,311 |
| Common | 433,119 |

**Table 4.** Details of count in binomial sets.

*3.1.2.2. Trinomial set vectors*

In the same way we did for binomial set vectors, we have produced three sets for the trinomial sets (see **Table 5**).

*3.1.2.3. Common sets*

In order to make the analysis more accurate, we removed all the common sets for both the binomial and trinomial sets. Since there are present at the same time both in malware and benign files, they do not provide meaningful information. As an additional advantage, we also reduce the size of the database and we spare time and memory (see **Tables 4** and **5**) [8].

### 3.2. The detection algorithm

We use a variant the $K$ -nn algorithm [17] whose aim is to compute the distance of a given vector (the file to analyze) to the sets of the training database. We then label the vector with respect to the set which is at the shortest distance. In practice, to classify an executable as a malware or a benign file, the detection algorithm consists in five tests. Three of them use directly the initial vectors extracted from its Import Address Table. The last two tests use the binomial and trinomial set vectors.

The detection algorithm is summarized in Algorithm 3 and implements several steps:

- The first test is a comparison with the blacklist vector. A simple bitwise AND is performed between both vectors. If the result is different from zero (characteristic malware functions are indeed shared by both vectors), then the executable is considered as a malware.
- The second test consists in performing a bitwise XOR between the file vector to classify and all vectors from the malware and legitimate file sets. The label (malware or benign file) will be determined by the shortest distance. We only keep the $2p + 1$ best values (usually $p = 15$) and apply a majority voting. Moreover, we also analyze whether there is gap in these $2p + 1$ distances. If we notice such a gap, we consider that the label for the file must be the same than that of the family of the vector for the gap. For example if the best value is 3 with malware label, and the second is 27 with the nonmalicious label, the file is considered as a malware (since 27 – 3 = 24 is far greater that generally observed).
- In the third test, we compare vectors with a bitwise AND test. The classification label is determined by the largest distance: the bigger the result, the closer is the vector to the corresponding vector set. In the same way we do the XOR test, we use gap criteria to discriminate a family in case of uncertainty. It is worth noticing that the AND and XOR test are not the same. While the XOR test enlightens the dissimilarities, the AND test favors similarities. In fact, both tests are complementary to each other.

| | Count |
|---|---|
| Goodware | 373,026,049 |
| Malware | 336,423,103 |
| Common | 283,4537 |

**Table 5.** Details of count in trinomial sets.

- The two last tests are based on the binomial and trinomial vector sets, we calculate which set yields the most common matches and hence we decide the label accordingly.

---

**Algorithm 3** IAT-based combinatorial detection algorithm (vectors and files are represented as GMP integers; binary operators are computed bitwise over GMP integers)

---

**Require:** File $f$ to analyze. Blacklit vector $\mathcal{B}$, malware vector set $\mathcal{M}$ and benign file vector set $\mathcal{G}$, malware binomial set vectors $\mathcal{MBS}$, malware trinomial sets vectors $\mathcal{MTS}$, benign file binomial set vectors $\mathcal{MBS}$, benign file trinomial sets vectors $\mathcal{MTS}$.

**Ensure:** File label (malware [1] or nonmalicious [0]).

type ← 0

compute $v = \mathcal{B}\ AND\ f$

**if** $v \neq 0$ **then**

type++

**end if**

compute the XOR distance of $f$ with vectors in $\mathcal{M}$ and $\mathcal{G}$

keep the 31 best vectors with their distance from $f$ and their label (malware or benign file)

**if** Malware labels are the most represented **then**

type++

**end if**

compute the AND distance of $f$ with vectors in $\mathcal{M}$ and $\mathcal{G}$

keep the 31 best vectors with their distance from $f$ and their label (malware or benign file)

**if** Malware labels are the most represented **then**

type++

**end if**

compute $d_{\mathcal{MBS}}$ and $d_{\mathcal{GBS}}$ (resp.)the distance of $f$ with vectors in $\mathcal{MBS}$ and $\mathcal{GBS}$)

**if** $d_{\mathcal{MBS}} > d_{\mathcal{GBS}}$ **then**

type++

**end if**

compute $d_{\mathcal{MTS}}$ and $d_{\mathcal{GTS}}$ (resp.)the distance of $f$ with vectors in $\mathcal{MTS}$ and $\mathcal{GTS}$)

**if** $d_{\mathcal{MTS}} > d_{\mathcal{GTS}}$ **then**

type++

**if** type ≥ 2 **then**

**return** 1 (malware)

**else**

**return** 0 (nonmalicious)

**end if**

---

### 3.3. Results and performances

In this section, we now detail the results of those different steps of the detection algorithm. With the initial sets only and without learning phase, we have a detection rate more than 98% and a really small false-positive rate (less than 3%). The false positive is mostly due to legitimate software, which uses packers that we can wrongly label as malware. However, by combining with white listing techniques (as we did in the French AV project DAVFI, see Section 4), the false-positive rate systematically tends toward zero. As explained before, only a very few legitimate software are using code packing as malware usually do.

#### *3.3.1. Blacklisted function vector*

The result for this test is generally zero (in more than 97% of the cases). But whenever this result is different (nonnull) we are certain that the file is a malware. This indicator about undocumented functions from the Windows API is discriminant only if the executable uses one of these functions.

#### *3.3.2. XOR & AND tests*

The tests for bitwise XOR and AND were the two first tests implemented (**Tables 6** and **7**). With a rather small database for each set (less than 30 Mb), we detect 99% of malware correctly. The following tables show the results using a part of the database only [8]. The aim is to determine whether a reduced database would provide significantly similar results thus enabling to spare memory.

To create the partial database, we keep only the most significant vectors in terms of information contained. This is directly connected to the sparsity of vectors. Another way to select the vectors to keep consists in computing their respective *Information Gain* [17]. Let us consider a vector $v$. Its information gain is given by the formula:

$$IG(v) = \sum_{v_j \in \{0,1\}} \sum_{C \in \{C_i\}} P(v_j, C)\log\left(\frac{P(v_j, C)}{P(v_j).P(C)}\right), \tag{3}$$

where $C$ is the class (malware or benign file), $v_j$ is the value of the $j$-th attribute, $P(v_j, C)$ is the probability that it has value $v_j$ in class $C$, $P(v_j)$, is the probability that it takes value $v_j$ in the whole training set (database) and $P(C)$ is the probability for the class $C$. With only 75% of the whole database, we detect 80% of the malware, while the rate of false positive is close to 0.

| % of original base | Size on disk | Detection rate | Time |
|---|---|---|---|
| 100 | 39 Mb | 99 | 2 s |
| 90 | 35 Mb | 93 | 2 s |
| 80 | 31 Mb | 86 | 2 s |
| 75 | 29 Mb | 81 | 2 s |

**Table 6.** Results for the AND test.

| % of original base | Size on disk | Detection rate | Time |
|---|---|---|---|
| 100 | 39 Mb | 98 | 2 s |
| 90 | 35 Mb | 97 | 2 s |
| 80 | 31 Mb | 87 | 2 s |
| 75 | 29 Mb | 80 | 2 s |

**Table 7.** Results for the XOR test.

### 3.4. Binomial and trinomial set vectors tests

With the binomial and trinomial set vectors we have built in the previous part, we detect 99% of malware containing an Import Address Table. Whenever an executable file has no IAT, it is strongly suspected to be a malware. Consequently it is labeled as such. However the size of database is relatively big: 121 Mb for the binomial set vectors and 34 Gb for the trinomial set vectors. To reduce the database sizes, once again we keep only the most significant vectors in each set. In this way, we reduce the time to analyze a file and the size of database. **Tables 8** and **9** give the best ratio to keep.

#### *3.4.1. General results*

The efficient approach consists in combining and chaining all the tests using different possible decision rules (one of the most efficient is the maximum-likelihood rule). The detection rate is then more than 99% while the false-positive rate is very close to 0 (without additional white listing techniques). **Tables 10** and **11** show the detection rate depending on the size of the initial database. The following table indicates us the rate of false positives on all our tests depending of the database size. As we can see, the rate of false positive is very good. False positive can be explained as follows:

- Software installers generally embed compressors and packers. Hence we observe the presence of a small IAT with many compression imports.
- The DotNet environment is developing more and more. DotNet files have really a small IAT. An optimization would be to analyze the internal imports.
- Update only programs. These programs are generally really near of webdownloaders (a functionality shared with malware), because they basically only try to connect on specific websites in order to check whether any new version is online.

| % of original base | Size on disk | Detection rate | Time |
|---|---|---|---|
| 100 | 121 Mb | 98 | 67 s |
| 90 | 109 Mb | 97 | 53 s |
| 80 | 96 Mb | 90 | 47 s |
| 50 | 60 Mb | 80 | 30 s |

**Table 8.** Results for binomial set vectors.

| % of original base | Size on disk | Detection rate | Time |
|---|---|---|---|
| 100 | 34 Gb | 99 | 287 s |
| 90 | 30 Gb | 98 | 240 s |
| 80 | 27 Gb | 93 | 223 s |
| 50 | 17 Gb | 82 | 153 s |

**Table 9.** Results for trinomial set vector.

| | | | | | | |
|---|---|---|---|---|---|---|
| % of original base | 100 | 90 | 80 | 70 | 60 | 50 |
| % of detection | 99 | 98 | 94 | 89 | 87 | 84 |

**Table 10.** Detection rate.

| | | | | | | |
|---|---|---|---|---|---|---|
| % of original base | 100 | 90 | 80 | 70 | 60 | 50 |
| % of false positive | 1 | 3 | 7 | 12 | 17 | 24 |

**Table 11.** False-positive rates.

In all three cases, white listing techniques and/or additional analysis routines (such as those presented in Section ??) will make the false-positive rate tends toward 0.

# 4. The DAVFI project

## 4.1. Presentation of the project

The DAVFI project [5] (standing for *Démonstrateur d'Antivirus Français et International* or *French and International Antiviral Demonstrator*) was a 2-year project (from October 2012 to September 2014) partially funded by the French Government (*National Fund for the Digital Society*). The objective of this project was to design, to implement and to test a proof-of-concept for a new generation, sovereign, multi-platform (Android, Linux, and Windows) open antivirus software.

The final proof-of-concept has been delivered in September 2014 and is based on a strongly multithreaded architecture. The latter is made of several modules which are chained and operate within two main resources: a resident notification pilot and an antiviral analysis service. The latter embeds two analysis streams, one for binaries and executable files, the other to process documents (and malware documents) specifically. In 2015, after a technical and operational validation by the French Directorate General of Armaments has been transferred to the private sector for the industrialization process. By now this project equips the French National Gendarmerie's computers (Linux version).

The DAVFI's general structure (we will focus on the Windows version) is summarized in **Figure 2**. The detailed internal structure of the executable analysis chain is depicted in **Figure 3**.

DAVFI/OpenDAVFI's detection architecture is based on several modules. Whenever a relevant file is accessed, antivirus' kernel drivers notify the analysis service for the file analysis. Then many possibilities are considered. First, the file may be already known by the analysis system to be a nonmalicious file. Such a file can be defined as part of the system or already scanned by the antivirus and therefore has not to be detected as malicious (**Figure 2**). For this purpose, dynamic white-listing and black-listing modules have been designed and implemented (modules 1.1, 1.2 and 1.3). Second, the file is a document file and must be analyzed by a specific module (module 4 in **Figure 3**) [9]. Third, if we deal with a script file, it must be analyzed by another specific module. In the last case of a binary executable file, the analysis involves the module 5. This module is in fact a chained sequence of sub-modules designed to filter the detection of a binary file (note that other modules are composed in the same way) as depicted in **Figure 3**.

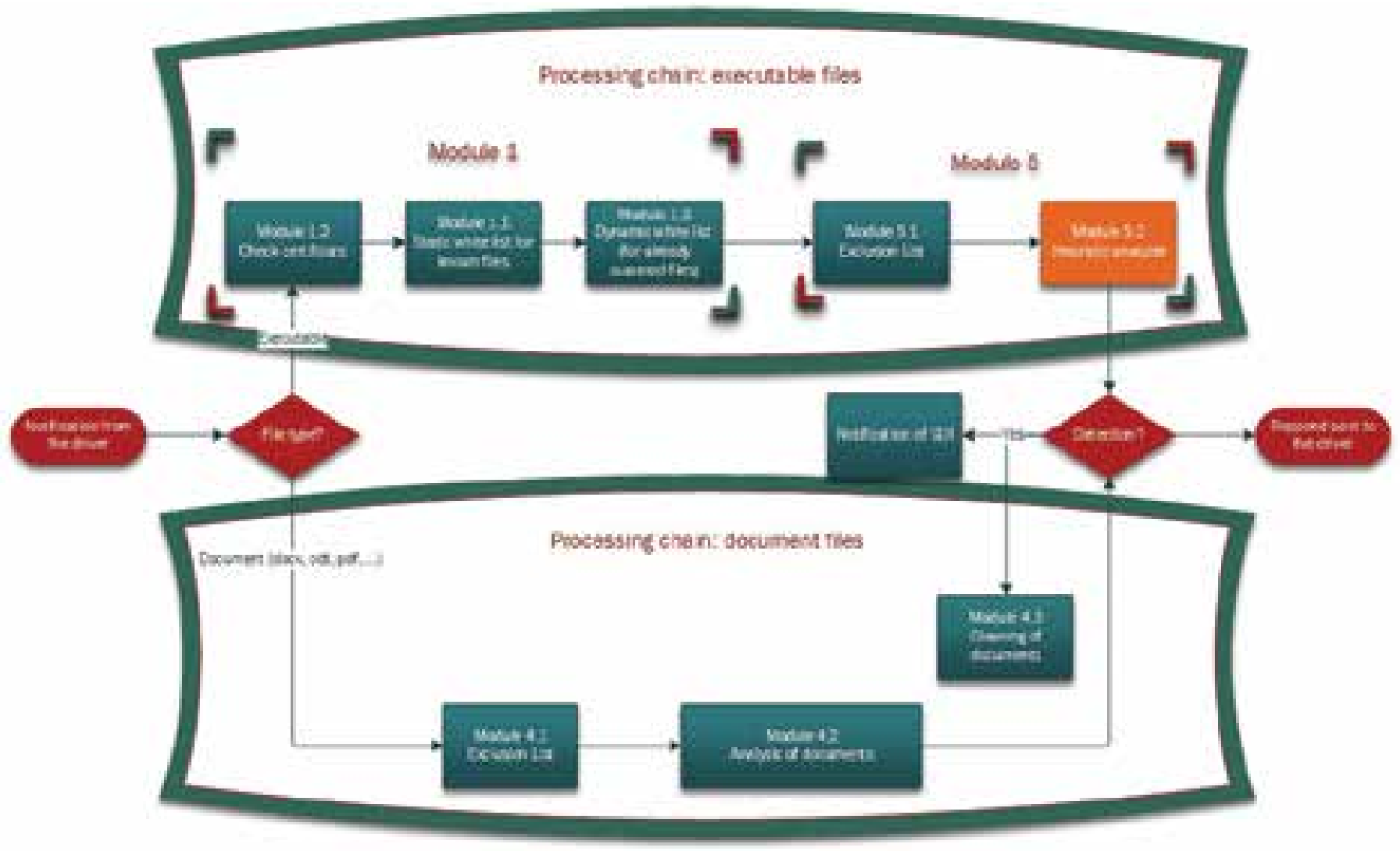

**Figure 2.** Overall structure of the windows DAVFI/OpenDAVFI application.

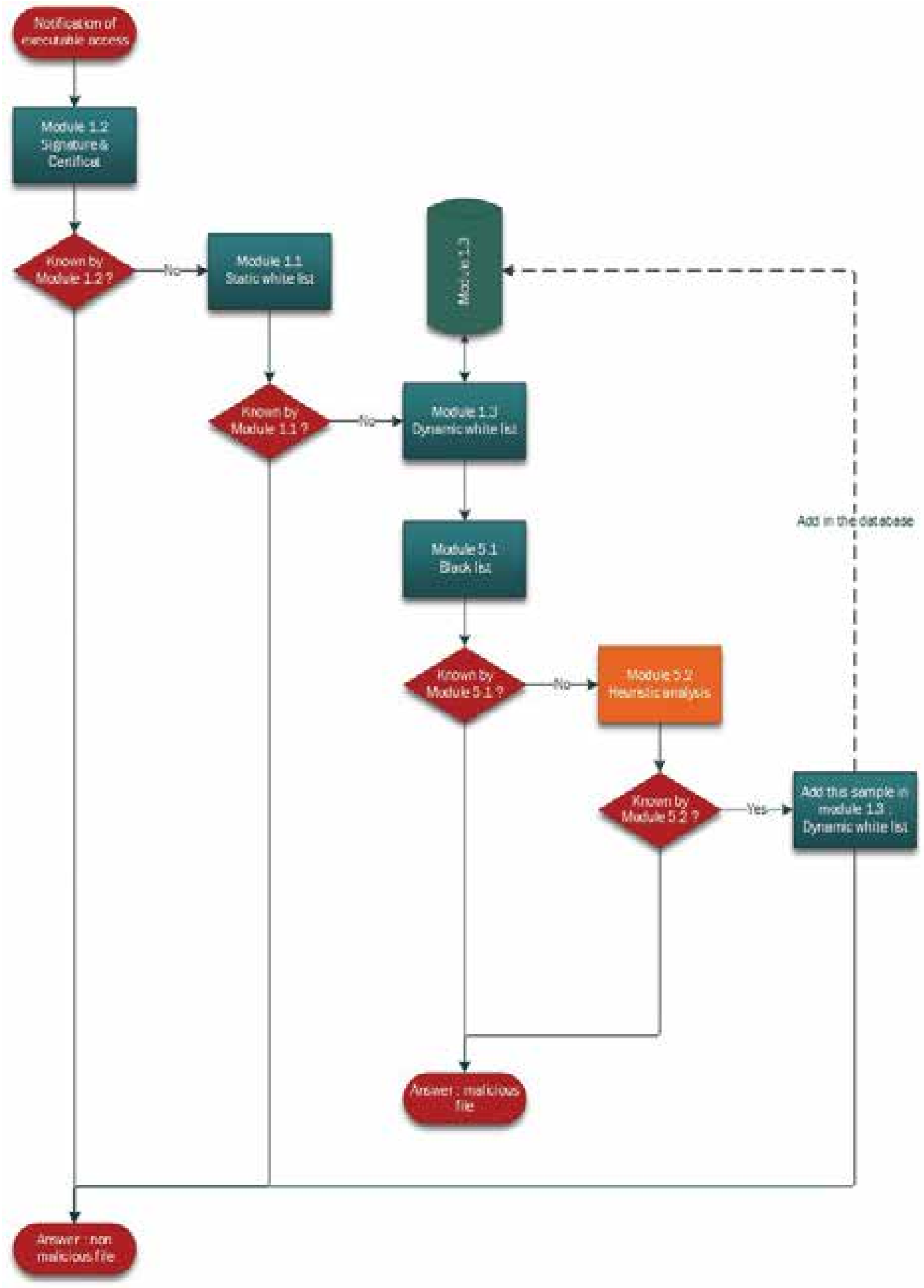

**Figure 3.** Overall structure of the windows DAVFI/OpenDAVFI executable file analysis module.

Whenever the module 5 starts, it checks with SEClamAV antivirus engine whether the file is a well-known malware or not. SEClamAv is used in our case for performance purposes to notify the heuristic detection module (module 5.2 which implements the detection algorithm presented in Section 2) with unknown files only. The heuristic module, which has been developed, is designed to detect both unknown malware and known malware. It is made of three parts: a header structural analysis [7] and two heuristic submodules which process the information contained in the *Import Address Table* (IAT) and in the *Export Address Table* (EAT) [6, 8]. It is also worth mentioning that the first filter in the DAVFI's analysis system is able to discard from detection all legitimate Windows kernel files (white-listing approach) or well-known benign files. This greatly reduces the false-positive detection rate.

Since heuristic detection is generally time-consuming, module 5.2 embeds a structural analysis chain which operates first [7]. Most AV software still uses detection techniques (either static or dynamic), which are however mostly based on the general concept of (static or heuristic) signature. However, we have observed that many malware do not comply to the Microsoft specifications with respect to the MZ-PE format [21]. Indeed, implementing malware techniques and tricks to fool a number of protection, detection or analysis techniques requires for the malware writer to take liberties with the file format specification. Consequently, a simple structural analysis with respect to this file format allows to identify executable that are indeed surely malware. As a consequence we avoid useless, time-consuming processing with the subsequent heuristic module.

### 4.2. Testing and technical evaluation

This project has been tested many times intensively during the two years of the project and then by the Directorate General of Armaments. A users committee (French Banks, French DoD, Prime Minister Office, and so on) has also been built for the DAVFI project. The aim was to involve end users, to have their operational feedback regarding antivirus software and to make them test a few modules in real-life conditions. Moreover, they feed us with unknown malware (usually manually detected in their respective CERT during the very first hours of the attack), most of them being not detected by commercial AV software (we use the VirusTotal [27] website for checking this point). Most of the samples provided related to targeted attacks. Final testing were organized as blind testing (we did not know which files were malware or benign files).

The performance results are very good and can be summarized as follows:

- The overall detection rate (true positive) is more than 97% while false-positive rate equal to 0.
- These overall results include unknown malware at the time of testing (the malware nature has been confirmed by manual malware forensics analysis). It is worth mentioning that the initial databases (presented in Sections 2 and 3 were not updated during the different testing phases).
- New tests in mid-2016 confirmed the previous results without database updating.

## 5. Conclusion and future work

In this chapter we have presented a different supervised detection algorithms working on data extracted from the IAT and EAT of binary executable files (Windows and Unices) and more broadly from their header. These particular pieces of information do not only describe the executable in a static way more precisely (use of far more complex and rich signatures) but also they capture the information related to program behaviors.

The overall performances which we have achieved show that is possible to detect unknown malware proactively and accurately. This yields enhanced detection capabilities while requiring far less database update. Beyond the experimental analysis, operational testing of those techniques has been performed on malware coming from the real world in real conditions. The results which have been observed fully satisfy the operational constraints and specifications with respect to unknown malware detection.

Future work will address the combinatorial modeling and processing of information contained in IAT/EAT. While we have considered mostly statistical aspects and initiated their combinatorial analysis in Section 3, it is possible to have a far more precise processing of this information when using combinatorial structures to synthetize the concept of behaviors and hence base a more accurate detection on the dynamical information contained in the code.

We also intend to extend the information used for detection. The study of data section or opcodes sections is a possible option in order to increase the number of detection criteria. These sections can provide correlations with the features we already consider.

As far as combinatorial techniques are concerned (Section 3), they can still somehow be time-consuming depending on the malware code to classify. The need for an important storage space when working with binomial and trinomial set vectors may also have an impact of the detection engine performances (mainly the computing time required for analysis). In case of a desktop computer, the user may not accept to wait more than a few seconds before he can access his data or resources. It is then better to use it upstream on a gateway, which would be dedicated to malware analysis and would check all the flow of incoming data.

As far as the size of the database is concerned, we can mitigate this point by considering that computer hard drives can store nowadays huge amounts of bytes. They also are large live memory (RAM) size. But it is still always unpleasant for the final user to let his antivirus software to be too much resource-consuming. Future work will consequently aim at reducing the database size by using suitable combinatorial designs [4]. The key approach lies in the ability to concentrate the information inside combinatorial design blocks while exhibiting correlation between IAT functions at a far higher order. We estimate that it is thus possible to reduce the database size at least by 75% without lessening the final detection performances.

Another future work deals with optimizing the detection with respect to binomial and trinomial vector-based detection. By adding or removing a few well-known combinations, it is possible to reduce the size of the database and the computing time.

The last work is to consider a limit for the Import Address Table size. In a few cases, malware writers are trying to fool the work performed by antivirus engines. As an example, they try to increase the malware size by loading and using too much external functions. It should then be rather easy to classify malware that are using more than a limited number of external functions but which actually only need and use less.

In the other hand, they may use a few stealth techniques to load and to use external functions without linking them in the Import Address Table. The improvement would then be to consider a file with a too small or too big Import Address Table as a malware.

## Author details

Eric Filiol

Address all correspondence to: filiol@esiea.fr

ESIEA – Operational Cryptology and Virology Lab (C+V), Laval, France

Chapter 2

# Cloud Cyber Security: Finding an Effective Approach with Unikernels

Bob Duncan, Andreas Happe and Alfred Bratterud

Additional information is available at the end of the chapter

**Abstract**

Achieving cloud security is not a trivial problem to address. Developing and enforcing good cloud security controls are fundamental requirements if this is to succeed. The very nature of cloud computing can add additional problem layers for cloud security to an already complex problem area. We discuss why this is such an issue, consider what desirable characteristics should be aimed for and propose a novel means of effectively and efficiently achieving these goals through the use of well-designed unikernel-based systems. We have identified a range of issues, which need to be dealt with properly to ensure a robust level of security and privacy can be achieved. We have addressed these issues in both the context of conventional cloud-based systems, as well as in regard to addressing some of the many weaknesses inherent in the Internet of things. We discuss how our proposed approach may help better address these key security issues which we have identified.

## 1. Introduction

There are a great many routes into an information system for the attacker, and while many of these routes are well recognized by users, many others do not see the problem, meaning limited work is being carried out on defense, resulting in a far weaker system. This becomes much more difficult to solve in the cloud, due to the multi-tenancy nature of cloud computing, where users are less aware of the multiplicity of companies and people who can access their systems and data. Cloud brings a far higher level of complexity than is the case with traditional distributed systems, in terms of both the additional complexity of managing new relationships in cloud, and in the additional technical complexities involved in running systems within the cloud. It runs on other people's systems, and instances can be freely spooled up and down, as needed.

Add to this the conception, or rather the misconception, that users can take the software, which runs on their conventional distributed systems network and run it successfully on the cloud without modification, thus missing the point that their solid company firewall does not extend to the cloud, and that they thus lose control over who can access their systems and data. Often, users also miss the point that their system is running on someone's hardware, over which they have limited or no control. While cloud service providers may promise high levels of security, privacy and vetting of staff, the same rigorous standards often do not apply to their subcontractors.

There are many barriers that must be overcome before cloud security can be achieved [1]. A great deal of research has been conducted toward resolving this problem, mostly through technical means alone, but this presents a fundamental flaw. The business architecture of an enterprise comprises people, process and technology [2], and any solution, which will focus on a technological solution alone, will be doomed to failure. People present a serious weakness to enterprise security [3], and while process may be very well documented within an organization, often it is out of date due to the rapid pace of evolution of technology [4]. Technology can benefit enterprises due to the ever improving nature and sophistication of software, which is a good thing, but at the same time can present a greater level of complexity, making proper and secure implementation within enterprise systems much more difficult. Another major concern is that the threat environment is also developing at a considerable pace [5].

Cloud computing has been around for the best part of a decade, yet we still have to see an effective, comprehensive security standard in existence. Those that do exist tend to be focussed on a particular area, rather than the problem as a whole, and as stated above, they are often out of date [4]. Legislators and regulators are not much further advanced. The usual practice is to state what they are seeking to achieve with the legislation or regulatory rules. Usually, they are very light on the detail of how to achieve these goals. To some extent, this is deliberate—if they specify the principles to apply to achieve their desired objective rather than the exact details, they do not have to keep updating the legislation/regulations as circumstances change. Often, they have no clue as to how to achieve these goals anyway, leaving it up to the users to work it out. This is the approach favored by the UK authorities, and it can be argued that it generally works well. In the US, they favor the rules-based approach, which, of necessity, requires far more work on the part of the government and regulators to keep the rules up to date. It also spawns an active industry of specialists who constantly probe the boundaries to see how far they can be pushed. Global enterprises often have to deal with both types of approach. In addition, the methodology deployed to achieve compliance is often flawed [6]. To this complex environment, we must now, of necessity, add the impact of both Industry 4.0, which encompasses mostly high-value targets, e.g. factories, and the Internet of things (IoT), which is likely to see a massive global explosion, to the mix.

The IoT has been around now for a considerable time, but it did not get much traction until the arrival of cloud computing and big data. In 2007, Gantz et al. [7] suggested that global data collection would double every 18 months, a prediction that looks like being very light when compared to the reality of data creation coming from the expansion of the IoT. Cisco noted that the IoT had really come of age in 2008, when there were now more things connected to the Internet than people [8]. The massive impact arising from this enabling of the IoT by cloud computing brings some exciting new applications and future possibilities in the areas of defense, domestic and home automation, eHealth, industrial control, logistics, retail, security and emergencies, smart airports, smart agriculture, smart animal farming, smart cars, smart cities, smart environment, smart metering, smart parking, smart roads, smart trains, smart transport and smart water, but also brings some serious challenges surrounding issues of security and privacy. Due to the proliferation of emerging and cheaply made technology for use in the IoT, it is well known that the technology is particularly vulnerable to attack. When we combine the IoT and big data, we compound this problem further. This area is poorly regulated, with few proper standards yet in place, which would suggest it might be more vulnerable than existing cloud systems, which have been around for some time now.

We are concerned with achieving both good security and good privacy, and while it is possible to have security without privacy, it is not possible to have privacy without security. Thus, our approach is to first ensure a good level of security can be achieved, and in Section 2, we discuss from our perspective how we have set about developing and extending this idea. In Section 3, we identify the issues that need to be addressed. In Section 4, we discuss why these issues are important, and what the potential implications for security and privacy are likely to be. In Section 5, we consider some current solutions proposed to address some of these issues and consider why they do not really address all the issues. In Section 6, we outline our proposed approach to resolve these issues, and in Section 7, we discuss our conclusions.

## 2. Development of the idea

The authors have developed a novel approach to addressing these problems through the use of unikernel-based systems, which can offer a lightweight, green and secure approach to solving these challenging issues. Duncan et al. [9] started by outlining a number of issues faced and explained how a unikernel-based approach might be used to provide a better solution. Bratterud et al. [10] provide a foundation for the development of formal methods, and to provide some clarity on the identification and use of good clear definitions in this space.

A unikernel is by default a single threaded, single address space mechanism taking up minimal resources, and [11] look at how the concept of single responsibility might be deployed through the use of unikernels in order to reduce complexity, thus reducing the attack surface and allowing for a better level of security to be achieved. Given the worrying expansion of security exploits in IoT, as exemplified by recent DDoS attacks facilitated by the inherent security weaknesses present in IoT architecture, Duncan et al. [12] looked at how the unikernel approach might be useful when used for IoT and big data applications. Duncan and Whittington [13] consider how to develop an immutable database system using existing database software, thus providing the basis for a possible solution for one of the major needs of the framework.

Unikernels use the concepts of both single address space and single execution flow. A monolithic application could be converted into a single large unikernel, but this would forfeit any real benefits to be gained from this architecture. To prevent this, we propose a framework that aids the deconstruction of business processes into multiple connected unikernels. This would allow us to develop complex systems, albeit in a much more simple, efficient, secure and private way. We must also develop a framework to handle the automated creation and shutting down of multiple unikernels, possibly carrying out a multiplicity of different functions at the same time. This concept is likely to be far more secure than conventional approaches. During runtime, the framework will be responsible for creation, monitoring and stopping of different unikernel services. While unikernels themselves do provide good functional service isolation, external monitoring is essential to prevent starvation attacks, such as where one unikernel effectively performs a denial-of-service attack by consuming all available host resources.

We have identified a number of other areas, which will need further work. We are currently working on developing a means to achieve a secure audit trail, a fundamental requirement to ensure we can retain as complete a forensic trail as possible, for which we require to understand how to properly configure an immutable database system, capable of withstanding penetration by an attacker. This work follows on from Ref. [13]. However, in order to run such a system, we will need to develop a control system to co-ordinate multiple unikernel instances operating in concert. We will also have to develop a proper access control system to ensure we can achieve confidentiality of the system and to maintain proper privacy. To help with the privacy aspects, we will also require to develop a strong, yet efficient approach to encryption.

In addition, the framework must provide means of input/output for managed unikernels, including facilities for communication and data storage.

Communication is both concerned with inter-unikernel communication as well as with providing interfaces for managed unikernels to the outside world. As we foresee a message-passing infrastructure, this should provide means for validating passed messages including deep packet inspection. This allows for per-unikernel network security policies and further compartmentalization, which should minimize the impact of potential security breaches.

In real-world use cases, we require the framework to be capable of handling mutable data, such as the ability to record temporary states, logging information or ensuring that persistent application and or user data can be maintained. Unikernels themselves by definition are immutable. In order to resolve this conflict, the framework must provide a means to persist and QUERY data in a race-free manner. It may be necessary to provide specialized data storage, depending on the use case. For example, system log and audit trail data require special treatment to prevent loss of a complete forensic record, thus requiring an append-only approach. Since persistent data storage is inherently contrary to our immutable unikernel approach, we do not enforce data storage to be implemented within unikernels. Being pragmatic, we defer this functionality to the framework, i.e. a means of storage is provided by the framework, rather than by the unikernels themselves.

We also believe it may be possible to develop a unikernel-based system to work with the serverless paradigm. With those frameworks, source code is directly uploaded to the cloud service. Execution is triggered in response to events; resources are automatically scaled. Developers do not have any system access except through the programming language and provided libraries. We see unikernel and serverless frameworks as two solutions to a very similar problem, reducing the administrative overhead and allowing developers to focus their energy on application development. Serverless stacks signify the "corporate-cloud" aspect: developers upload their code to external services and thus invoke vendor lock-in in the long run. Unikernels also allow users to minimize the non-application code, but in contrast to serverless architectures, this approach maintains flexibility with regard to hosting. Users can provide on-site hosting or move toward third-party cloud offerings. We expect serverless architecture providers to utilize unikernels within their own offerings. They are well suited to encapsulate the user provided applications and further increase the security of the host's infrastructure.

We are also developing penetration testing approaches, using fuzzing techniques, adapting tools and sanitizers, hardening tools and whatever else we can do to strengthen the user environment to achieve our aims. The ultimate goal is to make life so difficult for the attacker that they will be forced to give up and move on to easier pickings elsewhere. We have also been applying all the usual attack methods to our test systems to assess whether our approach will work. This should allow us to be sure that each component will be fit for purpose before we move on to the next component. In this way, by developing each component of the system to automatically integrate with the rest, the system should ultimately become far more robust as a result.

We now have a good idea of how the concept needs to be developed, and what future plans are needed to progress the development toward a highly secure and efficient system for cloud users. In the next section, we consider what exactly the issues are that we need to address in more detail.

## 3. What are the issues?

The fundamental concepts of information security are confidentiality, integrity, and availability (CIA), which is also true for cloud security. The business environment is constantly changing [14], as are corporate governance rules and this would clearly imply changing security measures would be required to keep up to date. More emphasis is now being placed on responsibility and accountability [15], social conscience [16], sustainability [17, 18], resilience [19] and ethics [20]. Responsibility and accountability are, in effect, mechanisms we can use to help achieve all the other security goals. Since social conscience and ethics are very closely related, we can expand the traditional CIA triad to include sustainability, resilience and ethics. These, then, must be the main goals for information security.

We now consider a list of ten key management security issues identified in Ref. [1], which provide detailed explanations for each of these items on the list. These items represent management-related issues, which are often not properly thought through by enterprise management.

The 10 key management security issues identified are:

- the definition of security goals,
- compliance with standards
- audit issues,
- management approach,
- technical complexity of cloud,
- lack of responsibility and accountability,
- measurement and monitoring,
- management attitude to security,
- security culture in the company,
- the threat environment.

These are not the only issues to contend with. There are a host of technical issues to address, as well as other, less obvious issues, such as social engineering attacks, insider threats (especially dangerous when perpetrated in collaboration with outside parties), state-sponsored attacks, advanced persistent threats, hacktivists, professional criminals, and amateurs, some of whom can be very talented. There are many known technical weaknesses, particularly in web-based systems, but the use of other technology such as mobile access, "bring your own device" (BYOD) access, and IoT can all have an adverse impact on the security and privacy of enterprise data.

In spite of what is known about these issues, enterprises often fail to take the appropriate action to defend against them, or do not understand how to implement or configure this protection properly, leading to further weakness. Staff laziness can be an issue. Failure to adhere to company security and privacy policies can also be an issue. Use of passwords, which are too simple, is an issue. Simple things, such as the use of yellow stickies can be a dangerous weakness when stuck on computer screens, with the user password in full view for the world to see.

Lack of training for staff on how to properly follow security procedures can lead to weakness. Failure to patch systems can be a serious issue. Poor configuration of complex systems is often a major area of weakness. Poor staff understanding of the dangers in email systems presents a major weakness for enterprises. Failure to implement simple steps to protect against many known security issues presents another problem. Lack of proper monitoring of systems presents a serious weakness, with many security breaches being notified by third-party outsiders, usually long after the breach has occurred.

We will take a look at some of these technical vulnerabilities next, starting with one of the most obvious. Since cloud is enabled through the Internet, and web-based systems play a huge role in providing the fundamental building blocks for enterprise systems architecture, it makes sense to look at the vulnerabilities inherent in web-based systems.

### 3.1. Web vulnerabilities

Security breaches have a negative monetary and publicity impact on enterprises, thus are seldom publicly reported. This limits the availability of empirical study data on actively exploited vulnerabilities. However, web vulnerabilities are well understood, and we can source useful information on the risks faced through this medium by using data from the work of the Open Web Application Security Project (OWASP) [21], who publish a top 10 list of web security vulnerabilities every 3 years.

The OWASP Top 10 report [21] provides a periodic list of exploited web application vulnerabilities, ordered by their prevalence. OWASP focuses on deliberate attacks, each of which might be based upon an underlying programming error—for example, an injection vulnerability might be the symptom of an underlying buffer overflow programming error. OWASP also provides the most comprehensive list of the most dangerous vulnerabilities and a number of very good mitigation suggestions. The last three OWASP lists for 2007, 2010 and 2013 are provided in **Table 1**.

This list, based on the result of analysis of successful security breaches across the globe, seeks to highlight the worst areas of weakness in web-based systems. It is not meant to be exhaustive, but instead merely illustrates the worst 10 vulnerabilities in computing systems globally. It is clearly concerning that the same vulnerabilities continue to recur year after year, which clearly demonstrates the failure of enterprises to adequately protect their resources properly.

| 2013 | 2010 | 2007 | Threat |
|---|---|---|---|
| A1 | A1 | A2 | Injection attacks |
| A2 | A3 | A7 | Broken authentication and session management |
| A3 | A2 | A1 | Cross site scripting (XSS) |
| A4 | A4 | A4 | Insecure direct object references |
| A5 | A6 | - | Security misconfiguration |
| A6 | - | - | Sensitive data exposure |
| A7 | - | - | Missing function level access control |
| A8 | A5 | A5 | Cross site request forgery (CSRF) |
| A9 | - | - | Using components with known vulnerabilities |
| A10 | - | - | Unvalidated redirects and forwards |

**Table 1.** OWASP top ten web vulnerabilities—2013 to 2007 [21].

Thus in any cloud-based system, these vulnerabilities are likely to be present. However, there are likely to be additional potential vulnerabilities, which will also need to be considered. We group the different vulnerabilities into three classes based on their impact on software development. Low-level vulnerabilities can be solved by applying local defensive measures, such as using a library at a vulnerable spot. High-level vulnerabilities cannot be solved by local changes, but instead need systematic architectural treatment. The last class of vulnerability is application workflow-specific and cannot be solved automatically but instead depends on thoughtful developer intervention.

Two of the top three vulnerabilities, A1 and A3, are directly related to either missing input validation or output sanitation. Those issues can be mitigated by consistently utilizing defensive security libraries. Another class of attack that can similarly be solved through a "low-level" library approach is A8. By contrast, "high-level" vulnerabilities should be solved at an architectural level. Examples of these are A2, A5 and A7. The software architecture should provide generic means for user authentication and authorization, and should enforce these validations for all operations. Some vulnerability classes, i.e. A4, A6 and A10, directly depend on the implemented application logic and are hard to protect against in a generic manner. Some other vulnerabilities can be prevented by consistently using security libraries, while other vulnerabilities can be reduced by enforcing architectural decisions during software development.

New software projects are often based upon existing frameworks. Those frameworks bundle both default configuration settings as well as a preselection of libraries providing either features or defensive mechanisms. Software security is mostly regarded as a non-functional requirement and thus can be hard to get funding for. Those opinionated frameworks allowed software developers to focus on functional requirements while the frameworks took care of some security vulnerabilities.

Over the years, those very security frameworks have grown in size and functionality, and as they themselves are software products, they can introduce additional security problems into otherwise secure application code. For example, while the Ruby on Rails framework, properly used, prevents many occurrences of XSS-, SQLi- and CSRF-attacks, recent problems with network object serialization introduced remotely exploitable injection attacks [22]. The affected serialization capability was not commonly used but was included in every affected Ruby on Rails installation. Similar problems have plagued Python and its Django framework [23]. All of these are further aggravated as, by design, software frameworks are generic—they introduce additional software dependencies, which might not be used by the application code at all. Their configuration often focuses on developer usability, including an easy debug infrastructure. Unfortunately, from a security perspective, everything that aids debugging also aids penetration.

OWASP acknowledged this problem in its 2013 report by introducing A9. The reason for adding a new attack vector class was given as: "the growth and depth of component based development has significantly increased the risk of using known vulnerable components" [21].

Of course, when it comes to the use of IoT with cloud, we need to look beyond basic web vulnerabilities. The IoT can also use mobile technology to facilitate data communication, as well as a host of inherently insecure hardware, and we look at this in more detail in the next section.

### 3.2. Some additional IoT vulnerabilities

OWASP now produces a list of the worst 10 vulnerabilities in the use of mobile technology, which we show in the list of **Table 2**.

Of course, it is not quite as simple as that the IoT mechanics extend beyond traditional web technology and mobile technology. In 2014, OWASP also developed a provisional top 10 list of IoT vulnerabilities, which we outline in **Table 3**.

An important point to bear in mind is that the above list represents just the OWASP top 10 vulnerability list. OWASP is currently working on a full list of 130 possible IoT vulnerabilities, which should be taken into account. OWASP also provides some very good suggestions on how to mitigate these issues.

While the above just covers security issues, we also have to consider the challenges presented by privacy issues. With the increase in punitive legislation and regulation surrounding issues of privacy, we must necessarily concern ourselves with providing the means to ensure the goal of privacy can be achieved. The good news is that if we can achieve a high level of security, then it will be much easier to achieve a good level of privacy [9]. Good privacy is heavily dependent on having a high level of security. We can have security without privacy, but we cannot have privacy without security.

| 2013 code | Threat |
|---|---|
| M1 | Insecure data storage |
| M2 | Weak server side controls |
| M3 | Insufficient transport layer protection |
| M4 | Client side injection |
| M5 | Poor authorization and authentication |
| M6 | Improper session handling |
| M7 | Security decisions via untrusted inputs |
| M8 | Side channel data leakage |
| M9 | Broken cryptography |
| M10 | Sensitive information disclosure |

**Table 2.** OWASP top ten mobile vulnerabilities—2013 [21].

| 2014 Code | Threat |
|---|---|
| I1 | Insecure web interface |
| I2 | Insufficient authentication/authorization |
| I3 | Insecure network services |
| I4 | Lack of transport encryption |
| I5 | Privacy concerns |
| I6 | Insecure cloud interface |
| I7 | Insecure mobile interface |
| I8 | Insufficient security configure-ability |
| I9 | Insecure software/firmware |
| I10 | Poor physical security |

**Table 3.** OWASP top ten IoT vulnerabilities—2014 [24].

While the IoT has progressed significantly in recent years, both in terms of market uptake and in increased technical capability, it has very much done so at the expense of security and privacy. For example, accessing utility companies, including nuclear in the US [25], damage caused to German steel mill by hackers [26], drug dispensing machines hacked in US [27], plane taken over by security expert mid-air [28], and a hack that switched off smart fridges if it detected ice cream [29]. While enterprises often might not care too much about these issues, they should. If nothing else, legislators and regulators are unlikely to forget, and will be keen to pursue enterprises for security and privacy breaches. In previous years, it was often the case that legislators and regulators had little teeth, but consider how punitive fines have become in recent years following the banking crisis in 2008. In the UK in 2014, the Financial Conduct Authority (FCA) fined a total of £1, 427, 943, 800 [30], during the year, a more than 40 fold increase on 5 years previously.

As we already stated in Section 1, there are no standards when it comes to components for the IoT. This means there is a huge range of different architectures vying for a place in this potentially massive market space. Obviously, from a technical standpoint, greater flexibility and power can be obtained through good use of virtualization. Virtualization is not new and has been around since 1973 [31]. Bearing in mind that dumb sensors do not have enough resources or lack hardware support for virtualization (or at least Linux-based virtualization), we will have a quick look at some of the most popular hardware in use in this space.

ARM [32] presented the ARM capabilities in 2009. ARM is one of the most used platforms in the IoT and has virtualization extensions. Columbia University has developed KVM/ARM, an open-source ARM virtualization system [33]. Dall and Nieh [34] have written an article on this work for LWN.net and for a conference [35]. Paravirtualization support in ARM Coretex A8 has been around since 2006, and ARM Coretex A9 since 2008, with full virtualization since approximately 2009. Virtualization is also in Linux Kernel 3.8. There are also MMU-less ARMs, although it is unlikely that these could be used, unless we were to forfeit the unikernel's protection.

Many modern smart devices can handle virtualization—devices such as play stations, smart automotive systems, smart phones, smart TVs and video boxes. This may not necessarily be the case for small embedded components, such as wear-ables, sensors and other IoT components. MIPS also supports virtualization [36, 37]. Some Intel Atom processors support virtualization (the atom range is huge). However, the low-power Intel Quark has absolutely no support for virtualization. The new open-source RISC-V architecture [38] does support virtualization.

Many current IoT systems in use do have the capability to handle virtualization. For example, most high-powered NAS systems now have virtualization and application support. Thus, we could potentially utilize NAS or other low-powered devices, many of which are ARM, MIPS or x86, to aggregate data on-site and then transport the reduced volume of data to the cloud.

Right now, we must carefully consider the current state of security and privacy in a massively connected world. It is clear that "big brother" is very capable of watching. Not just through the use of massive CCTV networks, but also through IoT-enabled devices, which will become embedded in every smart city. It is estimated that in smart cities of the future, there will be at least 5000 sensors watching as you move through the city at all times. How much personal information could leak as you walk? How much of your money could NFC technology in the wrong hands steal from you, without you being aware of it happening? Do you trust the current technology? We can read about more of these issues in Ref. [39].

### 3.3. Some basic enterprise vulnerabilities

Of course, there are some additional enterprise vulnerabilities that we also need to take into account. These are frequently exploited by the threat environment, and thankfully, we have access to some statistics collected by various security breach reports issued by many security companies [40–42], which will clearly demonstrate the security and privacy problems still faced today, including the fact that the same attacks continue to be successful year on year, as demonstrated by the six-year summary of the Verizon reports shown in **Table 4**. There is no figure provided by Verizon for 2015, as they changed the layout for that year.

We have been looking at an extensive range of management and technical issues above. Yet, there are some fundamental issues which impact directly on the people in the enterprise, as exemplified by the image below in **Figure 1**.

These attacks have been successfully used for decades, in particular the first three, which also happen to be the most devastating. It is no joke to state that in any organization "People are the weakest link", because the first three rely entirely on the inattentiveness and stupidity of users to succeed.

Thus, we can see that there are a considerable number of issues, which all enterprises will face when considering how to run a system, which can offer both a good level of security and privacy. It is necessary to raise awareness of these issues and reasons as to why they are important, and so, we take a look at this in the next section.

| Threat | 2010 | 2011 | 2012 | 2013 | 2014 | 2016 |
|---|---|---|---|---|---|---|
| Hacking | 2 | 1 | 1 | 1 | 1 | 1 |
| Malware | 3 | 2 | 2 | 2 | 2 | 2 |
| Misuse by company employees | 1 | 4 | 5 | 5 | 5 | 4 |
| Physical theft or unauthorized access | 5 | 3 | 4 | 3 | 4 | 6 |
| Social engineering | 4 | 5 | 3 | 4 | 3 | 3 |

**Table 4.** Verizon top 5 security breaches—2010 to 2014, 2016 (1 = highest) [40, 43–47].

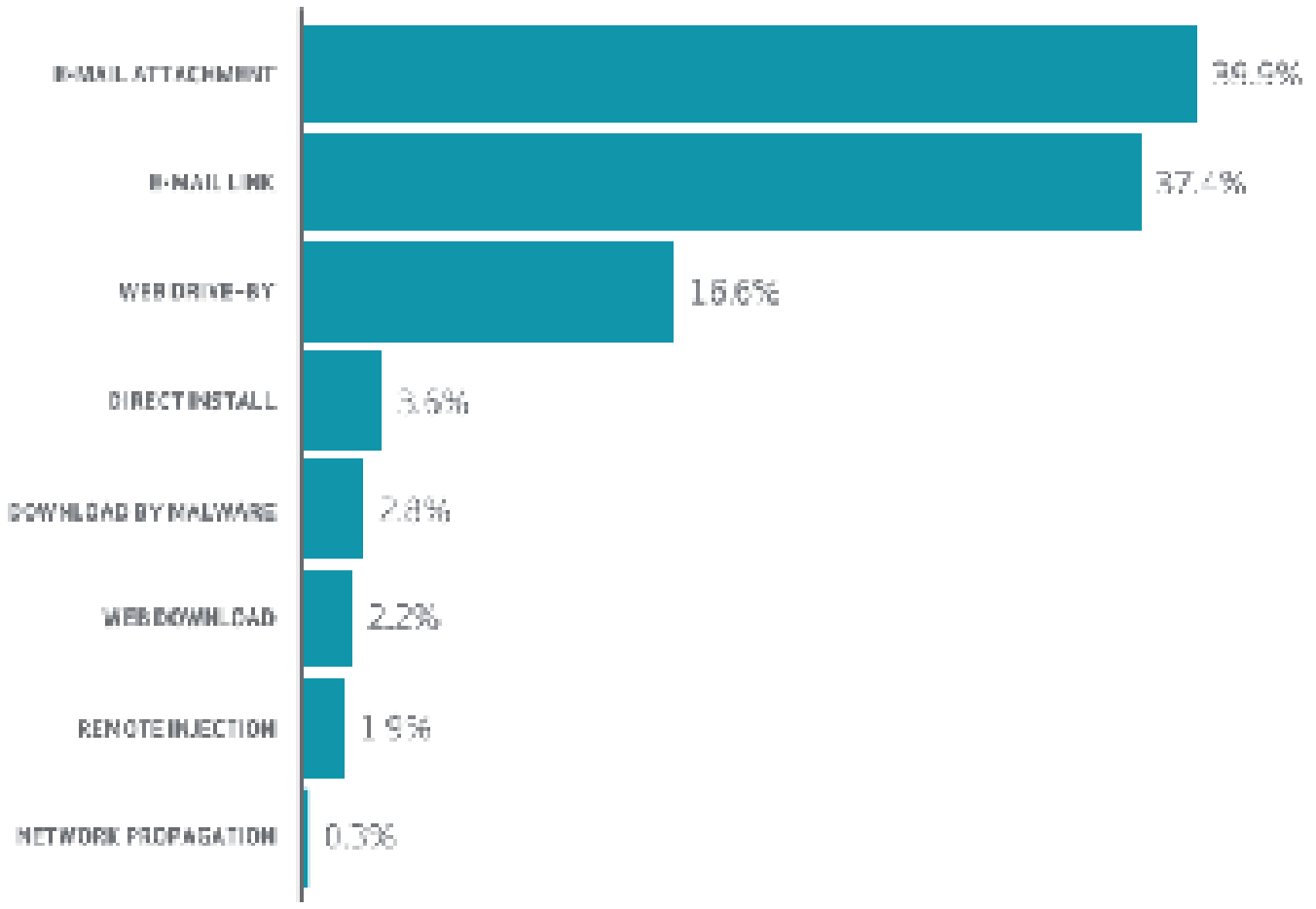

**Figure 1.** The most successful people attacks ©2015 Verizon.

## 4. Why this is important

This question is fairly obvious and easy to answer. Security breaches have a negative monetary and publicity impact on enterprises. In light of the increasing fine levels being applied by regulators, particularly in the light of the forthcoming EU General Data Protection Regulation (GDPR), which introduces the potential for a massive increase in fine levels, enterprises are starting to understand just how much of an impact this will have for them. The impact on an enterprise of a breach can be considerable, particularly, as often happens, where the breach is identified by third parties. This generally ensures that the impact on reputation will be much higher.

Staff often cannot believe that a breach has taken place, and by the time they realize what has happened, the smart attackers have eradicated all traces of their incursion into the system.

There will be virtually zero forensic evidence, so even bringing in expensive security consultants with highly competent forensic teams will be of little value. Quite apart from the possible financial loss to the company from any theft of money or other liquid assets, there will be the possibility of huge reputational damage, which can have a serious negative impact on their share price. Such enterprises are unlikely to have a decent business continuity plan either, which makes recovery even more difficult.

If an enterprise cannot tell how the attackers got in, what data they stole, or compromised, or how they got their hands on whatever was removed, it will be very difficult, time consuming and expensive to recover from such an event. Where an enterprise has a robust defensive strategy implemented, at least in the event of a breach, they will have a better chance of containing the impact of the breach, and a better chance of being able to learn from the evidence left behind.

Another good reason for considering this as an important issue is that it may well be a mandatory legislative or regulatory requirement for the enterprise to maintain such a defensive system. It is clear that legislators and regulators have started to take cyber security far more seriously, and therefore, it is likely that the level of fines levied for breaches will continue to rise.

We provide here a few examples to demonstrate how the authorities are starting to get tougher:

- In 2016, ASUS was sued by the Federal Trade Commission (FTC) [48], because they were not providing updates for their insecure routers.
- In January, the FTC started investigations into D-Link [49, 50], "Defendants have failed to take reasonable steps to protect their routers and IP cameras from widely known and reasonably foreseeable risks of unauthorized access, including by failing to protect against flaws which the Open Web Application Security Project has ranked among the most critical and widespread web application vulnerabilities since at least 2007".
- Schneier [51] recently spoke in front of the US House in favor of legal frameworks for IoT Security.
- In the EU, the General Data Protection Regulation (GDPR) will come into force in May 2018. It will also increase this (monetary) problem for companies with a maximum monetary penalty for a single breach of the higher of €10m or 2% of global turnover, and for a more serious breach involving multiple failings, the higher of €20m or 4% of global turnover.

Despite the fact that cyber security is often seen as falling into the "Cinderella" category of IT expenditure, there is abundant evidence that there are clear benefits to be derived from taking this matter seriously. For those who are not entirely convinced by these arguments, there are many security breach companies who compile annual reports showing the most successful breaches, providing some insight into the weaknesses that expose these vulnerabilities, and discussing the potential mitigation strategies that might be deployed in defense of these attacks. In the next section, we take a look at how current solutions approach these issues.

## 5. Current solutions

Over recent years, a great many solutions to these problems have been proposed, developed and implemented. The early works were generally directed toward conventional corporate distributed IT systems. By comparison to the cloud, these are generally well understood and relatively easy to resolve, and many companies have enjoyed good levels of protection as a result of these efforts. However, cloud changes the rules of the game considerably; because there is often a poor understanding of the technical complexities of cloud, and often a complete lack of understanding that the cloud runs on someone else's hardware and often software too, resulting in a huge issue from lack of proper control. With cloud, the lack of proper and complete security standards [6] also presents a major issue, as does the method of implementing compliance with these standards [52].

The other major issue, particularly when cloud is involved, is that implementing partial solutions, while very effective for the specific task in hand, will not offer a complete solution to the security problem. There are a great many potential weaknesses involved in cloud systems, and without taking a much more comprehensive defensive approach, it will likely prove impossible to effectively secure systems against all the threats faced by enterprises.

Any solution, which only addresses partially the overall security of the enterprise, will be doomed to failure. The only way to be sure a solution will work, will be to properly identify the risks faced by the organization and provide a complete solution to attempt to mitigate those risks, or to accept the particular risks with which the enterprise is comfortable. A fundamental part of this process, which is all too frequently forgotten, is the need to monitor what is happening in the enterprise systems after the software solution is installed, not whenever compliance time comes around, but day to day on an ongoing basis.

There are two major trends developing in the quest to tackle these issues:

1. preventing security breaches,
2. accepting that security breaches will happen and trying to contain breaches.

Traditionally, perimeter security such as firewalls is utilized to prevent attackers from entering the corporate network. The problem with this approach is that this model was well suited to network architectures comprising a few main servers with limited connections between clients and servers. However, as soon as the number of connections explodes, this approach can no longer cope with the demand. This first occurred with the influx of mobile devices and will again become problematic as new Industry 4.0 and IoT devices continue to be implemented and expanded.

We take a look at some of the current solutions available today and will discuss potential weaknesses which pertain to them. The following list shows the most common approaches to tackle cloud cyber security issues:

- Perimeter security (e.g. firewalls) problems with IoT, etc.

  Perimeter Security prevents malicious actors from entering the internal—private—network. Traditionally based on network packet inspection at important network connection points, e.g., routers, by now it includes endpoints such as client computers or mobile devices;

- Sandboxes, e.g. for analysis

  Sandboxes: traditionally anti-virus software utilized signature-based blacklists. Due to newer evasion techniques, e.g., polymorphic payloads, those engines provide insufficient detection rates. Newer approaches utilize sandboxes, e.g., emulate a real system within which the potentially malicious payload is executed and traced. If the payload's behavior seems suspicious, it is quarantined;

- Virtualization

  Virtualization solutions allow program execution within virtual machines. As multiple virtual machines can execute on the same physical host, this allows separation of applications into separate virtual machines. As those cannot interact, this creates resilience in the face of attacks—a successful attack only compromises a single application instead of the whole host. Virtualization is also used to implement sandboxes, easy software deployment, etc.;

- Containers

  Containers: originally used to simplify software deployments. Newer releases of container technologies allow for better isolation between containers. This solves similar problems as virtual machines while improving memory efficiency;

- Software development: secure lifecycles (testing, etc.)

  Software Development: recently security engineering has become a standard part of software development approaches, e.g., Microsoft Secure Development Life-cycle. Mostly, this focuses on early security analysis of potential flaws and attacks and subsequent verification of implemented mitigations;

- Software development: new "safe" languages

  New Programming Languages: as programs are implemented within a given programming language, their security features are very important for the security of the final software product. A good example are common memory errors in programs written in C-based programming languages. Recently, a new generation of programming languages aim to provide both performance and safety—examples of those are Rust, Go or Swift. Rust has seen uptake by the Mozilla community and is being used as part of the next-generation JavaScript engine. Go is used for systems programming and has been adopted by many virtualization management and container management solutions;

- Software development: hardening (i.e., fuzzing)

  Fuzzing: new tooling allows for easy fuzzing of application or operation system code. Automated appliance of this attack technique has yielded multiple high-level memory corruption errors recently;

- Software development: hardening (i.e., libraries)

  Hardened Libraries and Compilers: recent processors provide hardware support for memory protection techniques. Newer compilers allow for automatic and transparent usage of those hardware features—on older architectures, similar features can be implemented in software yielding slightly lower security and performance;

- Architecture: (micro)services

    Microservice Architectures encapsulate a single functionality within a microservice that can be accessed through standardized network protocols. If implemented correctly, each service contains minimal state and little deep interaction with external services. This implicitly implements the Separation of Concern principle and thus helps security by containing potential malicious actors.

The serverless paradigm is also gaining some traction and examples for this software stack can be seen in **Table 5**.

Another recent security battleground is the IoT. Recent examples are the 1Tbps attack against "Krebs on Security" or the 1.2Tbps DDoS attack taking out Dyn in October 2016 [53]. While the generic security landscape is already complex, IoT adds additional problems such as hard device capability restrictions, manifold communication paths and very limited means of updating already deployed systems. In addition, software is often an after-thought; for example, for usage within power-grids, all hardware must be certified with regard to their primary focus (i.e. power distribution). All later alterations, such as security-critical software updates, would void the certification and thus produce substantial costs for the device manufacturer. This leads to a situation where security updates are scarce in the best scenario. Unikernels can improve upon the current situation [12] on both the device as well as at the cloud-level. It should be understood that device level means on the local sensors/actors (problems with security and updates) and cloud level is for scale-out of backend processing of data generated by IoT devices.

| Product | Supported languages |
|---|---|
| Amazon Lambda | Java, Node.js, Python |
| Google Cloud Functions | Node.js |
| IBM BlueMix | Languages supported by CloudFountry, e.g. Java, Node.js, Go, C#, PHP, Python, Ruby |
| Microsoft Azure Functions | C#, F#, Node.js, Python, PHP |

**Table 5.** Example of serverless offerings ©2016 Happe, Duncan and Bratterud.

## 6. Proposed solutions

The most interesting aspect of unikernels from a security point of view is their inherent minimalism. For the purpose of this chapter, we define unikernels as:

- a minimal execution environment for a service,
- providing resource isolation between those services,
- offering no data manipulation on persistent state within the unikernel, i.e. the unikernel image is immutable,

- being the synthesis of an operating system and the user application,
- only offering a single execution flow within the unikernels, i.e. no multitasking is performed.

The unikernel approach yields an architecture that implicitly follows best software architecture practices, e.g. through the combination of minimalism and single execution flow, the separation of concern principle is automatically applied. This allows for better isolation between concurrent unikernels.

It is absolutely necessary to recognize the magnitude of the dangers posed by the threat environment. Enterprises are bound by legislation, sometimes regulation, the need to comply with standards, industry best practice, and are accountable for their actions. Criminals have no such constraints. They are completely free to bend every rule in the book, do whatever they want, manipulate, cajole, hack or whatever it takes to get to the money. They are constantly probing for the slightest weakness, which they are more than happy to exploit without mercy. It is clear that the threat environment is developing just as quickly as the technological changes faced by the industry [6, 54, 55]. We need to be aware of this threat and minimize the possible impact on our framework. While we have absolutely no control over attackers, we can help reduce the impact by removing as many of the "classic attack vectors" as possible, thus making their life far more difficult. The more difficult it becomes for them to get into the system, the more likely they will be to go and attack someone else's system.

In the interests of usability, many more ports are open by default than are needed to run a system. An open port, especially one which is not needed (and therefore not monitored) is another route in for the attacker. We also take the view that the probability of vulnerabilities being present in a system increases proportionally to the amount of executable code it contains. Having less executable code inside a given system will reduce the chances of a breach and also reduce the number of tools available for an attacker once inside. As Meireles [56] said in 2007 "… while you can sometimes attack what you can't see, you can't attack what is not there!". We address these issues by making the insides of the virtual machine simpler. We also propose to tackle the audit issue by making configuration happen at build time [57, 58], and then making services be "immutable" after deployment, making "configuration lapses" (i.e. through conflicts caused by unnecessary updates to background services etc.) unlikely.

Bearing in mind the success with which the threat environment continually attacks business globally, it is clear that many enterprises are falling down on many of the key issues we have highlighted in Section 3. It is also clear that a sophisticated and complex solution is unlikely to work. Thus, we must approach the problem from a more simple perspective.

### 6.1. Unikernel impact on efficiency

Cloud achieves maximum utilization and high energy efficiency through consolidating processing power within data centers. Multiple applications run on the same computing node, but often control on node placement or on concurrently running applications are not possible. This means for security, isolation between different applications, users, or services, is critical. A popular but inefficient solution is to place each application or service within a virtual machine [59]. This is very similar to the initial usage of virtualization within host-based systems; Madnick gives a good overview of the impact of virtualization in Ref. [60]. Containers could present a more efficient approach [61], but as they were originally developed to improve deployment, their security benefits are still open to debate [62].

A useful benefit of the applied minimalism is a reduced memory footprint [63, 64] plus a quick start-up time for unikernel-based systems. Madhavapeddy et al. utilize this for on-demand spin-up of new virtual images [65], allowing for higher resource utilization, leading to improved energy efficiency.

### 6.2. Unikernel impact on security

The most intriguing aspect of unikernels from a security perspective is their capability as a minimal execution environment. We can now define the attack surface of a system as:

**Definition 6.1** attack surface. The amount of bytes within a virtual machine [10].

When it comes to microcode, firmware and otherwise mutable hardware such as field-programmable gate arrays (FPGAs), physical protection can be seen as a gray area. This definition is intentionally kept general in order to allow further specifications to refine the meaning of "physically available" for a given context. The following example can serve to illustrate how the definition can be used for one of many purposes.

Building a classic VM using Linux implies simply installing Linux and then installing the software on top. Any reduction in attack surface must be done by removing unneeded software and kernel modules (e.g. drivers). Take TinyCore Linux as an example of a minimal Linux distribution and assume that it can produce a machine image of 24MB in size.

During the build of a unikernel, minimization is performed, meaning the resulting system image only includes the minimum required software dependencies. This implies that no binaries, shell or unused libraries are included within the unikernel image. Even unused parts of libraries should never be included within the image. This radically reduces included functionality and thus the attack surface. In addition, this can loosen the need for updates after vulnerabilities have been discovered in included third-party components—if the vulnerable function was not included within the set of used functions, an update can be moot.

The situation after a security breach with unikernels is vastly different to traditional systems. Assuming that the attacker is able to exploit a vulnerability, e.g. buffer overflow, he gains access to the unikernel system's memory. Due to having no binaries and only reduced libraries, writing shell code [66], a machine code that is used as a payload during vulnerability execution will not work. Common payloads spawn command shells or abuse existing libraries to give attackers access through unintended possibilities, which is complicated. Pivot attacks depending on shell-access are thwarted. But, all direct attacks against the application, e.g. data extraction due to insecure application logic, are still possible. A good example of this is the recent OpenSSL Heartbleed vulnerability [67]. A unikernel utilizing OpenSSL would also be vulnerable to this, thus allowing an attacker to access its system memory, including the private SSL key. We argue that functionality should be split between multiple unikernels to further compartmentalize breaches.

Next-generation hardware-supported memory protection techniques can benefit from minimalism. For example, the Intel Secure Guard Extensions [68, 69] allow for protected memory enclaves. Accessing these directly is prohibited and protected through specialized CPU instructions. The protection is enforced by hardware, so even the hypervisor can be prevented from accessing protected memory. Rutkowska has shown [70] that deploying this protection scheme for applications has severe implications. Just protecting the application executable is insufficient, as attacks can inject or extract code within linked libraries. This leads us to conclude that the whole application including its dependencies must be part of the secure-memory enclave. Simplicity leads to a "one virtual machine per application" model, which unikernels inherently support. We propose that unikernels are a perfect fit for usage with those advanced memory protection techniques.

Returning to the theme of "*software development frameworks providing sensible defaults but getting bloated, and thus vulnerable over time*", unikernels provide an elegant solution; while the framework should include generic defensive measures, the resulting unikernel will, by definition, only include utilized parts, thus reducing the attack surface.

### 6.3. Service isolation

A fundamental premise for cloud computing is the ability to share hardware. In private cloud systems, hardware resources are shared across a potentially large organization, while on public clouds, hardware is shared globally across multiple tenants. In both cases, isolating one service from the other is an absolute requirement.

The simplest mechanism to provide service isolation is *process isolation* in classic kernels, relying on hardware supported virtual memory, e.g. provided by the now pervasive x86 protected mode. This has been used successfully in mainframe setups for decades, but access to terminals with limited user privileges has also been the context for classic attack vectors such as stack smashing, root-kits, etc., the main problem being that a single kernel is being shared between several processes and that gaining root access from one terminal would give access to everything inside the system. Consequently, much work was done in the 1960s and 1970s to find ways to completely isolate a service without sharing a kernel. This work culminated in the seminal 1974 paper by Popek and Goldberg [71], where they present a formal model describing the requirements for complete instruction level virtualization, i.e. *hardware virtualization*.

Hardware virtualization was in wide use on e.g. IBM mainframes since that time, but it was not until 2005 that the leading commodity CPU manufacturers, Intel and AMD introduced these facilities into their chips. Meantime, paravirtualization had been reintroduced as a workaround to get virtual machines to run on these architectures, notably in Ref. [72]. While widely deployed and depended upon, the Xen project has recently evolved its paravirtualization interface toward using hardware virtualization in, e.g., PVH [73], stating that *"PVH means less code and fewer Interfaces in Linux/FreeBSD: consequently it has a smaller Trusted Computing Base (TCB) and attack surface, and thus fewer possible exploits"* [74].

Yet another mechanism used for isolation is operating system-level virtualization with containers, e.g. Linux Containers (LXC) popularized in recent years by Docker, where each container represents a userspace operating environment for services that all share a kernel. The isolation mechanism for this is classic process isolation, augmented with software controls such as cgroups and Linux namespaces. Containers do offer less overhead than classic virtual machines. An example where containers makes a lot of sense would be trusted in-house clouds, e.g. Google is using containers internally for most purposes [75]. We take the position that hardware virtualization is the simplest and most complete mechanism for service isolation with the best understood foundations, as formally described by Popek and Goldberg, and that this should be the preferred isolation mechanism for secure cloud computing.

**6.4. Microservices architecture and immutable infrastructure.**

Microservices is a relatively new term founded on the idea of separating a system into several individual and fully disjoint services, rather than continuously adding features and capabilities to an ever growing monolithic program. Being single threaded by default, unikernels naturally imply this kind of architecture; any need for scaling up beyond the capabilities of a single CPU should be done by spawning new instances. While classic VMs require a lot of resources and impose a lot of overhead, minimal virtual machines are very lightweight. As demonstrated in Ref. [76], more than 100,000 instances could be booted on a single physical server and Ref. [58] showed that each virtual machine including the surrounding process requires much less memory than a single "Hello World" Java program running directly on the host.

An important feature of unikernels in the context of microservices is that each unikernel VM is fully self contained. This also makes them immune to breaches in other parts of the service composition, increasing the resilience of the system as a whole. Adding to this, the idea of *optimal mutability* (defined below) and each unikernel-based can in turn be as immutable as is physically possible on a given platform. In the next paper in this series, we will expand upon these ideas and take the position that composing a service out of several microservices, each as immutable as possible, enables overall system architects and decision makers to focus on a high-level view of service composition, not having to worry too much about the security of their constituent parts. We take the view that this kind of separation of concerns is necessary in order to achieve scalable yet secure cloud services.

**6.5. No shell by default and the impact on debugging and forensics**

One feature of unikernels that immediately makes it seem very different from classic operating systems is the lack of a command line interface. This is, however, a direct consequence of the fact that classic POSIX-like CLIs are run as a separate process (e.g. bash) with the main purpose of starting other processes. Critics might argue that this makes unikernels harder to manage and "debug", as one cannot "log in and see what's happened" after an incident, as is the norm for system administrators. We take the position that this line of argument is vacuous; running a unikernel rather corresponds to running a single process with better isolation, and in principle, there is no more need to log in to a unikernel than there is to log in to, e.g., a web server process running in a classic operating system.

It is worth noting that while unikernels by definition are a single-address-space virtual machine, with no concept of classic processes, a read-eval-print loop (REPL) interface can easily be provided (e.g. IncludeOS does provide an example)—the commands just would not start processes, but rather call functions inside the program. From a security perspective, we take the view that this kind of ad-hoc access to program objects should be avoided. While symbols are very useful for providing a stack trace after a crash or for performance profiling, stripping out symbols pointing to program objects inside a unikernel would make it much harder for an attacker to find and execute functions for malicious and unintended purposes. Our recommendation is that this should be the default mode for unikernels in production mode.

We take the view that logging is of critical importance for all systems, in order to provide a proper audit trail. Unikernels, however, simply need to provide the logs through other means, such as over a virtual serial port or ideally over a secure networking connection to a trusted audit trail store.

Many UNIX period system administrators will require some mental readjustment due to the lack of shell access. On the other hand, the growing DevOps movement [77] abolishes the traditional separation into software development and system administration but places high importance on the communication between and integration of those two areas. Unikernels offer an elegant deployment alternative. The minimized operating system implicitly moves system debugging to application developers. Instead of analyzing errors through shell commands, developers can utilize debuggers to analyze the whole system, which might be beneficial for full-stack engineering.

Lastly, it is worth mentioning that unikernels in principle have full control over a contiguous range of memory. Combined with the fact that a crashed VM by default will "stay alive" as a process from the VMM perspective and not be terminated, this means that in principle the memory contents of a unikernel could be accessed and inspected from the VMM after the fact, if desired. Placing the audit trail logs in a contiguous range of memory could then make it possible to extract those logs also after a failure in the network connection or other I/O device normally used for transmitting the data. Note that this kind of inspection requires complete trust between the owner of the VM and the VMM (e.g. the cloud tenant and cloud provider). Our recommendation would be not to rely on this kind of functionality in public clouds, unless all sensitive data inside the VM is encrypted and can be extracted and sent to the tenant without decrypting it.

### 6.6. Why use unikernels for the IoT?

Why use unikernels for the IoT [78]? Unikernels are uniquely suited to benefit all areas (sensor, middleman and servers) within the IoT chain. They allow for unified development utilizing the same software infrastructure for all layers. This may sound petty, but who would have thought JavaScript could be used on servers (think node.js) a couple of years ago?

Using hardware virtualization as the preferred isolation mechanism, we take the view that there are three basic approaches we can use to deliver our requirements, namely the monolithic system/kernel approach, the microkernel approach and the unikernel approach. IaaS cloud providers will typically offer virtual machine images running Microsoft Windows or one or more flavors of Linux, possibly optimized for cloud by, e.g., removing device drivers that are not needed. While specialized Linux distributions can greatly reduce the memory footprint and attack surface of a virtual machine, these are general purpose multi-process operating systems and will by design contain a large amount of functionality that is simply not needed by one single service. We take the position that virtual machines should be specialized to a high degree, each forming a single purpose microservice, to facilitate a resilient and fault tolerant system architecture which is also highly scalable.

In Ref. [11], we discuss six security observations about various unikernel operating systems: choice of service isolation mechanism; use of a single address space, shared between service and kernel; no shell by default and the impact on debugging and forensics; the concept of reduced attack surface; and microservices architecture and immutable infrastructure. We argue that the unikernel approach offers the potential to meet all our needs, while delivering a much reduced attack surface, yet providing exactly the performance we require. An added bonus will be the reduced operating footprint, meaning a more green approach is delivered at the same time.

### 6.7. For IoT on the client

Unikernels are a kind of virtualization and offer all of its benefits. They provide application developers with a unified interface to diverse hardware platforms, allowing them to focus on application development. They provide the ability to mask changes of the underlying hardware platform behind the hypervisor, allowing for application code to be reused between different hardware revisions. Also, disparate groups within an enterprise often perform system and application development. Use of a unikernel decouples both groups, allowing development in parallel. Application developers can use a virtualized testing environment on their workstations during development, which will mirror the same environment within the production environment.

Unikernels can certainly produce leaner virtual machines compared to traditional virtualization solutions. This results in a much reduced attack surface, which creates applications that are more secure. Use of a resource efficient unikernel, such as IncludeOS, minimizes the computational and memory overhead that otherwise would prevent virtualization from being used. The small memory and processing overhead enables the use of virtualization on low-powered IoT devices and also aids higher capacity devices. Lower resource utilization allows for either better utilization (i.e. running more services on the same hardware) or higher usage of low power modes, reducing energy consumption, both of which increase the sustainability of IoT deployments.

A feature in high demand by embedded systems is atomic updates. A system supporting atomic updates either installs a system update or reverts back to a known working system state. For example, Google's Chrome OS [79] achieves this by using two system partitions. A new system upgrade is installed on the currently unused partition. On next boot, the newly installed system is now used, but the old system is preselected as a backup boot option if this boot does not work. If the new system boots, the new system becomes the new default operating system and the (now) old partition will be used for the next system upgrade.

This delivers high resilience in the face of potentially disrupting Chrome OS updates. A similar scheme is set to be introduced for the forthcoming Android Version 7. This scheme would be greatly aided by unikernels, as they already provide a clear separation of data and control logic. A system upgrade would therefore start a new unikernel and forward new requests to it. If the underlying hypervisor has to be upgraded, likely a very rare event, the whole system might incorporate the dual boot-partition approach.

### 6.8. For IoT on the server

Taking account of the large estimated number of IoT devices to be deployed in the near future, computational demand on the cloud can be immense. While IoT amplifies the amount of incoming traffic, it has some characteristics that should favor unikernel-like architectures.

Our envisioned unikernels utilize a non-mutable state and are event-based. This allows simplified scale-out, i.e. it allows for dynamically starting more unikernels if incoming requests demand it. We believe that many processing steps during an IoT dataflow's lifetime will be parallelizable, e.g. data collected from one household will not interact with data gathered by a different household from another continent during the initial processing steps, or possibly never at all. Since they do not interact, there is no chance of side effects, thus the incoming data can instantly be processed by a newly spawned unikernel.

Two recent trends in cloud computing are cloudlets and fog computing. The first describes a small-scale data center located near the Internet's edge, i.e. co-located near many sensors and acting as the upstream collection point for the incoming IoT sensors, while the second describes the overall technique of placing storage or computational capabilities near the network edges. A unified execution environment is needed to allow easy use of this paradigm. When the same environment is employed, application code can easily be moved from the cloud toward the networks' edge, i.e. into the cloudlets. Unikernels offer closure over the application's code, so the same unikernel can be re-deployed at a cloudlet or within a central data center.

The unikernel itself might place requirements on external facilities such as storage, which would need to be provided by the current execution environment. A consumer-grade version of this trend can already be seen in many high-powered NAS devices, which allow for local deployment of virtual machines or containers. This moves functionality from the cloud to a smallest-scale local processing environment. A good use case for this would be Smart Homes; here, a local NAS can perform most of the computations and then forward the compressed data toward a central data center. Also, this local preprocessing can apply various cryptographic processes to improve the uploaded data's integrity or confidentiality.

## 7. Conclusions

We have taken a good hard look at cyber security in the cloud, and in particular, we have considered the security implications of the exciting new paradigm of the IoT. While the possibilities are indeed exciting, the consequences of getting it wrong are likely to be catastrophic. We cannot afford to carry blindly on. Instead, we must recognize that if the issues we have outlined on security and privacy are not tackled properly, and soon, we will all be sleepwalking into a disaster. However, if we realize that we need to take some appropriate actions now, then we will be much better placed to feel comfortable in living in an IoT world. There are considerable potential benefits for everyone to be offered from using our unikernel-based approach. While we see security and confidentiality of data as paramount—and given the forthcoming EU's GDPA, we believe the EU agrees. Security and privacy do not directly translate into a direct monetary benefit for companies and thus are seldom given enough incentive for change to allow serious improvement to gain traction. To better convince enterprises, we offer the added benefit of increased developer efficiency. Experienced and talented developer resources are scarce at hand, so making the most of them is in an enterprise's best interest. The broad application of a virtualization solution allows them to better reuse existing knowledge and tools, as developers gain a virtual long-term environment that they can work in.

Virtualization in combination with the special state-less nature of many unikernels provides a solution for short-term processing spikes. Processing can be scaled-out to in-company or public clouds by deploying unikernels as they do not require external dependencies and as they do not contain state, deployments are simplified. After their usage, they can be discarded (no state also means that no compromising information is stored at the cloud provider). In the case of sensitive information, special means, e.g. homomorphic encryption or verifiable computing technologies need to be employed to protect data integrity or confidentiality.

Unikernels offer a high energy efficiency. This allows companies to claim higher sustainability for their solutions while reducing their energy costs. We view our proposed solution as taking a smart approach to solving smart technology issues. It does not have to be exorbitantly expensive to do what we need, but by taking a simple approach, sensibly applied, we can all have much better faith in the consequences of using this technology (as well as having the comfort of being able to walk through a smart city without having our bank account emptied).

## Author details

Bob Duncan[1]*, Andreas Happe[2] and Alfred Bratterud[3]

*Address all correspondence to: bobduncan@abdn.ac.uk

1 Computing Science, University of Aberdeen, Aberdeen, UK

2 Department of Digital Safety & Security, Austrian Institute of Technology GmbH, Vienna, Austria

3 Department of Computer Science, Oslo and Akershus University, Oslo, Norway

Chapter 3

# Machine Learning in Application Security

Nilaykumar Kiran Sangani and Haroot Zarger

Additional information is available at the end of the chapter

**Abstract**

Security threat landscape has transformed drastically over a period of time. Right from viruses, trojans and Denial of Service (DoS) to the newborn malicious family of ransomware, phishing, distributed DoS, and so on, there is no stoppage. The phenomenal transformation has led the attackers to have a new strategy born in their attack vector methodology making it more targeted—a direct aim towards the weakest link in the security chain aka humans. When we talk about humans, the first thing that comes to an attacker's mind is applications. Traditional signature-based techniques are inadequate for rising attacks and threats that are evolving in the application layer. They serve as good defences for protecting the organisations from perimeter and endpoint-driven attacks, but what needs to be focused and analysed is right at the application layer where such defences fail. Protecting web applications has its unique challenges in identifying malicious user behavioural patterns being converted into a compromise. Thus, there is a need to look at a dynamic and signature-independent model of identifying such malicious usage patterns within applications. In this chapter, the authors have explained on the technical aspects of integrating machine learning within applications in detecting malicious user behavioural pattern.

## 1. Introduction

Cybersecurity, a niche domain is likely to be compared in parallel to a cat and mouse game where sometime the offensive team (attacker/hacker) has an advantage and sometime the defensive team (cyber sec personnel). This never settling game has changed drastically over a period of time having born to various attack vectors targeting humans or what is largely known as the weakest link in the security chain.

Over the years, Information Technology (IT) has witnessed a massive paradigm shift. Initially, it was about mainframes, client-server model, closed group of systems, and the attacks were very limited and focused towards these only. Down the line of time, the former has been transformed completely into the web-based layer, clouds, virtualisation, and so on, thus adding greater complexity in the whole development-deployment architecture—of applications and infrastructure—thus making the attack surface further difficult for the hackers. What has remained constant is the human factor and the same is being exploited in large to circumvent the protection mechanisms which are in place.

Traditional signature-based solutions are functioning great in preventing against known attacks, but the paradigm shift of the technologies is making the signature-based systems inadequate against the newborn attacks and malicious exploits. Thus, the need of the hour is to implement which is a dynamic and signature—less thus evolved machine learning (ML).

Machine learning is not a new domain or technology. It has been in use in other areas since the 1950s. The missing link is the intersection of cybersecurity and machine learning. One of the best examples of early use of machine learning in security is the case of spam detection.

In this chapter, we cover how cybersecurity has evolved over a period of time and how attacks have become more tactical and sophisticated. We also talk about what is machine learning and its associated components. In this part, we cover how combination of machine learning and security adds value to an organisation. Later on, we focus on the application layer and web applications in specifics. And, finally, we talk about focusing on merging machine learning and applications to provide a pattern-based analytics of security within applications.

The second section covers in detail how cybersecurity has evolved over a period of time and how attacks have become more tactical and sophisticated. Section 3 focuses on the application layer and web applications in specifics. It will also cover how web applications have grown over time and the threats associated with them. Section 4 talks about what is machine learning and its associated components. This section, in addition, will also cover how combination of machine learning and security adds value to an organisation.

Section 5 targets on the merger of machine learning and application to provide a pattern-based analytics layer of security within applications.

## 2. Evolution of cybersecurity

By definition, Cybersecurity can be defined as 'the body of technologies, processes and practices designed to protect networks, computers, programs and data from attack, damage or unauthorised access'. One of the most challenging elements of cybersecurity is the quickly and constantly evolving nature of security risks. Adam Vincent pronounces the problem [1]:

*'The threat is advancing quicker than we can keep up with it. The threat changes faster than our idea of the risk. It's no longer possible to write a large white paper about the risk to a particular system. You would be rewriting the white paper constantly.'*

Initially cybersecurity used to be relatively simple. The enterprise network comprised of mainframes, client-server model, closed group of systems and the attacks were very limited with viruses, worms and Trojan horses being the major cyber threats. The focus was more towards malwares such as virus, worms and trojans with purpose of causing damage to the systems. It started with virus which needed to be executed in order to cause a malfunction or damage to the system. As this was something where manual intervention was required for propagation, a new type of malware came into existence, that is, 'Worm' similar to virus but with self-replicating feature, that is, they do not require a human intervention or a program to execute. These cyber threats randomly targeted computers directly connected to the Internet but posed little threat. Within the enterprise networks with firewalls on the perimeter and antivirus protection on the inside, the enterprise appeared to be protected and relatively safe. Occasionally an incident would occur and security teams would fight it.

The initial attacking methodology was attacking the infrastructure. This involved the traditional approach of compromising the systems by getting inside the network though loopholes such as open ports, unknown services, and exploiting system-related vulnerabilities in the infrastructure. At this time, the offensive teams started to recognise and closed these gaps as much as possible, reducing the attack surface. Over a period of time, as the infrastructure changed, the former has been transformed completely into the web applications, web-based layer, clouds, virtualisation, mobility, and so on, thus adding greater complexity in the whole development, deployment architecture of applications and infrastructure and changing the attack surface further as shown in **Figure 1**. Attackers started getting inside the enterprise networks, and once they were inside they operated in stealthy mode. By attaining access, they controlled the infected machines and managed them through command and control systems (C&C servers). Vulnerable systems were exploited within the enterprise for lateral movement among computers on the network, capturing user credentials and other critical information of more and more users within the organisation. The final nail in the coffin was privilege escalation, art to gaining elevated access to the machine, get control of the systems administrator accounts in charge of everything. Once these attackers got administrative control of the enterprise, they were able to do anything they wanted. It was like 'Keys to the Kingdom'.

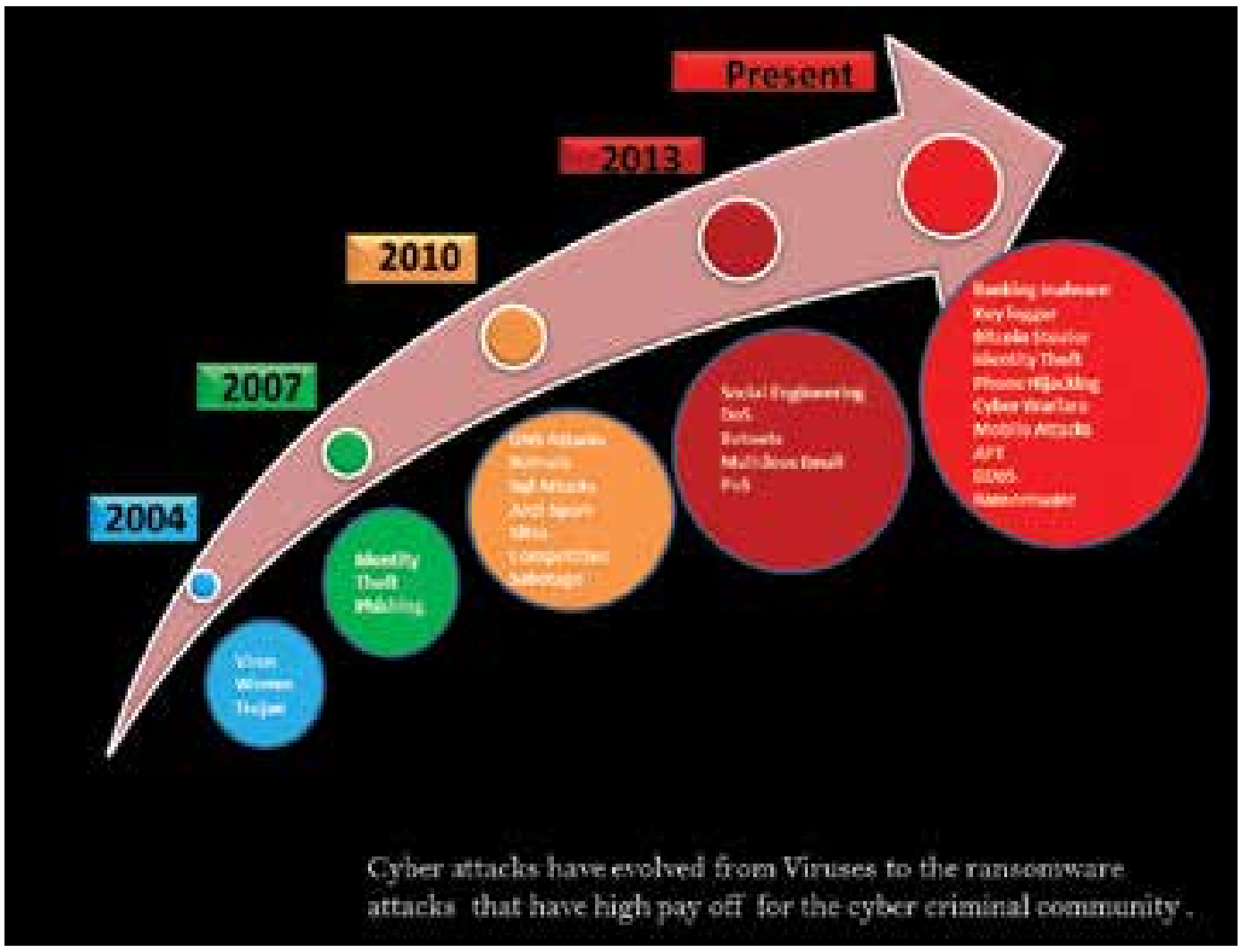

**Figure 1.** Cybersecurity attacks evolution over time.

Similarly, to the cat and mouse game or as we have seen in Tom and Jerry, to overcome each other as they used to change the tactics, the very same applies when we talk about attackers versus defenders in cyber space. Attackers take the advantage of zero-day exploits, vulnerabilities, and so on to compromise the systems, whereas defenders use secure mechanism such as hardening, patching, segmentation and other security controls to reduce the surface attack. This way the enterprise is locked down up to a certain extent, thus reducing the attack surface. For the attackers with less threat surface to attack due to the lock down, the only possibility seen by them towards a breach lies in web application.

### 2.1. Why web applications are vulnerable?

Before we begin, let us have a basic understanding of web applications. A web application or web app is a software application in which the client (or user interface) runs in a browser. Common web applications include webmail, online retail sales, online auctions, wikis, instant messaging services and many other functions [2]. For organisations, whether they are a private entity or government, to conduct business online, it has to provide services to the outside world. Over a period or so, the web has been embraced by millions of businesses as an inexpensive medium to communicate and exchange information with customers [3]. Therefore, they are vital to businesses for expanding their online presence, thus fashioning long-term and beneficial relationships with customers. There is no doubt in saying that web applications have become such a universal phenomenon over a period of time. Web applications are convoluted and multifarious in nature, and due to this behavior, they are widely mysterious and completely misinterpreted [3].

Regardless of the advantages, web applications do raise a number of security concerns. Severe weaknesses or vulnerabilities allow hackers to gain direct and public access to databases in order to extract sensitive data. Many of these databases contain critical information (personal, official, financial details, etc.) making them a frequent target of hackers. Although defacing corporate websites are still commonplace, nowadays, hackers prefer gaining access to the sensitive data residing on the database server because of the immense pay-offs in selling the data.

The greater complexity, including the web application code and underlying business logic, and their potential as a vector to sensitive data in storage, or in process, makes web application servers an obvious target for attackers [12].

In **Figure 2**, it is easy to see how a hacker can quickly access the data on the database through creativity and negligence or human error, leading to vulnerabilities in the web applications.

As mentioned, websites use databases to store and fetch the required information to the users. If a web application is vulnerable, that is, it can be exploited by the attackers, then the database associated with the web application is at serious risk, as it contains all the critical information that the attackers are looking for. Recent research shows that 75% of cyberattacks are done at web application level [3].

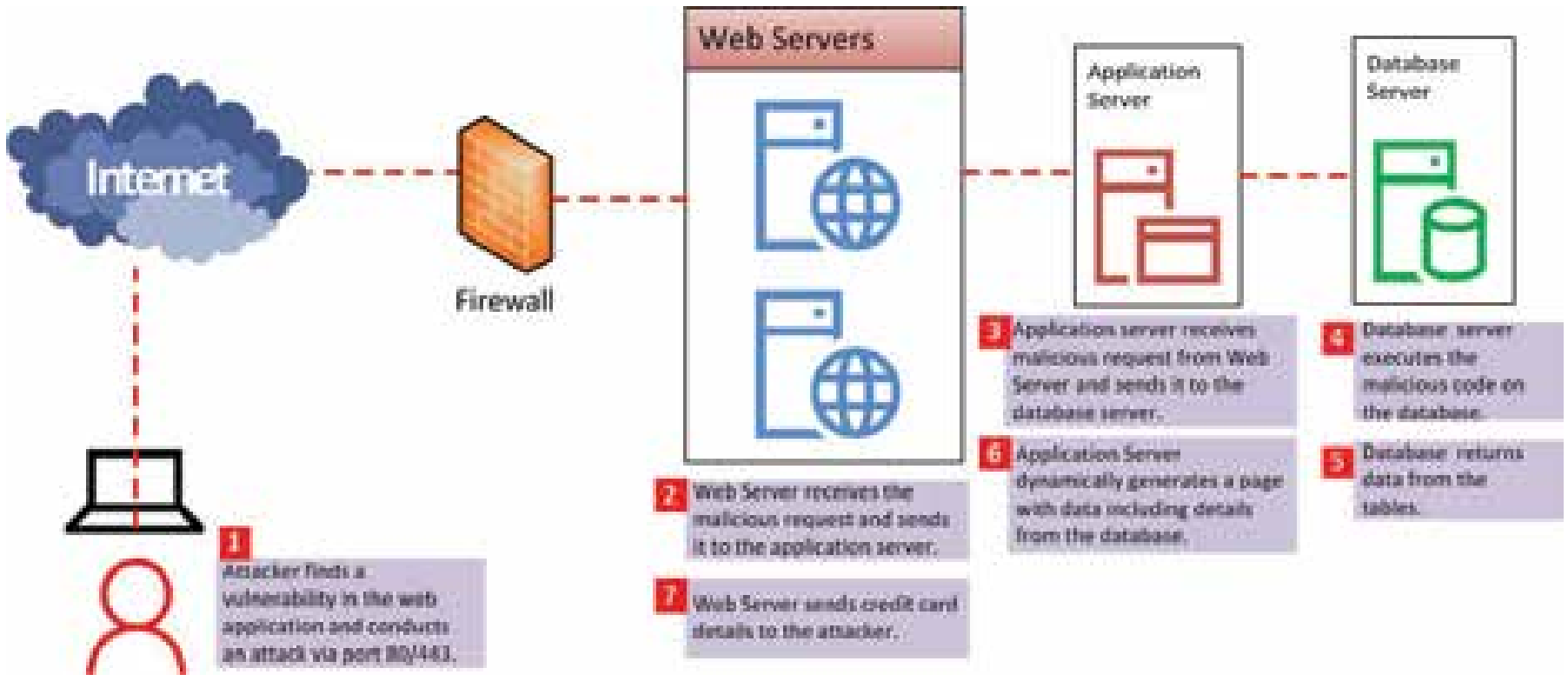

**Figure 2.** How an attacker exploits a web application.

Web application vulnerabilities have drastically increased over the past few years, as companies demand faster web application releases to fulfil the end-user requirements. Vulnerabilities associated with web applications are risky for organisations as they endanger not only brand and reputational damage but also loss of data, legal action and financial penalties associated with these incidents. The outcomes continue to confirm the majority understanding that the web application vector is a foremost and less protected path for attackers [11].

Web application scene is altering continuously over time. Evolution of web layer has enhanced rich experiences and functionalities directly within/from the browser. As a result of this flexibility and scalability that web applications provide, web applications and web services are rapidly replacing the legacy applications, and, as a result, broadening the surface attack which increases attacker's chances of exploitation, primarily since traditional network layer security controls such as firewalls and signature-based intrusion prevention and detection systems (IPS/IDS) have little or no role to play in detecting and preventing an attack occurring via the web application.

### 2.2. Cybersecurity attacks

In the past few years, the trend has played out in more and more breaches hitting the headlines. Some of the cyberattacks that shock the IT world include the following:

- RSA SecurID breach: Year 2011 [4]

  In 2011, RSA's enterprise was breached and the security keys for many of its customers were believed to have been stolen. This breach prompted RSA to replace millions of its SecurID tokens to restore security for its customers.

- Columbian Independence Day Attack: Year 2013

  In 2013, a large-scale cyberattack held on 20 July—Columbian Independence Day—against 30 Colombian government websites. As the most successful single-day cyberattack against a government, most websites were either defaced or shut down completely for the entire day of the attack. Attacks included both web and network vectors including web application and network Distributed Denial of Service (DDoS) attacks. [5]

- eBay Data Breach: Year 2014 [6]

  eBay went down in a blaze of embarrassment as it suffered this year's biggest hack so far. In May 2014, eBay revealed that hackers had managed to steal personal records of 233 million users.

- Sony Picture Entertainment: Year 2014 [7]

  On 25 November 2014, something new happened in the history of data theft activity. A group calling itself GOP or The Guardians of Peace hacked into Sony Pictures, causing severe damage to the network for days and leaked confidential data. The data included personal information about employees and their families, e-mails and copies of then-unreleased Sony films and other information.

- Dyn Cyber Attack: Year 2016 [9]

  The largest cyberattack in recorded history happened on 21 October 2016, causing temporary shutdown of websites such as Twitter, Netflix, Airbnb, Reddit and SoundCloud. The threefold hack caused mass Internet outage for large parts of the USA and Europe.

These incidents are a few of the numerous cybersecurity breaches and attacks that have occurred over the past few years [8]. The trend indicates that the attacks are more towards personal identities, financial accounts and healthcare information and getting such information on millions or tens of millions of people. Looking at the trend here, these types of cyberattacks are moving down market over time. In simple terms, the techniques that nation states were using few years back are being used by cyber criminals currently [10]. In the real-world scenario, we have to expect that these types of less known attacks will become more public in the near future as exploits and techniques will surge and become available to larger communities. These types of threats may be affecting a small group of organisations at a given time, but progressively they will become more common. Organisations have to be regularly evolving their defences [10].

### 2.3. Web application threat trend

As per Verizon's [12] recently released Data Breach Investigation Report (DBIR) for 2016 which is constructed on real-world investigations and security incidents:

1. When we compared this year's data to last year's data, the total number of attacks this year was significantly higher than last year (see below).
2. Conventional web attacks rose by 200 and 150%, respectively, continuing the trend from last year, with larger numbers and larger volumes of scanning campaigns across the Internet.

3. The volume and persistency of attacks indicate industrialisation of and automation behind organised efforts.

4. Ninety-five per cent of confirmed web app breaches were financially motivated [12].

5. Web application attacks are the #1 source of data breaches [12].

6. Data breaches caused by web application attacks are rapidly rising. The percentage of data breaches that leveraged web application attacks has increased rapidly in the last. This indicates that the web applications in many organisations are not just exposed but are also extremely susceptible compared to other points of attack [12].

**Figure 3** illustrates the occurrence rates of different attack methods that resulted in data loss. The grey bars indicate the corresponding figure for the past year, that is, 2015. It clearly shows that web application attacks accounted for the highest proportion of attacks that resulted in breaches.

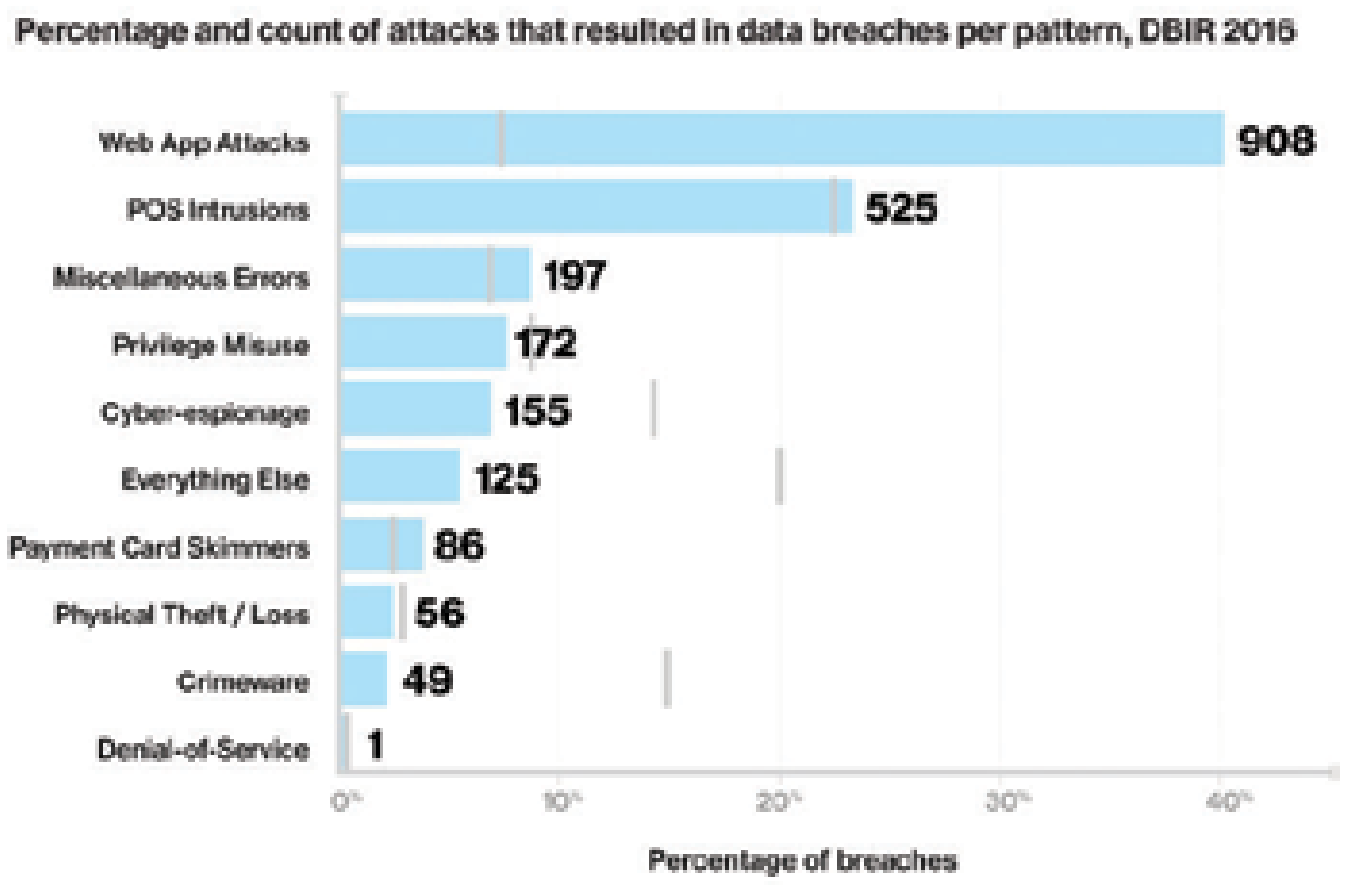

**Figure 3.** DBIR statistics report.

## 3. Web application threats

### 3.1. Web application security: a new boundary break

In the recent era, each and every business has web applications to showcase its online presence, conduct business online, and so on. These applications are hosted on multiple online servers, databases, infrastructures, and so on. And, thus, inherits security risks from the underlying technologies and its associated components. An interesting fact, in 2012 alone, there has been reported around more than 800 hacking events and around 70% of them where via issues in web applications, thus, making the web the new boundary for security, which is not as easy to pull a kill switch like the network [28].

These days we see application development is more focused towards the web, creating applications for every business and personal needs. Looking at an increase in web applications, hackers are more focused to alter their threat attack model targeting these applications instead of a complete protected infrastructure, networks, and so on [29]. Web applications are susceptible to attack from the time they go online. As more inventive attack strategies and structures seem on the Internet, end users and the organisations that provide web services need to shield their systems from being compromised. According to Gartner, around 75% of all external attacks occur at the application level [18]. Web 2.0 helps enterprises in conducting their business; however, an understanding needs to be adhered to that it also introduces a surfeit of damaging risks [18]. The beauty about web application is that in the past the applications were created with scripts as there were no frameworks to support a web developer. But these days, rise in various web development languages, such as Java, NET, WordPress, PHP, Ajax and JQuery, allows a developer to create web application with delivering wide range of functionality in less than no time. Thus come the security issues with the underlying frameworks.

### 3.2. Security risks in a web application

Application security risks are universal and can pose an unswerving threat to business availability. Business world works using web-based applications and web-based software. Because of the propagation of web-based apps, vulnerabilities within the web applications are the new attack path for malicious actors/hackers. An attack of a web-based application possibly will produce information that should not be available, browser spying, identify theft, theft of service or content, damage to corporate image or the application itself and the feared Denial of Service (DoS). The nature of http, hackers find it very easy and lucrative to modify the parameters and execute functionality that was not envisioned to be performed as a function of the application [30, 32, 33].

Businesses and organisations are anticipating large amount of capital in expenditure to safeguard and secure their complete networks (internal/online) and servers. And yet, when it comes to web application security, there is a huge ignorance towards its protection, or, at the very least, considered as an undervalued aspect within the threat model architecture. This notion of thought is considered to be ill-fated, as it has been seen that most security attacks occur online via applications. As per Gartner Group, '75% of cyber-attacks and internet security violations are generated through internet applications'. It is just that organisations are unable to comprehend the security loopholes which exists in web applications [31].

Open Web Application Security Project (OWASP) Top 10 increases cognizance of the challenges organisations face safeguarding web application security in a swiftly fluctuating application security environment. Let us focus on the OWASP Top 10 from 2013 as described below [34].

Injection: Injecting aka inserting code to trick an application in triggering unplanned activities which serve as a deviation from the business functional logic. One of the most preferred injection hacks, which is being actioned by the hackers, is a SQL Injection (SQLi) attack. In SQLi type of attack, malicious actor (hacker) injects a SQL declaration within the application to perform malicious actions like deleting the database, retrieving sensitive database records, and so on.

Broken Authentication and Session Management: Hackers can take over user identities, unauthenticated pages and hide behind a genuine user ID to gain easy admittance to your data and programs.

Cross-Site Scripting (XSS): XSS inserts malicious scripts within the web applications. This malicious script lies within the client side code (browser) and targeted for a different user of the application.

Insecure Direct Object References: Most websites store user records based on a key value that the user controls. When a user inputs their key, the system regains the equivalent information and presents it to the user. An Insecure Direct Object Reference occurs when an authorisation system fails to avert one user from gaining access to another user's information.

Security Misconfiguration: Security misconfiguration is a reference to application security systems that are half-finished or ailing managed. Security misconfiguration can occur at any level and in any part of an application and, thus, is both highly common and effortlessly noticeable.

Sensitive Data Exposure: Inadvertent data leak is a grave problem to everyone using a web application that contains user data.

Missing Functional Level Access: Wrongly configured user access control system can allow users the capacity to achieve functions above their level.

Cross-Site Request Forgery (CSRF): The attack functions on a web application in which the end users' client (browser) has performed an undesired action (user has no knowledge until the task has been performed) in which the very same user is authenticated.

Using Components with Known Vulnerabilities: Open source development practices drive innovation and reduce development costs. However, the 2016 Future of Open Source Survey found that momentous encounters remain in security and management practices. It is critical that organisations gain perceptibility into and control of open source software in their web applications.

Unvalidated Redirects and Forwards: When a web application accepts unverified input that affects URL redirects, malicious actors/hackers can redirect users to malicious websites. In addition, hackers can alter automatic forwarding routines to advance access to sensitive information.

### 3.3. Associated motive in a web application hack

Users' accessing web application(s) are indirectly accessing the critical resources such as the web server and database server (if applicable). Software developers intend to spend vast amount of their project allocated time in developing the functionality and ensuring a timely release thus binding less or no time to security requirements. The reason for this can be due to lack of understanding/implementing security measures/controls in a web application [19]. For whatever reason, applications are often peppered with vulnerabilities that are used by attackers to advance access to either the web server or the database server. Some of the aspects what an attacker seeks for [19]—defacement, redirect the user to a malicious website, inject malicious code, steal user's information, steal bank account details, access unauthorised and restricted content, and so on.

## 4. Machine learning (ML)

### 4.1. What is ML?

An ardent subset of artificial intelligence dedicated to the formal study of learning systems. Machine learning is a methodology of performing data analysis which automates an analytical model [13]. In other words, machine learning is all about learning to do a task better in the future based upon its previous learned patterns in the past [14]. ML being a subsection of artificial intelligence provides systems/computers with the power to learn without being explicitly programmed [14]. One of the reasons why ML is picking up traction in the IT world is because as and when patterns are developed with new data, ML algorithms has the ability to independently adapt and learn from the data and information. With ML, computers are not being programmed but are altering and refining algorithms by themselves [13, 14].

Looking at other definitions, ML discovers the study and construction of algorithms that can learn and make predictions of data. In other words, it focuses on prediction-making through the use of computers [13–15].

With the rise in new-generation-technologies being witnessed in the twenty-first century, ML today is something that cannot be compared to what it used to be in the historical past. The past has been witnessed in the rise of various ML algorithms and the complexity of the calculations being carried out; however, it is just during the recent times, the recent ML algorithms have been tuned in such fashion that the whole complex mathematical calculations, analysing big data at a much greater and faster—a very recent development [16, 17]. Some of the underneath examples (but not limited to) have adapted ML within their service space:

- Google's Self-Driving Cars: ML algorithms are used to create models in classifying various types of objects in different situations [18].
- Netflix: ML is used in improving the member experience [19].
- Twitter: ML is being applied to enhance its video strength [20].

### 4.2. Rise in ML

An immense amount of popularity is being gained over Data Mining and Bayesian analysis due to a fast-pace adaptation of ML in solving business problems. Computational power, availability and various type of data, cheap and powerful data storage is ever growing which is some of the few attractive factors towards adapting ML. What this mean is that it is quickly possible to fire up an automated predictive model which can analyse larger and complex datasets and deliver accurate final outcomes [25]. This results in an additional value towards predictions which leads in creating smarter real-time decisions without human intervention [25]. Within the software vertical, artificial intelligence is being a popular technology to integrate within a service as the mandate for analytics is motivated more by growth in type of both structured and unstructured data [21].

In the 1930s and 1940s, the pioneers of computing, such as Alan Turing, began framing and playing with the most basic aspects of ML such as a neural network which has made today's ML probable [27].

As per [25], humans create couple of models every week; with ML, thousands of models are created within a week. Upsurge in computing power is one of the prime reason from a transformational shift from theoretical to practical implementation. High number of researchers and industry expertise are contributing towards the advancements in this space as it is constantly being used in solving some real issues across industries including (but not limited to) healthcare, automotive, financial service, cloud, oil and gas, governments, and so on. Data (be it small or large) residing within these types of industries contain a large number of patterns and insights. ML creates the ability to discover various patterns and trends within these giving rises to substantial results.

The rise of cloud computing, massive data storage, devices connecting with each other (Internet of Things [IoT]), and mobile devices play a huge role in the adaption of ML.

### 4.3. ML methodologies

Supervised Learning: Algorithms are trained on labelled data, essentially leading to its meanings where an input having a looked-for output. In other words, a supervised learning algorithm with an input variable denoted as P and an output variable denoted as Q and algorithms are used to create and learn a mapping function (f) via the input to the output.

$$Q = f(P)$$

The goal of a supervised learning algorithm is to achieve an estimate mapping function so that for every new input (P), a new predicted output (Q) is created. In other words, the learning algorithm receives a set of inputs with their corresponding outputs, and the algorithm learns by equating its concrete output with correct outputs in order to find errors and have the learning model modified accordingly. Supervised learning algorithms make use of patterns to predict the values of the label on unlabelled data. This is achieved by classification, regression, prediction, and so on [25, 22].

Supervised learning is used to predict probable future events within applications having vast amount of historical data [25]. An example is detecting likely fraud patterns in credit card transactions.

Unsupervised Learning: Unsupervised learning is where only an input data (P) is available with no equivalent output variables. The aim of unsupervised learning is to model the construction of the data in order to learn more about the data. Algorithms are required to discover a structure, an inference and meaning within the data in order to arrive to a conclusion. These algorithms do not have any type of historical data in order to predict the output unlike supervised algorithms [25]. Unsupervised learning does not have any explicit outputs and nor exists a dependency environment factor within the input variables; it brings to accept preceding predispositions as to what aspects of the structure of the input should be seized in the output [23]. In an aspect, unsupervised learning locates patterns in the data which succours in arriving to a constructive meaningful decision.

### 4.4. Adaption of ML in industry

ML has been widely adopted across various sectors within the industry to solve real-life business statements. Data is available within the whole global space and to derive a deep understanding from it, ML is the methodology to be consumed for such derivations. We live in the golden era of innovative technologies and ML is one of them [24]. ML has created an ability to solve problem declarations horizontally and vertically across aviation, oil and gas, finance, sales, legal, customer service, contracts, security, and so on, due to its greatest capability of learning and improving. ML algorithms has been a great stimulus in creating applications and frameworks to analyse data which brings in a great predictive accuracy and value to enterprise's data, leading to a sundry company-wide strategy ensuing faster and stimulating more profit [25].

One such example is of the revenue teams across the industry, they are converting the practical aspect of ML in augmenting promotions, compensations and rebates driving the looked-for behaviour across various selling streams. **Figure 4** draws a mind of ML applications within some of the industries [25].

ML has been a chosen integration within the industries for its skill to constantly learn and improve. As we have seen, ML algorithms are very iterative in nature, having the flexibility to make it learn towards a vision of achieving an optimised and a useful outcome [25]. ML's data-driven acumen is infusing every corner of every industry and it's starting to disrupt the way business is done worldwide. Leveraging ML has enabled processes to be re-calibrated inevitably and improved for reduced cycle times, created a higher quality of delivered goods and allowed for new products to be established and tested. The ability to influence data for more accurate decision-making in place of instinctive feel [26]. According to a representative from Gartner, as quoted, 'Ten years ago, we struggled to find 10 machine learning – based business applications. Now we struggle to find 10 that don't use it' [26].

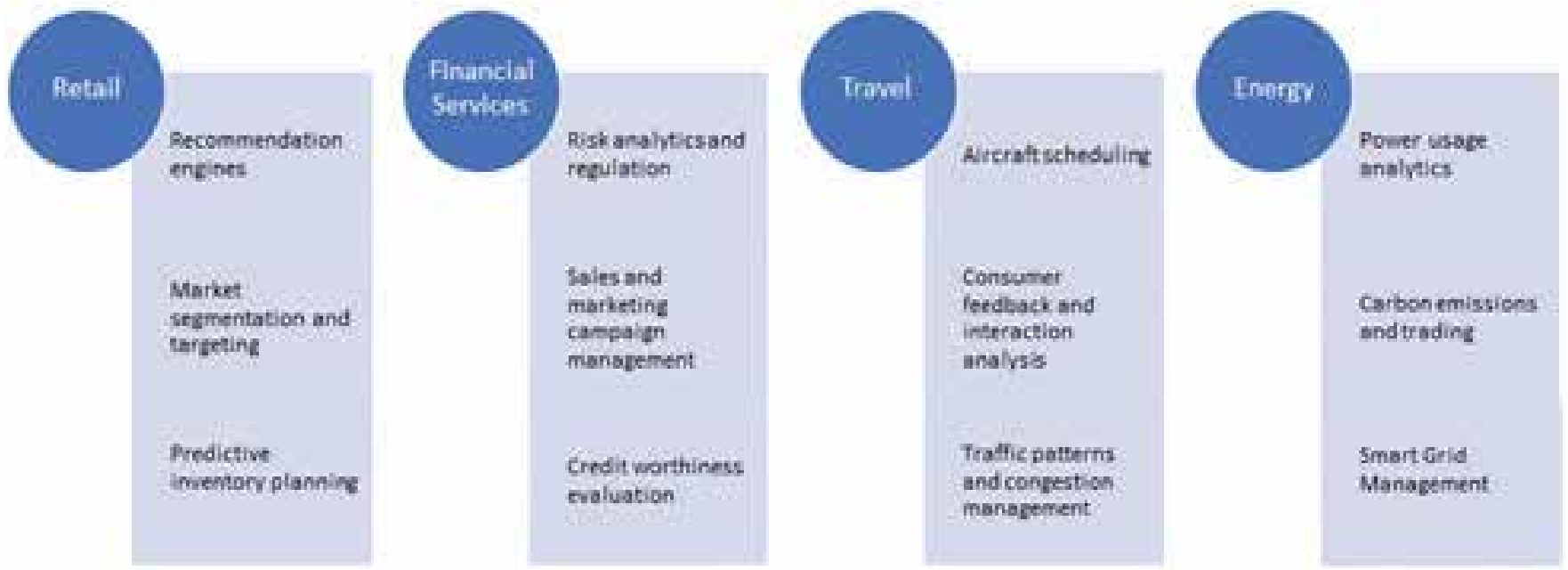

**Figure 4.** ML applications in industries.

ML's rise in the industry makes data a growing vital part of how a business makes decisions. Because of this, data scientists will take up a complete central focused role in organisational strategies as data is becoming a core agent of change within a business. It is forecasted that with a wealth of data in business given the occurrence of sensors and IoT implementation, the wide ability to influence data will be critical to building competitive advantage [26].

ML should not be understood as a technology component, and, by the rise within the current era, it confidently is not a short-term trend. With its impact within the industries and various business sectors by bringing out toppling business models and with its rising maturity in terms of sophisticated algorithms being advanced, it will continue to be the solitary driver in shifting the complete viewpoint in decision making and having a truly workable conclusion [26].

### 4.5. Machine learning usage in cybersecurity

Machine learning (ML) is not something new that security domain has to adapt or utilise. It has been used and is being used in various areas of cybersecurity. Different machine learning methods have been successfully deployed to address wide-ranging problems in computer security. Following sections highlight some applications of machine learning in cybersecurity such as spam detection, network intrusion detection systems and malware detection [39].

#### *4.5.1. Spam detection*

Traditional approach of detecting spam is usage of rules also known as knowledge engineering [39]. In this method, mails are categorised as spam- or genuine-based set of rules that are created manually either by the user. For example, a set of rules can be:

- If the subject line of an email contains words 'lottery', its spam.
- Any email from a certain address or from a pattern of addresses is spam.

However, this approach is not completely effective, as a manual rule doesn't scale because of active spammers evading any manual rules.

Using machine learning approach, there is no need specifying rules explicitly; instead, a decent amount of data pre-classified as spam and not spam is being used. Once a machine learning model with good generalisation capabilities is learned, it can handle previously unseen spam emails and take decisions accordingly [40].

#### *4.5.2. Network intrusion detection*

Network intrusion detection (NID) systems are used to identify malicious network activity leading to confidentiality, integrity or availability violation of the systems in a network. Many intrusion detection systems are specifically based on machine learning techniques due to their adaptability to new and unknown attacks [39].

#### *4.5.3. Malware detection*

Over the last few years, traditional anti-malware companies have stiff competition from new generation of endpoint security vendors that major on machine learning as a method of threat detection [41]. Using machine learning, machines are taught how to detect threats, and, with this knowledge, the machine can detect new threats that have never been seen before. This is a huge advantage over signature-based detection which relies on recognising malware that it has already seen.

#### *4.5.4. Machine learning and security information and event management (SIEM) solution*

Security information and event management (SIEM) solutions have started leveraging machine learning into its latest versions, to make it quicker and easier to maximise the value machine data can deliver to organisations [42]. Certain vendors are enabling companies to use predictive analytics to help improve IT, security and business operations.

## 5. Uniting machine learning and application security

In the last few sections, we have seen that the web application attacks are constantly evolving, and building protection mechanism on the fly has been a complex task. So, with all of the recent threats and attack trends on web application, one may ask what exactly is machine learning and how is it applied in these situations.

Inferring from a much wider scope and having it elucidated, machine learning imparts the understanding as a line of drills where the algorithm would 'train' a machine in cracking a problem. In order to understand the above statement, we need to comprehend it via an example, let us imagine a task to determine if the animal in the photo is a lion or an elephant. Prior to coming out with this conclusion, it is imperative to train the machine by providing 'n' photos of elephants and 'm' photos of lion. As the machine trains, a picture can be supplied and the output will be predicted if the supplied picture is a lion or an elephant.

The effectiveness of a machine learning model is determined in the accuracy of its predictions; in other words, a predictive analytical model needs to be derived. In order to explain this, let us now provide the model with around 10 pictures of elephants and the output imparts eight being elephants and two depicted as lions. In this case, we derive the model to be 80% precise. Looking at this being on the brink of accuracy, there is a way to improve the model. And the improvement will be by providing more data; in other words, deliver knowledges to improve its proficiencies meaning to provide a large number of photos to train the machine as increase in data volume rises large developments aiming at an acceptable accuracy of the model. An implausible frequency of growth of web applications over the years produces large sum of logs which leads to a methodology in improving the precision over a period of time.

Let's explain the above perspective in web application scenario. Any three-tier web application consists of web traffic logs, application logs (normally terms as business layer) and the database logs (normally termed as data access layer). When we look at the logs, let's say we look at one category, that is, login attempts on the application. The output of the login can be either a successful attempt or a failure attempt. Compared to our example of elephants and lions, to train the failure or successful attempts we provided it with 100 logs of successful attempts and 100 logs of failed attempts. Once the model or the machine is trained, we can provide a log and it can tell me if it is a failed attempt or a successful attempt.

Now for predictive analysis, if we provide the model with 10 web logs of successful login attempts, out of that it says that seven are successful attempt and three are failed attempts, we can say that the model is 70% accurate. One way to improve a machine learning system is to provide more data, essentially provide broader experiences to improve its capabilities and with the application logs this is not a challenge. Any application which is accessed my thousands of users can generate huge number of logs on daily basis, thus increasing the accuracy of the machine learning model or algorithm.

### 5.1. ML detecting application security breaches

Researchers are constantly working on implementing ML techniques in detecting various application security level hacks. But the authors have proposed an extraction algorithm, which is based upon various ML algorithms. The authors [36] adapted various ML algorithms, such as SVM, NB and J48, to develop the vulnerability prediction model. They have emphasised on vulnerability prediction prior releasing an application. In an environment where time and resource are very minimal, web application security personnel require an upper-level support in identifying vulnerable code. A complete practical methodology in bringing out predicted vulnerable code will surely assist them in prioritising the secure code vulnerabilities.

Inferring from this thought process, authors [37] have worked towards bringing out a substantial pattern that illustrates both input validation and sanitisation code which are expected to be the predicted vectors of web application vulnerabilities. They have applied both supervised and semi-supervised learning when building vulnerability predictors based on hybrid code attributes. Security researchers are utilizing ML towards web application vulnerability detection. SQL Injection being one of the most preferred attack vector of hackers, authors [38] have displayed their work by coming with a classifier for detection of SQL Injection attacks. The classifier implements Naïve Bayes ML algorithm in conjunction with application security principle of Role-Based Access Control implementation for detection of such attacks.

### 5.2. Anomaly detection and predictive analysis

Anomaly detection is the documentation of items, events or observations which do not conform to an expected pattern or other items within a dataset. Anomalies are also termed as outliers. These outliers will detect an issue which is not normal compared to its learned model. Industries are adapting anomaly detection techniques in identifying medical problems, financial frauds, and so on.

Anomaly detection is not limited just to security but it is being utilised in various other domains such as financial fraud uncovering, fault detection systems for structural defects, event detection in sensor networks used in petroleum industries and many other. It is used in preprocessing the data, to eliminate any abnormal data from the dataset. By eliminating the abnormal data in supervised learning results in a statistically significant increase in accurateness.

Looking at the vast amount of cyberattacks increasing on web applications, the authors of Azane were inspired by the complete study of anomalies and patterns which led them to present a research towards the implementation of an ML engine comprising of an anomaly detection and predictive analysis framework at an application level to detect certain user behaviour in order to predict if it is a normal usage or an attack. The authors have explained a prototype model that will describe the Azane which is a machine learning framework [35] for web applications. Azane as a proof-of-concept algorithm designed by the authors has played a major role at the applications log level to detect anomalies at the application workflow level and also serve as a prediction base for any future events. The workflow in **Figure 5** comprises multiple stages: Application Logs → Pre-processing → Training Data → ML Algorithm → Test Phase → Predictive Model Output.

Let us understand each phase in general:

**1.** Logs

This is the first and the foremost phase upon which the whole model depends. We need to understand that in order for the model to work, logs are necessary. The authors have taken the dual aspects of logging into consideration and have applied their algorithms in order to derive a meaningful context. This phase is more about collection of logs and verifying whether the log contains the parameters that are required for analytical purpose.

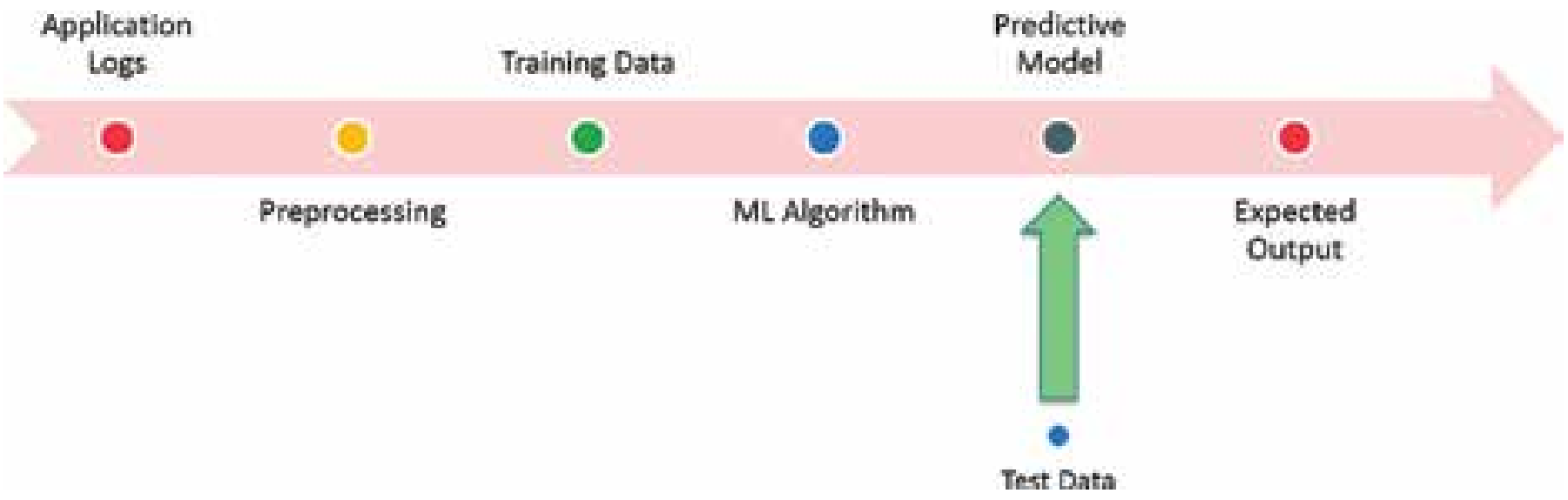

**Figure 5.** Anomaly detection workflow.

**2.** Pre-processing

This phase is more of transforming a given dataset into a format in which the ML algorithm can deduce and learn from it. This phase emphasises on making your data compatible with the machine learning algorithms. A challenge which can occur is that various algorithms make different assumptions about the data which may require to conduct individual analysis to see which algorithm is more suitable to the business needs. Further, when you follow all of the rules and prepare your data, sometimes algorithms can deliver better results without the pre-processing.

Pre-processing in general includes the following steps:

**a.** Load the data

**b.** Split the dataset into the input and output variables for machine learning.

**c.** Apply a pre-processing transform to the input variables.

**d.** Summarise the data to show the change.

**3.** Training data

Training data phase is more of a data that is an outcome of pre-processing and will be used to train the algorithm. This data plays a vital role to feed in the ML algorithm as it has the right amount of input and output data. Training data is more about making the machine learning algorithms aware about the data attributes and their values.

**4.** Machine learning algorithm

This phase is used to identify the algorithm which suits your dataset and final outcome.

**5.** Test phase (Learning: Predictive Model)

In simpler terms, training data is bought in to create a learning set which will service as a predictive model against the selected algorithm to validate the prediction or the accuracy of the model. A training set is learnt and this particular set of learnt data is used to discover potentially predictive relationships. This whole analysis is based on the training data which forms the baseline of the predictive analysis model.

**6.** Predictive model output

Now the final step is to test the predictive model with the accuracy with the new data known as the test data. A test set is a set of data used to assess the strength and utility of a predictive relationship.

Azane was developed to unite machine learning and application security in order to protect web applications from sophisticated type of attacks by predicting the application's user pattern.

ML algorithms yield a near-to-accurate result when huge amount of data is fed and trained in order to aid in spotting malicious patterns. Data should be consistent for the ML algorithm to work to its fullest. Combining ML output with other infrastructure devices, such as IPS and firewall, will strengthen the correlation and assist in drilling down to its correctness and validation of a web application attack. Finally, once the patterns are identified and analysed and blocked, this can be integrated to a SIEM solution for a complete centralised management metrics reporting.

In this chapter, we have seen how machine learning can be integrated within application security in order to prevent attacks on web-driven applications.

## 6. Conclusion

Our daily life, economic growth and a country's security is highly dependable on a safe and secure cyber universe. Hackers are always on a lookout in breaching the cyber universe in identifying vulnerable loopholes to steal information, data, money, having services disrupted, and so on. The inventiveness of hackers has led to the advance of new attack vectors and new ways of exploiting bugs in a web application. Web application breaches have evolved the cyber war between application owners and hackers. As companies cope with a more urbane threat landscape, they will have no choice but to innovate, automate and predict hacking identifications, attempts and breaches in their web applications.

Talking about prediction, machine learning as a technology has erupted vastly in the whole cyber implementation space. These decision-making algorithms are known to solve several problems as seen in the above illustrations. Following a simple principle of prediction, machine learning has shown itself as the problem solver for any given type of problem occurring within the complete technology space. Looking at the in-depth capability of machine learning, the cybersecurity industry started its adaptation. The collection and storage of large amount of data points is rapidly rising in cybersecurity where machine learning plays a huge role in analyzing different use case patterns. Another facet where machine learning is being utilized is in identifying and defending against vulnerabilities in the complete cyber eco-system and web application being a part of it.

Integrating machine learning in web applications are proving to serve as an identification and prevention against web hacking breaches by analysing the usage patterns of the web application. As seen within the above sections, machine learning has been a success in identifying various attacks, and research works have been carried out. Future of web application security lies in the hands of machine learning as we are stepping in the space of large data residing in web applications, logs being written every millisecond and attacks being witnessed at large.

## Conflict of interest

All work presented in this chapter is our own research/views and not those of our present/past organizations and institutions. It does not represent the thoughts, intentions, plans or strategies of our present/past organizations and institutions.

## Author details

Nilaykumar Kiran Sangani[1]* and Haroot Zarger[2]

*Address all correspondence to: sanganinilay@hotmail.com

1 BITS Pilani-Dubai Campus, Dubai, United Arab Emirates

2 Abu Dhabi Company for Onshore Petroleum Operations Ltd., Abu Dhabi, United Arab Emirates

# Advanced Access Control to Information Systems: Requirements, Compliance and Future Directives

Faouzi Jaidi

Additional information is available at the end of the chapter

**Abstract**

The swift cadence of Information and Communication Technologies (ICT) is at the origin of a new generation of open, ubiquitous, large-scale, complex, and heterogeneous information systems (IS). Inextricably linked with this evolution, a number of technical, administrative, and social challenges should be urgently addressed. Security and privacy in critical IS are recognized as crucial issues. The access control is well adopted as a typical solution for securing sensitive resources and ensuring authorized interactions within IS. The chapter deals mainly with the thematic of advanced access control to IS and particularly to relational databases. We present a synthesis of the state of the art of access control that encloses a study of research advancements and challenges. We introduce and discuss requirements and main characteristics for deploying advanced access control infrastructures. Then, we discuss the problem of the conformity of concrete access control infrastructures, and we propose a conformity management scheme for monitoring the compliance between low-level and high-level policies. Finally, we provide and discuss proposals and directives to enhance provably secure and compliant access control schemes as a main characteristic of future IS.

## 1. Introduction

Nowadays, Information and Communication Technologies (ICTs) developments bring out significant security concerns related to the deployment and operation of IS. In fact, in today's IS infrastructures characterized by their criticality, openness, complexity and heterogeneity (such as e-commerce, e-government, and e-health care), security and privacy are recognized as crucial issues. Ensuring a high level of security with a minimal overhead is the main goal of research activities. Among several security mechanisms, it is commonly agreed that the access control is a strong driving force and is well adopted as a typical solution for ensuring a high level of protection of critical infrastructures. This mechanism is fundamental to ensure higher confidentiality and integrity of sensitive data and services within IS. It helps in a structured manner—*generally enforced according to an access control policy with reference to a security policy*—to define and organize accesses and interactions within a specific system. The standing of the access control in the protection of critical resources has been well studied in literature. Several mechanisms, approaches, and models have been proposed to structure, specify, and enforce access rights. As a part of this chapter, we review in an exhaustive manner and discuss access control advancements. We highlight the evolution of access control infrastructures from traditional solutions to fine-grained solutions.

Despite the great advancements, several requirements and concerns need to be addressed for defining and setting up efficient and reliable access control infrastructures. Indeed, specifying and enforcing a trustworthiness access control infrastructure, ensuring its coherence, and monitoring its conformity have now become complicated and even puzzling activities. Moreover, it is commonly agreed that effective and proficient administration and management of access control infrastructures are recognized as main issues, while mastering these tasks is crucial as it would help to guarantee a higher security of IS. We study and discuss basic requirements for deploying advanced access control infrastructures. We discuss the problem of the conformity of concrete access control infrastructures, and we propose future directives to enhance provably secure and compliant access control schemes as a main characteristic of future IS.

The remainder of this chapter is structured as follows. In Section 2, we introduce and review access control advancements. In Section 3, we study and discuss main access control challenges for IS. In Section 4, we focus on advanced access control to IS. We study the main requirements for deploying advanced solutions, and we propose a compliance management solution. We present future directives to enhance provably secure and compliant access control schemes. Finally, Section 5 concludes the chapter.

## 2. Access control advancements: from traditional approaches to fine-grained access control

### 2.1. Introduction to access control

The access control is defined as any physical/logical mechanism by which a system controls and manages the access and the use of its resources. This mechanism allows the system to grant or revoke privileges for active entities (subjects) to access to or to perform some actions on passive entities (objects). The mechanism is also identified as authorization service or reference monitor that generally enforces access control policies.

An access control policy—*in general defined in the context of a security policy*—corresponds to the sets of rules and practices that regulate within a specific system how different resources (data and services) are operated, managed, and distributed. A security policy has to identify for a specific system the security objectives and the associated threats [1]. The policy acts mainly on three levels or aspects: administrative, physical, and logical. In the administrative level, we focus on the organizational security and the corresponding administrative procedures within the organization. In the physical level, we need to define the necessary procedures and means to protect the set of resources from physical risks and accesses. Finally, in the logical level, we define legitimate and authorized actions a user can perform. This level contains a set of security mechanisms such as the identification, authentication, and access control.

### 2.2. The generic model of access control

The generic model to control access to resources in the context of databases (as an example) is defined according to **Figure 1**. (**1**) A subject (an active entity) requests access to a database object (a passive entity) to perform some actions. (**2**) The authorization service checks the set of rights granted to the subject by the defined access control policy. (**3**) Then, an access decision (grant or revoke) is accorded to the subject.

### 2.3. Access control models

The emphasis on access control for ensuring high protection of critical infrastructures has been extensively justified in literature. The three main reference models have been defined: discretionary, mandatory, and role-based access control (RBAC) models. The wide deployment and great success of the standard role-based access control model have initiated several research works leading to the definition of advanced models for a fine-grained access control.

#### *2.3.1. The discretionary access control (DAC) model*

The discretionary access control (DAC) model has appeared mainly with Lampson [2] in the 1970s who defined the structure of the access control matrix based on the subject, object, and action notions. The model allows to restrict the access to objects on the basis of the identity of the subjects and/or the groups of subjects. In the DAC model, the owner of a specific resource fixes itself the access rights to the resource for all users of the system. Moreover, a subject who has an access authorization is able to pass this permission (perhaps indirectly) to other subjects. The controls are discretionary in the sense that the management of access rights is left to the discretion of the users.

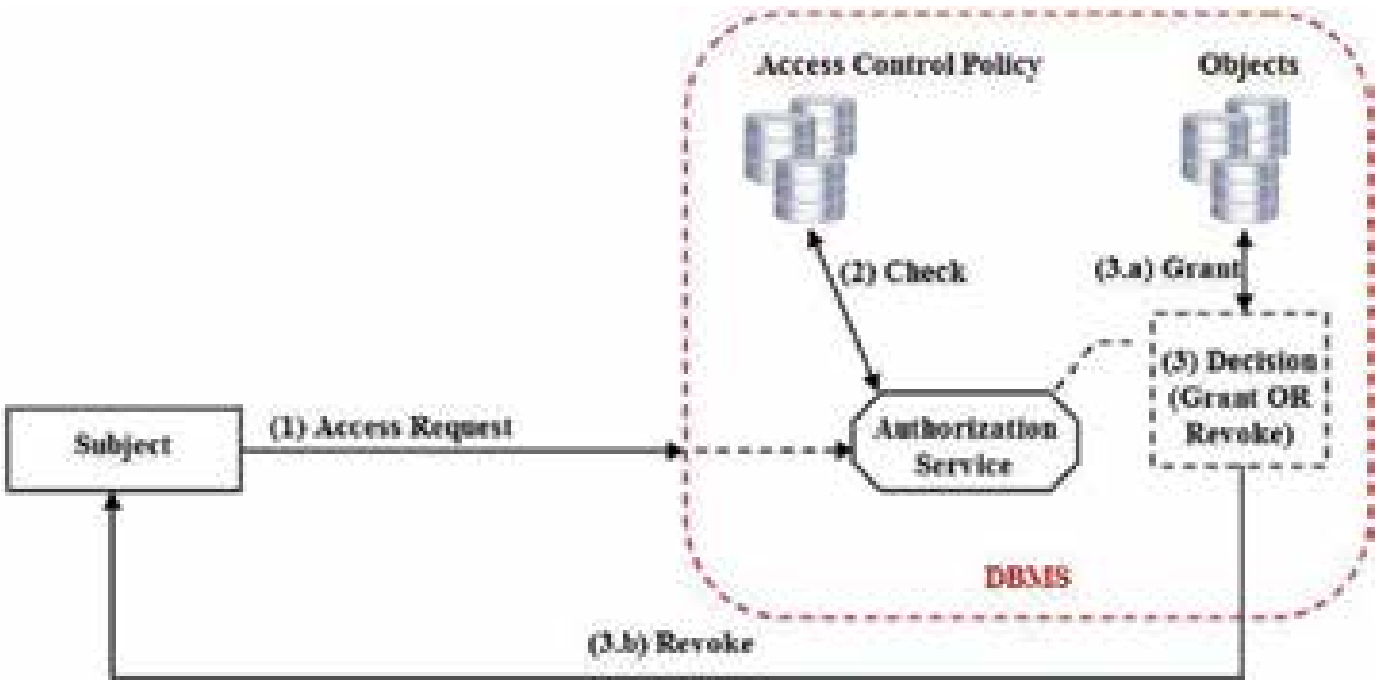

**Figure 1.** The generic model of access control.

*2.3.2. The mandatory access control (MAC) model*

The mandatory access control (MAC) [3, 4] model controls access to resources on the basis of the notion of security level. It associates a confidentiality level to each subject and object. The set of subjects and objects is classified according to corresponding confidentiality levels, and authorized actions are derived based on associated levels. In this model, the system manages directly user rights based on the corresponding security information (confidentiality levels) assigned to users and objects. This model was motivated by the need of providing higher security by preventing unauthorized accesses and protecting resources from Trojan horses.

*2.3.3. The role-based access control (RBAC) model*

The role-based access control (RBAC) model [5, 6] controls access to resources on the basis of the concept of roles assigned to users. A role represents a function within an organization. The model defines access rights to resources based on the roles assigned to users. A variety of RBAC conceptual models had been defined. The core model (noted $RBAC_0$) defines, for any system, the minimum requirements to support the access control based on roles. The *hierarchical RBAC* (noted $RBAC_1$) extends $\mathrm{RBAC}_0$ with the concept of role hierarchy. The *constrained RBAC* (noted $RBAC_2$) extends $\mathrm{RBAC}_0$ with the notion of constraints. Constraints are a set of restrictions defined on RBAC components, such as static separation of duty, dynamic separation of duty, prerequisite, and cardinality constraints. The global model called *consolidated model* (noted $RBAC_3$) contains both of the hierarchy and the constraint concepts. It includes $\mathrm{RBAC}_1$ and $\mathrm{RBAC}_2$ models and consequently the $\mathrm{RBAC}_0$model.

*2.3.4. Fine-grained access control (X-BAC) models*

To structure access rights and to meet specific application needs, numerous extensions have been proposed for the RBAC family, called X-based access control models. The ORBAC [7] model represents a main extension defined as a conceptual and industrial framework to meet security needs for sensitive healthcare communications. RBAC+ is a dynamic model to enforce fine-grained access control to web databases. The model extends the standard model with the notions of application, application profile, and sub-application session. The GEO-RBAC [8] model has been proposed to take into account spatial contextual information. It is motivated by security requirements of location-based services and mobile applications as well as the increased concern for the management and sharing of geographical information in strategic applications like environmental protection and homeland security [8]. The temporal RBAC and the generalized temporal-RBAC extensions [9] are defined to take into account temporal contextual information and to constrain the use of permissions to specific temporal periods. Several other extensions that define the concepts of teams (Team-BAC), tasks (task-LAC), lattice (Lattice-LAC), organizations (Organization-BAC), or contexts (Context-LAC) have been also defined to structure rights. The concept of attribute-based access control (ABAC) organizes access based on the evaluation of attributes. RBAC extensions are defined for a fine-grained access control policy specification to meet new security challenges.

Content-based access control is a particular case of fine-grained access control specific to database systems. In this type of access control, access decisions (to authorize or to deny an access request) are based mainly on the content of the data to be accessed.

### 2.4. Synthesis

The main discretionary access control model expressed through access matrices has been widely used and implemented in several applications in the commercial and academic fields, such as in Unix/Linux operating systems. Nonetheless, the management of permissions list is tedious and prone to errors in the case of a large number of users and permissions. Moreover, this model is difficult to administer in the case of large infrastructures and is vulnerable to Trojans.

In order to improve and strengthen access control to critical resources, several research works gave rise to other models, mainly the MAC model based on the definition and exploitation of security levels. In this context, the early defined approaches have proposed to organize the set of security levels in a strict order. Then, the obligation to refine and relax this constraint made it possible to organize the security levels according to lattices. The notion of lattice allows structuring the security levels according to a hierarchy that gives more freedom to administrators in the modeling of access control policies. This makes the MAC models well adapted to highly structured IS that require a high confidentiality, namely, military systems or sensitive enterprise systems. However, in the case of large-scale organizations, it is too rigid to apply the MAC model since it is difficult to classify a huge number of objects and subjects in a predefined number of security levels. Moreover, database management systems (DBMS) that adopt only the MAC model as a unique access control solution had a little commercial success due to their rigidity and the strict hierarchy they impose on users and objects.

Even though historical access control models have introduced the concepts and generic principles of access control, the concern to impose strict access control rules has led to the definition of the notion of structured intermediate levels between subjects and permissions. The general principle of the new models is to introduce a new level of indirection (defined by roles) between users and permissions. The main purpose of this new concept is to facilitate the administration of access control policies. Indeed, in role-based access control models, it is not necessary to update the whole access control policy when adding a new subject (user), and it is sufficient to assign it rights through one or more roles. Thus, the use of this intermediate entity (role) reduces considerably administrative errors and contributes to master difficulties and complexities in the management of access rights through the mechanism of inheritance between roles. This model of access control has received a particular attention by the research community and has become in its simplest or most complex form the most used model [10]. This huge success has made from this model a standard for access control [11]. This led to the elaboration of multiple models and derivations for this family of access control named X-BAC for the definition of a fine-grained access control.

In order to highlight the main advancements of access control, we present in **Table 1** a comparative study and analysis of the discussed access control models and mechanisms with reference to their approaches, applications, and capabilities.

From a security perspective, we present in **Table 2**, a summary of security analysis of the discussed access control models and mechanisms that identifies the relative strengths and weaknesses of existing approaches and their security vulnerabilities.

| **Features** | **DAC** | **MAC** | **RBAC** | **Fine-grained models (exp. ABAC)** |
|---|---|---|---|---|
| Context of application | Commercial and academic fields: Unix/Linux operating systems | Suitable to highly structured IS: military systems and intelligent environments | Various IS | Specific to particular applications |
| Implementation | Access control matrix\access control and capability lists | Security levels for subjects and objects | Roles and authorization constraints | Specific mechanisms (exp. ABAC: attributes) |
| Sensitivity | No fine-grained access control | No fine-grained access control | No fine-grained access control | Fine-grained access control specific to particular applications |
| Policy updates | The policy update is costly | The policy update is costly | The policy update is simple | The policy update is simple |
| Policy management and administration | Tedious management, prone to errors, and difficult to administer | Requires a higher management to account and update security levels | Easier than previous models | Easier than previous models |
| Advantages | Flexibility of usage<br>Enforces the sharing of information | Multilevel security<br>Ensures higher integrity<br>Well-adapted to highly structured IS | Includes the advantages of historical models<br>Intermediate levels between subjects and permissions<br>Hierarchy of roles and constraints<br>Adapted to complex and distributed areas | Fine-grained controls<br>Respond to specific access control needs<br>Solve RBAC limits<br>Focus on other concepts than roles (exp. attributes in ABAC)<br>Higher flexibility<br>Adapted to complex, distributed, open, and dynamic areas |
| Disadvantages | Problem of scalability<br>No distinction between users and subjects<br>Security problems | Too rigid to apply<br>Problem of scalability | Not suitable to dynamic environments<br>Static access control<br>Requires roles engineering<br>Does not support contextual rules | Difficulty in compliance management<br>More complex to implement than RBAC |

**Table 1.** Comparative study of access control models.

| Features | DAC | MAC | RBAC | Fine-grained models (exp. ABAC) |
|---|---|---|---|---|
| Performance and integrity | Low: possibility to settle insecure rights | High: based on security level | High | Very high |
| Access decision | Ownership | Centralized | Centralized | Centralized |
| Vulnerability | Vulnerable to malicious programs such as Trojan horses and covert channels | Vulnerable to covert channels | Vulnerable to inner threats, particularly administrative threats | Vulnerable to inner threats, particularly administrative threats |
| Flexibility | Flexible | Rigid | Flexible | Higher flexibility |
| Security separation of duties | Does not support | Does not support | Static and dynamic separation of duties | Static and dynamic separation of duties |
| Constraints and conditions | Does not support | Does not support | Constraints and condition enforcement | Constraints and Conditions enforcement |
| Inference (indirect access) | Fail to deal with inferences | Fail to deal with inferences | Fail to deal with inferences | Requires specific study to each model |
| Transitivity | No control on transitive access flow | Transitivity is controlled | Transitivity is controlled | Transitivity is controlled |
| Least privilege and delegation of rights | Supports | Does not support | Supports | Supports |

**Table 2.** Security analysis of access control models.

## 3. Access control to databases

### 3.1. Mechanisms

In a database context, a number of mechanisms can be enforced in a cooperative manner for ensuring the control of legitimate accesses and preventing unauthorized accesses. The diversity of access control mechanisms for database systems illustrates on the one hand the importance of the access control for protecting sensitive data and services and on the other hand the difficulty and the complexity of defining a reliable access control solution. We present in the following a list of the principal access control mechanisms for database systems.

- *Passwords*: a database management system allows to associate passwords for the identification of users and to enforce passwords for the activation of roles.
- *Privileges*: a database management system allows defining a set of privileges for managing the empowerment of users. It provides system privileges and object privileges that allow users performing specific actions across the system and accessing database objects.
- *Views*: a view represents an important and very useful mechanism for restricting access to data. It is a most common mechanism adopted by database management systems to support content-based access control.

- *Triggers*: triggers allow to automatically enforce access restrictions as well as security rules. They especially allow enforcing authorization constraints mainly pre- and post-authorization constraints.
- *Stored* procedures: stored procedures may be used in order to define privileges associated with a user's job functions and to ensure that access to data and services are performed according to the defined rules.
- *Encryption mechanisms*: a number of encryption mechanisms contribute to access control for databases. They concern several applications, such as password encryption, data encryption, digital signature, authorization tickets, etc.
- *Access control policies*: it consists on enforcing access control policies based on different models such as DAC, MAC, RBAC, etc. Indeed, database management systems provide mechanisms based on different access control models allowing the management of access rights.
- *Specific mechanisms*: this type of mechanisms is context dependent and specific to every DBMS. As an example, we cite the component Oracle Database Vault that allows restricting access to sensitive data even for database administrators.

### 3.2. Policy enforcement

In a database context, enforcing an access control policy consists in deploying within a database system, generally in a distributed manner, the schema of the access control policy. This distribution of the policy between different active components of the system ensures a better management of access requests and operations of the database resources. In fact, an access control policy is often spread over several levels. (i) The first level is defined by the database management system itself. Indeed, a DBMS makes it possible to define and store in its depository (data dictionary) a set of information and access rules allowing it to control and manage access to data. (ii) The second level is defined by application servers that allow restricting access to applications and data. (iii) The third level is relative to the application part. Indeed, the software application can manage itself the level of accreditation of users, and it connects to the database under its own logical identity and decides which information the users can consult, modify, etc. (iv) The fourth level belongs to the set of privileges associated to different actors of the system and defined in the directories of users and operating systems of the IS.

### 3.3. Challenges

Even though, the access control is becoming increasingly important for protecting sensitive data in critical systems; several issues are recognized as crucial challenges in today's access control infrastructures. In fact, the efficient and secure administration and management of access control infrastructures, the safety analysis of access control models, and the risk assessment in access control systems are recognized as fundamental issues. Moreover, setting up a trustworthiness environment of access control and monitoring its compliance and coherence have emerged as complicated and confusing tasks. Indeed, in unreliable and untrustworthy environment, the administration of access control policies considered as a fundamental security aspect generally raises a critical analysis problem when the process of administration is distributed and/or potentially untrusted users contribute to this process. As a consequence, collusion attempts and inner threats may take place to generate crucial and invisible breaches to circumvent the access control policy. Moreover, an administrator via its administrative privileges has power increasingly disputed when the safety of data is threatened [12]. Given that administrative roles are naturally powerful, if they are not used in cautious manner, a malicious administrator or a powerful user can corrupt the policy and create other breaches difficult to identify.

In database context, most of existing DBMS use the *closed-world policy* as a main authorization model. Under this circumstance, whenever a user tries to access a database object and no authorization rule is found, the access is denied. Hence, the lack of authorizations is interpreted as no authorization. Nonetheless, this policy has a major drawback since it does not prevent the user from receiving the required authorization some times in the future [13]. Moreover, most of existing DBMS act as a *black box*, and it is difficult for administrators and security architects to identify the actual state and the compliance level of the concrete policy enforced by the DBMS. Recently, researchers are convinced of the urgency of this topic given the challenges of securing data. A few attempts addressed the topic of reverse engineering of access control policies to externalize the low-level schema of an access control policy enforced by a relational DBMS [14, 15].

## 4. Advanced access control

Advanced access control solutions should provide an efficient and flexible access control with a reasonable (minimal) overhead for ensuring a higher protection of private data and sensitive resources. From a security perspective, we consider that a reliable and trusted access control infrastructure for future information systems should take into account several requirements and should provide a minimum of security features.

### 4.1. Requirements

#### *4.1.1. Confidentiality and integrity*

In critical systems, it is compulsory to consider and treat sensitive data and services as confidential resources which integrity should be preserved. In this context, the confidentiality and the integrity of the IS resources must be preserved, and a main requirement for the access control system is that it should not allow in any way illegal accesses and unauthorized exploitation of the system resources.

#### *4.1.2. Privacy*

In open and untrustworthy environments, preserving the privacy of different users and actors in a critical system is highly required. The access control solution should consider the protection of private data as well as the preservation of a high level of secrecy as main objectives.

*4.1.3. Authenticity*

The authenticity of the actors of the system is an important aspect that needs to be taken into account by the access control solution. Ensuring the authenticity in the system relies principally on the validation of the origins and the identities of the different actors of the system. In order to ensure a high level of authenticity, the access control system should dispose of reliable identification and authentication mechanisms such as cryptographic and digital signature.

*4.1.4. Robustness*

The access control system must ensure a high level of robustness in the sense that inappropriate and unauthorized accesses should not be expected. The access control rules have to be rigorous enough to authorize desired accesses and prevent illegal accesses. The robustness of the access control as a crucial question is described with the levels of confidentiality and privacy provided by the system.

*4.1.5. Flexibility*

In critical infrastructures, such as healthcare and commercial systems, emergency accesses should be preserved by the access control system. In this case, the access control system must integrate flexible controls as regards to emergency cases. The flexibility needs to be integrated in a smooth and transparent manner to permit emergency accesses based on a delegation of rights or an overriding of access privileges.

Moreover, revoking access privileges is an important aspect that should be considered by access control solutions. Indeed, the access control system should be flexible in the sense that it should allow revoking access rights in an easy manner especially in critical situations when users abuse the trust and threaten the IS.

*4.1.6. Non-repudiation*

In order to pass up incidents linked to user's irresponsibility and negligence, it is recommended that the access control system has to integrate non-repudiation mechanisms such as auditability. This helps mainly in auditing illegal access and collusion attempts that allows strengthening the system with the corresponding prevention rules and controls.

*4.1.7. Administration, management, and compliance*

An access control system has to remain compliant and coherent with regard to the validated requirements without alterations. The necessity of integrating flexible controls in the system should not be at the origin of the non-compliance situations. In this context, we verify that a secure and efficient management of the access control infrastructure is a main requirement that has a wide impact on the quality of the system. Indeed, a faulty access control policy, a miss-configuration of the policy, or flaws in the policy and system deployment can result in serious vulnerabilities. A reliable management helps to precisely capture the security properties and needs that access control should adhere to bridge the gap of abstraction between the access control policy and the corresponding mechanisms.

### 4.2. Compliance management

The traditional life cycle of an access control policy defines three main phases: the specification, the verification, and the implementation of the policy. Then, the policy evolves with reference to maintenance and administrative tasks, following the evolution of security needs. Throughout its life cycle, the policy can undergo confused alterations: (i) It may record illegal updates and non-compliant changes with regard to its original specification. This generally occurs following an intrusion attempt or an illegal delegation of rights. (ii) It may contain incoherent and conflicting access control rules. This generally occurs following inner threats and collusion attempts and particularly in case the policy is defined by using more than a unique model of access control that leads to redundancy, inconsistency, and contradiction in the expression of the policy. We consider that the identification of the discrepancies between the abstract level of the access control policy (the specification) and its concrete level is crucial since correct operation and enforcement of access control policies by corresponding applications rely on the hypothesis that the specifications are correct and valid.

In large-scale, open and untrustworthy environments, the administration and management of an access control policy (considered as main security aspects) generally raise a critical analysis problem in case of a distributed administration of the policy, and/or potentially untrusted users (in most cases represent malicious administrators) contribute to the administration process. As a consequence, collusion attempts and inner threats may take place to generate crucial and invisible breaches to circumvent the policy. In fact, a faulty access control policy, a miss-configuration of the policy, or flaws in the policy and system deployment can result in serious vulnerabilities. In a database management system (DBMS) context, we easily check that as business and private data is exposed to several security threats and attacks, an access control policy is also subject to the same dangers [16]. According to *Imperva Application Defense Center* reports in 2013 and 2015, *Excessive and Unused Privileges* and *Privilege Abuse* are identified as most critical threats in top 10 database security threats.

Moreover, in the context of healthcare and e-healthcare systems (as a typical critical infrastructures), access control solutions should be rigorous to ensure a higher protection and flexible to treat emergency cases. We check that the simultaneous coupling of two necessary but contradictory objectives (robustness and flexibility) has a direct influence and a wide impact on the compliance of the deployed access control policy [17].

Our proposal to address issues related to the deployment and management of access control policies extends the traditional life cycle of access control policies with pertinent phases that we consider as necessary activities for ensuring the trustworthiness and the compliance of security policies. We consider three main levels of compliance management of access control infrastructures like illustrated in **Figure 2**. The first level concerns the management of the conformity between security and functional needs and the specification and design of the access control system. Indeed, a main requirement in the deployment of the access control infrastructure is specifying security and business needs, mastering, and validating its basics and expressiveness. Then, evolving the policy according to new security and business needs is highly required to maintain coherence between security needs and the high-level policy.

The second level concerns the management of the compliance between the specification of the policy and its concrete implementation. This level ensures the identification of faulty access control policy, miss-configurations or flaws in the policy and the system that can result in serious vulnerabilities. To ensure a high level of trustworthiness, it is highly recommended to proceed with verifying the conformity between the specification and the first implemented instance of the policy and particularly before the concrete exploitation of the system. The third level concerns the management of the compliance between the high-level policy as a reference model and any concrete instance of the policy. This helps detecting illegal updates and non-compliant changes in the concrete instance with regard to its original specification. It allows also identifying incoherent and conflicting access control rules occurred following inner threats and collusion attempts.

To ensure an efficient and secure deployment and management of reliable access control policies, we cover three key security aspects like illustrated in **Figure 3**. (i) The specification, verification, and implementation of the policy invariants, (ii) the validation of a concrete (implemented) instance of the policy regarding its original specification, and (iii) the adjustment and optimization of the access control policy schema. In fact, the goal during the specification phase is to capture the maximum of security needs and to distinguish the invariants that must meet any concrete instance of the access control policy. *Security architects* dispose, during this phase, of security modeling languages that extend classical application modeling languages. Verifying the exactitude of the specification and adopting a *model-driven architecture* (MDA) approach are highly interesting in system development and allow especially reaching the implementation via successive refinements of the verified specification. During the validation phase, the reference stage (the specification) and the concrete instance are facing in a logical framework allowing formal reasoning and compliance demonstration. To do so, two preliminary phases are necessary: a reverse engineering phase that allows generating the schema of the implemented policy and a formalization phase for representing the extracted policy in our formal framework. The optimization phase corrects the redundancy anomalies and helps check the properties of the graph of roles, calculate the power of a role, etc. Obtained results allow the adjustment and the up to date of the corresponding policy.

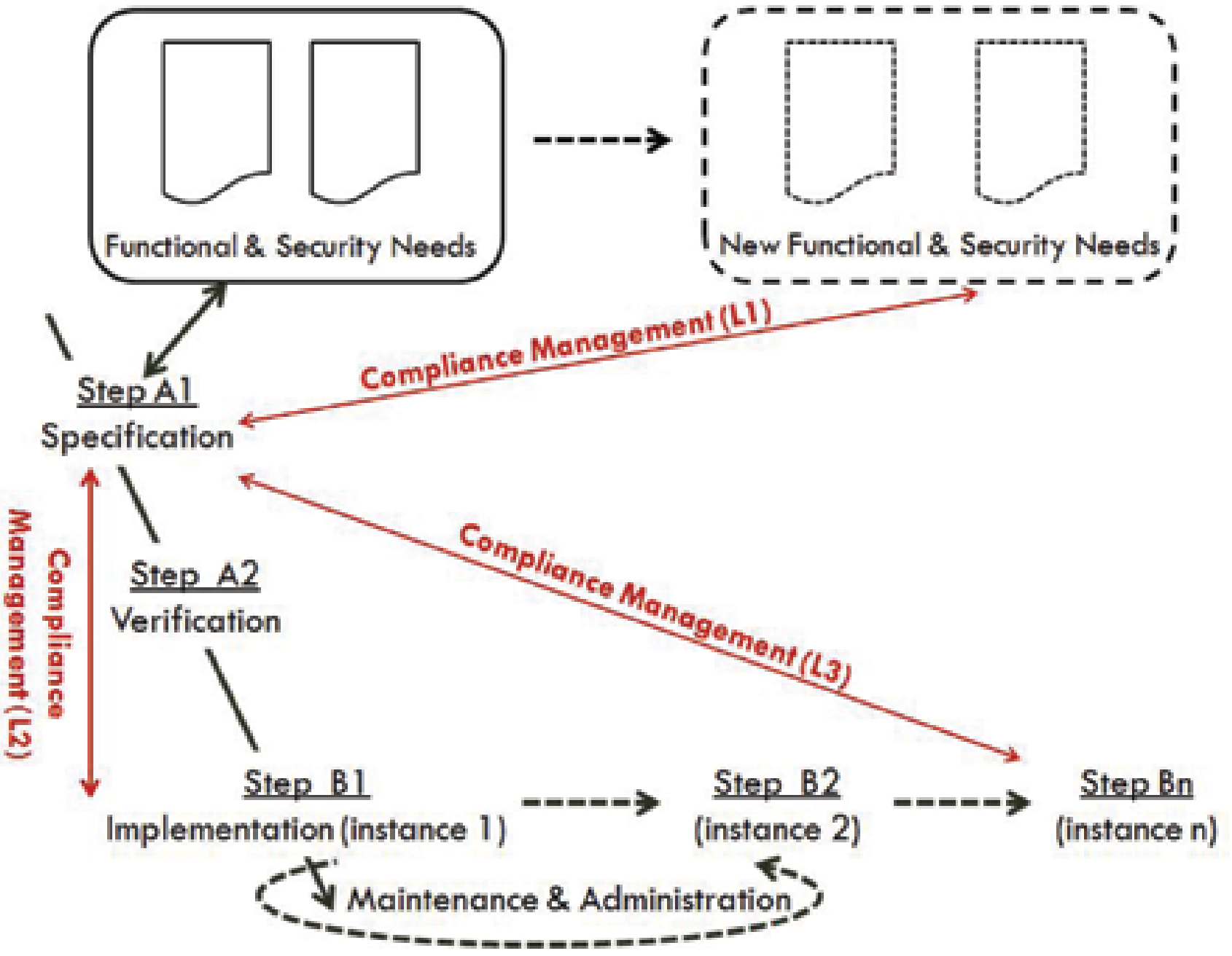

**Figure 2.** Compliance management levels.

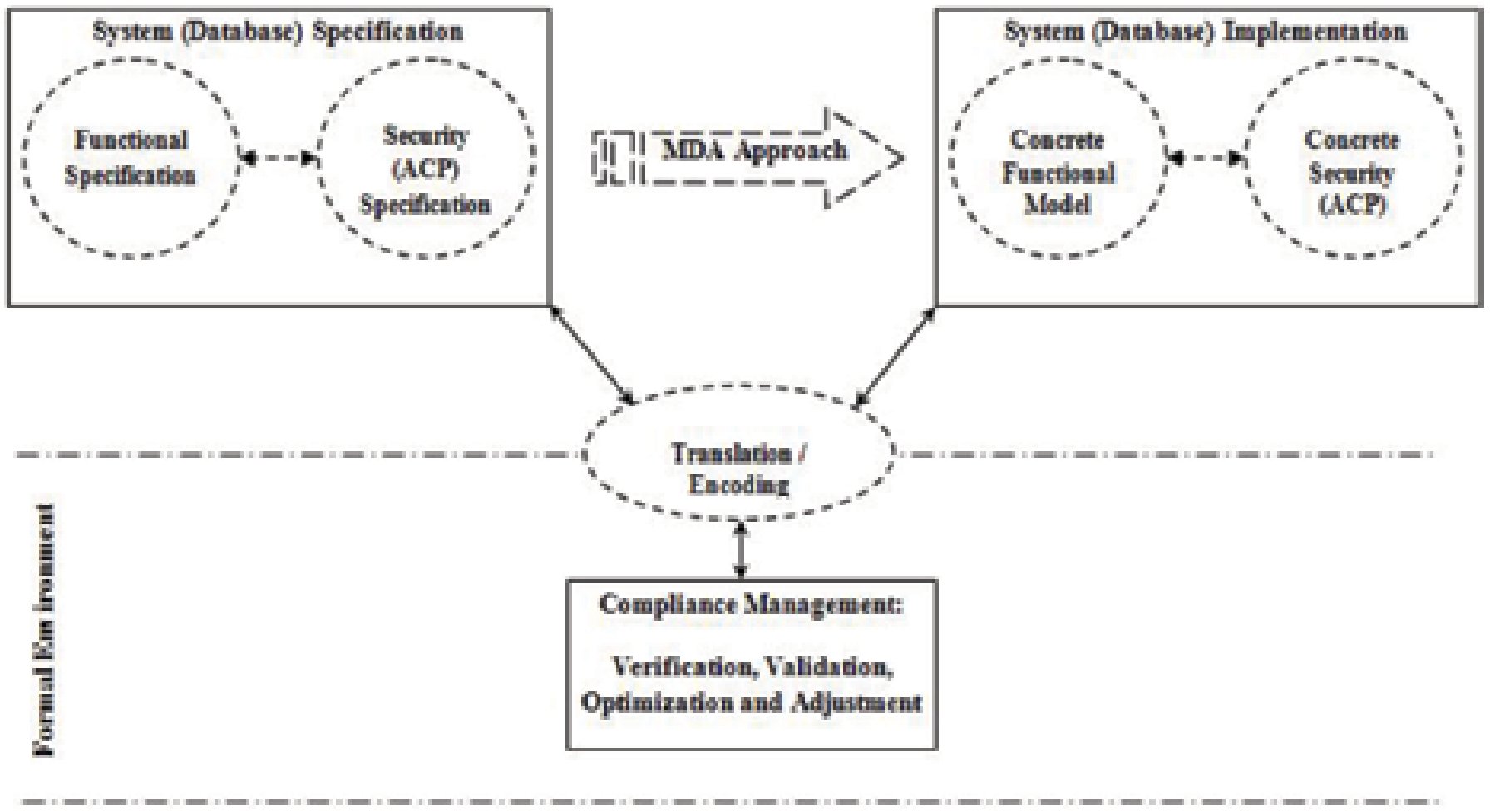

**Figure 3.** Formal approach of compliance management.

### 4.3. Future directives

#### *4.3.1. Compliance management of distributed policies*

Today's IS generally comprises several heterogynous components. Securing the IS requires mainly defining and enforcing a global security policy to be distributed on several active components that participate—in a collective manner—to the system security. The *security architect* is responsible for defining a global access control policy (GACP) and for defining the sub-policies ($ACP_1$, $ACP_n$) relative to each active component in the system such as firewalls (FW), database management systems (DBMS), application servers (AS), operating systems (OS), enterprise directory services (LDAP), etc.

A global management of the compliance of the global access control policy enforced by the IS consists in monitoring the conformity of all sub-access control policies enforced by active components separately and in verifying the conformity of the global policy taking into account the interactions between those components, like illustrated by **Figure 4**. As a future research direction, a global compliance management process—which integrates compliance management of sub-policies—should be defined and investigated for a global check of the conformity of the global policy taking into account interactions between active components of the system.

*4.3.2. Reverse engineering access control policies*

The management of the compliance of access control policies within databases relies on a reverse engineering step for externalizing the concrete implementation of a policy from the DBMS. In literature, numerous research works addressed the thematic of retro-conception or reverse engineering in the context of relational databases. This was at the origin of the development of professional tools for databases reverse engineering. Existing tools allow to generate the functional model of a concrete database, while they do not allow to generate the complete security model. In other words, they do not offer the opportunity to extract all the components of persistent access control policies. The reverse engineering procedure is based on the exploitation of the DBMS data dictionary.

Actually, a few research works addressed this important topic and defined reverse engineering techniques for extracting concrete policies from the Oracle DBMS [14, 15]. Even though, defined reverse engineering approaches can be generalized, we need specific reverse engineering procedures relative to each DBMS. A main contribution in this thematic is to define specific reverse engineering techniques relative to other familiar database management systems such as Informix, MySQL, etc.

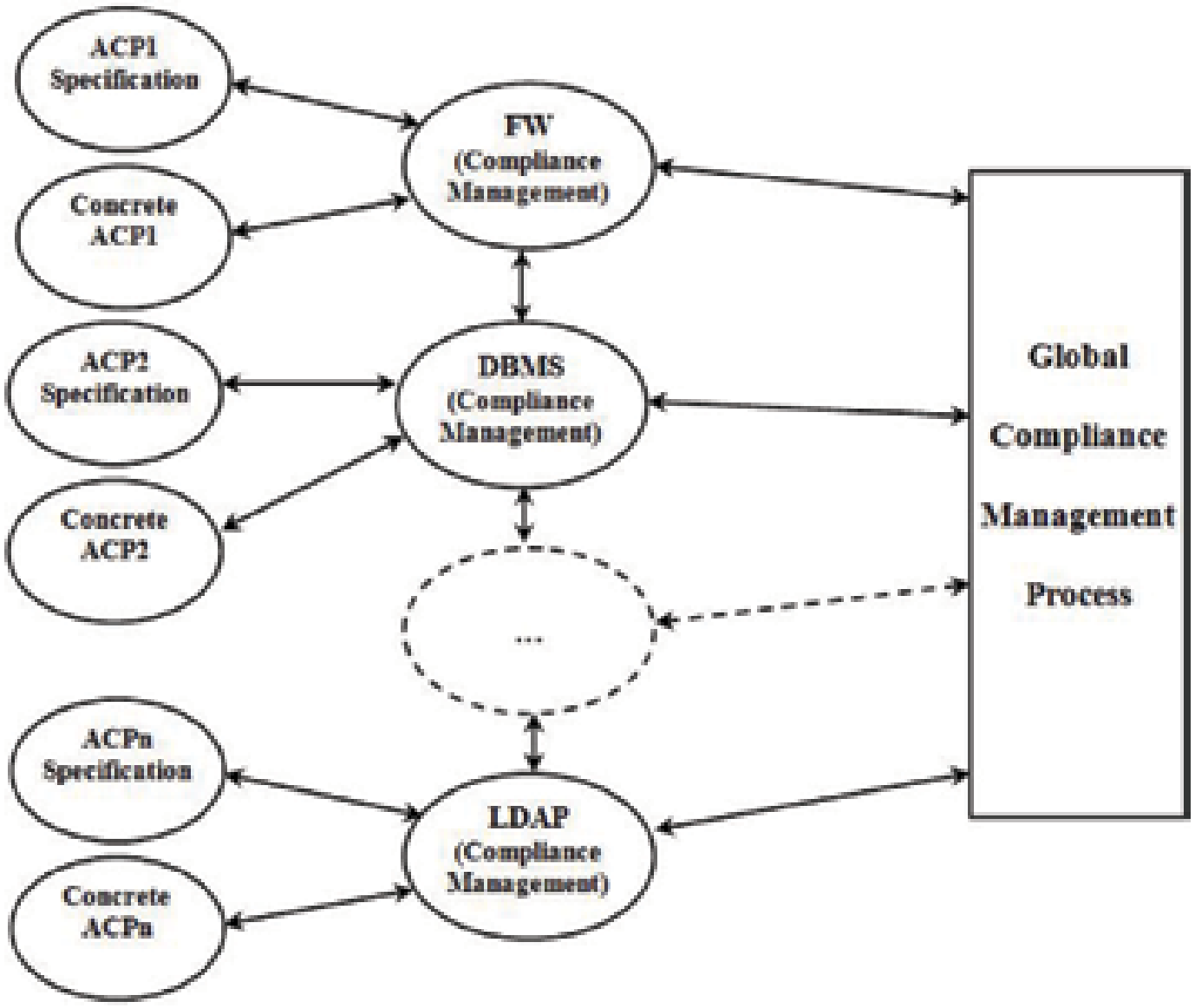

**Figure 4.** Distributed compliance management.

*4.3.3. Compliance management of access control policies in the context of object and NoSQL databases*

The fact that the concepts and principles of object-oriented and NoSQL databases are completely different from traditional databases (mainly relational databases), the application of traditional security controls is not adequate for providing effective security measures for object-oriented and NoSQL databases. Due to the specificity of object databases, we need—in a future research directive—to define the necessary techniques for monitoring the compliance of implemented access control policies in object-oriented DBMS.

## 5. Conclusion

Addressing security issues in today's information system requires mainly defining a trustworthy environment of access control. In this chapter, we review the actual state of research in access control to highlight the main advancements and challenges. We introduce and discuss requirements and the main characteristics for deploying advanced access control infrastructures. We illustrate that the management of the compliance of today's access control infrastructures represents a main issue for the deployment of secure systems. To address this thematic, we discuss the problem of the conformity of concrete access control infrastructures, and we propose a conformity management scheme for monitoring the compliance between low-level and high-level policies. Moreover, we highlight some future research directives that comply with the discussed thematic.

## Author details

Faouzi Jaidi

Address all correspondence to: faouzi.jaidi@gmail.com

Digital Security Research Unit, Higher School of Communication of Tunis (Sup'Com), University of Carthage, Tunisia

Chapter 5

# Protection of Relational Databases by Means of Watermarking: Recent Advances and Challenges

Javier Franco Contreras and Gouenou Coatrieux

Additional information is available at the end of the chapter

**Abstract**

Databases represent today great economical and strategic concerns for both enterprises and public institutions. In that context, where data leaks, robbery as well as innocent or even hostile data degradation represent a real danger, and watermarking appears as an interesting tool. Watermarking is based on the imperceptible embedding of a message or watermark into a database in order, for instance, to determine its origin as well as to detect if it has been modified. A major advantage of watermarking in relation to other digital content protection mechanisms is that it leaves access to the data while keeping them protected by means of a watermark, independent of the data format storage. Nevertheless, it is necessary to ensure that the introduced distortion does not perturb the exploitation of the database. In this chapter, we give a general overview of the latest database watermarking methods, focusing on those dealing with distortion control. In particular, we present a recent technique based on an ontological modeling of the database semantics that represent the relationships in between attributes—relationships that should be preserved in order to avoid the appearance of incoherent and unlikely records.

## 1. Introduction

The evolution of multimedia technologies and communications has resulted in a remarkable increase in the construction, transfer and sharing of databases. As a consequence, data gathering and management into databases or data warehouses or the scaling up to big data become important economical and strategic concerns for enterprises and public administrations in decision making. The expansion of data-mining and assisted analysis tools are just two examples that highlight the growing value of these databases. In that context, information leaks, thefts (confidentiality, traceability) or even degradations (integrity/authenticity), intentional or not, represent a real menace. This has recently been proved by the Wikileaks [1] or the Falciani cases [2], where large amounts of sensitive data have been exposed publicly on the Internet due to internal leaks.

Several protection mechanisms have been proposed so as to protect digital contents. A nonexhaustive list encompasses user authentication, access control and encryption which are helpful for confidentiality, digital signatures that can support data integrity and non repudiation and logs that can help to trace data. However, these security solutions offer an *a priori* protection in the sense that once they are bypassed, or more simply when the data access is granted, data are no longer protected.

On the contrary, watermarking can complementarily provide an *a posteriori* data protection. By definition, watermarking lies on the insertion of a message (some security attributes) into a host document (e.g., an image, an audio signal or, in our case, a database) by slightly modifying it based on the principle of controlled distortion. Watermarking leaves thus access to the data, which can be manipulated or consulted, while staying protected by means of a watermark. This watermark or the message it corresponds to (or equivalently the embedded security attributes) may serve as the protection of the owner rights, data integrity, data traceability, etc. Its versatility makes watermarking a really attractive solution for sensitive data protection.

While there is vast knowledge in the field of multimedia watermarking [3, 4], the interest in database watermarking has been limited, to date, with about 100 publications since the seminal method of Agrawal and Kiernan, which dates to 2002 [5]. In particular, and as we will see in the sequel, relational database watermarking differs from multimedia contents watermarking in several points. Among them, two are worth highlighting—i) records in a database can be reorganized without changing the meaning of the database, in opposition to highly correlated neighbor samples in a signal or pixels in images and ii) the existence of specific manipulations a database may undergo like tuple suppression and insertion which will modify the database structure. At the same time, depending on the nature and on the sensitivity of the data, more or less strict distortion constraints have to be considered and managed or at least modeled.

This chapter addresses the latest advances on the protection of relational databases by means of watermarking. We focus, in particular, on methods that aim at preserving the informative content of database. If in the past distortion control techniques preserved database statistics, a recent one suggests taking into account the semantic meaning of database records.

This chapter is divided into five main sections. First, in Section 2, we present the main applications of database watermarking. In Section 3, we come back and sum up the basic principles of database watermarking, highlighting the main differences with watermarking of multimedia contents (i.e., images, video). Section 4 gives an overview of the existing database watermarking techniques, putting in evidence "how" distortion control in database watermarking is most of the time achieved. In this section, we describe, in more detail, a semantic distortion control by means of ontologies—a modeling that is much more general. This solution is illustrated considering a practical case with a medical database containing inpatient stay records.

## 2. Applications of database watermarking

As depicted above, watermarking stands for the insertion or dissimulation of a message (a watermark) into the records or the attributes' values of a database. Depending on the relationship between the host database and the embedded message, different applications have been proposed.

### 2.1. Copyright and ownership assertion

As in the case of multimedia content watermarking, the first developed and most-studied watermarking application corresponds to database copyright protection. It relies on the insertion of an identifier associating the host document to its owner (creator or buyer) [6]. This identifier or watermark should be imperceptible and resistant to any operations, especially those aiming at removing the watermark. The first database watermarking technique, introduced by Agrawal and Kiernan [5], focused on copyright protection.

### 2.2. Traitor tracing and database traceability

In some cases, the identification of the recipient of one database can be a priority so as to trace a possible illegal redistribution. Watermarking is referred in that context as "fingerprinting" [7]. Herein, each distributed copy of the content is marked with an identifier or fingerprint which uniquely identifies an individual. If one of the receivers decides to illegally reroute or redistribute the database, it becomes possible to identify him or her [8]. The way these fingerprints are built has received a lot of research effort in order to make them resistant to collusion attacks in which several users owning copies of the same content cooperate in order to obtain an unwatermarked version. Such fingerprints or user identifiers are anticollusion codes [9, 10] and have, as an objective, the identification of at least one or several colluders in a coalition of users.

In the same vein, such traitor tracing solutions can serve as the identification of a dishonest user at the origin of a data leak. As previously exposed, a message identifying the user is embedded when he/she accesses the content. If the information is retrieved online, it will be possible to identify the responsible person by extracting the message. Contrary to the previous problem, the collusion attack is of less concern as such data leaks are usually the result of one user.

### 2.3. Integrity control (tamper detection)

Integrity or authenticity control represents the third main application of database watermarking. Indeed, it is essential to ensure data integrity, especially when they acquire a legal value or if they contribute to sensitive decision making. That is especially the case of the medical domain in case of litigations.

Fragile or semifragile watermarking constitutes attractive alternatives. In opposition to robustness, the fragility of the mark to databases' manipulations can herein be useful. The absence or the incorrect detection of a mark will indicate a loss of data integrity. Depending on the applicative context, the mark can be designed to resist some specific manipulations but not to all. If all manipulations have to be detected, we will talk about fragile watermarking [11, 12]. Such techniques are usually very sensitive, like a digital signature or message authentication code, and can indicate which parts of the database have been altered [13]. On the contrary, a semi-fragile watermark will be designed to be robust to some innocent manipulations, that is allowed in the applicative framework, and fragile to hostile attacks [14, 15].

## 3. Database watermarking: Specificities and a general chain of watermarking

A database *DB* is composed of a finite set of relations $\{R_i\}_{i=1,\ldots,NR}$. Hereon, for sake of simplicity and without loss of generality, we will consider one database based on one single relation constituted of *N* unordered tuples $\{t_u\}_{u=1,\ldots,N}$, each of *M* attributes $\{A_1, A_2, \ldots, A_M\}$. The attribute $A_n$ takes its values within an attribute domain, and $t_u.A_n$ refers to the value of the $n^{th}$ attribute of the $u^{th}$ tuple. Each tuple is uniquely identified by either one attribute or a set of attributes, and we call its primary key $t_u.PK$. Tuples or attributes in such a database can be reorganized, removed and added by the user. In this section, we expose the fundamentals of how this kind of structure can be watermarked and with which purposes.

The application of existing signal or image watermarking techniques to databases is not a straightforward process. Relational databases differ from multimedia contents in several aspects that must be taken into account when developing a watermarking scheme.

### 3.1. Database structure and watermark insertion/detection synchronization

One of the main differences is that samples in a multimedia signal are sorted into a specific order, in a temporal (e.g., audio signal samples) and/or spatial domain (e.g., pixels of an image or video), giving a sense of the content itself to the user. Close samples are strongly correlated with usually important information redundancy. This is not the case of relational databases, the purpose of which is to provide efficient storage of independent elements within a common structure. Thus, tuples in a relation are not stored in any specific order. At the same time, because tuples or records can be stored and reorganized in many ways in a relation without impacting the database information, questions arise between the synchronization of the watermark insertion and the watermark reading/extraction processes. Indeed, with signals or images, one can count on their intrinsic structure, working, for instance, on blocks or groups of consecutive samples or on transformed coefficients so as to conduct the insertion of one symbol of the message. The same strategy is not so easy to apply in the case of relational databases where tuples can be displaced, added and removed. Identification of watermarked elements in a database (records or attributes) and consequently the synchronization between the watermark insertion and detection stages require specific solutions. In order to make the watermark insertion/reading independent of the database structure, or more clearly of the way this one is stored, a preprocessing step is usually applied before message insertion/reading and (see **Figure 1**) following different possible strategies.

The first approach [5] consists of secretly constituting two groups of tuples based on a secret key. One group contains the tuples to be watermarked while the tuples in the second are not modified. In order to obtain the group index of a tuple $t_u$ in the relation $R_\gamma$, it makes use of a HASH function (H) modulo a parameter $\gamma \in N$ which controls the number of tuples to modify. If we define $t_u.PK$ as the primary key of a tuple, $K_S$ as the secret watermarking key, mod as the modulo operator and ∥ as the concatenation operation, the condition $H\left(K_S \| H\left(t_u.PK \| K_S\right)\right) \bmod \gamma = 0$ indicates whether a tuple must be watermarked or not. In [16], the HASH operation is replaced by a pseudo-random generator initialized with the tuple primary key concatenated with the secret key. Notice that these methods allow for embedding a message of one bit only. This consequently restricts the range of possible applications. In order to increase the capacity, Li *et al.* [8] proposed an evolution of the previous method in which one bit of the message is embedded per selected tuple. To do so, the watermark bit to embed in the tuple $t_u$ is also selected taking into account the tuple primary key $t_u.PK$ and the secret key $K_s$. This allows the insertion of a multi-bit watermark offering more applicative options.

A more advanced solution consists of a "tuple grouping operation," which outputs a set of $N_g$ that is nonintersecting groups of tuples $\{G_i\}_{i=1,\ldots,Ng}$. This allows spreading each symbol of a message *S* (or equivalently of the watermark) over several tuples, increasing, then, the watermark robustness against tuple deletion or insertion (i.e., the capability to detect/extract the message even if the database is modified).

The first strategy proposed in [17] is based on the existence of special tuples called "markers" which serve as a boundary or frontier between groups of tuples organized in a user-dependent order. A group corresponds to the tuples between two group markers. More clearly, tuples are ordered according to the result of a cryptographic HASH operation applied to the most significant bits (*MSB*) of the tuples attributes concatenated to a secret key $K_S$ such as $HASH(Ks \| MSB \| Ks)$. Then, tuples for which $H(K_s \| t_u . PK) \bmod e = 0$, where *e* is a parameter that fixes the size of groups, are chosen as group markers. The main drawback of this approach stands on the deletion of some of the markers which will induce a loss of watermark symbols. The extracted message will be shorter than the one originally embedded. To overcome this issue, the most common strategy consists of calculating the group index number $n_u \in [0, N_g - 1]$ of $t_u$ as in Eq. (1) [18]. Using a cryptographic hash function, such as the secure hash algorithm (SHA), ensures the secure partitioning and the equal distribution of tuples into groups.

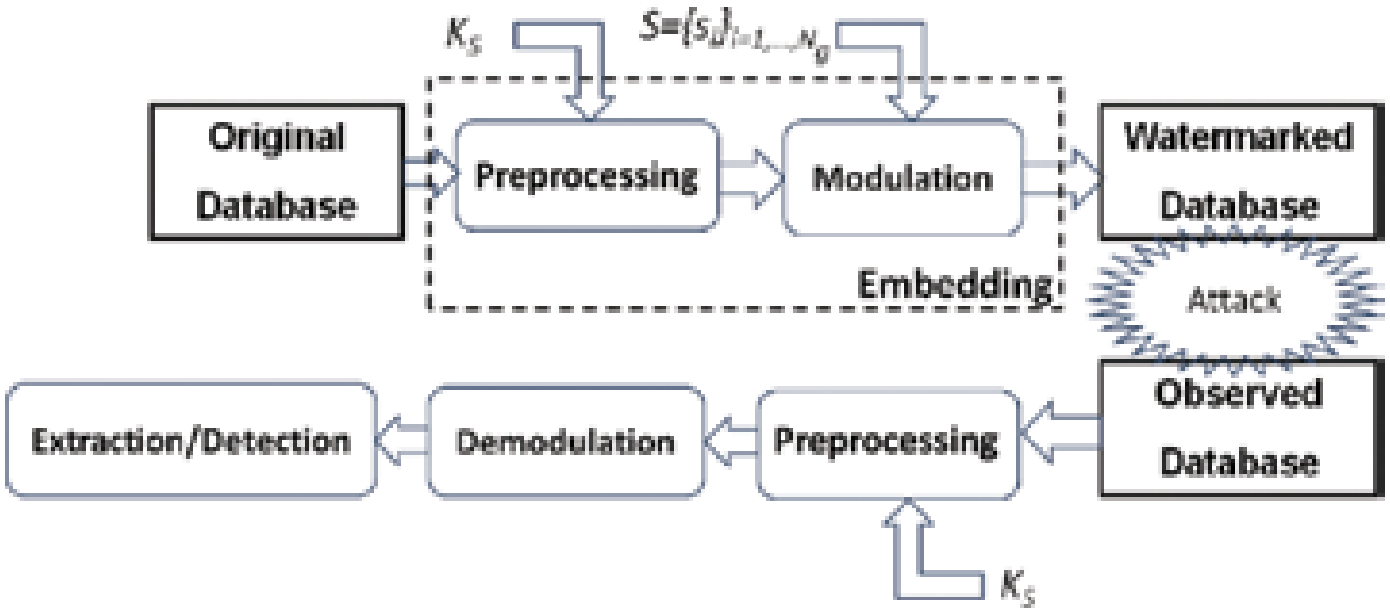

**Figure 1.** A common database watermarking chain. The message S is the concatenation of different symbols independently inserted into groups of tuples secretly constituted based on a secret watermarking key $K_S$.

$$n_u = H(Ks \| H(Ks \| t_u . PK)) mod Ng \tag{1}$$

Once this preprocessing task is conducted, one bit or symbol of the message is then embedded per group of tuples by modulating or modifying the values of one or several attributes according to the rules of watermarking modulation (e.g., modifying the attribute's statistics as in Ref. [17] or the tuple order as in Ref. [11]). Thus, with $N_g$ groups, the inserted message corresponds to a sequence of $N_g$ symbols—$S = \{s_i\}_i=1,\ldots,Ng$.

Some other approaches that do not make use of the primary key for group construction have also been proposed. Shehab *et al.* [18] regroup tuples depending on the MSB of some attributes. The main disadvantage of this strategy stands on the fact that groups can have very different sizes as MSBs do not usually follow uniform distribution. In a similar way, and with the same disadvantage, Chang *et al.* [19] propose to construct a virtual primary key from a fragment of some categorical or textual attributes.

### 3.2. Database manipulations

Another important difference between multimedia and database watermarking is associated with the frequency and the nature of manipulations over the data. In the multimedia case, filtering and compression operations are common. They modify the signal samples' values but do not change the signal structure (a filtered image or of a part of it will be close to its original version). In databases, insertion and deletion of tuples are frequent. They can be seen as sub-sampling and oversampling operations but with irregular distribution, a quite rare situation in signal processing, especially if the process output should keep an image structure. Moreover, databases may be queried so as to extract pieces of information that present an interest to the user.

### 3.3. Numerical and categorical attributes

Beyond the database structure and manipulation, one must also consider that the information contained in a database may come from different sources, for example, different services in a hospital. Hence, the attributes of the database can be of very heterogeneous nature while having semantic logic. In particular, one may have to handle numerical and categorical attributes or complex data such as images and so on. Categorical attributes differ from numerical attributes in the absence of order relationships in between the values of their domain. For example, considering the attribute "*eye colour,*" no rule states *a priori* that "*blue*" is greater or smaller than "*green.*" It is then difficult to apply mathematical operations in this context. We cannot say what will be the result of "*blue*" plus "*brown.*" This is not the case in multimedia signals where all the samples are numerical with the same dynamic. Nevertheless, solutions have been proposed to handle such categorical attributes even though it appears more difficult to control the distortion and preserve the meaningful value of the database. In a more general way, if image or video watermarking makes use of perceptual models of the human perception defects in order to mask the watermark, database watermarking requires other kind of distortion control solutions. As we will see, they are based on statistical and semantic aspects, some of which will be exposed in Section 4.

## 4. Overview of database watermarking schemes

This section presents an overview of the state of the art in database watermarking. Marking modulations are classified according to four criteria. The first two correspond to the robustness of the watermark or its fragility against database modifications. As stated earlier, robustness is the capability to retrieve the watermark after the protected database has been innocently (i.e., modifications that are authorized in the applicative framework) or malevolently (i.e., modification where the purpose is to remove the watermark) modified. Robustness is an important property in traitor tracing and copyright protection (ownership proof) frameworks. On the contrary, fragility of the watermark to some or all database modifications is a property that is at the basis of integrity control (tamper detection) applications.

The other two criteria are watermark imperceptibility and the database information the watermarking modulation uses so as to embed the watermark (e.g., categorical or numerical attributes, tuples' order, etc.). The former is a fundamental issue in database watermarking. This is why we propose a second classification level which depends on the way methods deal with data distortion. We will thus distinguish methods with or without distortion control, "distortion free" methods and lossless or reversible methods.

### 4.1. Robust methods

In the sequel, these methods are presented depending on the pieces of information they modulate in a database. We propose to distinguish three categories. *Distortion-based methods* modify or alter the values of some attributes of the database, these attributes being numerical or categorical, satisfying or not distortion constraints. The second class we suggest to consider regroups *lossless or reversible distortion-based methods*. The reversibility property ensures that it is possible to remove the watermark and to restore the original attributes' values of the database. The last class of methods modulates the database structure for message embedding. These schemes are referred to as *distortion-free methods*, due to the fact that they do not modify the record attributes values.

#### *4.1.1. Distortion-based methods*

##### *4.1.1.1. Modification of numerical data*

The first database watermarking method proposed by Agrawal and Kiernan [5] inserts a watermark by bit substitution into the least significant bits (LSB) of the database attributes' values. The tuples, attributes and bits to be modified are secretly selected by means of a hash operation (see Section 3). In this scheme, the watermark bit sequence depends on the database content and is not known by the user, that is, it corresponds to a database "footprint." At the detection, if the database has been watermarked, the expected number of bit correspondences (or equivalently the correlation) in between the recomputed database footprint and the extracted watermark should be near to 100%, while this number logically falls down to 50% if it has not.

Li *et al.* [8] extended the previous method so as to allow the insertion of a sequence of bits: a multi-bit message. Considering thus a multi-bit message $m$, the $j$th bit of the $t$th attribute $A_t$ of the record $t_u$, which we call $b_j$, is set to a value $b_j' = b \oplus m[q]$, where $b \in \{0, 1\}$ is a mask bit obtained from a random sequence generator $S$ such as $b = S(K_S \| t_u.PK) \bmod 2$, $m[q]$ is the $q$th bit of m secretly selected based on a function of $K_S$ and $t_u.PK$ and $\oplus$ is the xor operator. The message is inserted several times in the database. At the detection, for each secretly selected tuple, the operation $b \oplus b_j'$ is computed so as to extract the binary value inserted in $t_u.A_t$. This bit extraction is followed by a majority vote strategy so as to determine the final value of the extracted message. Such repetition increases the watermark robustness. Since Li *et al.*, different approaches following the same embedding strategy have been proposed with as objective to increase the complexity of message extraction or tampering an attacker [20, 21].

#### *4.1.1.2. Modification of categorical data*

Categorical attributes differ from numerical data in the absence of order relationships in between the values of their attribute domain. Sion *et al.* [22] were the first to propose a method for this kind of data. Let us consider an attribute $A_t$ which takes its values in the finite value domain $\{a_1, a_2, a_3, \ldots, a_{Na}\}$. These different values do not have a predefined order. However, a numerical value can be arbitrarily assigned to each categorical value creating thus a virtual attribute dynamic as for numerical attributes. By doing so, they can then apply a numerical attribute modulation, for instance LSB substitution. The main problem of this method is the strong distortion it can introduce when the meaning of the new value is considerably different from the original one.

#### *4.1.1.3. Introduction of "fake" tuples and/or attributes*

Another type of method is based on the insertion of new pieces of information (e.g., tuples or attributes) into the database. In that case, even though the original information has not been modified, one can consider that a certain distortion results from the additional data. Indeed, they can bias the result of database queries or of some statistical analysis. Pournaghshband [23] presents a method that inserts false tuples. In order not to impact the database integrity or coherence, it constructs primary key values for each new tuple so as to respect the key integrity constraint (there should not be duplicated primary key values). The detection seeks for fake tuples. The presence of one of them indicates that the database has been watermarked as they are only known to the database owner.

In a scenario not too different of this one which focuses on the watermarking of ontologies, we find the method of Suchanek and Gross-Amblard [24] based on the same strategy of false information insertion in order to identify the ontology owner.

*4.1.2. Distortion control-based methods*

In order to increase the watermark imperceptibility and to not modify the normal use of the data, distortion control techniques have been considered. All of them work on numerical attributes. Gross-Amblard published in 2003 a theoretical work [25] oriented to distortion minimization in the case of *a priori* known aggregation queries. Minimal distortion is considered to be obtained if the result of these queries is exactly the same as the one obtained with nonwatermarked original data. In this framework, Gross-Amblard modulates pairs of tuples involved in the result of the same query with distortion of identical amplitude but of opposite sign for each tuple in the couple so as to compensate introduced perturbation in average. This algorithm has been extended and implemented in the Watermill method proposed by Lafaye *et al.* [26]. A limitation of this approach is that queries should be *a priori* known. Moreover, only aggregation queries are considered. Regarding other kind of queries (e.g., selection of a set of tuples), Lafaye *et al.* apply the method of Sion *et al.* [17] which is based on the modification of attribute's values statistics under information quality constraints, defined by means of the mean squared error (MSE). Once groups of tuples are constructed, Sion *et al.* compute a reference value *ref* that is calculated in each group according to the mean (*avg*) and the standard deviation ($\sigma$) of the attribute to the watermark such as: $ref = avg + c\sigma$, where $c \in (0, 1)$ is a user-defined parameter. The embedded bit value depends on the number of attributes' values in a group $v_c$ that are over this reference. More clearly, for a group of $Nt$ tuples, insertion relies on two parameters, $v_{true}$, $v_{false} \in (0, 1)$, in a way that a bit "0" is embedded if $v_c < Nt \times v_{false}$ and a bit "1" is embedded if $v_c > Nt \times v_{true}$. At the same time, if the modification exceeds the quality constraints, a fixed threshold or a rollback operation is applied, that is, all the operations performed onto the tuples of a group are undone.

Shehab *et al.* [18] enhanced the method of Sion *et al.* with a more efficient management of distortion constraints, while solving, at the same time, some issues linked to the group creation strategy (see Section 2.2). Watermarking is presented as a constrained optimization problem, where a dissimulation function $\Theta$ is maximized or minimized depending on the bit value to embed. The optimization space is limited by the quality constraints set. In the example given by the authors, $\Theta$ represents the number of elements which exceed a certain reference value (same value as in the method of Sion *et al.*). At the detection, the value of $\Theta$ is calculated and the detected bit is a 1 (resp. 0) if its value is greater (resp. smaller) than a threshold $T$. The value of $T$ is calculated from the embedding information so as to minimize the probability of a decoding error.

Lately, Kamran *et al.* [27] have proposed the concept of "once-for-all" usability constraints. Considering a database that should be transferred to several users, they proved that if the detection threshold is fixed in order to ensure a correct detection for the most restrictive set of constraints, then detection reliability is independent of the constraints. This most restrictive set of constraints can be named "once-for-all" usability constraints. One drawback of this method is that its robustness and the lowest distortion it induces stand on a very short mark embedded into a few number of tuples. If the "good" tuples are altered, that is the watermarked tuples, the database is unprotected. Notice also that the modulation on which this scheme is based has some security issues allowing detecting any chosen watermark even if it has not been embedded into the database. The same authors propose in another work a watermarking scheme that preserves classification results of an *a priori* known data-mining process [28]. To do so, attributes are first grouped according to their importance in the mining process. Then, some local (i.e., for a set of attributes) and global constraints are derived from some dataset statistical characteristics that are relevant to the mining process. Finally, the allowed perturbation for a set of attributes is determined by means of optimization techniques.

As it can be seen, all the above methods aim at preserving statistical properties of the database. They do not consider the existence of strong semantic links in between attributes values in a tuple, links that should be preserved when modifying attributes values. Indeed, tuples must remain semantically coherent in order to: (i) assure the correct interpretation of the information without introducing incoherent or unlikely records and (ii) keep the introduced perturbations invisible to the attacker. In order to solve these issues, Franco-Contreras and Coatrieux propose to consider an ontological modeling of the semantic relations between attributes values in the database so as to guide the watermark embedding [29]. Being the most recent approach in dealing with attributes distortion control, we will present it in more detail in the next section.

### *4.1.3. Lossless or reversible methods*

#### *4.1.3.1. Lossless watermarking of numerical data*

In some cases, there is an interest or even a need of being able to recover the original database from its watermarked version. For instance, one may want to perform some operations on the original data or update the watermark. The reversibility property is herein of great interest. Robust lossless watermarking has been recently considered in the context of relational databases. Most of the existing methods are an adaptation of techniques proposed for image watermarking [30] and, as these, they are predominantly fragile with some exceptions.

Let us start by the latter, that is, robust methods. In Ref. [31], Gupta and Pieprzyk propose a zero-bit watermarking method where a binary meaningless pattern is embedded into secretly chosen tuples with attributes which are real numbers. To do so, a secretly chosen LSB from the integer part of an attribute is replaced by a pseudo-random-generated bit. To make the scheme reversible, the original LSB value is inserted into the space left by shifting the LSB representation of the fractional part of the attribute. The presence of the binary pattern is checked by the detector, indicating if the database has been watermarked or not. In order to reduce data distortion, Farfoura *et al.* [32] suggest watermarking the fractional part of one numerical attribute by means of prediction-error expansion modulation originally proposed by Alattar in [33] for images. Although this method is robust against common database manipulations (e.g., tuple addition or removal), the watermark will not survive a simple rounding integer operation. Beyond, it is important to notice that difference expansion modulation has not been designed for being robust to attributes' values modifications (this is the same for images). Indeed, Farfoura *et al.* [32] achieve watermark robustness with the help of a majority vote strategy, repeating thus several times the message into the database.

On its side, the method by Li *et al.* [34] constructs groups of tuples according to a clustering technique. The maximal modification that can be introduced into a tuple ensures that it will remain in the same group from the detector point of view. The watermarked value of an attribute is then calculated from an expansion of the polar angle of the attributes to the watermark. However, and as reported by its authors, this method is not fully reversible as some little errors can be found in the recovered data.

Recently, Franco-Contreras *et al.* [35] adapted the lossless watermarking scheme based on circular histogram modulation, originally proposed for images by De Vleeschouwer *et al.* [36], to the watermarking of relational databases. More precisely, this scheme modulates the relative angular position of the circular histogram center of mass of one numerical attribute in the relation. This scheme allows the embedding of a robust sequence and a fragile sequence at the same time and it can be thus considered for ownership control and traceability as well as for integrity control. Details on experimental and theoretical performance of this scheme can be found in Ref. [35].

*4.1.3.2. Reversible watermarking of categorical data*

In the method proposed by Chang *et al.* [19], one bit of a message is embedded by replacing the last letter of the last word of a textual attribute with another one from previously constructed reference sets. More precisely, before message embedding, two reference sets are constructed, one for each possible bit value, "0" or "1," which simply correspond to a secret ordering of letters in the alphabet, that is $\{a, b, \ldots, z\}$. At the detection, the knowledge of these reference sets allows for extracting the embedded bit as well as the restoring of the original letters' values. If high robustness against classic attacks is achieved, the use of a spelling checker will help erase the embedded message.

*4.1.3.3. "Attribute distortion-free" methods*

In the above methods, it is assumed that a slight distortion can be carried out for message insertion without perturbing the interpretation or any *a posteriori* uses of data. However, if one may consider that no data perturbation can be introduced, methods that do not modify attributes values can represent as interesting alternatives. These attribute distortion-free robust embedding strategies play on the way textual or categorical attributes values are encoded.

Al-Haj and Odeh [37] embed a binary image by modifying the number of spaces between words. In the same vein, Hanyurwimfura *et al.* [38] take advantage of the Levenshtein distance between words in order to select the words between which the space can be modified, those at the smaller distance. **Figure 2** illustrates the application of this modulation on a textual attribute. It is considered that such kind of modification does not induce any information quality loss. Instead of modifying the spaces in between words, Shah *et al.* suggest to alter the encoding of attributes values and work with capital and small letters of secretly selected attributes [39]. According to the bit to embed the complete word (or phrase) or only the first letter is capitalized. Notice that even if their authors present these methods as being robust, the watermark will be easily identified and erased by means of simple manipulations changing the way the attributes are encoded (e.g., fixing the number of spaces or changing words' capitalization).

| | | | |
|---|---|---|---|
| 155124 | Inflammation of right knee | 155124 | Inflammation of    right knee |

**Figure 2.** An example of distortion-free watermarking of categorical attributes considering the method by Al-Haj and Odeh [37].

## 4.2. Fragile methods

In contrast to robust methods, fragile methods are designed so as to make the watermark disappear after database manipulations. This makes them of interest for verifying the integrity of a database (tamper detection), which is the main objective of the following methods.

### *4.2.1. Distortion-free methods*

As stated in Section 4.1.3.3, by definition, such kind of methods do not modify the values of the attributes and basically consist of introducing “virtual attributes” or the modulation of the tuples’ (or attributes’) organization in the relation.

Prasannakumari [40] proposed the addition of one or several virtual attributes into the relation, which will contain the watermark information. They propose the following steps. In a first time, groups of tuples are constructed. One or several attributes of NULL value are next inserted in all tuples of the relation. For each group, the value of the virtual attribute is replaced by an aggregate of the values of a chosen numerical attribute in the group. The aggregate can be the sum, the mean value, the median, etc. Then, for each tuple, the checksum of each attribute is calculated and concatenated to the virtual attribute value. At the verification stage, the same steps are followed. Integrity of data is verified if recomputed checksums correspond to the extracted ones. Working at the tuple level allows the identification of the records of the database that have been modified.

Regarding “attribute distortion-free” strategies playing with the tuple organization, that is to say reorganizing the way tuples are ordered in the database, they have been originally introduced by Li *et al.* [11]. Their scheme works as follows. In order to embed the watermark, tuples are grouped and ordered into a group depending on the value of a hash function calculated on some attributes concatenated with the tuple primary key and the owner secret key. Database is then rewritten or reorganized so as to store tuples according to the increasing order of their hash. For a group $i$, the watermark is a sequence $W_i$ of length $l_i = N_i/2$ with $N_i$ the number of tuples in the group. Insertion consists of reorganizing the order of pairs of tuples in the group depending on the bit to embed. One bit of $W_i$ is embedded in a pair by interchanging the position of the tuples in the database so as to encode “1” or left unchanged in order to encode “0”. Thus at the detection, if in a pair the hashes of the tuples do not respect their hash ordering, a bit value “1” will be extracted, “0” on the contrary. Other approaches that have been proposed later on allow the identification of the manipulations the database underwent (e.g., tuple suppression) [12, 41] or the increase of the embedding capacity, that is, the number of watermark bits that can be embedded [13].

It is important to notice that methods based on tuple reordering are extremely fragile and constrain database handling by the database management system (DBMS). Tuple order should be preserved. As a consequence, their application context remains limited.

#### 4.2.2. Lossless or reversible methods

Lossless watermarking is well adapted for verifying the integrity or authenticity of a database. For instance, it allows the embedding of a digital signature or a message authentication code like SHA [15] computed over the whole database. At the verification stage, one just has to extract the watermark, restore the database and compare the extracted signature with the recomputed one. If signatures do not match, the database has been modified. Such protection is based on fragile lossless watermarking and different strategies have been proposed. Again, these methods have been derived from reversible watermarking scheme or modulation proposed for images.

##### 4.2.2.1. Methods working on numerical data

The histogram shifting (HS) modulation, a well-known lossless modulation for images, has been applied by Zhang *et al.* [14] to partial errors in a relation, that is in the differences between the values of one attribute of two consecutive tuples. The histogram of one digit of these differences is computed. Classes on the right side of the maximum class of the histogram are shifted to the right, creating thus an empty class close to the histogram maximum (see **Figure 3**). The attributes, the value of which belong to the maximum class, are then shifted to empty class value so as to code a "1" or left unchanged to code "0." The main drawback of this approach stands on the fact that consecutive tuples in the relation are not necessarily correlated values (contrary to contiguous pixels in an image). As a consequence, the less significant digits of the calculated differences may follow an almost uniform distribution, which seriously reduces the embedding capacity. Notice that in an image, the pixel difference distribution is close to a Gaussian distribution of small standard deviation.

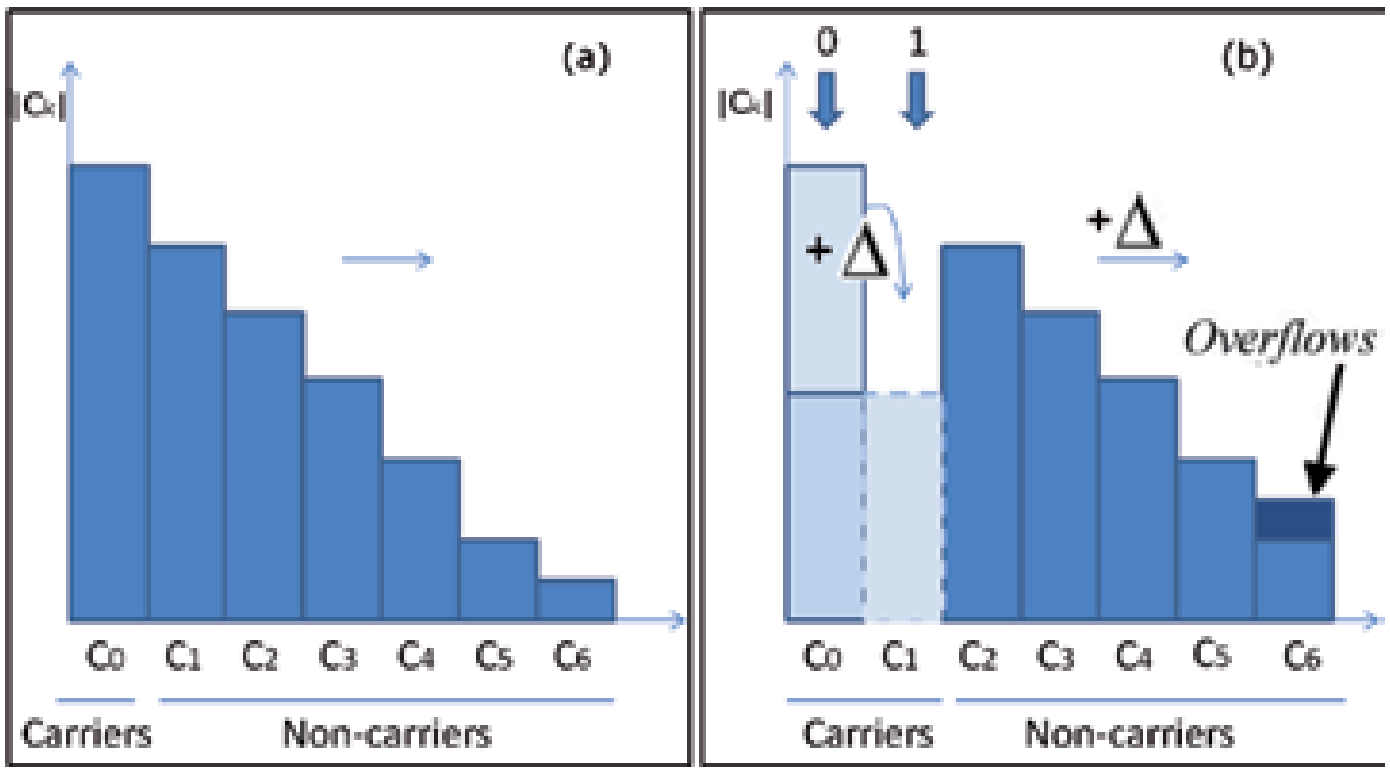

**Figure 3.** Histogram shifting modulation. (a) Original histogram and (b) histogram of the watermarked data.

Another approach proposed by Chang and Wu [42] considers the use of a support vector machine (SVM) classifier. One SVM is trained with a set of tuples selected so as to obtain a classification function *f*(*V*) used by the next to predict the values of one numerical attribute. Then, they apply the difference expansion modulation, another well-known lossless watermarking modulation, for message embedding. Basically, this modulation expands the differences between original and predicted values adding one virtual least significant bit that is used for message embedding. The distortion magnitude is unpredictable and, as underlined by its authors, it can be high in some cases.

*4.2.2.2. Methods working on categorical data*

Coatrieux *et al.* [15] adapted the histogram shifting modulation to categorical data, being the first lossless watermarking method for *this* kind of attributes. Following the general watermarking chain exposed in **Figure 1**, one group of tuples is secretly divided in two subgroups, $SG_1$ and $SG_2$. The number of occurrences of each value of the attributes considered for embedding in $SG_1$ is used to construct a virtual dynamic. More clearly, values are organized depending on their cardinality, as exposed in **Figure 4**. Attributes of $SG_2$ are then watermarked accordingly to this virtual dynamic based on histogram shifting modulation (see **Figure 3**). The embedded watermark can be a signature of the database used to verify its integrity.

### 4.3. Comparative view of database watermarking methods

A synthetic classification of the methods described in Sections 4.1 and 4.2 is given in **Table 1**, depending on: the type of the watermarking modulation, the applicative context, the type of the watermarked data, the type of watermark distortion control and, finally, the robustness or fragility of the watermark against common attacks.

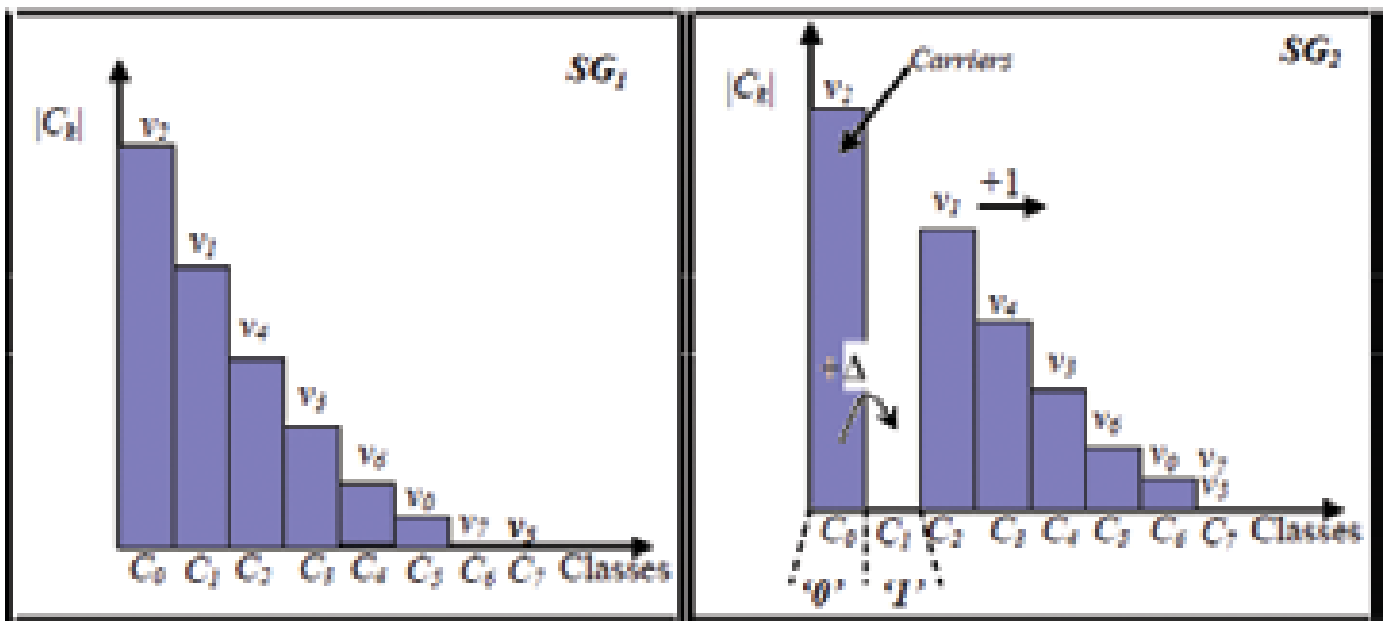

**Figure 4.** Histogram shifting applied to one categorical attribute with values $\{c_i\}_{i=0,\dots,7}$. $SG_1$- tuple subgroup use to derive the histogram x-axis (i.e. “virtual dynamic”). $\{c_i\}$ are sorted depending of their occurrences in $SG_1$. $SG_2$ – tuple subgroup use for embedding applying HS based on $SG_1$ virtual dynamic.

| Authors | Method | Application | Data type | Distortion control | Robustness |
|---|---|---|---|---|---|
| Agrawal and Kiernan [5] | LSB substitution | Ownership proof | Numerical | No | Robust |
| Li et al. [8] | LSB substitution | Traitor tracing | Numerical | No | Robust |
| Wang et al. [20] | Insertion of an image into LSB | Ownership proof | Numerical | No | Robust |
| Wang et al. [21] | Insertion of speech into LSB | Ownership proof | Numerical | No | Robust |
| Sion et al. [22] | MSB of the frequency histogram | Ownership proof | Categorical | No | Robust |
| Pournaghshband [23] | Insertion of fake tuples | Ownership proof | – | No | Robust |
| Lafaye et al. [26] | Modification of pairs of tuples | Traitor tracing | Numerical | Yes, respecting query constraints | Robust |
| Sion et al. [17] | Modification of statistics | Ownership proof | Numerical | Yes, rollback if required | Robust |
| Shehab et al. [18] | Modification of statistics | Ownership proof | Numerical | Yes, controlled by optimization techniques | Robust |
| Kamran et al [27] | Modification of LSB | Ownership proof | Numerical | Yes, parameters respecting constraints | Robust |
| Kamran et al [28] | Modification of LSB | Ownership proof | Numerical | Yes, controlled by optimization techniques | Robust |
| Gupta and Pieprzyk [31] | Difference expansion | Ownership proof | Numerical | Yes, reversible | Robust |
| Farfoura et al. [32] | Insertion into the fractional part | Ownership proof | Numerical | Yes, reversible | Robust |
| Li et al. [34] | Polar angle expansion | Ownership proof | Numerical | Yes, reversible | Robust |
| Franco Contreras et al. [35] | Circular histogram shifting | Ownership proof | Numerical | Yes, reversible | Robust |
| Chang et al. [19] | Modification of letters | Ownership proof | Categorical | Yes, reversible | Robust |
| Al-Haj and Odeh [37] | Modification of spaces between words | Ownership proof | Text | No but distortion has no impact | Robust |
| Haryunwinfura et al. [38] | Modification of spaces between words | Ownership proof | Text | No but distortion has no impact | Robust |
| Shah et al. [39] | Capital/noncapital letters | Ownership proof | Text | No but distortion has no impact | Robust |
| Prasannakumari [40] | Insertion of attributes | Tamper detection | – | Distortion free | Fragile |

| Authors | Method | Application | Data type | Distortion control | Robustness |
|---|---|---|---|---|---|
| Li et al. [11] | Reordering of order | Tamper detection | – | Distortion free | Fragile |
| Kamel and Kamel [12] | Reordering of tuples | Tamper detection | – | Distortion free | Fragile |
| Bhattacharya and Cortesi [41] | Reordering of tuples | Tamper detection | – | Distortion free | Fragile |
| Guo [13] | Reordering of tuples | Tamper detection | – | Distortion free | Fragile |
| Zhang et al. [14] | Histogram shifting of partial errors histogram | Tamper detection | Numerical | Yes, reversible | Fragile |
| Chang and Wu [42] | SVM used to predict values in detection | Tamper detection | Numerical | Yes | Fragile |
| Coatrieux et al. [15] | Histogram shifting of categorical attributes | Tamper detection | Categorical | Yes, reversible | Fragile |

**Table 1.** A synthetic overview of database watermarking methods.

# 5. Preserving semantic data quality in database watermarking

As exposed in the previous section, distortion control-based watermarking methods mainly focus on minimizing the distortion of the database statistics. Such consideration does not necessarily take into account the semantic relationships that exist in between the attributes' values in a tuple. Database semantics should not be neglected; it will avoid the introduction of incoherent or impossible records by the watermarking process, records that give clues about the presence of a watermark to an attacker.

In this section, we propose the use of an ontological model of the semantic links in between the attributes' values in a relational database in order to minimize the distortion [29].

## 5.1. Definition and main components of ontology

By definitions in the literature [43–45], ontologies allow defining shared concepts in some specific area of knowledge and how these are related by means of a common vocabulary in order to overcome the intrinsic heterogeneity and complexity of the real world. An important feature of ontologies is that they are interpretable by both human operators and computer programs, representing a gateway between human and artificial knowledge.

Even though authors do not come to an agreement in terms of what components ontology should have, most definitions contain the following elements defined by Gruber [45]: classes, relations, axioms and instances.

Concepts or classes correspond to abstract groups, sets or collections of objects. Examples of concepts could be: *Person*, *Car*, *Thing*, etc. The notion of classes depends on ontology.

For instance, one can define the class "*Thing*" that (in the abstract sense of the word) may contain anything one could imagine (e.g., *Person, Car, Book*, etc.). As exposed, classes can contain other classes and a universal class may contain every other class.

An individual or instance corresponds to the ground level concept of ontology; it is a concrete instantiation of an element or an object (e.g., a person named *Peter* or a car *Renault Clio*). Notice that the frontier between an individual and a class is quite blurred. It relies on the considered ontology. Individuals are described in ontology by a set of attributes. Examples of attributes can be *has-name*, *has-age* and so on. The value of an attribute is defined by a data type, for example, integer, string.

Objects in the domain are associated by means of relations specifying interactions between them. We can have relations between classes, between an individual and a class, between individuals, etc. For example, we know that one person *is-child-of* another person or that Batman *fights-against* the Joker.

We invite the reader to consult [43–45] for more information about ontological modelling of knowledge.

### 5.2. Relational databases and ontologies

A relational database consists of a finite set of relations $\{R_i\}_{i=1,\dots,NR}$ where one relation $R_i$ contains a set of $N$ unordered tuples $\{t_u\}_{u=1,\dots,N}$, each of which having $M$ attributes $\{A_1, A_2, \dots, A_M\}$ (see Section 3). Defined as such, this data structure lacks semantic information about the attributes meaning and links between different attributes' values in a tuple. Considering as an example the database of inpatient stay records, with a relation to the records which include the attributes *Gender and Diagnosis*, one should take care that the watermarking process does not turn a record such as {*Gender* ="*female*", *Diagnosis* ="*pregnant*"} into { *Gender* ="*male*", *Diagnosis* ="*pregnant*"}.

To overcome this issue, we propose to use the concepts and relations of ontology associated with the database in order first to model the knowledge one can have of the semantic relationships in between attributes of tuples and second to identify the maximum tolerated distortion for numerical attributes that will be watermarked. To do so, one question to consider is how to *make* interact these two database and ontology structures.

As stated above, concepts in an ontology are linked by means of relationships that specify hierarchical or associative interactions between them. From this standpoint, each domain value, subset or a range of values of an attribute $A_t$ can be associated with one ontology concept. We depict in **Figure 5** an illustrative extract of such mapping considering the example of one database containing pieces of information related to in-patient stay records and its associated ontology, ontology one must *a priori* know or elaborate. Notice that such relations cannot be easily identified by means of a simple statistical analysis of the database.

In this example, the value "*Alzheimer*" in the domain of the attribute "*diagnosis*" can be associated to a concept "*Alzheimer*" in medical ontology. This concept is related to another concept "*≥ 60 years old,*" which can be mapped into a range of possible values for the attribute "age."

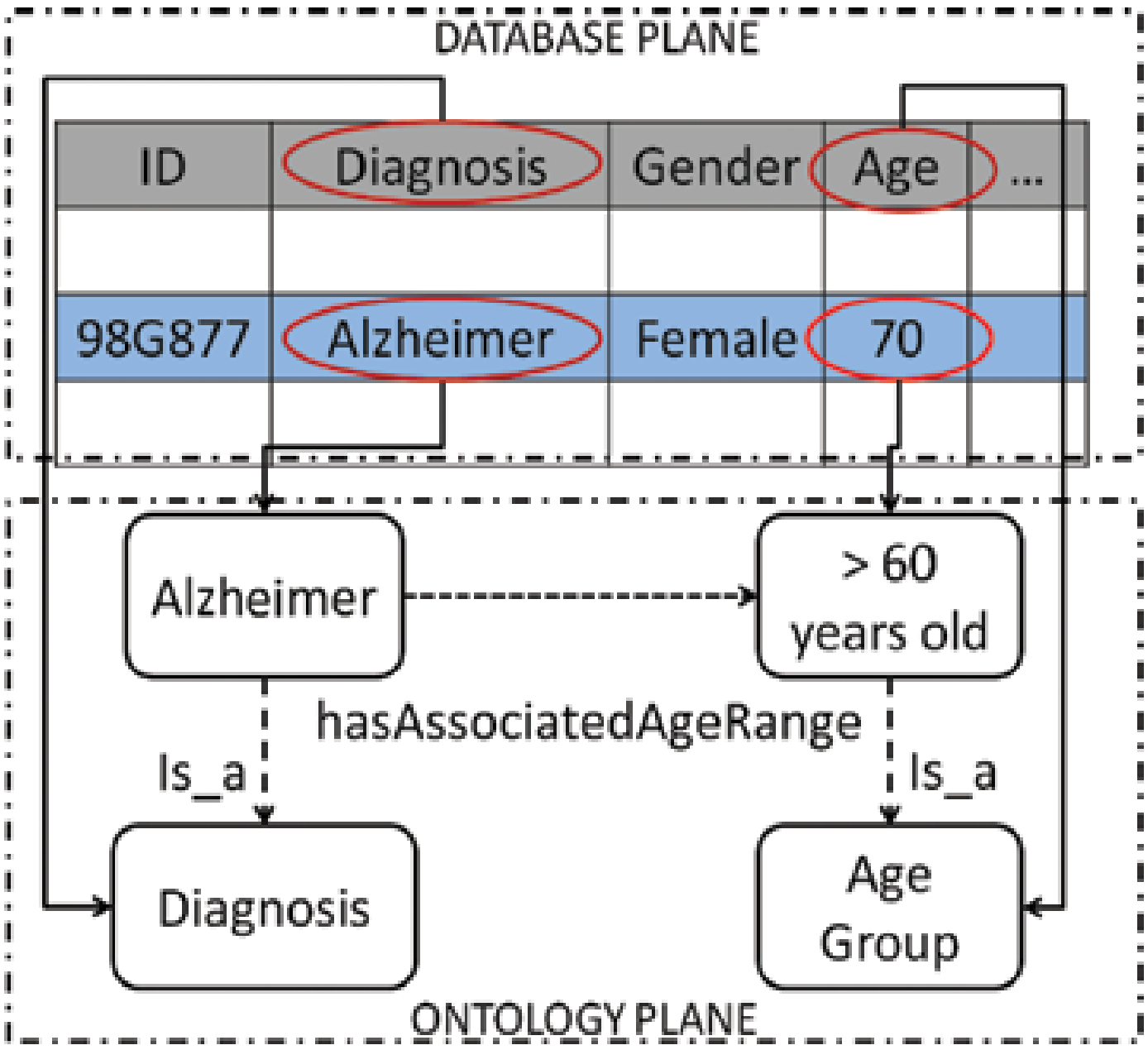

**Figure 5.** An existing connection between a relational database (database plane) and ontology (ontology plane) in the case of a database of in-patient stay records. Dotted and dashed arrows represent ontological relations between concepts in the ontology. Solid arrows represent connections between attributes or attributes values and ontological concepts [29].

From a watermarking point of view, this semantic relationship informs us that one attribute age value should not be turned into a value smaller than 60 in a tuple where the "*diagnosis*" attribute value is "*Alzheimer.*" If we generalize, assuming the numerical attribute $A_t$ is considered for watermarking, its value in the $u^{th}$ tuple, that is $t_u.A_t$, semantically depends on the set of values $S_{tu.At}$ of the other attributes of $t_u$, that is $t_u.\{A_1, \ldots, A_{t-1}, A_{t+1}, \ldots, A_M\}$ or a subset of them.

The distortion limits of $t_u.A_t$ can be defined as $Rg_{tu.At}$, that is, the allowable range of values $t_u.A_t$ can take after the watermarking process under the semantic constraint set $S_{tu.At}$ in order not to introduce incoherent or unlikely tuples in the watermarked database (see **Figure 6**). If we come back to the previous example, where $A_t$ = "*age*" is an integer, the value $t_u.age$ belongs to an integer range $Rg_{tu.age}$ imposed by the set $S_{tu.age}$ = "*Alzheimer.*"

Let us now consider that a categorical attribute $A_c$ has been selected for message embedding in the relational database DB. We recall that $t_u.A_c$ corresponds to the value of the attribute $A_c$ in the $u^{th}$ tuple.

As above, the range of values $Rg_{tu.Ac}$ that $t_u.A_c$ can take is semantically linked as $S_{tu.Ac}$. Because $A_c$ is a categorical attribute, $Rg_{tu.Ac}$ is a set of categorical values $Rg_{tu.Ac} = Val_1, \ldots, Val_{Nvals}$. Again, $Rg_{tu.Ac}$ can be identified by querying the ontology. For example, if we consider $A_c$ = "*diagnosis*" in regard to the attribute $A_{c+1}$ = "*gender,*" we know that for a tuple where $tu.A_{c+1}$ = "*Male,*" $t_u.A_c$ cannot be equal to "*Multiple gestation.*"

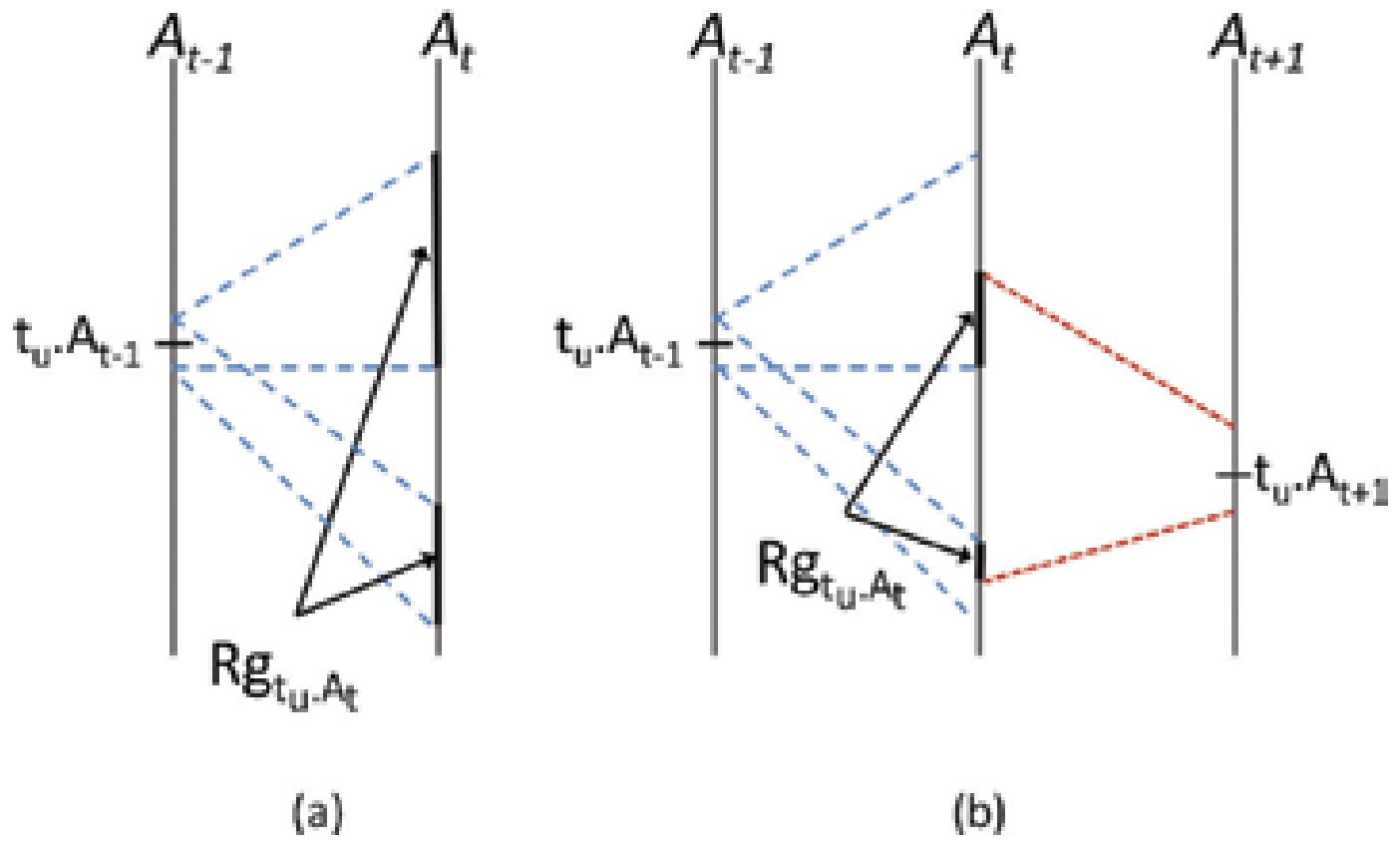

**Figure 6.** Identification of the allowable range $Rg_{tu.At}$ of values the attribute $t_u.A_t$ of the *u*th record can take for watermarking purpose. $Rg_{tu.At}$ is derived from the relationship in between the values of $t_u.A_t$ with those of other attributes of $t_u$: (a) when $A_t$ is only related to $A_{t-1}$ (in that case, the set of constraints is $S_{tu.At} = t_u.A_{t-1}$) and (b) when $A_t$ is only related to $A_{t-1}$ and $A_{t+1}$ (in that case, the set of constraints is $S_{tu.At} = \{t_u.A_{t-1}, A_{t+1}\}$). In the first case, $Rg_{tu.At}$ corresponds to the union of different intervals while in the second it is at the intersection of allowable ranges imposed by the two values in $S_{tu.At}$.

It is important to notice that this semantic distortion control is complementary to any other statistical distortion control method. Indeed, additionally to the semantic constraints, one may aim at preserving the correlation or the mutual information between attributes. Then, a global solution associating semantic distortion control and statistics distortion control, such as the technique suggested by Kamran *et al*. [27] can be constructed. This was done in Ref. [29], where it was also shown that the semantic control does not reduce the watermark robustness performance.

# 6. Conclusion

In this chapter, we gave an overview of most recent database watermarking techniques as well as of the latest advances in terms of database distortion control. As for multimedia data, like images or video, this aspect is important due to the fact that: (i) watermarking should preserve data quality and should not interfere with the *a posteriori* use and interpretation of data and (ii) induced distortion should not betray the presence of a watermark. To do so, semantic and statistic distortion controls have both to be considered. As illustrated, ontology appears as a good candidate so as to model the semantic relationships *and* will help to avoid the occurrence of incoherent or unlikely tuples.

Several issues are still to be considered in database watermarking and have to be addressed. Regarding the control of the database distortion, the joint application of statistical and semantic constraints should be analyzed. Indeed, in order to minimize the risk of incorrect data interpretation and to ensure the correct result of data-mining operations both criteria must be considered. Moreover, the automation of distortion control is still a challenge, especially in terms of complexity.

## Acknowledgements

This work has received a French government support granted to the CominLabs excellence laboratory and managed by the National Research Agency in the "Investing for the Future" program under reference ANR-10-LABX-07-01, and to the ANR project INSHARE, ANR-15-CE19-0024-02.

## Author details

Javier Franco Contreras[1]* and Gouenou Coatrieux[1,2,3]

*Address all correspondence to: javier.francocontreras@watoo.tech

1 WaToo, Plouzané, France

2 Institut Mines-Telecom, IMT Atlantique Bretagne-Pays de la Loire, Brest, France

3 Institut national de la santé et de la recherche médicale, Laboratory of Medical Information Processing(LaTIM), Brest, France

Chapter 6

# Implementing Secure Key Coordination Scheme for Line Topology Wireless Sensor Networks

Walid Elgenaidi, Thomas Newe, Eoin O'Connel, Muftah Fraifer, Avijit Mathur, Daniel Toal and Gerard Dooly

Additional information is available at the end of the chapter

**Abstract**

There has been a significant increase in the implementation of wireless sensor networks (WSNs) in different disciplines, including the monitoring of maritime environments, healthcare systems and industrial sectors. WSNs must regulate different sorts of data transmission such as routing protocols and secure key management protocols. An efficient WSNs' architecture must address the capability for remote sensor data management, for example encrypted transmitting data between nodes. This system demonstrates the capability to adapt its sensor members in the network in response to environmental changes or the condition of sensor nodes. The key management technique for any secure application must minimally provide security services such as authenticity, confidentiality, integrity, scalability (S), and flexibility. This chapter studies and analyzes different key management schemes that are implemented in WSN applications and evaluates the performance of secure key coordination algorithm for line topology WSNs. This scheme provides traveling packet for a source to end user via an individually encrypted link between authenticated sensor nodes. It will be shown how security algorithms are applied on a network, such as advanced encryption standard (AES)-based WSNs in real time, e.g., Waspmote sensor platform at the University of Limerick Campus.

# 1. Introduction

Key management schemes should provide a level of security requirements, low-energy consumption, and low-key storage overhead (KSO). However, some requirements are conflicting and difficult to provide at the same time. For these reasons, an efficient key management scheme should be applied to meet the requirements of the deployed applications. Moreover, distribution techniques that are applicable employ assorted key management schemes such as asymmetrical key cryptography and require numerous communication and computation capabilities. Thus, it is substantial to examine the different requirements, constraints, and evaluate various key management schemes that are applied to wireless sensor network (WSNs). The key management schemes based on WSNs must meet particular criteria for efficiency in light of vulnerability to attackers, including scalability, authenticity, flexibility, resistance value against node capture, key connectivity, and key storage overhead. This chapter investigates the core mechanism in the security of WSNs, which is managing the security keys in the network. Although wireless sensor nodes have limited resources and vulnerable against malicious attacks, different key management schemes in WSNs have been proposed in order to secure the communications between network members.

This chapter is organized as following; after briefly introduced, a detailed review about the existing security key management techniques based on symmetric encryption techniques in WSNs will be presented in Section 2. This chapter will classify schemes into different classes relying on key distribution approaches into three categories; probabilistic predistribution technique, deterministic predistribution technique and combined (probabilistic and deterministic) predistribution technique. The most common security schemes implemented in each technique will be explained and evaluated in detail in Sections 3–5. In Section 6, the performance evaluation of predistributed schemes will be presented in terms of distribution technique (DT), node type (NT), scalability (S), resistance value against node capture (RV), key connectivity (KC), and key storage overhead (KSO). Section 7 will explain the design, outdoor implementation, and evaluation of a secure and efficient key coordination algorithm for line topology-based WSNs. The performance evaluation of the outdoor implementation measurements will be discussed in terms of the received single strength indicator (RSSI) and current consumption. Section 8 highlights how to connect Libelium motes to IoT sensors via Intel Galileo in order to provide a secure solution for an end-to-end IoT system via the cloud. The chapter is concluded in Section 9.

# 2. Symmetric cryptography key management schemes in WSNs

Most of the key management schemes work in phases including key generation, key transmission, generation of transport key, and revocation of compromised keys. In the symmetric cryptography predistribution scheme, keys are stored in the sensor nodes that are used in a transport key (session key) generation phase. The predistribution key management techniques can be classified based on key distribution mechanism to two types: probabilistic and deterministic scheme. In this section, key management schemes are classified into classes that are relying on key distribution approach as illustrated in **Figure 1**. For a better understanding, we have classified these existing key management schemes into three different categories according to their distribution techniques.

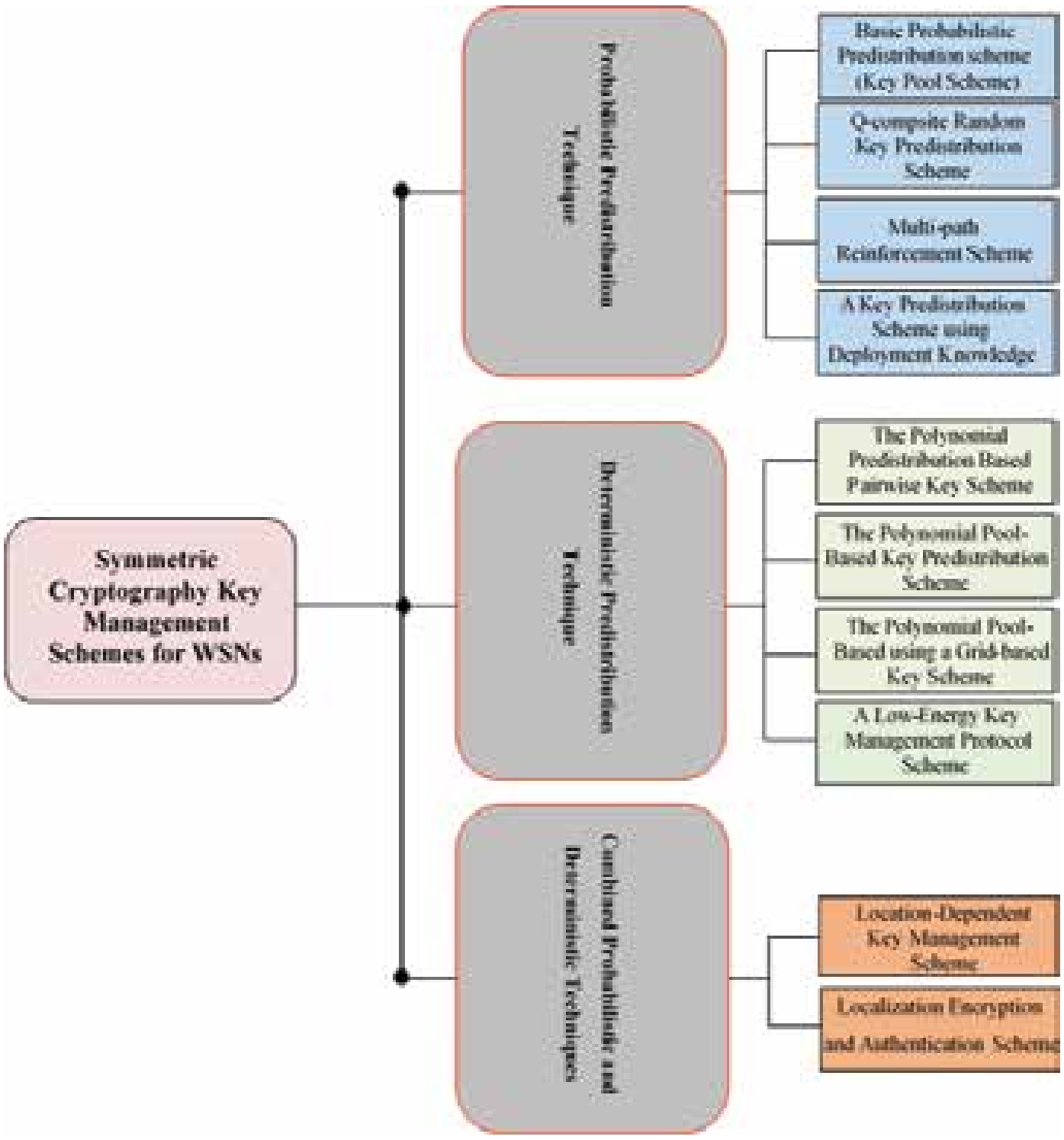

**Figure 1.** Classification of key management schemes based on symmetric cryptography for WSNs.

## 3. Symmetrical probabilistic key management based on predistribution technique

This technique is based on an offline distribution of a number of keys where the identities to each sensor node (key ring) are drawn from a large pool of keys. The main idea in this scheme is that any given pair of nodes in the network will share a common key with a certain probability. The initial step in the deployment phase is the exchanging of keys identifiers between sensor nodes in order to discover their common keys. A secure channel will be established between nodes when two nodes notice that they have the same common key [1].

### 3.1. The basic probabilistic key predistribution scheme

**Figure 2** illustrates the basic probabilistic key predistribution scheme. The overlaps between key rings present the shared keys between sensor nodes. Thus the configuration of the sensor network topology is established relying upon the secure channels between sensor nodes. After the deployment phase, and as a result of the exchange of key identifiers messages between node A and node B, they will use $K_4$ as their shared common key in order to secure the communication channel between them as shown in **Figure 2**. ON the other hand; if one sensor node is captured, the attacker could reveal the key ring which leads to compromised one or more secure channels in the network.

### 3.2. Q-composite random key predistribution scheme

Chan et al. [2] proposed a solution to improve the basic probabilistic key predistribution scheme against node capture attack called $q$-composite random key predistribution scheme. This solution used $q$ number of common keys to establish a secure channel between sensor nodes, which is boosting the network performance against node capture attack. As the number of keys overlap between two nodes is increased, the scheme performs more resistance against communication link break attack. Therefore, the scheme provides security under small-scale attacks.

### 3.3. The multipath reinforcement scheme

The multipath reinforcement scheme offers good security [2]. This scheme relies on updating links between nodes after the key setup phase. Since security is more of a concern than bandwidth or power drain, the scheme offers good security with additional sensor network communication overhead. The communication links established between sensor nodes after the key discovery phase in the basic scheme are based on the random selection of keys from the key pool. This allowed sensor nodes in the network to share a subset of the same keys and, thereby, possibly threaten multiple nodes when only one is compromised. Firstly, Anderson and Perrig [3] applied a multipath to reinforce links in a random key establishment scheme, then Chan et al. [2] used the multipath key reinforcement scheme in order to establish a communication path between every two sensor nodes in WSNs. This technique makes the multipath reinforce scheme stronger with communication links better than that the original basic scheme.

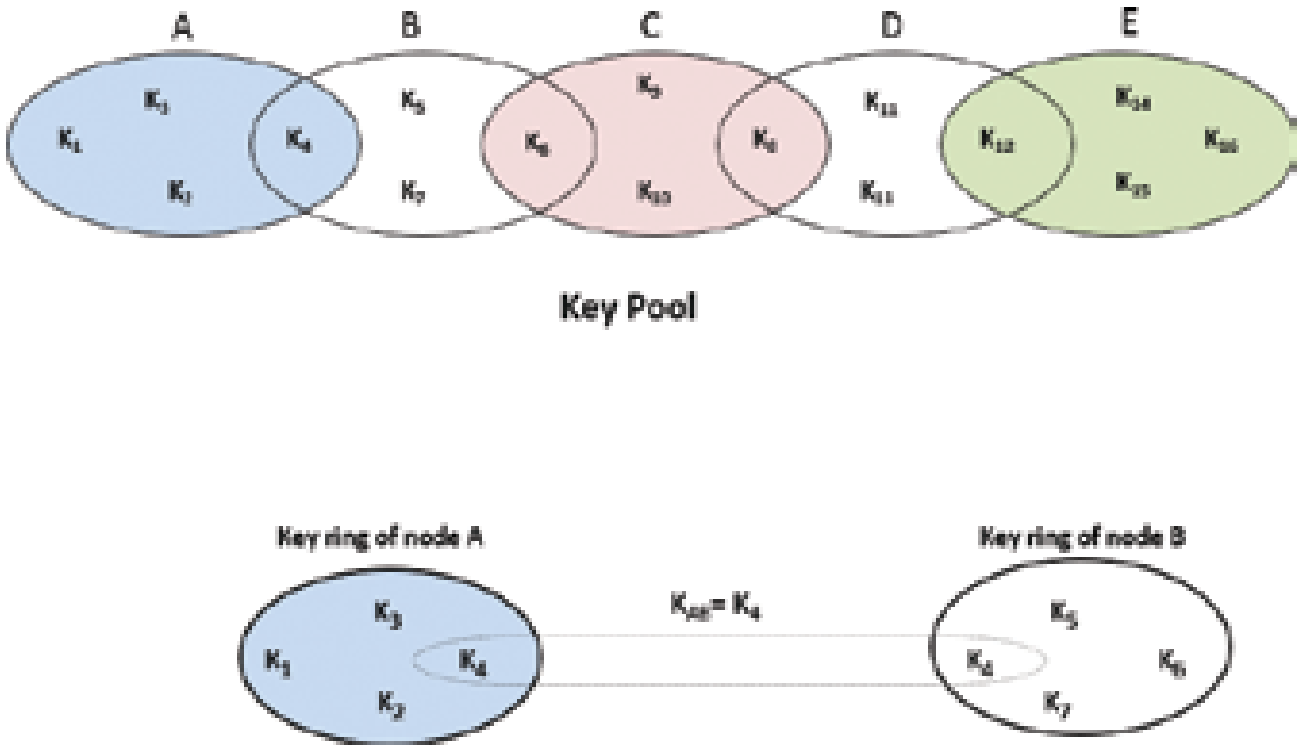

**Figure 2.** Key pool scheme of key pool size 16 with 4 size of key ring.

### 3.4. The basic probabilistic key predistribution scheme using deployment acknowledge

To ensure that the attacker cannot compromise secure channels between noncompromised nodes in Refs. [3, 4] proposed the basic probabilistic key predistribution scheme using deployment acknowledge. This scheme uses non-uniform probability density functions (pdfs) in order to perform deployment acknowledge. In other words, the locations of sensor nodes should be in certain positions. This scheme allowed sensor node to store only the keys that are shared with the other nodes in the networks.

## 4. Deterministic technique based on predistribution scheme

Generally, in deterministic technique, only the individual keys are predistributed offline. These keys are shared between sensor nodes and the base-station. Thus, after deployment, each node broadcasts only its node identifier using the shared individual keys in order to establish a secure channel with other nodes in the network. However, there is no demand to exchange key identifiers. In predistribution deterministic scheme, the pairwise keys are established after the deployment phase.

### 4.1. The polynomial-based pairwise key deterministic predistribution scheme

In Ref. [5], polynomial-based pairwise key deterministic predistribution scheme is generated dependent on a bivariate $t$-degree polynomial.

$$f(x,y) = \sum_{i,j=0}^{n} q_{ij} x^i y^j, \quad \text{where } q_{ij} = q_{ji} \tag{1}$$

where $q$ is a prime number above a finite area of $Fq$, where that should be large enough for a secret cryptographic key, and it has the property of $f(x, y) = f(y, x)$. Initially, every sensor node in the network loaded with a polynomial share of $f(x, y)$, such as $f(i, y)$ in node $i$. Thus, nodes $i$ and $j$ can generate the common key via evaluating the $f(i,y)$ at point $j$ or $f(j,y)$ at point $i$.

For example, we assume that sensor nodes 4 and 5 are loaded with a bivariate two-degree polynomial:

$$f(x,y) = x^2 + 3xy + y^2 \tag{2}$$

So the bivariate two-degree polynomial in node 4 is as follows:

$$f(4,y) = 4^2 + 3(4y) + y^2 \tag{3}$$

Also, the bivariate two-degree polynomial in node 5 is as follows:

$$f(5,y) = 5^2 + 3(5y) + y^2 \tag{4}$$

After nodes exchanged their identifiers, sensor node 4 calculates the share pairwise key with node 5:

$$f(5,4) = 5^2 + 3(5)(4) + 4^2 = 101 \tag{5}$$

And sensor node 5 calculates the shared pairwise key with node 4:

$$f(4,5) = 4^2 + 3(4)(5) + 5^2 = 101 \tag{6}$$

This scheme is secure and does not expose the link information of other nodes, if $n$ nodes are not compromised. This is making $n$ collusion resistant where the $n$ value relies on the sensor memory storage availability. Since each sensor node in this technique needs to store an $n$-degree polynomial, which occupies $(n + 1) \log(q)$ memory space. Therefore, this scheme is suitable only for nodes with big size of memory storage.

### 4.2. The polynomial pool-based key predistribution scheme

Modifications based on the basic scheme can address the issues of the scheme. Thus, instead of using a single $t$-degree polynomial, the polynomial pool-based key predistribution scheme is used. This scheme has three phases:

**Step 1. Setup phase:** the setup server randomly selects a subset of polynomials Fi from a set of a generated bivariate $t$-degree polynomials $F$ over finite field $Fq$ which are assigned with a particular ID for the server, and deployed into each node's memory, e.g., node $i$.

$$F_i \subseteq F \tag{7}$$

**Step 2. Direct key establishment:** each sensor node in this stage finds the same share of the bivariate polynomial with the other nodes. Next process is that each sensor node establishes the pairwise key.

**Step 3. Path key establishment:** This phase is based on the key predistribution scheme aforementioned in the previous section.

Initially, randomly selected polynomials are uploaded into each node. However, this scheme will functionally resemble the polynomial pool-based key distribution, if only one polynomial remaining in the pool, or all of the polynomials are 0-degree. While polynomial deployment is the main issue of the setup phase, the complex issue is in the direct key establishment phase, where each sensor node needs to find the other sensor nodes that share the same polynomial. In this stage, sensor nodes use two techniques: deterministic pairwise key based on predistribution scheme [6] and real-time discovery scheme [7]. In the deterministic pairwise key based on predistribution scheme, the knowledge of the nodes with which each node will share a polynomial is preloaded. In this approach, each node carries the node IDs of sensor nodes, which share the same polynomial. The drawbacks of this approach are that sensor node does not offer the flexibility of joining new nodes into the network. In fact, this approach is predistributed information of sensor nodes, which helps an adversary to gain access to the stored data in the case of a compromised node.

In the real-time discovery approach, source node aims to find a communication path with other sensor nodes that did not share polynomial in order to establish a pairwise key. The source node sends a message with the destination node ID to the intermediate nodes that already share a common key. Since the communication paths between sensor nodes that share common keys is secure, we assume that a node that is located between the source and destination shares the same secret key with the destination node. Thus, a communication link between the source and destination nodes has been discovered through which they may discover a common key. However, the issue of real-time discovery approach is the overhead of communication.

### 4.3. The polynomial pool-based using a grid-based key scheme

Ref. [8] provides all the properties of the polynomial pool-based key predistribution. In this scheme, each two sensors can establish a pairwise key in case of uncompromised nodes and the nodes can communicate with each other. Also if some sensor nodes are compromised, uncompromised nodes still have a great opportunity for the establishment of a pairwise key. By using grid-based key predistribution, a sensor node can determine the other sensor nodes, which can establish a pairwise key with, and it can choose which proper polynomial should be used for key establishment.

This scheme consists of constructing a $m \times m$ grid, where the number of sensor nodes in network can be at most $\mathrm{m}^2$ and a set of $2m$ polynomials is constructed as follows:

$$\{f_i^c(x,y), f_i^r(x,y)\}, \quad i = 0, \ldots m-1 \tag{8}$$

where $c$ and $r$ represent the column and row. Thus, each row $i$ in the grid is associated with a polynomial $f_i^r(x,y)$ where $i$ = 0 to $m - 1$, and each column $j$ in the grid is associated with polynomial $f_j^c(x,y)$ where $j$ = 0 to $m - 1$. The setup server assigns to each node a unique intersection point ($i$, $j$) in the grid and allocated polynomial shares of $f_i^r(x,y)$ and $f_j^c(x,y)$. Using this information, each node can perform key discovery and path key establishment. Each sensor node has two bivariate polynomials, while each polynomial (assigned to either the row or the column) shared by $m$ different sensor nodes in the network. Therefore, each node can directly establish two ($m - 1$) pairwise keys with other sensor nodes.

In this scheme, the attacker can prevent the two nodes from establishing a shared key, or by attacking the communication link between two nodes by either compromising the pairwise key. Moreover, the adversary can attack the entire sensor network by attacking a pair of nodes and finding their common key without actually compromising the nodes. This can be done through compromising the sharing polynomial of the two nodes. Thus, the attacker must compromise at least $t$+1 nodes in order to discover the polynomial share. In case the attacker successfully discovers the polynomial that is shared between two nodes, sensor nodes can avoid such an attack by establishing a common key through path key establishment.

### 4.4. A low-energy key management protocol for wireless sensor networks scheme

The main idea in Ref. [9] is that sensor nodes do not generate keys, and there is no sensor-to-sensor secure communication. Thus, each sensor node needs to store particularly two secret keys to secure the communication links with the base-station and the cluster gateways. Since the sensor nodes are not trusted in this scheme, storing a small number of keys is advantages for the memory constrained. These keys are stored in the flash RAM of sensor nodes before deployment. The gateways must have big size of memory resources in order to store a huge number of keys. Each gateway in the cluster stores all of the keys that are shared with the sensor nodes, keys shared with the command node, and key shared with one other gateway in the network (**Figure 3**). In addition, the command node stores all of the secret keys in the network.

The number of keys stored by the command node is equal to the total number of the gateways and the sensor nodes. Before the deployment phase, each sensor is preloaded with the identifier and the shared key of its cluster gateway. Then after deployment, each sensor node broadcasts "hello" message included in the identifier of its gateway. Therefore, the gateway shapes the cluster and sends assigned message to all sensor nodes in its own gateway cluster. Subsequently, each gateway broadcasts the identifiers of its group to the other gateways in the network. This broadcast stage helps minimize the volume of inter-gateway traffic.

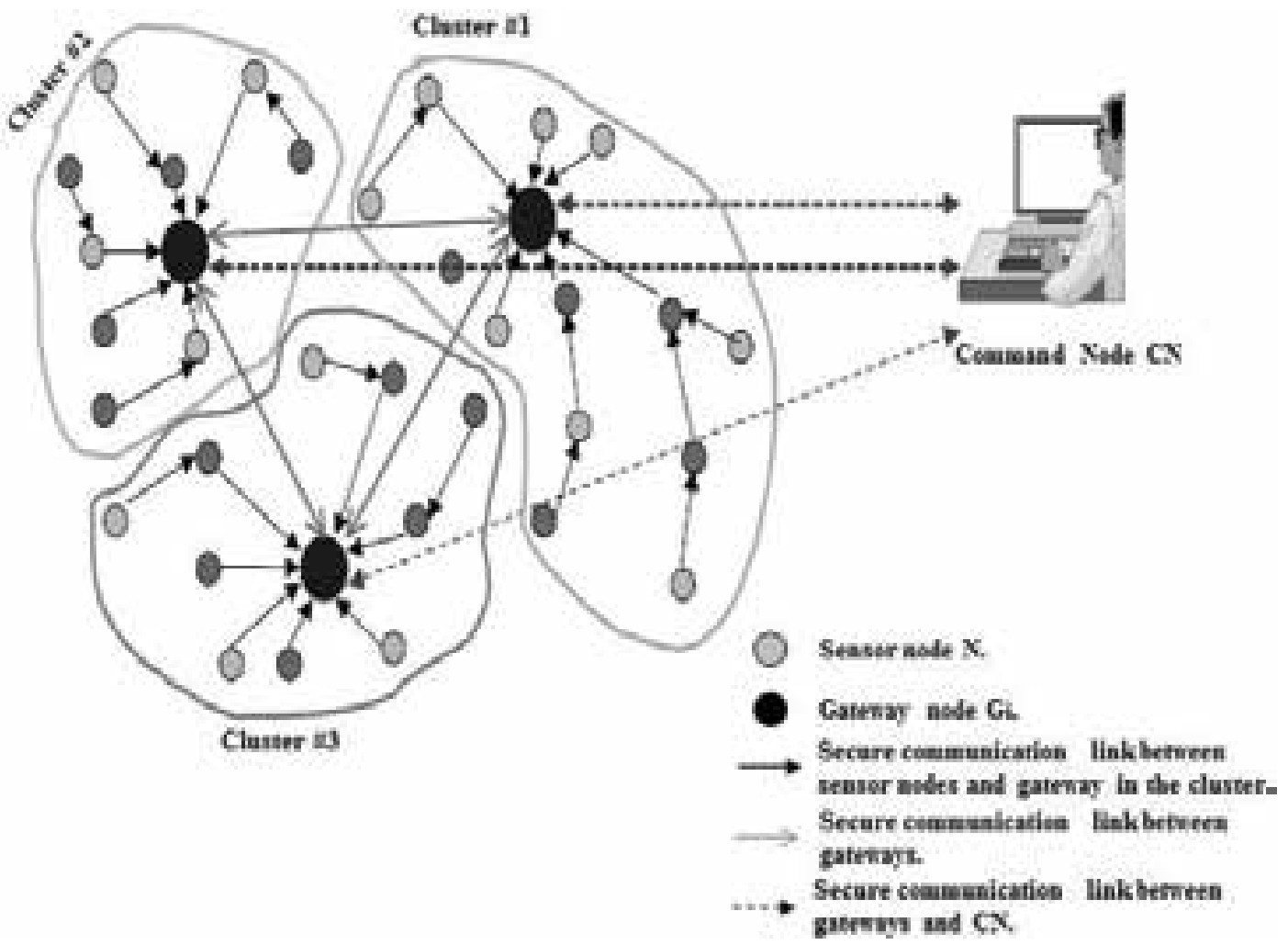

**Figure 3.** A low-energy key management protocol for WSNs.

The scheme provides a cost-effective key management infrastructure for security wireless sensor networks and the sensor nodes are not required to store large numbers of keys. Moreover, the scheme supports key revocation and renewal mechanisms, an energy conserving technique to provide key management for sensor networks presented as well.

## 5. Combined techniques based on predistribution scheme

### 5.1. Location-dependent key management scheme

Relying on their environment positions of nodes, the location-dependent key management scheme in Ref. [10] decides which keys to load on each node. All sensor nodes in this approach are considered to be static and communicate through encrypted links, also new nodes can be added at any time. Additionally, this scheme is assumed that nodes are transmitting at different power levels and different ranges. In this scheme, the nodes that transmit at different power levels and are tamper proof are called anchors. Each sensor node in the location-dependent key management scheme is preloaded with a subset of keys along with a single common key that is shared between all nodes. These keys are computed by a key server and placed into a key pool. However, the anchor nodes do not get keys from the key pool [11].

### 5.2. Localization encryption and authentication protocol

Zhu et al. [12] innovated a key management scheme called localization encryption and authentication protocol (LEAP). This scheme offers the main WSNs security requirements such as confidentiality and authentication. Also, LEAP supports diverse communication patterns, including unicast which is addressing a single node, a local broadcast which is addressing a group of nodes in a neighborhood, and global broadcast by addressing all the nodes in the network. One of the crucial points in LEAP is survivability, where compromising of some sensor nodes does not affect the entire network. Furthermore, LEAP is based on the theory that different types of messages exchanged between sensor nodes, where all the packets transferred in the network always need to be authenticated and encryption of packets carrying routing information is not always needed. In LEAP, each sensor node stores four types of keys as illustrated in **Figure 4** including individual, pairwise, cluster, and group.

- *Individual key:* This is a unique key that is shared between each sensor node and the base station in the network. Sensor nodes use this key to secure the communication link with the base station. The individual key is generated and preloaded into each sensor node before the deployment process. The individual key $K_{IN}$ for node *N* (each node has a unique id) is generated as follows:

$$K_{IN} = fK_m(N) \tag{9}$$

where *f* is a pseudo-random function and $K_m$ is the master key known only to the controller. Thus, the controller might only keep its master key in order to generate any individual key rather than store all the individual keys. In fact, the computational efficiency of pseudorandom functions is negligible, thus the controller generates individual keys on the fly.

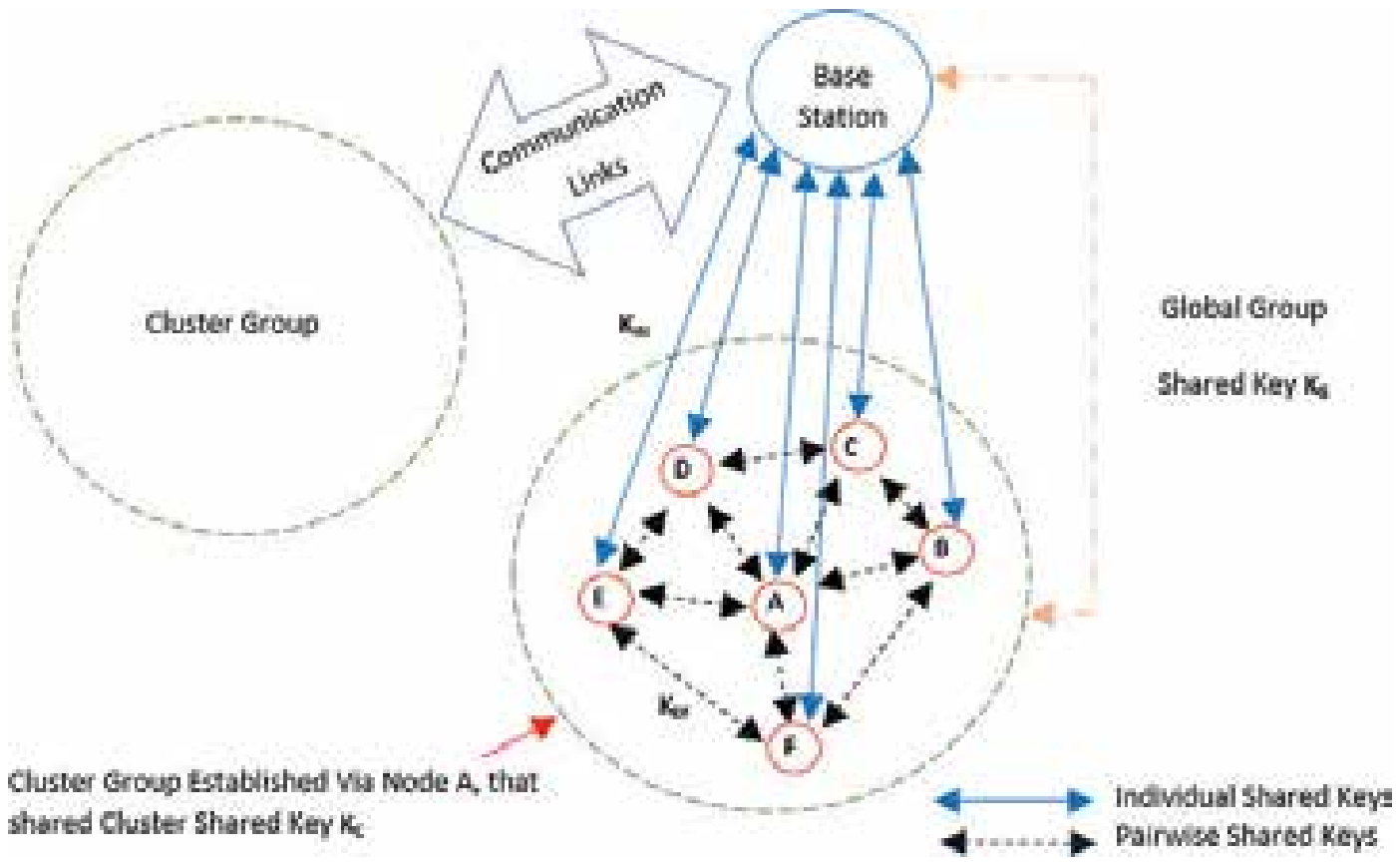

**Figure 4.** Communication links in LEAP.

- *Pairwise key:* This key is shared between each sensor node and its immediate neighbors (one-hop neighbors) in the network. Pairwise Key is used to secure communication links between sensor nodes which are used to transfer individual messages like sharing a cluster key or sending data to an aggregator node. Therefore, this type of pairwise key is most commonly used in the network. This scheme assumes that neighbors of a node are relatively static and a sensor node that is being added to the network will discover most of its neighbors at the time of its initial deployment ($T_{est}$). Moreover, [12] showed in practice that $T_{est}$ is smaller than the time interval ($T_{min}$) for an attacker to compromise a sensor node. There are four stages that describe the demonstration of the pairwise key establishment of new deployed node (e.g., *N*) in the network. These stages are key predistribution, neighbor discovery, pairwise key establishment, and key erasure.

- *Key predistribution:* During the initial stage of key predistribution, the $K_{IN}$ is generated by the controller and loaded into each node. Then each node (e.g., *N*) computes a master key.

$$K_N = fK_{IN}(N) \tag{10}$$

- *Neighbor discovery:* The first stage in deployment phase is discovering node's neighbors. By broadcasting a "hello" message that contains its ID, Node *N* starts to discover its neighbors, and in the same time starts a timer of $T_{est}$. ACK response of the "hello" message contains neighbor ID ( e.g., *V*). Node *N* authenticates the ACK using master key $K_V$ as follows:

$$K_V = fK_{IN}(V) \tag{11}$$

Since node *N* knows $K_{IN}$, it can also derive $K_V$ and then verify node *V*'s identity.

- Pairwise key establishment: Both Nodes $N$ and $V$ compute their pairwise key as the following:

$$K_{NV} = fK_V \tag{12}$$

- *Key erasure:* In the case a timer expires, node $N$ erases $K_{IN}$ and all the master keys (e.g., $K_v$) of the other neighboring nodes, which it generated in the neighbor discovery stage, and keeps its own master key $K_N$. Subsequently, the communication links between node $N$ and another node cannot be decrypted without the key $K_{IN}$.

  - Cluster key: This is a shared key ($K_c$) between multiple neighboring nodes, which is very important since it is mainly used for securing locally broadcast messages. In case, two neighboring nodes intend to send the same message to the base station, the node with a better message signal may be elected to send the message to the base station. A discovery of this process is only possible to implement if a node shares a common key with its neighboring nodes. In order to reduce the network communication overhead, LEAP uses cluster key to select which messages to transfer. The establishment of the cluster key between neighbors in the network relies on the shared pairwise key. Where node $N$ generates $K_c$ and before it transmits this key to its neighbors, it needs to encrypt that key using the pairwise key that is shared with each neighbor ($V_1$, $V_2$, $V_3$, …, $V_n$). Next step, node $V_1$ decrypts and stores the $K_c$, and then it sends back its cluster key to node $N$. In node revocation, the new cluster key will be generated and swapped between all the neighboring nodes in the cluster via the same aforementioned method.

  - A global group key: All nodes in the network shared this key including the base station. This key is used by the base station to encrypt messages which are broadcasted to all nodes in the network. A basic approach to establish a global key $K_G$ for a sensor network is to preload each node with the global key in key predistribution stage. However, this key needs to securely update in case a compromised node is revoked. Ref. [12] proposed an efficient mechanism that relied on cluster keys ($Kc$), where the transmission will only be one key. The mechanism was employed in Ref. [13] for broadcast authentication. The sensor nodes in Ref. [13] are organized into a breadth-first spanning tree, where each node remembers all the neighbors including the parent and children of a spanning tree [14].

## 6. Evaluation of the performance of key predistribution scheme

This section evaluates the performances of aforementioned predistribution schemes in terms of distribution technique (DT), node type (NT), scalability (S), resistance value against node capture (RV), key connectivity (KC) and key storage overhead (KSO). **Table 1** illustrates the comparison between the schemes.

| Scheme | KDT | NT | S | RV | KC | KSO | Ref |
|---|---|---|---|---|---|---|---|
| The basic probabilistic key predistribution scheme | Probabilistic | All | Good | $1-\frac{k_{\text{ring}}}{k_p}$ | Medium | $K_{\text{ring}}$ | [1] |
| Q-random key predistribution Scheme | Probabilistic | All | Medium | $1-\frac{k_{\text{ring}}!(k_{\text{ring}}-q)!}{(k_{\text{ring}}-q)!k_p!}$ | Poor | $K_{\text{ring}}$ | [2] |
| The basic multipath reinforcement scheme | Probabilistic | All | Good | $1-\frac{P(2P-P^2)^{N_N}}{0.5865nd^2N_N}$ | | | [2, 3] |
| The basic probabilistic key predistribution scheme using deployment acknowledge | Probabilistic | All | Good | $1-\frac{\frac{m}{\lvert S\rvert}}{\left(1-\frac{m}{\lvert S\rvert}\right)^x}$ | Good | $\le K_{ring}$ | [3, 4] |
| The polynomial based pairwise key deterministic predistribution scheme | Deterministic | All | Excellent | $1-\frac{\left(\frac{k}{\lvert S\rvert}\right)}{\left(1-\frac{k}{\lvert S\rvert}\right)^x}$ | Good | $ck$ where ($K_{\text{ring}} \le s$) | [5] |
| The polynomial pool-based using a grid-based key scheme [++] | Deterministic | All | Good | $1-\frac{12m-6}{N}$ | Good | $2(t+1)\log(q)+2(t+1)l$ | [6–8] |
| A low-energy key management protocol for wireless sensor networks scheme | Deterministic | 1. Cluster gateways | Excellent | $1-\frac{G-1+CN_1+C_1}{C_{i+1}+CN_{i-1}+\sum_1^t G_i-1}$ | Poor | $G+C_i+1$ | [9] |
| | | 2. Command nodes | | $1-\frac{G}{n+G}$ | | $G$ | |
| | | 3. Ordinary nodes | | $1-\frac{1}{n-1}$ | | $2\,K$ | |
| Localization encryption and authentication protocol | Combined | All | Excellent | $1-\frac{1+2N_n+n}{\frac{n(n-1)}{2}+2N_n}$ | Excellent | $2N_n+3$ | [11–13] |

$k_{\text{ring}}$ : Key ring.
$k_p$ :Key Pool.
$q$: composite random key.
$p$: Basic propabilty of sompromise link.
$d$: Probability of sharing sufficient communication keys.
$n$: Number of deplied sensor nodes.
$\lvert S\rvert$: Size of the global key pool.
$m$: Number of keys carried by each sensor node.
$N_n$: Number of node neighbors.
$x$: Number of compromised nodes.
$G_k$: Keys generation.
$C$: Total number of independently generated.
Keys via a unique generation key $G_k$.
$S$: Polynomials in the sensor node.
$G$: Number of gateways
$C_i$: Number of nodes in the cluster.
$CN_i$: command node.

**Table 1.** Comparison between predistribution schemes.

### 6.1. Key distribution technique (KDT)

The aims of the key management protocol are to establish secure communication links that are used to exchange data between sensor nodes and update the encryption keys in case of some node compromised. KDT refers to the approach of distributing secret keys between authorized sensor nodes in WSN in order to establish secure communication links.

### 6.2. Node type (NT)

As we defined the KDT above, sensor nodes are classified into different categories based on their function. In some cases in WSNs, nodes could work as supervised nodes or supervisor nodes. Thus, in this section, the NT indicates to node classification in the networks.

### 6.3. Scalability (S)

The term of scalability in WSNs means each key management scheme must allow for the variation in the size of the network. In other words, distributed scheme must be able to add more nodes in case of new nodes request to join the network.

### 6.4. Resistance value against node capture (RV)

As adversary might attack the network by compromising one or more nodes in the network and reveal the secure communication links between sensor nodes, the attacker can compromise the entire network. Therefore, key management scheme must resist against such attacks. In this section, we evaluate the RV as the ratio between number of compromised communication links ($N_{CCL}$) and the total communication links ($N_{TCL}$) in the network when one sensor node is compromised:

$$RV = 1 - \frac{N_{\mathrm{CCL}}}{N_{\mathrm{TCL}}} \tag{13}$$

where $RV$ = 1 implies that capturing one sensor node in a key management scheme will not affect the other secure communication links in the networks. However, RV = 0 implies that capturing one sensor node in a key management scheme will lead to compromise the rest of the security communication links in the network.

### 6.5. Key connectivity (KC)

KC means the probability of neighboring nodes sharing some common keys or nonneighboring nodes securely communicating via secure intermediate nodes sharing common keys. Therefore, the communication between sensor nodes in the network must be high.

### 6.6. Key storage overhead (KSO)

Due to the limitations on sensor nodes in term of memory space, KSO is considered as the most important constraint in designing suitable key management scheme in WSNs. In this section, we evaluate KSO as the total number of keys stored in each sensor node in the network.

## 7. Secure and Efficient Key Coordination Algorithm for Line Topology Network maintenance for use in maritime wireless sensor networks

The Secure and Efficient Key Coordination Algorithm for Line Topology Network (SEKCA) is a symmetric security scheme for a maritime coastal environment monitoring-based WSN [15, 16].

The scheme provides security for travelling packets via individually encrypted links between authenticated neighbors, thus addressing the main issues in security mechanisms based on the symmetric encryption algorithm, including memory, storage space, key generation, and a reiteration of a global rekeying process. Furthermore, SEKCA proposes a dynamic update key based on a trusted node configuration, called a Leader node ($L_n$), which works as a trusted third party. The technique has been implemented in real time on a Waspmote test bed sensor platform [16]. The deployment of a Waspmote sensor node integrated with the XBee 802.15.4 pro module in an outdoor environment using IEEE 802.15.4 /2.4GHz standard will be used to demonstrate the implementation of the algorithm. This mechanism relies upon the concepts of location-based routing [17]. Packets travel between nodes based on the information of the next repeater node. Subsequently, the static position of the sensor node encourages proposing a strategy for managing and controlling the transmission of secure packets between wireless nodes. In addition, maintaining network connectivity when removing or joining nodes makes it necessary to know the identification of the neighboring node/nodes in order to exchange the cryptographic keys and create a secure communication link. The main idea behind this work is to allow an ordinary node to determine its authenticated neighbor without the use of complex computations. Nodes will make their decision depending on the recommendation message from the $L_n$ [16]. The $L_n$ is located at a calculated distance from the line topology of ordinary nodes. The packet travels from the source to the destination based on the location of the nodes in the neighboring node's list of members, which is coordinated from/by the $L_n$. Each ordinary node must forward the packet to its next authenticated neighbor through link encryption with its adjacent key, "$k_{nj}$". The fundamental motivation behind this strategy is to configure a network line topology with simple and scalable security algorithms [16]. **Figure 5** illustrates the deployment of the nodes in our scheme and **Table 2** provides details on the notation used [16].

### 7.1. Implementation outline of SEKCA

The Waspmote sensor node as shown in **Figure 6** is provided with different frequency radio and protocols. In SEKCA, the XBee-Pro protocol is used for communication between nodes. This provides for a maximum communication distance of 7000 m between nodes which is ideal for the line topology used in a maritime coastal environment monitoring.

SEKCA is based on Advanced Encryption Standard (AES) with a key length of 128 bits, where AES encrypts a block of elements using the electronic codebook (ECB) encryption mode as shown in **Table 3**.

Sensor data "M" is encrypted in the application layer via software with AES 128 using the source key "$k_s$", which is shared exclusively between the source and the destination nodes. Then, the encrypted frame is encrypted again with the shared adjacent key "$k_{nji}$" (AES-128), which is shared exclusively between every set of two neighbors as in Eq. (14). The repeater node that forwards the sensor data to the destination in the network will decrypt the information once using the shared adjacent key, "$k_{ni}$" [16]. Then, to ensure complete confidentiality and privacy, before forwarding the data to the next repeater, the node will encrypt it via its adjacent key, "$k_{nj}$". Thus, the repeater will not be able to see the original sensor data transmitted due to the encryption with the source key, "$k_s$" [16]. Equation below shows this process where the red text indicates decryption with a shared adjacent key which is shared with the neighbor of the node performing the encryption [16].

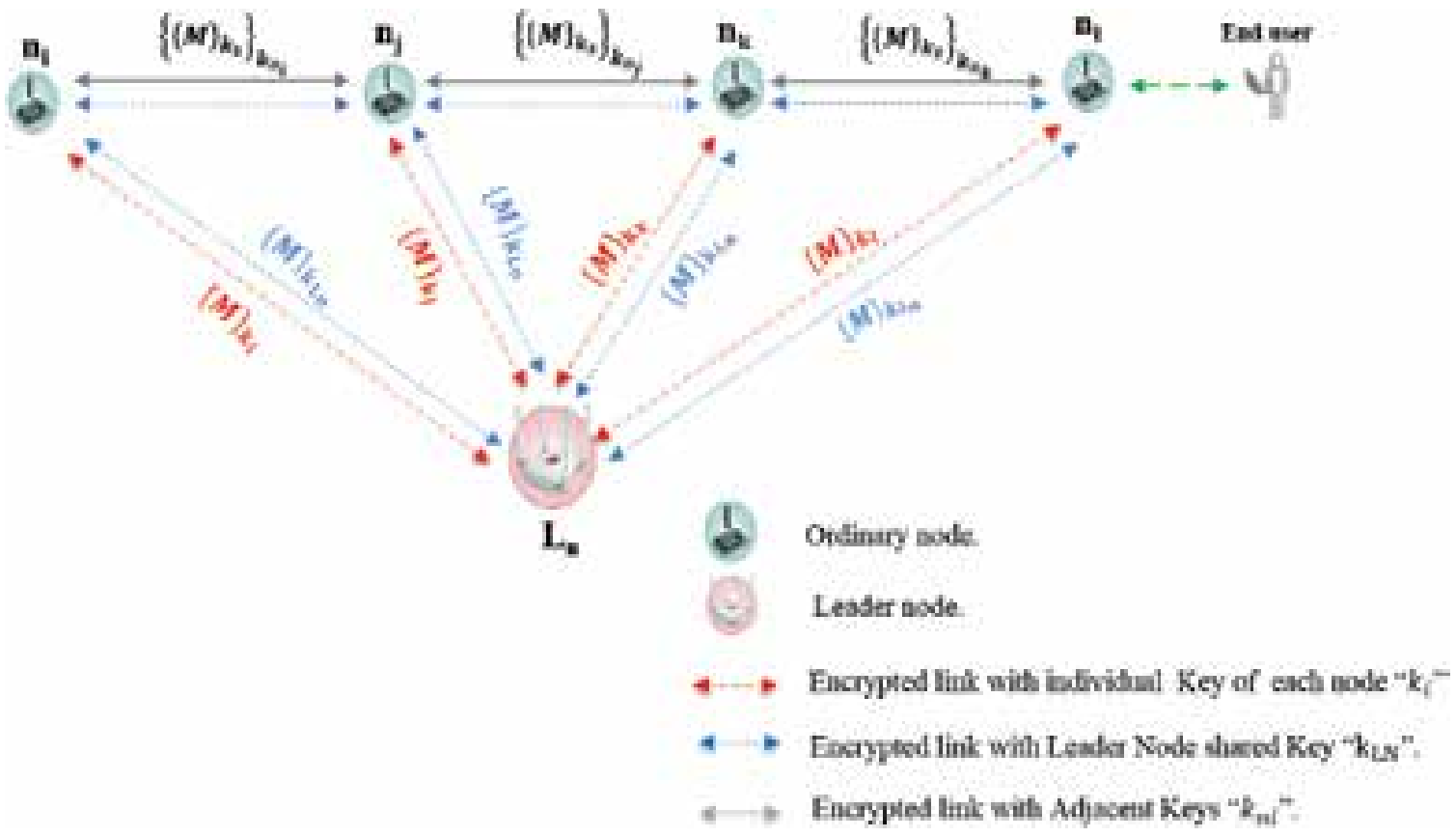

**Figure 5.** Nodes in line topology “originally published in Ref. [16], under a CC BY license”.

$$\rightarrow nj{:}\left\{\{M\}_{k_s}\right\}_k \text{then at } nj{:}\left\{\left\{\{M\}_{k_s}\right\}_k\right\}k \text{ then } nj \rightarrow nk{:}\left\{\{M\}_{k_s}\right\}_{k_{nj}} \text{etc.} \tag{14}$$

All ordinary nodes and the $L_n$ must share a master key called the Leader node key, “$k_{Ln}$”. This key is used for all the confidential communication between network members in processes such as when a new member joins the network, and for monitoring the behavior of the ordinary nodes [16]. The individual key “$k_j$” is a unique predistributed key between every ordinary node and the leader node. This key is used in the re-keying phase during the revocation process [16].

| Symbol | Description |
|---|---|
| $n_i$, $n_j$, $n_k$, $n_l$ | Ordinary nodes in the line topology. |
| $Ln$ | Leader node. |
| $\{\ \}_k$ | Symmetric encryption/decryption with key $K$. |
| $M$ | Transmitted sensor data. |
| $k_s$ | Secret encryption source key. |
| $k_{ni}$ | Secret encryption adjacent key of node $i$. |
| $k_i$ | Secret encryption individual key of node $i$. |
| $k_{Ln}$ | Secret encryption leader node key. |
| $\{\{M\}_{ks}\}_{kni}$ | Sensor data encrypted with source and adjacent keys. |
| $\{M\}_{ki}$ | Message encrypted with individual key of node $i$. |
| $\{M\}_{kLn}$ | Message encrypted with leader nod key. |

**Table 2.** Explanation of notation used in **Figure 1**.

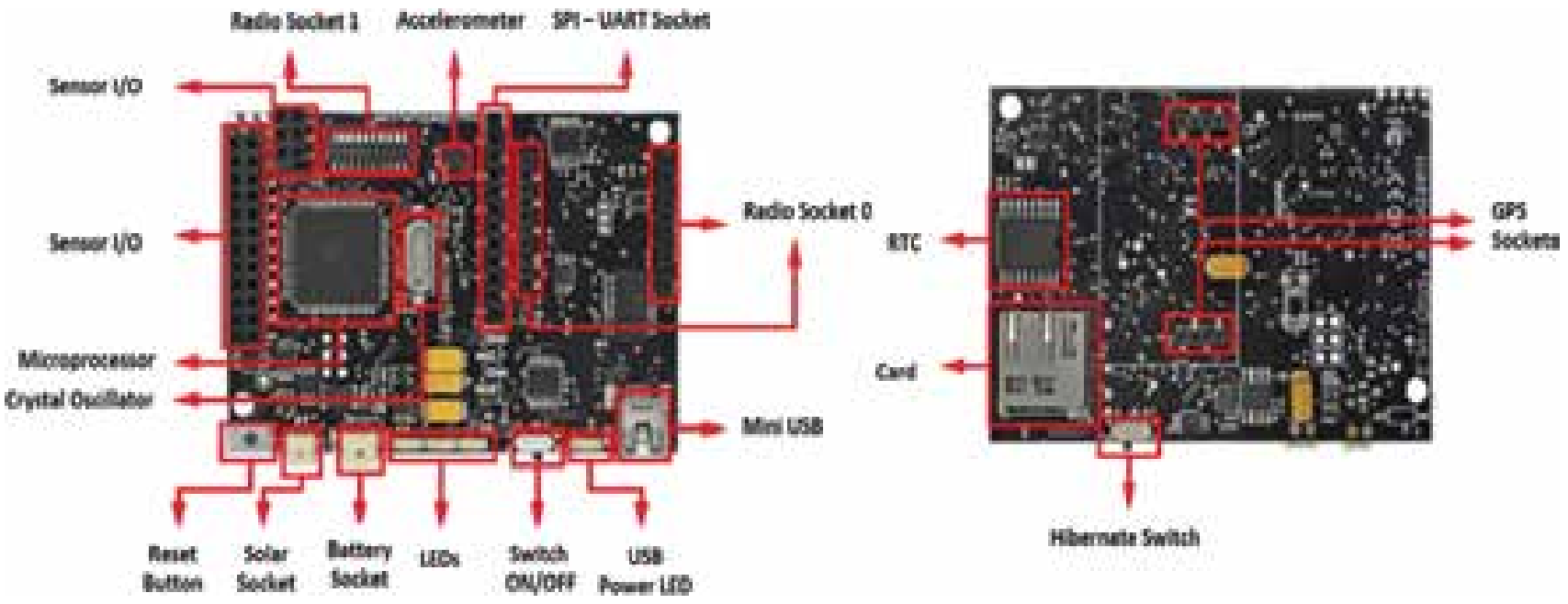

**Figure 6.** Waspmote components top and bottom sides “originally published in Ref. [16], under a CC BY license”.

| Algorithm | Key size | Data block size | Mode cipher | Padding |
|---|---|---|---|---|
| AES-128 | 128 bits | 16 bytes | ECB | ZEROS |

**Table 3.** AES-128 with ECB cipher mode and zeros padding “originally published in Ref. [16], under a CC BY license”.

In the case of a network member being revoked, only the leader node key ‘$k_{\text{ln}}$’, and one adjacent key needs to be renewed [16]. Subsequently, every node in the network has a key re-generation mechanism to create a new key [16]. This mechanism relies upon the Message Digest 5 algorithm (MD5) outlined in **Table 4** and a hash of the Real Time Clock (RTC) value as shown below.

$$k = H(\text{RTC}) \tag{15}$$

| Algorithm | Output size (bits) | Internal state size (bits) | Block size (bits) | Max message size (bits) | Word size (bit) |
|---|---|---|---|---|---|
| MD5 | 128 | 128 | 512 | 264-1 | 32 |

**Table 4.** MD5 hash algorithm “originally published in Ref. [16], under a CC BY license”.

Due to the straight line network topology, only the node located before the revoked node must update its own adjacent key "$k_{nj}$". This node then needs to share its new key with the new neighbor that replaces the revoked node in the authenticated neighbors list [16]. This stage is coordinated by the $L_n$ via a method of unicasting a revoked message that is encrypted with a predistributed individual key “$k_j$” [16].

### 7.2. Practical outdoor implementation of SEKCA

The outdoor implementation involves four Waspmote nodes, a Waspmote Gateway, four XBee 802.15.4 Pro modules with antennae, an MC1322x USB ZigBee dongle, an Agilent 66321D mobile communication DC Source, a Waspmote Pro IDE version 04 with Waspmote Pro API version 013 software based on Arduino, X-CTU provided by Digi, and the Wireshark network analyser [16]. Each Waspmote was placed on a fixed pole at a height of 80 cm from the ground. The effects of temperature and humidity on RSSI in WSNs as in Ref. [18] were considered. In this scenario the temperature was between 20 and 21°C and the humidity was 70%. SEKCA was implemented into three scenarios [16]:

- Four nodes and the gateway at a distance of 80, 120, or 160 m between end points in line topology.
- Three nodes and the gateway at a distance of 80, 120, or 160 m between end points in the line topology.
- One repeater node between sender and the gateway at a distance of 80, 120, or 160 m between ends in the line topology [16].

In this experiment, the received single strength indicator (RSSI) was measured at the gateway. **Figure 7** shows the average value of RSSI which was measured after receiving 300 encrypted packets of 79 bytes in size at a baud rate of 115,200 bps [16]. These values were measured in the three aforementioned scenarios of distances and repeaters. The signal strength in the 80 m scenario was the strongest in the all case of one, two, and three repeaters at exactly—48, −41 and −36 dB, respectively. However, in the case of one repeater the 160 m scenario had the minimum signal strength value at −64 dB. While, in the case of all possible scenarios the maximum achievable signal strength is −36 dB, which is the 80 m distance with three repeaters scenario. These measurements are then used by the leader node to determine the positions of future new joining nodes into the line topology [16].

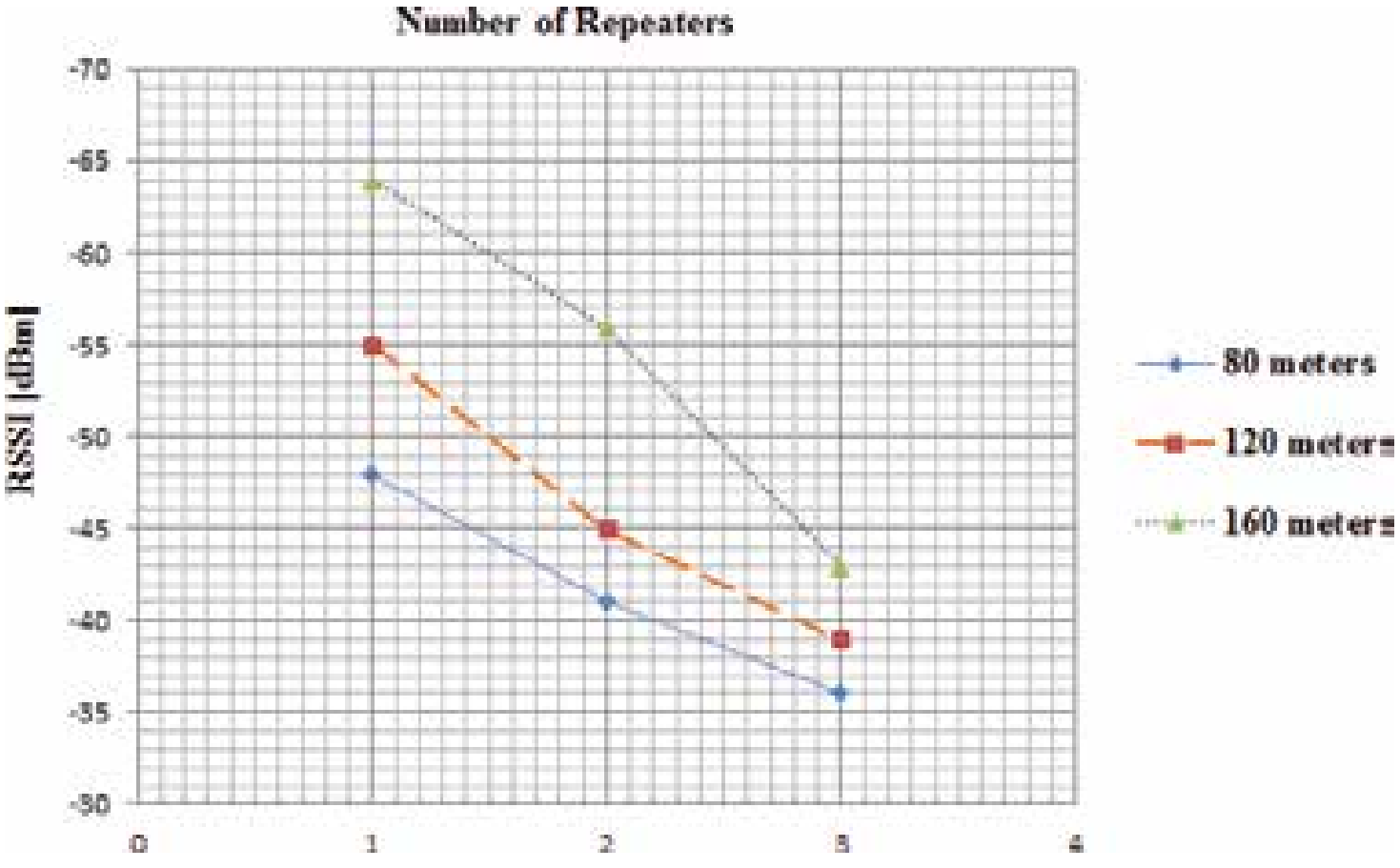

**Figure 7.** Average RSSI in three different scenarios of outdoor implementation "originally published in Ref. [16], under a CC BY license".

Data processing relies on the size of the data and the approach used in processing this data. Furthermore, SEKCA has been adapted to use minimum current consumption for data processing. The current consumption of the transmission data has been improved by using sleeping schedules for the receiving/transmitting modules (the XBee current consumption is from 37 to 64 mA with the mode ON fully operational). The measurements were taken using the 66321D-Agilent with input 3.689 V and 0.19999 Ω resistance. The average current consumption of the XBee module is 64.9123 mA when the leader node is operational. The measured current consumption has been improved from 64.9123 to 7.4355 mA, when the system returned to sleeping mode after transmitting/receiving data, which is a reduction of over 88%. **Table 5** provides the current consumption of the scheme modes, this includes when the XBee module is fully functional and sleeping [16].

| Scheme mode | Max measured current consumption (mA) | Min measured current consumption (mA) | Average measured current consumption (mA) |
|---|---|---|---|
| Fully functional system | 66.732 | 63.5297 | 64.9123 |
| XBee module (sleeping) | 9.4567 | 5.6248 | 7.4355 |
| XBee module current consumption (awake) | 57.2753 | 57.9049 | 57.4768 |

**Table 5.** Measured current consumption of scheme modes "originally published in Ref. [16], under a CC BY license".

SEKCA has improved some common security issues in WSNs [16], including:

- *Efficiency:* 88.55% of current consumption was reduced using the XBee sleeping mode. This will improve the lifetime of all network sensor nodes.
- *Data confidentiality:* By doubling the encryption of the messages, we ensure that only the gateway in the network can decrypt the original data (using AES 128) and after that, we establish peer-to-peer encryption between repeaters [16].
- *Authentication:* Each node has an individual key shared with the leader node, which is used to "sign" the messages in order to ensure the authenticity of the new neighbor and the rekeying. This signing is performed when the leader node reconfigures the network topology in the case of joining or revoking network members [16].
- *Memory capacity storage:* By using the re-generation function, there is no need for each node to store as many keys as are required in some key predistribution schemes where a large pool of keys is located in each node before deployment. In our scheme, each node has only four keys. Therefore, every node requires only a small amount of memory for key storage, which is four times 128 bits [16].
- *Non-repudiation:* by signing the XBee acknowledgment messages, there is now verifiable proof that the information sent has really been received by a specific sensor node. This signing using a shared source/destination key, *Ks*, that provides weak nonrepudiation as a public key scheme is not yet available in our system.

- *Forward security:* In the reconfiguration of the scheme, new neighbors could not communicate directly before authentication process. The authentication process coordinates via a leader node [16].
- *Backward security:* When a node is revoked from the network, the system ensures that the node cannot receive any new data from the network [16].

## 8. Toward connecting Libelium Motes to IoT sensors via Intel Galileo

Currently, the system being proposed uses Libelium motes, which connect sensors to the Intel Galileo board. Different Libelium motes show Arduino wireless modules that may be connected to Intel Galileo. These modules are RFID, NFC, Bluetooth, BLE, Wi-Fi, GPRS, 3G, and GPS, which may be connected as Libelium's open hardware division. Now, the Intel Galileo is based on the single core 32bit, 400MHz Quark SoC X1000 processor. In addition, it supports 3.3 or 5 volt shields and has an Ethernet port and two USB ports. Moreover, Galileo software is compatible with Windows, Mac OS and Linux, and the Intel Galileo development board is a prodigious tool for quickly prototyping simple, interactive designs for different types of IoT projects. According to the related works such as Refs. [18–20], it is noticeable that Galileo board is a promising and great tool for any potential IoT projects. When integrating it with sensors it can make big gains thereby bringing several IoT projects to life. Various IoT objects can access and control components such as temperature sensors, humidity sensors, gas sensors and light sensors, and so on. Essentially, sensors are considered the eyes and ears of potential IoT objects by many engineers and researches. This paradigm is proposed to be an end-to-end two-way communication that can be set-up with MQTT protocol as it is a secure solution for an end-to-end IoT system via the cloud. This includes encrypted data in a cloud to keep the data safe in the cloud.

## 9. Conclusion

This chapter presents an overview of existing key management distribution schemes of a WSN running a symmetric encryption security scheme. A review of probabilistic and deterministic basic key predistribution techniques has shown that the key management-based symmetric encryption algorithm requires less computation and energy consumption; however, these techniques have weaknesses in scalability, resistance value against node capture, and key storage overhead. These weaknesses make them nonapplicable to large-scale wireless sensor networks. Furthermore, different probabilistic and deterministic schemes' basic key predistribution technique was compared in terms of key distribution technique (probabilistic/deterministic), scalability, resistance value against node capture, key connectivity, and key storage overhead in order to show an efficient symmetric-key based scheme for WSNs. In addition to the key management schemes for WSNs, Secure and Efficient Key Coordination Algorithm for Line Topology Network (SEKCA) was also described. The algorithm tackled number of security services such as that made the re-keying process a local operation and minimized the use of key memory storage. The technique was implemented on the Waspmote platform in real time and analysed in terms of a received single strength indicator (RSSI) and current consumption. As cloud computing addresses the issues of data storage, it is also providing a secure solution for an end-to-end IoT system; whereas, security keys could be generated and stored in the cloud.

## Acknowledgements

This work was part supported by the SFI Centre for Marine Renewable Energy Ireland (MaREI). Grant References: 12/RC/2302 and 14/SP/2740. In addition, part of the chapter is taken from our previous work as referenced in [16].

## Author details

Walid Elgenaidi[1], Thomas Newe[1,3], Eoin O'Connel[1,3]*, Muftah Fraifer[2], Avijit Mathur[1], Daniel Toal[3] and Gerard Dooly[1,3]

*Address all correspondence to: eoin.oconnell@ul.ie

1 Department of Electronic and Computer Engineering, Optical Fibre Sensors Research Centre, Mobile & Marine Robotics Research Centre, University of Limerick, Limerick, Ireland

2 Department of Computer Science and Information System, Interaction Design Centre, the University of Limerick, Limerick, Ireland

3 Department of Electronic and Computer Engineering, Mobile and Marine Robotics Research Centre, University of Limerick, Limerick, Ireland

Section 2

# Communications Security

Chapter 7

# Energy-Secrecy Trade-offs for Wireless Communication

Ruolin Zhang

Additional information is available at the end of the chapter

**Abstract**

This chapter investigates the secrecy-energy trade-offs for communication in wireless networks. It is shown that privacy requirements for applications such as image or video transmissions do not require perfect secrecy, and the level of privacy can be quantified using a Rate-Distortion metric. Using information theoretic analysis, the chapter formulates analytic secrecy trade-offs for various communication channel models. In particular we analyze the advantage of partial secrecy for the Gaussian and Rayleigh fading channel models and the MIMO channel. The impact of secrecy requirements and the inherent secrecy-energy-connectivity trade-offs are also analyzed for networks of wireless nodes.

## 1. Introduction

The secrecy of wireless communication was studied from an information theoretic perspective, and equivocation rate was defined to characterize secrecy capacity. Equivocation rate measures the uncertainty that the eavesdropper decodes about the transmitted message with a certain rate during a transmission. Perfect secrecy can be achieved if the equivocation rate is equal to the transmitted rate. In this chapter, the notion of partial secrecy is introduced, for the scenario that perfect secrecy cannot be achieve. The reason for studying partial secrecy is based on the consideration about another key performance metric of communication systems, energy efficiency. The energy savings can be obtained when a partial secrecy requirement is enforced.

For many practical applications, such as voice and image transmission, a certain percentage of loss will be accepted, since decoded data is useless for the eavesdropper. The quality of the source data reconstruction for those applications is typically characterized by the rate-distortion function. For example some certain scenarios will keep the eavesdropped message useless for the eavesdropper, such as some images for which all the key objects in the image have been transmitted with perfect secrecy, while the other objects and background in the image that are not important are transmitted without secrecy constrains.

To characterize the partial secrecy, the physical layer and application layer secrecy requirement was analyzed jointly. At the physical layer, the equivocation rate characterizes the uncertainty remaining at the eavesdropper after receiving the transmitted message. At the application layer, the application layer secrecy would be achieve if there is a close to no message which is decoded by the eavesdropper at application layer, and from the information theoretic view, the application layer secrecy is characterized by the rate-distortion function. The application lay metric is the rate-distortion function, and physical layer secrecy metrics is equivocation, which directly translates to the secrecy capacity.

In the remaining, the energy-secrecy trade-offs was studied for one point to one point communication, such as the Gaussian fading channels, MIMO channel and network level communication. In Section 2, the background was mentioned and in Section 3 system model was introduced, and in Sections 3 and 4 the partial secrecy in one point to one point communication and in network level communication has been studied. In Section 5, the further work is depicted.

## 2. Background

In the chapter, the work is based on the rich literature about information theoretic results for the secrecy capacity for various channel models. The perfect secrecy was in introduced and the wire-tap channel model was also first introduced in Ref. [1]. Later, the classic Gaussian wire-tap channel was introduced in Ref. [2], and the perfect secrecy capacity was first derived. The perfect secrecy capacity was analyzed for fading channels and multiple antennas, for example, in Refs. [3, 4]. An extension of the wire-tap channel model when a common message is transmitted was first proposed in Ref. [5], and it is known as the Broadcast Channel with Confidential Messages (BCC). In Ref. [5] capacity-equivocation region and the perfect secrecy capacity of the discrete memory-less BCC was characterized, while in Refs. [6–8] a more general channel model for the fading BCC was considered. Further results of the secrecy capacity of the BCC was studied [9–11].

Also more recently, multi-users extensions of the wiretap channel and broadcast channel have been investigated in Refs. [1, 12]. The multi-receivers wire-tap channel was considered, and its perfect secrecy capacity was derived for an arbitrary number of users in Refs. [13–15], and for two users in Ref. [16]. The perfect secrecy capacity of the Gaussian scalar multi-receivers wiretap channel, a degraded multi-receivers wiretap channel, was also derived in Refs. [17, 18]. Outage capacity analysis was also proposed for Gaussian MIMO channel in several papers in the literature, such as in Ref. [19].

About secrecy graph for network connectivity, based on a secrecy graph framework [20], earlier work in the literature has proposed network connectivity models, based on this eavesdropper worse than receiver channel condition, showing that network connectivity can be greatly affected by secrecy requirements, such as in Refs. [21, 22].

## 3. System model

Partial secrecy can be achieved when energy per bit cost need to be reduced, and thus for suitable applications the source may trade-off energy expenditure for secrecy rate. The common definition of secrecy capacity as the maximum rate at which the transmitted message can be transmitted with perfect secrecy was adopted. For this scenario this chapter considers that the initial stream of data can be split into private and non-private information sub-streams, but without imposing the condition that the non-private sub-stream needs to correctly decoded by the eavesdropper as in the previous work in Ref. [23]. $R_0$ and $R_s$ was denoted as the transmission rate of the non-private sub-stream and the private sub-stream, respectively, with the total transmission rate being defined as $R_t = R_0 + R_s$.

In partial secrecy, the equivocation rate may be smaller than the transmission rate at physical layer. From information theoretic, transmission rate, $R$, is measured in bits per transmission, if $R$ bits information are transmitted by the transmitter. Assume that a maximum transmission power $P$ can be used by the transmitter, and that transmitter has limited battery energy. Hence a performance metric, the energy per bit, is considered in partial secrecy. Therefore, the new metric of partial secrecy is characterized by energy per bit, $E_b$. Although the actual transmission rate is related to the system bandwidth, energy per bit is beyond the scope of that discussion. To characterize the energy efficiency of the transmission, the average energy per bit consumption was defined as

$$E_b = \frac{P}{R}\left(Joules/bit\right), \tag{1}$$

where $P$ is the average power transmission constraint, and $R$ is the number of bits per transmission.

In partial secrecy, the minimum Distortion, $D_{min}$, is the new metric at the application layer. The new metric is measured the transmitted message, which can be guaranteed to achieve at the eavesdropper. Also, the minimum Distortion is related to the transmission rate, and it accurately represent the transmitted message. For illustration purposes, a very simple rate-distortion function was characterized the distortion in the source reconstruction of transmitted message at the eavesdropper is defined as the ratio between the secrecy rate and the total transmission rate. The minimum guaranteed distortion metric, $D_{min}$, was defined as the percentage of the source rate that achieves secrecy, and thus will not be available at the eavesdropper:

$$D = \frac{R_s}{R_t}. \tag{2}$$

## 4. Partial secrecy in one point to one point communication

In Gaussian fading channel model, the partial secrecy applied in two cases: the known full channel CSI (Channel Statement Information) and known partial channel CSI, in which only the main channel CSI is known and channel from source to eavesdropper is not known.

Assume that the block length is large enough such that coding over one block can achieve a small probability of error. The channel input-output relationship is given by

$$Y = h_1 X + W, Z = h_2 X + V, \tag{3}$$

where $X$ is the channel input, and $Y$ and $Z$ are channel outputs at the legitimate receiver and eavesdropper, respectively. The channel gain coefficients $h_1$ and $h_2$ are proper complex random variables. We assume that hi is a stationary and ergodic random process. The noise processes $W$ and $V$ are zero-mean independent and identically distributed (i.i.d.) proper complex Gaussian random variables with variances $\mu^2$ and $\sigma^2$, respectively. The input $X$ is subject to the average power constraint $P$. This chapter assumes different noise power levels and different constant fading gain coefficients for the intended receiver and the eavesdropper. In the previous work [23], the ergodic partial secrecy capacity for this channel model is characterized as:

$$C_{ps}^{GuPS} = \left\{ \begin{array}{c} (R_0, R_1) \\ R_0 \le \frac{1}{2}\log\left(1 + \frac{(1-\beta)P|h_1|^2}{\mu^2 + \beta P|h_1|^2}\right) \\ R_1 \le \frac{1}{2}\log\left(1 + \frac{\beta P|h_1|^2}{\mu^2}\right) - \frac{1}{2}\log\left(1 + \frac{\beta P|h_2|^2}{\sigma^2}\right) \\ R_s \ge R_1 \end{array} \right. . \tag{4}$$

where $1-\beta$ is the fraction of the total power allocated to the common message, and $\beta$ is the fraction of the power budget that is allocated to the confidential message, which can be determined optimally [23], such as to maximize the partial secrecy capacity Eq. (4).

As in Ref. [23], given the power budget $P$, $\beta$ can be optimized to achieve the secrecy-capacity boundary, under the observation that the secrecy-capacity region is convex:

$$\max_{\beta\in[0,1]}\{\gamma_0 R_0(\beta) + \gamma_1 R_1(\beta)\}, \tag{5}$$

$$\beta\left(\frac{\gamma_1}{\gamma_0}\right) = \min\left\{\left(\frac{\gamma_1}{\gamma_0}\left(\frac{\sigma^2}{P|h_2|^2} - \frac{\mu^2}{P|h_1|^2}\right) - \frac{\sigma^2}{P|h_2|^2}\right)^+, 1\right\} \tag{6}$$

For $\beta \in [0,1]$, the condition should be satisfied

$$\frac{\sigma^2}{\sigma^2 - \mu^2|h_2|^2/|h_1|^2} \le \frac{\gamma_1}{\gamma_0} \le \left(1 + \frac{\sigma^2}{P|h_2|^2}\right)\frac{P|h_1|^2|h_2|^2}{\sigma^2|h_1|^2 - \mu^2|h_2|^2} \tag{7}$$

The choice of the ratio $\gamma_1/\gamma_0$, characterizes the energy-secrecy trade-off, by influencing both the achievable $R_s$, as well as the achievable $R_t$.

For applications more sensitive to transmission delay constraints, the outage partial secrecy capacity was investigate, defined as the maximum transmission and secrecy rate that can be supported for a given outage probability constraint, under the assumption that both the transmitter and the receiver have perfect channel-state information for both the regular transmission, as well as for the eavesdropper channel. Here the assumption is that the transmitter will defer transmission for the states that are associated with outage, and hence incur transmission delay.

$\left(\breve{R}_0, \breve{R}_s\right)$ was used to represent a target rate pair. The non-private sub-stream is transmitted at the rate $R_0$, while the private sub-stream is transmitted at the rate $R_s$. If the target rate pair is not achieved, an outage is claimed. The outage probability was defined as

$$P_{out} = \Pr\left\{\left(\breve{R}_0, \breve{R}_s\right) \notin C_s\left(\bar{h}, p\left(\bar{h}\right)\right)\right\}, \tag{8}$$

where $C_s\left(\bar{h}, p\left(\bar{h}\right)\right)$ is the secrecy capacity region for the channel with fading state, $\bar{h} = \left(h_1, h_2\right)$, and $p\left(\bar{h}\right)$ is the transmission power used by the source node. The transmission power is adapted to the CSI. The outage probability is considered under a long-term average power constraint $P$, so we have

$$E\left[p\left(\bar{h}\right)\right] \le P \tag{9}$$

For this scenario, the Rayleigh-fading broadcast channel with confidential messages was considered. Therefore $|h_1|^2$ and $|h_2|^2$ are exponentially distributed with parameters $\delta_1$ and $\delta_2$, respectively.

From the partial secrecy capacity region Eq. (4) we have the following two conditions:

$$\begin{aligned} R_0 &< \log\left(\frac{1+\dfrac{p(\underline{h})|h_1|^2}{\mu^2}}{1+\dfrac{\beta p(\underline{h})|h_1|^2}{\mu^2}}\right), \\ R_s &< \log\left(\frac{1+\dfrac{\beta p(\underline{h})|h_1|^2}{\mu^2}}{1+\dfrac{\beta p(\underline{h})|h_2|^2}{\sigma^2}}\right). \end{aligned} \tag{10}$$

Rewriting these conditions to reflect the constraints on the link gain values, such that the required transmission rate ($R_0$) and secrecy rates ($R_s$) are achievable, we obtain

$$\frac{\mu^2\left(2^{R_0}-1\right)}{p(\underline{h})\left(1-\beta 2^{R_0}\right)}<|h_1|^2$$
,
$$|h_2|^2<\frac{\sigma^2\left[\left(1+\frac{\beta p(\underline{h})|h_1|^2}{\mu^2}\right)2^{-R_s}-1\right]}{\beta p(\underline{h})}. \tag{11}$$

Hence, the outage probability can be computed as the probability that the above conditions hold,

$$\widehat{P}_{out}=1-\int_{f(R_0)}^{\infty}\int_{0}^{f(R_s)}\frac{1}{\delta_1}e^{-\frac{|h_1|^2}{\delta_1}}\frac{1}{\delta_2}e^{-\frac{|h_2|^2}{\delta_2}}d|h_2|^2d|h_1|^2=1-\exp\left(-\frac{1}{\delta_1}\frac{\mu^2\left(2^{R_0}-1\right)}{p(\underline{h})\left(1-\beta 2^{-R_0}\right)}\right)$$
$$+\frac{\delta_2}{\delta_2+\frac{\sigma^2}{\mu^2}2^{-R_s}\delta_1}\exp\left(-\frac{\sigma^2\left(2^{-R_s}-1\right)}{\beta p(\underline{h})\delta_2}-\frac{\delta_2+\frac{\sigma^2}{\mu^2}2^{-R_s}\delta}{\delta_1\delta_2}\frac{\mu^2\left(2^{R_0}-1\right)}{p(\underline{h})\left(1-\beta 2^{-R_0}\right)}\right), \tag{12}$$

where $f(R_0)=\left[\mu^2\left(2^{R_0}-1\right)\right]/\left[p(\underline{h})\left(1-\beta 2^{R_0}\right)\right]$ and $f(R_s)=\sigma^2\left[\left(1+\beta p(\underline{h})|h_1|^2/\mu^2\right)2^{-R_s}-1\right]/\left[\beta p(\underline{h})\right]$.

Similar as the known full channel CSI case, the partial secrecy-capacity region for the guaranteed partial secrecy channel can be determined as:

$$C_{ps}^{GuPS}=\begin{cases}(R_0,R_1)\\ R_0\le\frac{1}{2}\log\left(1+\frac{p_0|h_1|^2}{\mu^2+p_1|h_1|^2}\right)\\ R_1\le\frac{1}{2}\log\left(1+\frac{p_1|h_1|^2}{\mu^2}\right)-\int_0^{|h_1|^2}\frac{1}{2}\log\left(1+\frac{p_1|h_2|^2}{\sigma^2}\right)f\left(|h_2|^2\right)d|h_2|^2\\ R_s\ge R_1\end{cases}, \tag{13}$$

where $p_0$ is the power allocated to the no-private message, and $p_1$ is the power allocated to the private message. This partial secrecy channel region relies on the assumption of large coherence intervals and ensures that when $h_2 < h_1$. And the transmitter is under the power constraint

$$E\left[P(|h_1|)^2\right] \leq P. \tag{14}$$

For this scenario, the Rayleigh fading channel was considered. Therefore, in the Rayleigh fading channel, the partial secrecy capacity region could be changed to

$$C_{ps}^{GuPS} = \begin{cases} (R_0, R_1) \\ R_0 \leq \frac{1}{2}\log\left(1 + \frac{p_0|h_1|^2}{\mu^2 + p_1|h_1|^2}\right) \\ R_1 \leq \frac{1}{2}\log\left(1 + \frac{p_1|h_1|^2}{\mu^2}\right) - \frac{1}{2}\log\left(1 + \frac{p_1|h_1|^2}{\sigma^2}\right)\exp\left(-\frac{p_1|h_1|^2}{\delta_2}\right) + \\ \exp\left(\frac{1}{\delta_2 p_1/\sigma^2}\right)\left[Ei\left(\frac{|h_1|^2}{\delta_2} + \frac{1}{\delta_2 p_1/\sigma^2}\right) - Ei\left(\frac{1}{\delta_2 p_1/\sigma^2}\right)\right] \\ R_s \geq R_1 \end{cases} \tag{15}$$

It is easy to check that the objective function is concave in $p_1$ and hence, by derivation approach for maxing $R_s$, we get the following optimality condition

$$\frac{\partial(\gamma_0 R_0(p_1) + \gamma_1 R_1(p_1))}{\partial p_1} = \frac{|h_1|^2 \Pr\left(|h_2|^2 \leq |h_1|^2\right)}{\mu^2 + |h_1|^2 p_1} - \int_0^{|h_1|^2}\left(\frac{|h_2|^2}{\sigma^2 + |h_2|^2 p_1}\right) f\left(|h_2|^2\right) d|h_2|^2 = 0$$
$$p_0 = P - p_1. \tag{16}$$

For any main channel fading state is $|h_1|^2 \sim e^{x/\delta_1}/\delta_1$, and eavesdropper channel state is $|h_2|^2 \sim e^{x/\delta_2}/\delta_2$. The optimal transmit power level $p_1$ is determined from the above equation. If the obtained power level turns out to be negative, then the optimal value of $p_1$ is equal to 0.

In the Rayleigh fading channel, the condition can be rewrite as

$$(\gamma_1 - \gamma_0)\frac{|h_1|^2}{\mu^2 + |h_1|^2 p_1} - \gamma_1 \exp\left(-\frac{|h_1|^2}{\delta_2}\right)\frac{|h_1|^2}{\sigma^2 + |h_1|^2 p_1} - \gamma_1 \frac{\left(1 - \exp\left(-\frac{|h_1|^2}{\delta_2}\right)\right)}{p_1}$$
$$+ \gamma_1 \frac{\exp\left(\frac{1}{\delta_2 p_1^2/\sigma^2}\right)}{\delta_2 p_1^2/\sigma^2}\left[Ei\left(\frac{1}{\delta_2 p_1/\sigma^2}\right) - Ei\left(\frac{|h_1|^2}{\delta_2} + \frac{1}{\delta_2 p_1/\sigma^2}\right)\right] = 0 \tag{17}$$
$$p_0 = P - p_1.$$

If there is no positive solution to this equation for a particular, then we set $p_1 = 0$.

From the partial secrecy capacity region formula, the two conditions is abstained as following:

$$R_0 < \log\left(\frac{1+P|h_1|^2/\mu^2}{1+\beta P|h_1|^2/\mu^2}\right), R_s < \log\left(\frac{1+\beta P|h_1|^2/\mu^2}{1+\beta P|h_2|^2/\sigma^2}\right). \tag{18}$$

Rewriting these conditions to reflect the constraints on the link gains values such as given transmission rate ($R_0$) and secrecy rates ($R_s$) are achievable, we obtain:

$$\frac{\mu^2\left(2^{R_0}-1\right)}{P\left(1-\beta 2^{R_0}\right)} < |h_1|^2 \quad , |h_2|^2 < \sigma^2\left[\left(1+\frac{\beta P|h_1|^2}{\mu^2}\right)2^{-R_s}-1\right]/\beta P \tag{19}$$

Hence, the outage probability could be computed as the probability that the above conditions Eq. (19) do not hold.

$$\begin{aligned} P_{out} &= 1-\int_{\frac{\mu^2\left(2^{R_0}-1\right)}{P\left(1-\beta 2^{R_0}\right)}}^{\infty}\int_0^{\frac{\sigma^2\left[\left(1+\frac{\beta P|h_1|^2}{\mu^2}\right)2^{-R_s}-1\right]}{\beta P}} \frac{1}{\delta_1}\exp\left(\frac{|h_1|^2}{\delta_1}\right)\frac{1}{\delta_2}\exp\left(\frac{|h_2|^2}{\delta_2}\right)d|h_1|^2\, d|h_2|^2 \\ &= 1-\exp\left(-\frac{1}{\delta_1}\frac{\mu^2\left(2^{R_0}-1\right)}{P\left(1-\beta 2^{R_0}\right)}\right)+\frac{\delta_2}{\delta_2+\frac{\sigma^2}{\mu^2}2^{-R_1}\delta_1}\exp\left(-\frac{\sigma^2\left(2^{-R_s}-1\right)}{\beta P\delta_2}-\frac{\delta_2+\frac{\sigma^2}{\mu^2}2^{-R_1}\delta_1}{\delta_1\delta_2}\frac{\mu^2\left(2^{R_0}-1\right)}{P\left(1-\beta 2^{R_0}\right)}\right). \end{aligned} \tag{20}$$

In order to determine the partial capacity region for the MIMO fading channels, this chapter start from the results derived in Ref. [24] for perfect secrecy he MIMO Gaussian broadcast channel with non-private and confidential message. It expands the analysis in Ref. [24] to account for the fading model presented in partial secrecy, and consider that the private sub-stream will be transmitted as a private message (with rate $R_s$) and the no-private sub-stream will have no secrecy constraints and only the constraint that it will be correctly received by the intended receiver. The non-private stream will be transmitted with rate $R_0$.

The partial secrecy capacity formula can be then be determined as:

$$C_{ps} = \bigcup_{\beta\in[0,1]}\left\{\begin{array}{c} R_0 \le \frac{1}{2}\log\left|\mathrm{I}_M + \frac{(1-\beta)\frac{P}{N}}{\mu^2+\beta\frac{P}{N}}\mathrm{H}_1\mathrm{H}_1^H\right| \\ R_1 \le \frac{1}{2}\log\left|\mathrm{I}_M + \frac{\beta P}{\mu^2 N}\mathrm{H}_1\mathrm{H}_1^H\right| - \frac{1}{2}\log\left|\mathrm{I}_M + \frac{\beta P}{\sigma^2 N}\mathrm{H}_2\mathrm{H}_2^H\right|, \\ R_s > R_1 \\ R_t = R_0 + R_1 \end{array}\right. \tag{21}$$

where (·)H denotes the Hermitian transpose. In a full-rank system, Eq. (21) can be simplified by using singular value decomposition as

$$C_{ps} = \bigcup_{\beta_i \in [0,1]} \left\{ \begin{array}{l} R_0 \le \frac{1}{2}\sum_{i=1}^{M} \log \left| I_M + \frac{(1-\beta_i)\frac{P}{N}}{\mu^2 + \beta_i \frac{P}{N}} \lambda_i (H_1 H_1^H) \right| \\ R_1 \le \frac{1}{2}\sum_{i=1}^{M} \log \left| I_M + \frac{\beta_i P}{\mu^2 N} \lambda_i (H_1 H_1^H) \right| - \frac{1}{2}\sum_{i=1}^{M} \log \left| I_M + \frac{\beta_i P}{\sigma^2 N} \lambda_i (H_2 H_2^H) \right|, \\ R_t = R_0 + R_1 \\ R_s = R_1 \end{array} \right. \tag{22}$$

where $\lambda_i(H_1H_1^H)$ or $\lambda_i(H_2H_2^H)$ are the eigenvalues of $H_1H_1^H$ and $H_2H_2^H$. At the *i*th antennas, the power allocation is $(1-\beta_i)P$ for the non-private sub-stream, and the power allocation is $\beta_i P$ for the private sub-stream.

As in Eq. (22), given that the power budget $P$, $\beta_i$ can be optimized to achieve the secrecy-capacity boundary, and under the observation that the secrecy-capacity region is convex, we have the optimization condition:

$$\max_{\beta \in (0,1)} \{\gamma_0 R_0(\beta_i) + \gamma_1 R_s(\beta_i)\}. \tag{23}$$

By taking the derivative and setting it to 0 we get the optimal power allocation between the non-private and private sub-streams:

$$\beta_i = \frac{\gamma_1}{\gamma_0}\left( \frac{\sigma^2 N}{P\lambda_i(H_2H_2^H)} - \frac{\mu^2 N}{P\lambda_i(H_1H_1^H)} \right) - \frac{\sigma^2 N}{P\lambda_i(H_2H_2^H)}. \tag{24}$$

When there are stringent delay constraints, the transmitter cannot defer transmission and consequently, for some states of the channel, and outage may occur. Outage the event is defined as which the secrecy rate Rs is not contained in the partial secrecy capacity,

$$R_s \notin C_{ps}(H_1, H_2, P). \tag{25}$$

In case of an outage, the private sub-stream is not secured against eavesdropping. The probability of this event happening is referred to as the secrecy outage probability.

Formula (22) shows the partial secrecy capacity, as a function of the eigenvalues of matrix $H_kH_k^H$, which are random variables. The joint probability density function (pdf.) of these eigenvalues, after being ordered according to their amplitude, has been shown in Ref. [19] to be

$$p_{order}(\lambda_{k1},\ldots,\lambda_{kM}) = K_{M,N}^{-1}\left(\prod_i \lambda_{ki}^{N-M}\right)\cdot\left(\prod_{i>j}(\lambda_{ki}-\lambda_{kj})^2\right)\exp(-\sum_i \lambda_{ki}) \tag{26}$$

,

where $K_{M,N}$ is a normalizing factor.

Using the pdf in Eq. (26), the secrecy outage probability could be derived as follows

$$\begin{aligned}
P_S &= \Pr\left(\sum_{i=1}^{M}\left(\log_2\left(1+\frac{\beta P}{\mu^2 N}\lambda_i\left(\mathrm{H}_1\mathrm{H}_1^H\right)\right)-\log_2\left(1+\frac{\beta P}{\sigma^2 N}\lambda_i\left(\mathrm{H}_2\mathrm{H}_2^H\right)\right)\right)<R_s\right)\\
&= \Pr\left(\log_2\left(1+\frac{P}{\mu^2 N}\frac{1}{M}\sum_{i=0}^{M}\lambda_i\left(\mathrm{H}_1\mathrm{H}_1^H\right)\right)-\log_2\left(1+\frac{P}{\sigma^2 N}\frac{1}{M}\sum_{i=0}^{M}\lambda_i\left(\mathrm{H}_2\mathrm{H}_2^H\right)\right)<R_s\right)\\
&= \Pr\left(tr\left(\mathrm{H}_2\mathrm{H}_2^H\right)>\frac{1-2^{R_s}+M\frac{\beta_i P}{\mu^2 N}\mathrm{E}\left(\lambda_i\left(\mathrm{H}_1\mathrm{H}_1^H\right)\right)}{2^{R_s}\frac{\beta_i P}{MN\sigma^2}}\right)\\
&= 1-P\left[\frac{1-2^{R_s}+M\frac{\beta_i P}{\mu^2 N}\mathrm{E}\left(\lambda_i\left(\mathrm{H}_1\mathrm{H}_1^H\right)\right)}{2^{R_s}\frac{\beta_i P}{MN\sigma^2}},MN\right]
\end{aligned} \tag{27}$$

,

where tr(A) denotes the trace of A, and $\mathrm{E}\left(\lambda_i\left(\mathrm{H}_1\mathrm{H}_1^H\right)\right)=tr\left(H_1H_1^H\right)/M$ is the expectation of $\lambda_i\left(\mathrm{H}_1\mathrm{H}_1^H\right)$. P[$x$,$a$] is the normalized incomplete gamma function defined as

$$P[x,a]=\frac{1}{\Gamma(a)}\int_0^x u^{a-1}e^{-u}du,\ x\geq 0\,. \tag{28}$$

## 5. Partial secrecy in network level communication

In the wireless network, the partial secrecy-capacity region for transmission between nodes $i$ and $j$, with eavesdropper $e^*$ is given similarly as Eq. (4)

$$C_{ps}^{GuPS}=\bigcup_{\beta\in[0,1]}\begin{cases}(R_0,R_S):\\ R_0\leq\frac{1}{2}\log\left(1+\frac{(1-\beta)Ph(x_i,x_j)}{\mu^2+\beta Ph(x_i,x_j)}\right)\\ R_1\leq\frac{1}{2}\log\left(1+\frac{\beta Ph(x_i,x_j)}{\mu^2}\right)-\frac{1}{2}\log\left(1+\frac{\beta Ph(x_i,e^*)}{\sigma^2}\right)\\ R_S\geq R_1\end{cases} \tag{29}$$

Similar as Eq. (4), $\mu^2$ and $\sigma^2$ represent the channel noise levels at the receiver and eavesdropper, respectively. Also, $P$ represents the transmission power, with a power fraction $\beta$ which is power allocation fraction to the private stream.

The average link gain in Eq. (29) can be determined based on the distance between the receiving and the transmitting nodes:

$$h(x_i x_j) \approx \frac{1}{\left\| x_i - x_j \right\|^{\alpha}}, \tag{30}$$

where $\alpha$ is the amplitude loss exponent.

Then $e^*$ is denoted the eavesdropper with the nearest to the transmitter $i$,

$$e^* = \arg\max_{e} Ph(x_i, e). \tag{31}$$

To better analyze the privacy requirements on the network connectivity, we define the distance ratio $\xi$, as the ratio of the distance between the transmitter and eavesdropper versus the transmitter and receiver, as

$$\xi = \frac{\left\| x_i - e^* \right\|}{\left\| x_i - x_j \right\|} = \frac{r}{R}, \tag{32}$$

$R$ is the maximum distance for transmission achievable between transmitter and receiver at given rate and secrecy constraints. $r$ is the minimum distance requirement between transmitter and eavesdropper.

Using Eqs. (29) and (30) in conjunction with Eq. (32), we rewrite the partial secrecy capacity as follows

$$C_{ps}^{GuPS} = \bigcup_{\beta \in [0,1]} \begin{cases} (R_0, R_S): \\ R_0 \le \frac{1}{2}\log\left(1 + \frac{(1-\beta(\xi))\frac{P}{R^{\alpha}}}{\mu^2 + \beta(\xi)\frac{P}{R^{\alpha}}}\right) \\ R_1 \le \frac{1}{2}\log\left(1 + \frac{\beta(\xi)\frac{P}{R^{\alpha}}}{\mu^2}\right) - \frac{1}{2}\log\left(1 + \frac{\beta(\xi)\frac{P}{(\xi R)^{\alpha}}}{\sigma^2}\right) \\ R_S \ge R_1 \end{cases}. \tag{33}$$

The legitimate nodes and potential eavesdroppers are randomly located in the space according to a Poisson Point Process. To capture Eq. (29), the distance between the transmitter to closest eavesdropper node is further than the distance from transmitter to the receiver, which is equivalent to a geometrical condition: $r \geq R$.

The distance was derived to meet the constraints about the transmission rate, energy and secrecy requirement, according to Eqs. (29) and (32). A family of graphs is characterize by the secrecy and rate requirement.

Let $\Pi = \{x_i\} \subset \Re^d$ denote the set of legitimate nodes, and $\Pi_E = \{e_i\} \subset \Re^d$ denote the set of eavesdroppers. We define the rate secrecy family of graphs $G = \{\Pi, \mathrm{E}\}_{(E_b, R_S)}$ characterized by energy per bit and secrecy rate requirements, as the graph with vertex set and edge set

$$E = \left\{ \overrightarrow{x_i x_j} : R_0 \geq \eta', R_s \geq \eta' D, R_0, R_s \in C_{ps} \right\} \tag{34}$$

where $C_{ps}$ is the partial secrecy capacity of the link between the transmitter $x_i$ and the receiver $x_j$, such as in Eqs. (29) and (30). $\eta'$ is a threshold of the minimum required transmission rate for the communication link, and $\eta' D$ is a threshold is the required minimum secrecy rate for the communication links.

The condition of the edge in Eq. (34) can be rewrote as a geometrical relationship between the requirements for the distance from transmitter to receiver, and to the eavesdropper, as a function of distortion and energy per bit based on Eqs. (29) and (33), with $\eta = \eta'/2$:

$$E = \left\{ \overrightarrow{x_i x_j} : \begin{array}{l} R \leq \dfrac{r}{\left[ \left(2^{\eta} - 1\right) \dfrac{\mu^2}{\eta E_b} r^{\alpha} + \dfrac{\mu^2}{\sigma^2} 2^{\eta} \beta(\xi) \right]^{1/\alpha}} \\ R \leq \dfrac{r}{\left[ \left(2^{\eta D} - 1\right) \dfrac{\mu^2}{\beta(\xi) \eta E_b} r^{\alpha} + \dfrac{\mu^2}{\sigma^2} 2^{\eta D} \right]^{1/\alpha}} \end{array} \right\}. \tag{35}$$

**Figure 1** shows the dependence of the achievable secrecy rate, the overall transmission rate, the non-private stream rate and the distortion at the eavesdropper, as functions of the distance ratio metric. Numerical results were obtained for $\alpha = 2$ and $\mu^2 = \sigma^2 = 1$. Unless otherwise specified, $R = 1$.

**Figure 1** shows us that the highest transmission rate, yielding the most energy efficient transmission, is obtained for the case in which the eavesdropper is more further than the legitimate receiver, larger $\xi$. This case also related to perfect secrecy, but it will enforce more stricter constraints for the link availability at the network level. Hence it will negatively impact connectivity. If the eavesdropper is closer to the transmitter, the network connectivity is improved, while the higher energy expense is also much higher.

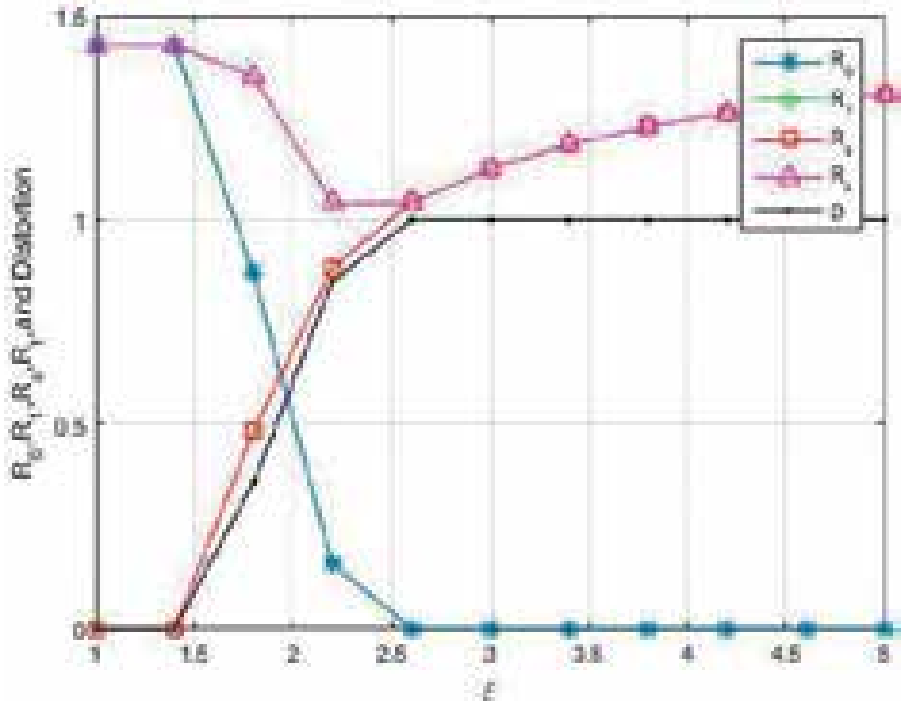

**Figure 1.** Secrecy-energy-connectivity trade-offs.

As in Ref. [20], the Poisson rate-secrecy graph is defined. In the Poisson rate-secrecy graph, $\Pi$, $\Pi_E$ are mutually independent, homogeneous Poisson point processes with densities $\lambda$ and $\lambda_E$, respectively, and consider $\lambda = 1$. But the rate-secrecy graphs are redefined by the transmission range and the distance ratio requirements by incorporate energy and secrecy constraints.

The edge condition in Eq. (35) is rewrite based on definition Eq. (34):

$$\xi \geq \left( \frac{\mu^2 2^{\eta D_{\min}} \beta(\xi) \eta E_b}{\sigma^2 \left( \beta(\xi) \eta E_b - R^\alpha \left( 2^{\eta D_{\min}} - 1 \right) \right)} \right)^{1/\alpha} \geq \left( \frac{\mu^2 2^{\eta D_{\min}}}{\sigma^2 \left( 1 - \frac{R^\alpha}{\eta E_b} \left( 2^{\eta D_{\min}} - 1 \right) \right)} \right)^{1/\alpha} . \tag{36}$$

Eq. (37) shows that a rate-secrecy capacity feasibility condition can be obtained by requiring to be positive, and it is given by

$$R < \left( \frac{\beta \eta E_b}{\left( 2^{\eta D_{\min}} - 1 \right) \mu^2} \right)^{1/\alpha} \leq \left( \frac{\eta E_b}{\left( 2^{\eta D_{\min}} - 1 \right) \mu^2} \right)^{1/\alpha} . \tag{37}$$

The bounds in Eqs. (36) and (37) hold for $\beta \in [0,1]$.

Eq. (37) gives a bound on the maximum transmission range that can be achieved, given transmission rate, energy per bit consumption, and eavesdropper's distortion constraints. For a given transmission rate and distortion requirements, the range of transmission can be made infinitely large by allowing infinitely large transmission power, with the energy per bit consumption going to infinity.

From Eqs. (36) and (37), it shows that secrecy can be achieved when we impose a range and a distance ratio constraint. The connectivity of the network is analyzed by determining the out-degree distributions of the nodes, and the probability of out-isolation, and the average out-degree for an arbitrary node in the network are studied.

There are two cases, which are worth to study: (a) the range limited case – for which distance from transmitter and receiver, $R$, is limited to a maximum value; (b) the unlimited case - $R \to \infty$, which corresponds to the unlimited transmission power case.

For the range limited case, unlimited transmission power and no range transmission limit is imposed. To calculate the out-degree of a vertex, we follow a derivation similar to that in Eq. (35), where it replaces the condition $R < r$. The out-degree probability can be wrote as:

$$P[N_{out} = n] = \frac{\lambda_E \xi^2}{1+\lambda_E \xi^2} \left( \frac{1}{1+\lambda_E \xi^2} \right)^n . \tag{38}$$

The probability that the origin node cannot communicate with another node in $\vec{G}_{1,\lambda_E,\infty,\xi}$ (out-isolation) is then determined to be

$$P_{out-isolation} = P[N_{out} = 0] = \frac{\lambda_E \xi^2}{1+\lambda_E \xi^2} . \tag{39}$$

When a transmission power constraint is imposed, a maximum transmission range $R$ can be determined as in Eq. (37).

As in Ref. [20], this thesis distinguishes two cases:

1. There is no eavesdropper inside a circle with radius $r = \xi R$. This case occurs with probability

$$P_0 = \exp\left(-\lambda_E \pi r^2\right) = \exp\left(-\lambda_E \pi \xi^2 R^2\right) . \tag{40}$$

   For this case, the number of good nodes inside the radius $R$ is given by a Poisson distribution, with the mean, $\pi R^2$.

2. There is an eavesdropper at the distance, $\rho$. Then the number of good nodes is given by a Poisson distribution restricted to a radius $R' = \rho/\xi$, with the mean of $\pi R'^2 = \pi\rho^2/\xi^2$.

Averaging cases (1) and (2) and making the change of variable $r = \rho^2/\xi^2$, we obtain an out-degree probability expression similar to that in Ref. [20], but for an enhanced equivalent arrival rate for the eavesdropper, $\lambda_E^* = \lambda_E \xi^2$:

$$P[N_{out} = n] = \frac{\lambda_E^* \left(1 - \frac{\Gamma(n,a)}{\Gamma(n)}\right) + \exp(-a)\frac{a^n}{n!}}{(\lambda_E^* + 1)^{n+1}} , \tag{41}$$

in which a is $\pi R^2(\lambda^*_E + 1)$, and $R$ is transmission range radius. Also, $\Gamma(.,.)$ is the upper incomplete gamma function.

Hence, the probability of out-isolation is derived as

$$P[N_{out} = 0] = \frac{\exp(-\pi R^2(\lambda^*_E + 1)) + \lambda^*_E}{1 + \lambda^*_E}. \quad (42)$$

The mean out-degree or in-degree can be determined to be

$$E[N^{out}] = E[N^{in}] = \frac{1}{\lambda^*_E}(1 - \exp(-\lambda^*_E \pi R^2)). \quad (43)$$

To achieve numerical results, we assume $\lambda = 1\ \mathrm{m}^{-2}$ and $\lambda_E = 0.08\ \mathrm{m}^{-2}$. **Figure 2** analyzes the out-degree probability on different secrecy level, $D$. The secrecy level is selected for perfect secrecy ($D = 1$) and for a value of distortion that gives a good level of privacy ($D = 0.6$). In **Figure 2**, the significant impact on network connectivity works when the secrecy constraint imposes. Note also that $\xi = 1$, which is the secrecy constraint imposed in Eq. (33), yields non-secrecy ($D = 0$), when transmission rate and energy constraints are also imposed.

In **Figure 3**, it shows the probability of out-isolation and the mean out degree distribution at the case of limited transmission range. When the secrecy level increases, the probability of out-isolation significantly increases. At the perfect secrecy situation, under transmission rate requirements, a minimum value of $\xi$, 2.5, is required (see **Figure 1**). And at the perfect secrecy, it leads to a 3.4 times increasing in the out-isolation probability and a 58% decreasing in the average mean degree, compared to the perfect secrecy studied in Ref. [20]. We also note that by relaxing the secrecy requirements, the out-isolation probability will reduce and the mean out-degree will improve. For example, choosing a distortion at the eavesdropper of $D = 0.6$ (obtained for $\xi = 2$), there is a 73% decreasing for the out-isolation probability and a 72% increasing in the mean out-degree, compared to the case of perfect secrecy (obtained for $\xi = 2.5$).

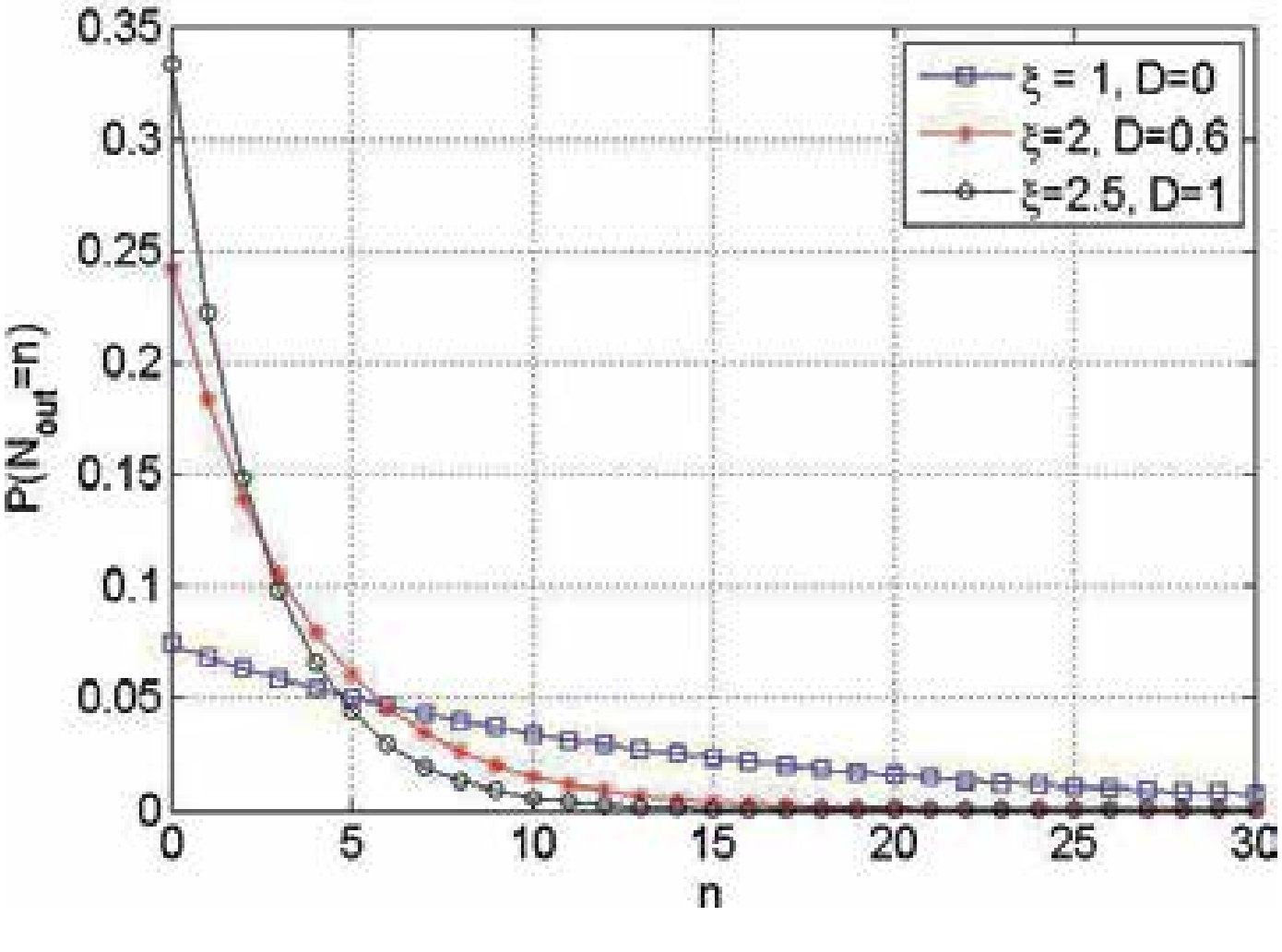

**Figure 2.** Out-degree probability—unlimited range scenario.

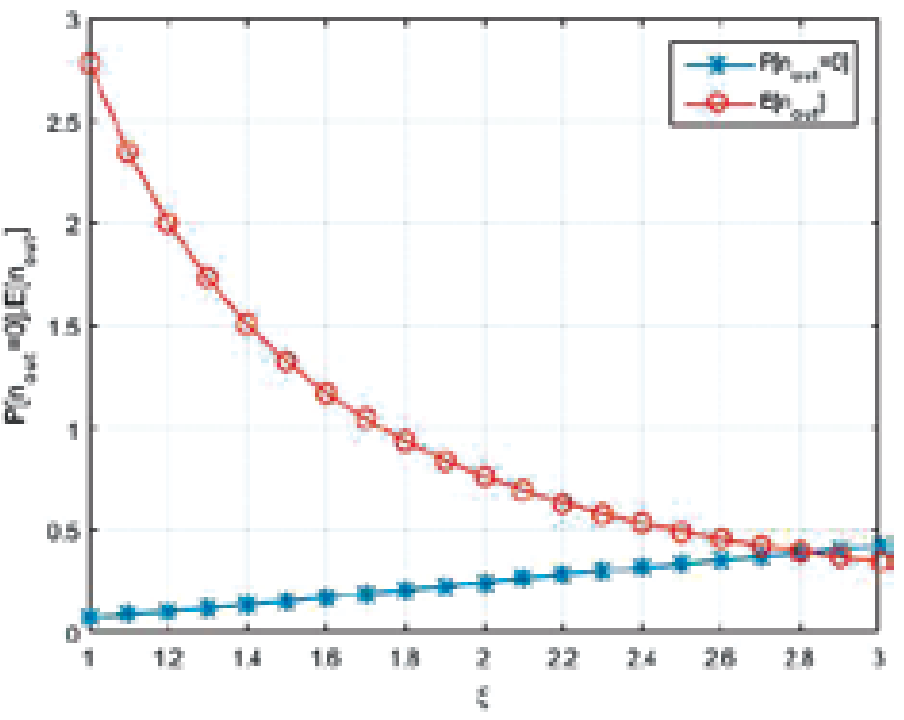

**Figure 3.** Out-isolation probability and mean out degree—limited range scenario.

## 6. Further works

The partial secrecy could also applies in the multi-nodes communication, such as relay-eavesdropper channel. In the no relay case, the source and receiver cannot have secure communication when the eavesdropper's channel is the same or better than the receiver's channel holds. In the relay case, although the source and receiver cannot have secure communication, while they can achieve the partial secrecy. Assuming the nodes of eavesdropper is the randomly distributed in the area, the probability of achieving at least a secrecy level of medium secrecy level could be characterized based on the properties of the random distribution.

Also, for the network level, the partial secrecy could be studied in different distribution of nodes. For example, the case that users are uniformly distributed in area could be considered.

## Author details

Ruolin Zhang

Address all correspondence to: rzhang2@stevens.edu

Stevens Institute of Technology, Hoboken, USA

Chapter 8

# Implementation of a Multimaps Chaos-Based Encryption Software for EEG Signals

Chin-Feng Lin and Che-Wei Liu

Additional information is available at the end of the chapter

**Abstract**

In the chapter, we adopted a chaos logic map and a quadratic map to develop the chaos-based multi-maps EEG encryption software. The encryption performances of the chaos-based software were studied. The percent root-mean-square difference (PRD) is used to estimate the accuracy of a correctly decrypted EEG signal with respect to the original EEG signal. Pearson correlation coefficient (PCC) is used to estimate the correlation between the original EEG signal and an incorrectly decrypted EEG signal. The seven encryption aspects were testing, the average PRD value of the original and correctly decrypted EEG signals for the chaos-based multi-maps software is $2.59 \times 10^{-11}$, and the average encryption time is 113.2857 ms. The five error decryption aspects were testing, the average PCC value of the original and error decrypted EEG signals for the chaos-based multi-maps software is 0.0026, and the average error decryption time is 113.4000 ms. These results indicate that the chaos-based multimaps EEG encryption software can be applied to clinical EEG diagnosis.

## 1. Introduction

Chaos-based encryption is an important research topic in the field of multimedia information communication and storage [1–8]. Compared to the advanced encryption standard (AES), data encryption standard (DES), and the Rivest, Shamir, & Adleman (RSA) algorithm, chaos-based cryptography can exhibit higher levels of security and strong anti-attack ability [7, 8]. The use of chaos-based encryption schemes has expanded steadily over the last few years. Sankpal and Vijaya [9] provided the insights on chaos-based image encryption. Chaotic encryption mechanisms with infinite precision and unpredictability are sensitive to initial conditions and chaotic parameters. Complex chaotic maps have higher levels of security. Chaos-based multimedia encryption can be used in an open access network, and the internet. Zhou et al. [10] proposed a cascade chaotic system using two one-dimension (1-D) chaotic maps in series. The 1-D chaotic maps included logistic, tent, and sine maps. Compared to the use of one 1-D chaotic map, the simulation results showed that the proposed cascade chaotic system had higher robustness and randomness, more unpredictable parameters, and improved chaotic properties and chaotic performance.

Babu and Ilango [11] integrated chaos-based look-up tables using a higher-dimensional Arnold's cat map (ACM) to achieve high encryption sensitivity with respect to the secret key space for audio encryption. The coefficients of the original and encrypted audio signals were employed to evaluate encryption robustness. Mostaghim and Boostani [12] proposed a chaotic visual cryptography (CVC) algorithm to increase steganography in security applications. The key space, key sensitivity, and correlation coefficient of the proposed CVC encryption method were demonstrated. Munir [13] proposed a chaos-based image encryption method using discrete cosine transform (DCT) in the frequency domain. The size of the encrypted image block was $8 \times 8$. The ACM was used to permute the encrypted image block and achieve visual image encryption. The 2D Henon chaotic map and skew tent map play a significant role in the design of permutation and diffusion image encryption mechanisms [14]. The Henon chaotic map generates two different chaotic addresses to permute the row and column of encrypted values in the shuffling process. Furthermore, the unimodal skew tent map was used to scramble the pixel values of the encrypted image using exclusive or (XOR) operations in the diffusion process. Liu et al. [15] proposed a pseudorandom bit generator (PRBG) using parameter-varying logistic map. The change mechanism of the parameters was designed, and the proposed PRBG displayed non-stationary behavior. The parameter-varying logistic map disrupted the phase space of the chaotic system, and could overcome phase space reconstruction to withstand attacks.

Awad et al. [16] investigated chaos-based encryption and transformation approaches using fuzzy keyword search for a mobile cloud storage system. The comprehensive tests showed that the proposed technology obtained a significantly more efficient solution to the searchable encryption problem compared to existing solutions. Huang et al. [17] developed an image cryptosystem using permutation architecture with block and stream ciphers. Ricardo and Alejandro [18] modified a 32-bit chaotic Bernoulli map PRBG using an 8-bit microcontroller. The multiplication, accommodation, addition, and shifting operations were integrated. Jolfaei et al. [19] indicated that permutation-only image ciphers have been used to protect multimedia information in recent years. In the permutation-only image encryption algorithm, the multimedia information is scrambled using a permutation mapping matrix generated by a PRBG. In previous studies [20–23], a chaos-based visual encryption mechanism, 2D chaos-based visual encryption scheme, C# based chaotic single map encryption system, and chaotic visual cryptosystem using empirical mode decomposition algorithm for clinical electroencephalogram (EEG) signals have been proposed. In the chapter, chaotic multimaps of one-channel clinical

EEG encryption software were developed to enhance the encryption robustness. The rest of this paper is organized as follows. The encryption algorithm focusing on the encryption software is investigated in Section 2. Section 3 provides implementation and experimental results of chaotic multimaps of visual clinical EEG encryption software. The conclusion and future work are presented in Section 4.

## 2. Encryption algorithm

The encryption algorithm consists of two main components: chaotic permutation address index approach (CPAIA) and chaotic clinical EEG signal generator approach (CCESGA) for the proposed encryption software. The encryption parameters are inputted to CPAIA, and the chaotic permutation address index sequence is generated. The chaotic permutation address index sequence is integrated to CCESGA, and the chaos-based encryption clinical EEG signal is generated. The proposed CPAIA is shown in **Figure 1**. The CPAIA algorithm proceeds as follows:

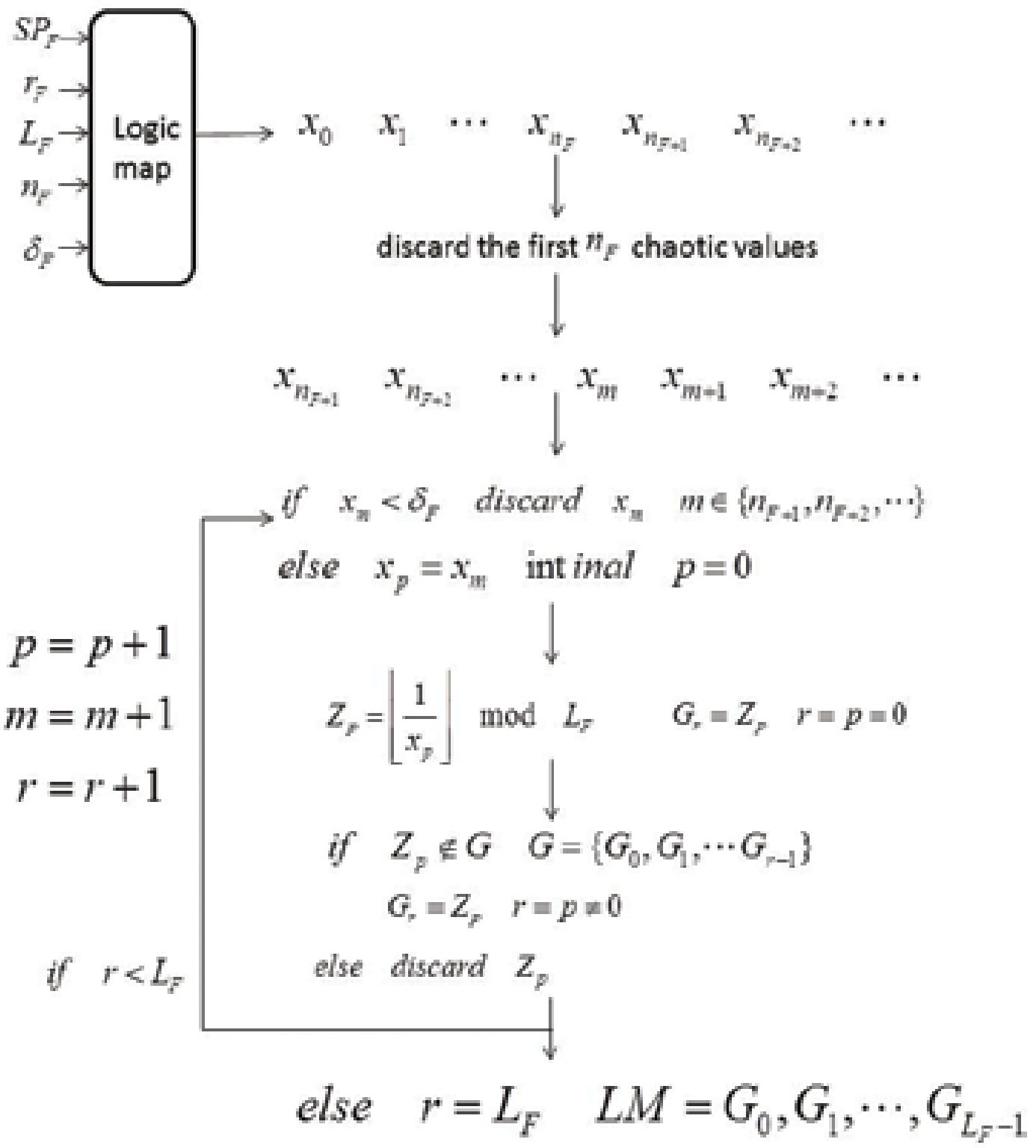

**Figure 1.** The proposed chaotic permutation address index approach.

Step 1: Input the encryption parameters, $SP_F \quad r_F \quad L_F \quad n_F \quad \delta_F$ , into the CPAIA.

$SP_F$: The initial value of the chaotic logic map.

$r_F$: The bifurcation parameter of the chaotic logic map.

$L_F$: The length of encryption clinical EEG signal.

$n_F$: The level of parameter 1 of security.

$\delta_F$: The level of parameter 2 of security.

The chaotic logic map was adopted in CPAIA, described as following.

$x_{n+1} = r_F x_n(1 - x_n)$.................................................................(1)

$x_o = SP_F$

Step 2: Discard the first $n_F$ chaotic logic map values; the chaotic logic map sequence is $x_{n_F+1} \quad x_{n_F+2} \quad \cdots \quad x_m \quad x_{m+1} \quad \cdots$ .

Step 3: If $x_m < \delta_F$, discard $x_m$, $m \in \{n_{F+1}, n_{F+2}, \cdots\}$

else

$x_p = x_m \quad initinal \quad p = 0$

Step 4:

$$Z_p = \left\lfloor \frac{1}{x_p} \right\rfloor \quad \text{mod} \quad L_F \quad G_r = Z_p \quad r = p = 0 \quad \text{..............................(2)}$$

Step 5: If $Z_p \notin G \quad G = \{G_0, G_1, \cdots, G_{r-1}\}$

$G_r = Z_p \quad r = p \neq 0$ .................................................................(3)

else discard $Z_p$.

Step 6: If $r = L_F \quad LM = G_0, G_1, \cdots, G_{L_F-1}$

$LM$ is the chaotic permutation address index sequence.

else

m = m + 1;

p = p + 1;

r = r + 1; and go to Step 3.

The proposed CCESGA is shown in **Figure 2**. The CCESGA algorithm proceeds as follows:

Step 1: Input the encryption parameters, $SP_G \quad r_G \quad L_F$, into the CCESGA.

$SP_G$: The initial value of the chaotic quadratic map.

$r_F$: The bifurcation parameter of the quadratic logic map.

$L_F$: The length of encryption clinical EEG signal.

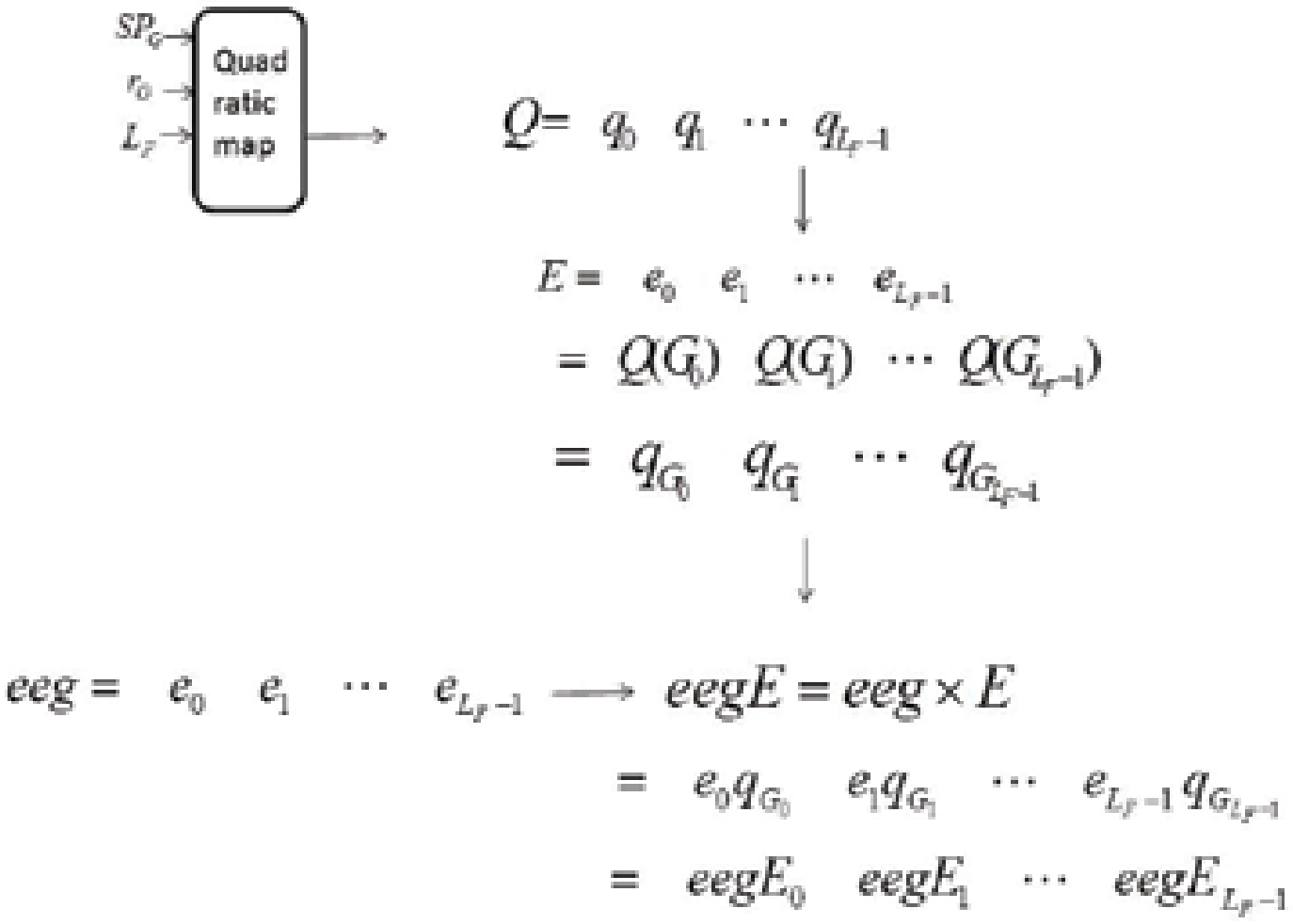

**Figure 2.** The proposed chaotic clinical EEG signal generator approach.

The chaotic quadratic map was adopted in CCESGA, and described as follows:

$q_{n+1} = 1 - r_G \cdot q_n^2$……………………………………………… …………(4)

$q_o = SP_G$

The generation the chaotic quadratic map sequence was $Q$,

$$Q = q_0 \quad q_1 \quad \cdots \quad q_{L_F-1}$$

Step 2: Input the chaotic permutation address index sequence, $LM$.

Step 3: Generate the chaos-based encryption sequence, $E$.

$E = e_0 \quad e_1 \quad \cdots \quad e_{L_F-1}$ …………………………… …………(5)

$= Q(G_0) \quad Q(G_1) \quad \cdots \quad Q(G_{L_F-1})$

$= q_{G_0} \quad q_{G_1} \quad \cdots \quad q_{G_{L_F}-1}$

Step 4: Input the clinical EEG signal, $eeg$.

$eeg = eeg_0 \quad eeg_1 \quad \cdots \quad eeg_{L_F-1}$ ……………………………………………..(6)

Step 5: Generate the chaos-based encryption clinical EEG signal, $eegE$.

$eegE = eeg \times E$…………………………………………………………….....(7)

$= e_0 q_{G_0} \quad e_1 q_{G_1} \quad \cdots \quad e_{L_F-1} q_{G_{L_F}-1}$

$= eegE_0 \quad eegE_1 \quad \cdots \quad eegE_{L_F-1}$

## 3. Chaotic multimaps visual clinical EEG encryption software

**Figure 3** shows the developed chaotic multimaps visual clinical EEG signal encryption software. The software was developed using C# language and Microsoft Visual Studio integrated development environment (IDE). The software includes input, encryption, decryption, storage, and display modules. One-channel clinical EEG signals were inputted in the software through an input module, and were encrypted using an encryption module; these encrypted clinical EEG signals were decrypted using a decryption module. Furthermore, the encryptions and decryptions were stored and displayed using storage and display modules, respectively. The ranges of encryption parameters $SP_F = x1$, $r_F = R$, $SP_G = x2$, $r_G = \alpha$, $n_F$, and $\delta_F$ are 0-1 real numbers (RNs), 3.6-4 RNs, 0-1 RNs, 1.4-2 RNs, 0-100000 integrate number, and 0-0.2 RNs, respectively. **Figure 4** displays the original one-channel clinical EEG signal, whose length is 10 s, and sample rate is 256 samples/s.

**Figure 5** shows the encrypted chaotic multimaps visual one-channel clinical EEG signal. The medical features of the encrypted EEG signal were visually unrecognizable and could not be applied to clinical EEG diagnosis. The encryption parameters $SP_F = x1$, $r_F = R$, $SP_G = x2$, $r_G = \alpha$, $n_F$, and $\delta_F$ are 0.6, 4, 0.6, 1.4, 100, and 0.05, respectively. The robustness of the developed visual chaotic multimaps encryption software was excellent. **Figure 6** shows the correctly decrypted one-channel clinical EEG signal. The percent root difference (PRD) value is adopted to evaluate the difference between the original and correctly decrypted clinical EEG signals and is defined as

$$PRD = 100 \times \sqrt{\frac{\sum_{i=1}^{L_F} (EEG_{ori}(i) - EEG_{dec}(i))^2}{\sum_{i=1}^{L_F} EEG_{ori}^2(i)}} \tag{8}$$

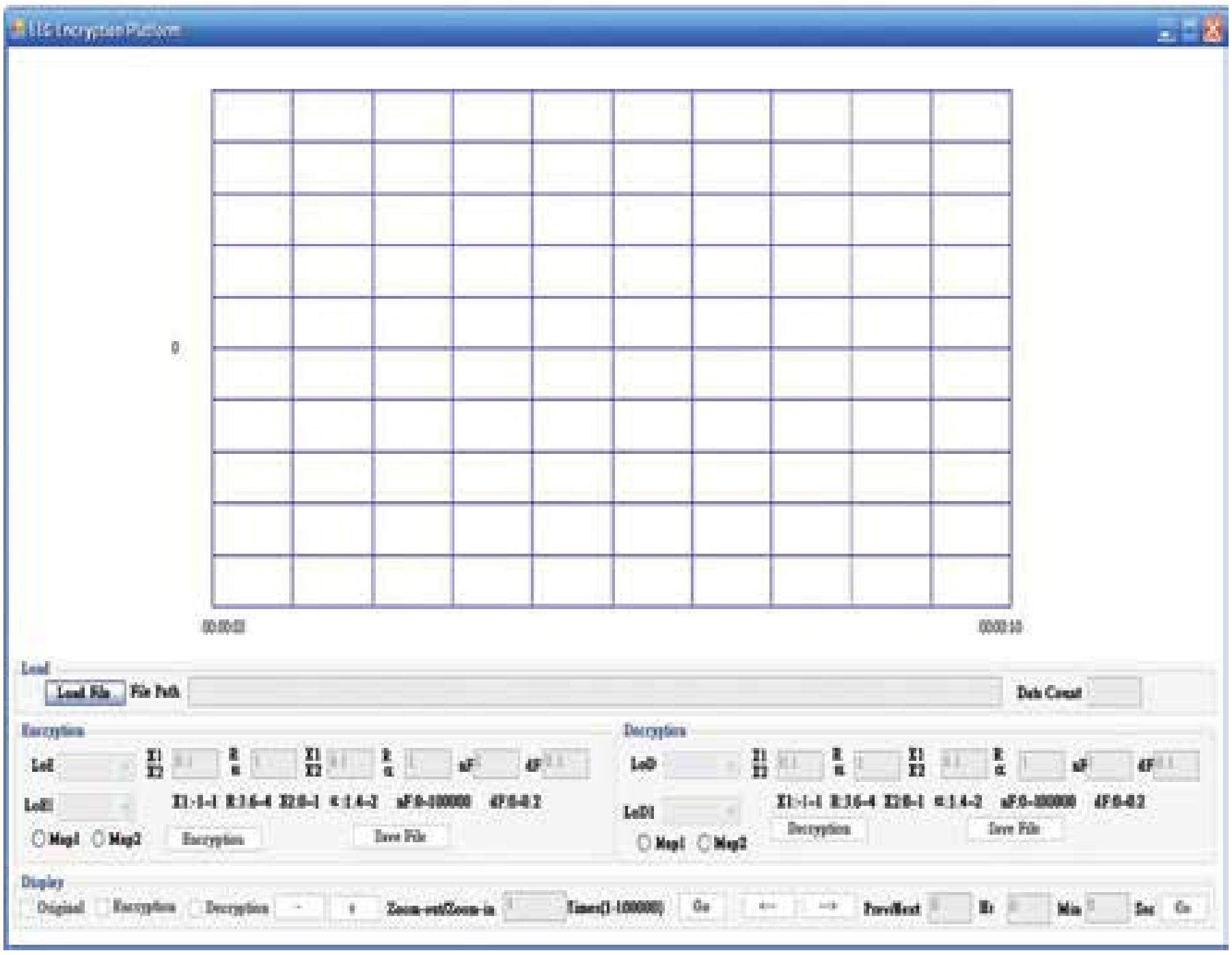

**Figure 3.** The developed chaotic multimaps visual clinical EEG signal encryption software.

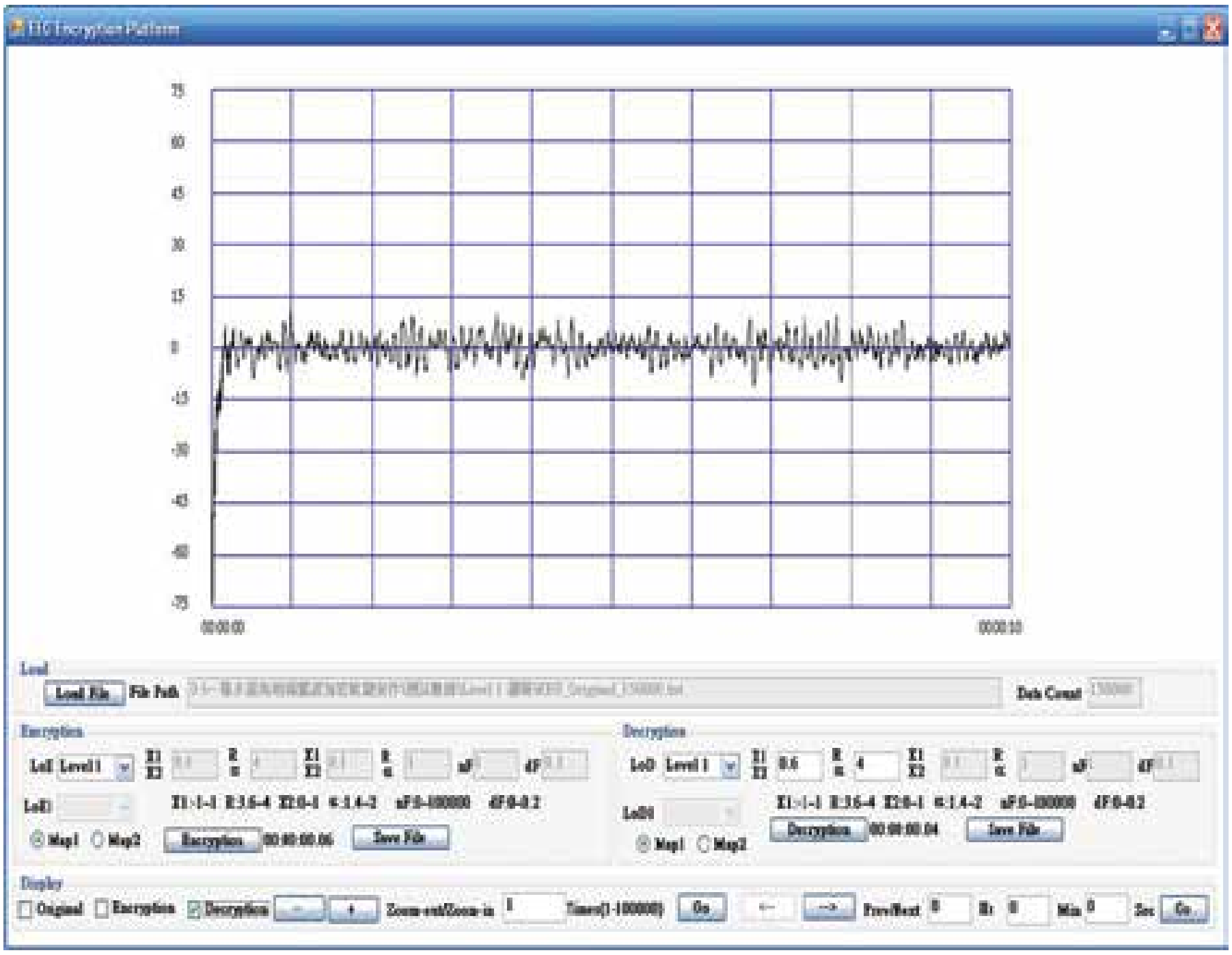

**Figure 4.** The original one-channel clinical EEG signal.

*$EEG_{ori}$: amplitudes of original clinical EEG signal.*

*$EEG_{dec}$: amplitudes of decrypted clinical EEG signal.*

The parameter $L_F$ is 2560. The PRD value of the original and correctly decrypted clinical EEG signals is $3.87 \times 10^{-11}$. **Table 1** lists the encryption parameters, PRD values of correct decryption, and encryption time of the proposed chaotic multimaps visual encryption mechanism for a clinical EEG signal. Seven encryption aspects were tested, and the average PRD value of original and correctly decrypted clinical EEG signal is obtained as $2.59 \times 10^{-11}$ and with the encryption time as 113.2857 ms. From **Figure 6** and **Table 1**, the accuracy of a correctly decrypted EEG signal was excellent, and the correct decryption speed was acceptable. **Figure 7** shows the decrypted one-channel clinical EEG signal with error decryption parameters; the encryption parameters $SP_F = x1$, $r_F = r$, $SP_G = x2$, $r_G = \alpha$, $n_F$, and $\delta_F$ are 0.6, 4, 0.6, 1.4, 100, and 0.05, respectively. In addition, the decryption parameters $SP_F = x1$, $r_F = r$, $SP_G = x2$, $r_G = \alpha$, $n_F$, and $\delta_F$ are 0.6, 4, 0.60001, 1.4, 100, and 0.05, respectively. The Pearson correlation coefficient (PCC) was adopted to evaluate the difference between the original and error decryption clinical EEG signals and is defined as

$$r = \frac{\sum_{i=1}^{L_F} EEG_{ori}(i) EEG_{errdec}(i) - \frac{\sum_{i=1}^{L_F} EEG_{ori}(i) \sum_{i=1}^{L_F} EEG_{errdec}(i)}{L_F}}{\sqrt{\left( \sum_{i=1}^{L_F} EEG_{ori}^2(i) - \frac{\left( \sum_{i=1}^{L_F} EEG_{ori}(i) \right)^2}{L_F} \right) \left( \sum_{i=1}^{L_F} EEG_{errdec}^2(i) - \frac{\left( \sum_{i=1}^{L_F} EEG_{errdec}(i) \right)^2}{L_F} \right)}} \tag{9}$$

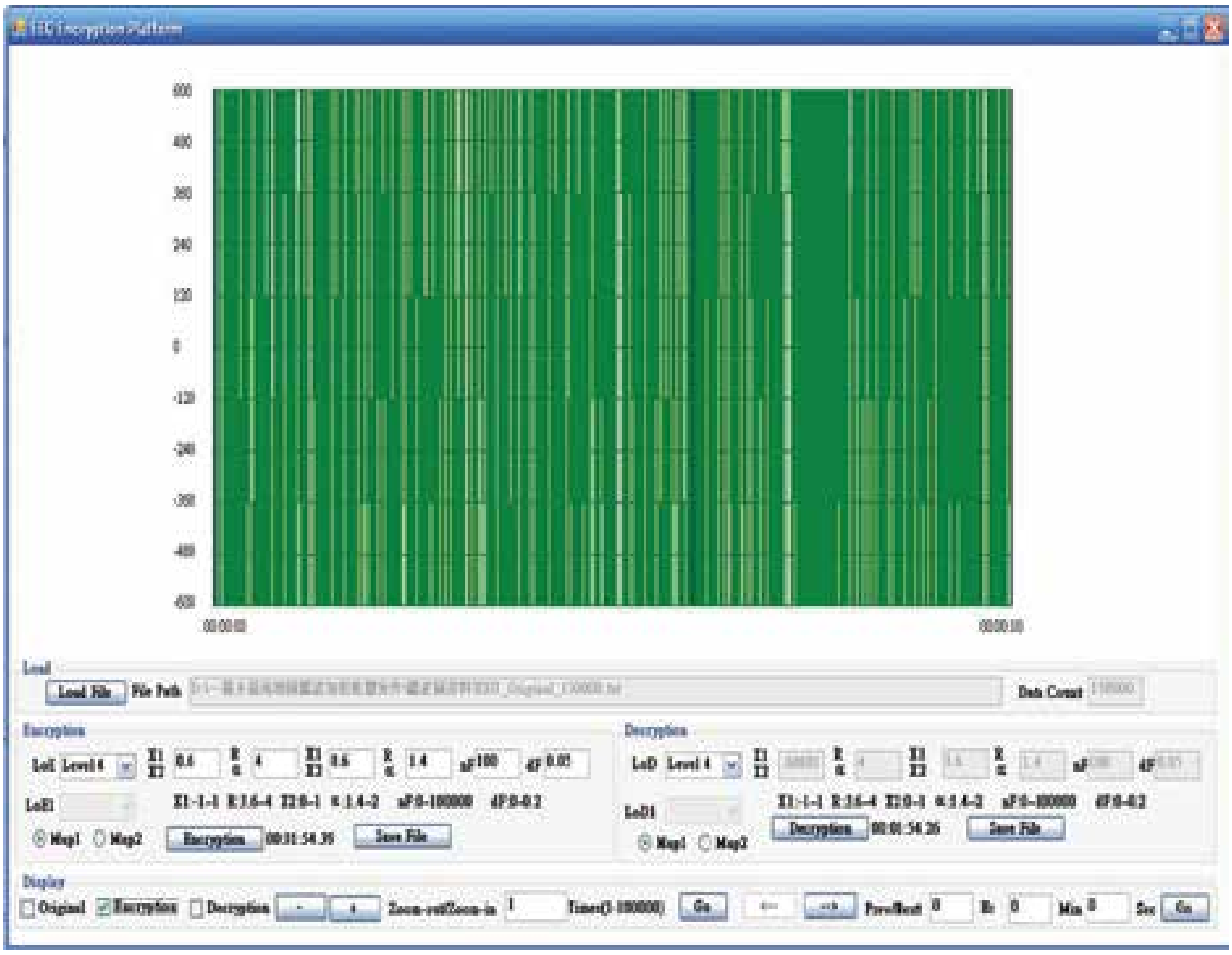

**Figure 5.** The encrypted chaotic multimaps visual one-channel clinical EEG signal.

$EEG_{ori}$: *amplitudes of the original clinical EEG signal.*

$EEG_{errdec}$: *amplitudes of error decrypted clinical EEG signal.*

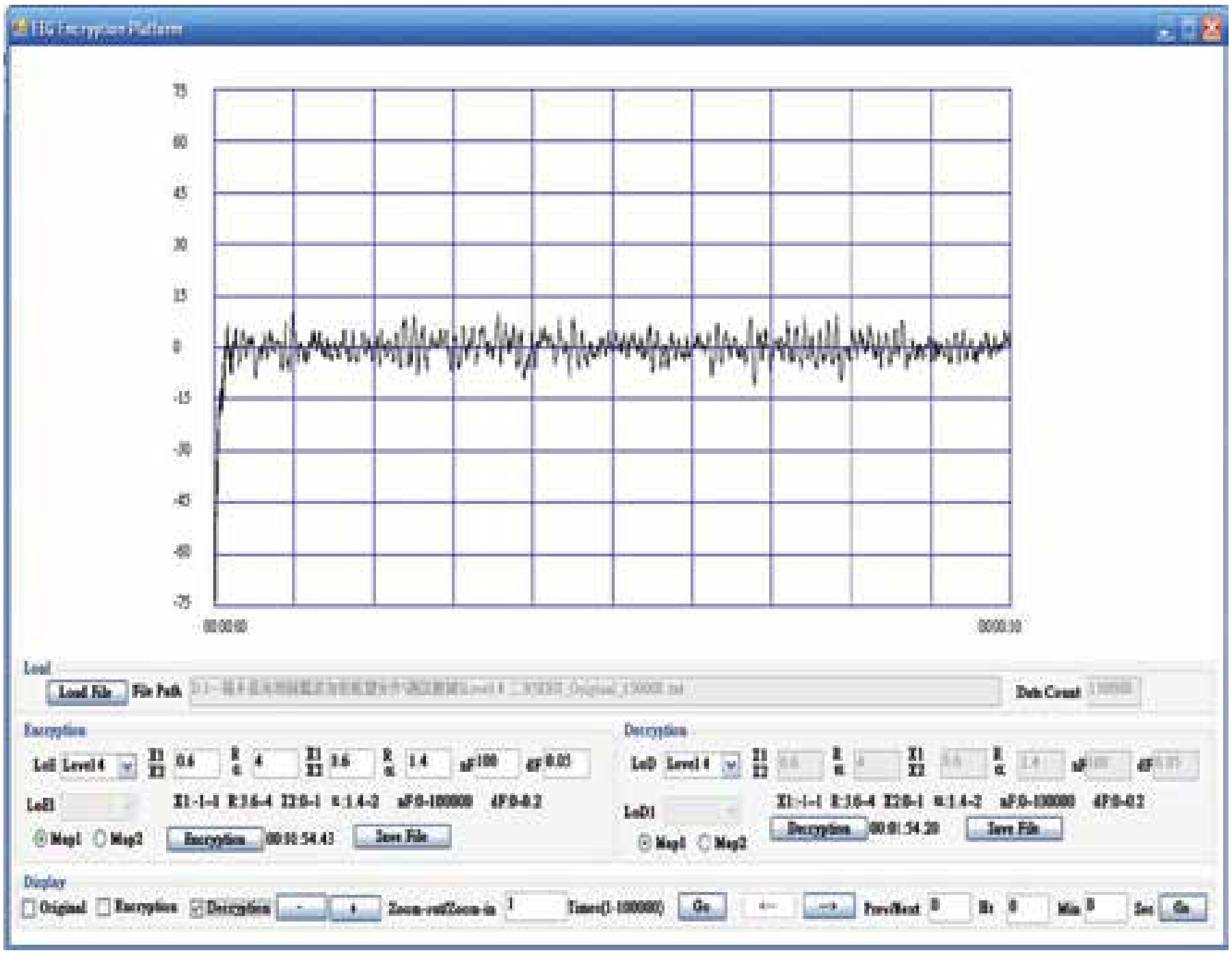

**Figure 6.** The correctly decrypted one-channel clinical EEG signal.

| Encryption | | | | | | PRD | Encryption time (ms) |
|---|---|---|---|---|---|---|---|
| $x_1$ | r | $x_2$ | $\alpha$ | $n_F$ | $\delta_F$ | | |
| 0.6 | 4 | 0.6 | 1.4 | 100 | 0.05 | $3.8734 \times 10^{-11}$ | 113 |
| 0.60001 | 4 | 0.6 | 1.4 | 100 | 0.05 | $1.2630 \times 10^{-11}$ | 113 |
| 0.6 | 3.9999 | 0.6 | 1.4 | 100 | 0.05 | $1.2362 \times 10^{-11}$ | 115 |
| 0.6 | 4 | 0.601 | 1.4 | 100 | 0.05 | $3.3033 \times 10^{-11}$ | 114 |
| 0.6 | 4 | 0.6 | 1.399 | 100 | 0.05 | $1.1655 \times 10^{-10}$ | 113 |
| 0.6 | 4 | 0.6 | 1.4 | 200 | 0.05 | $3.2141 \times 10^{-11}$ | 113 |
| 0.6 | 4 | 0.6 | 1.4 | 100 | 0.15 | $4.0856 \times 10^{-11}$ | 112 |

**Table 1.** The encryption parameters, PRD values of correct decryption, and encryption time of the proposed chaotic multimaps visual encryption mechanism for clinical EEG signal.

**Table 2** lists the decryption parameters, PCC values of error decryption, and error encryption time of the proposed chaotic multimaps visual encryption mechanism for the clinical EEG signal. For this, five error decryption aspects were tested, the average PCC value of original and error decrypted clinical EEG signal was obtained as 0.0026, and the error decryption time was 113.4000 ms. From **Figure 7** and **Table 2**, the original and error decrypted clinical EEG signal was extremely uncorrelated, and the error decryption time was accepted.

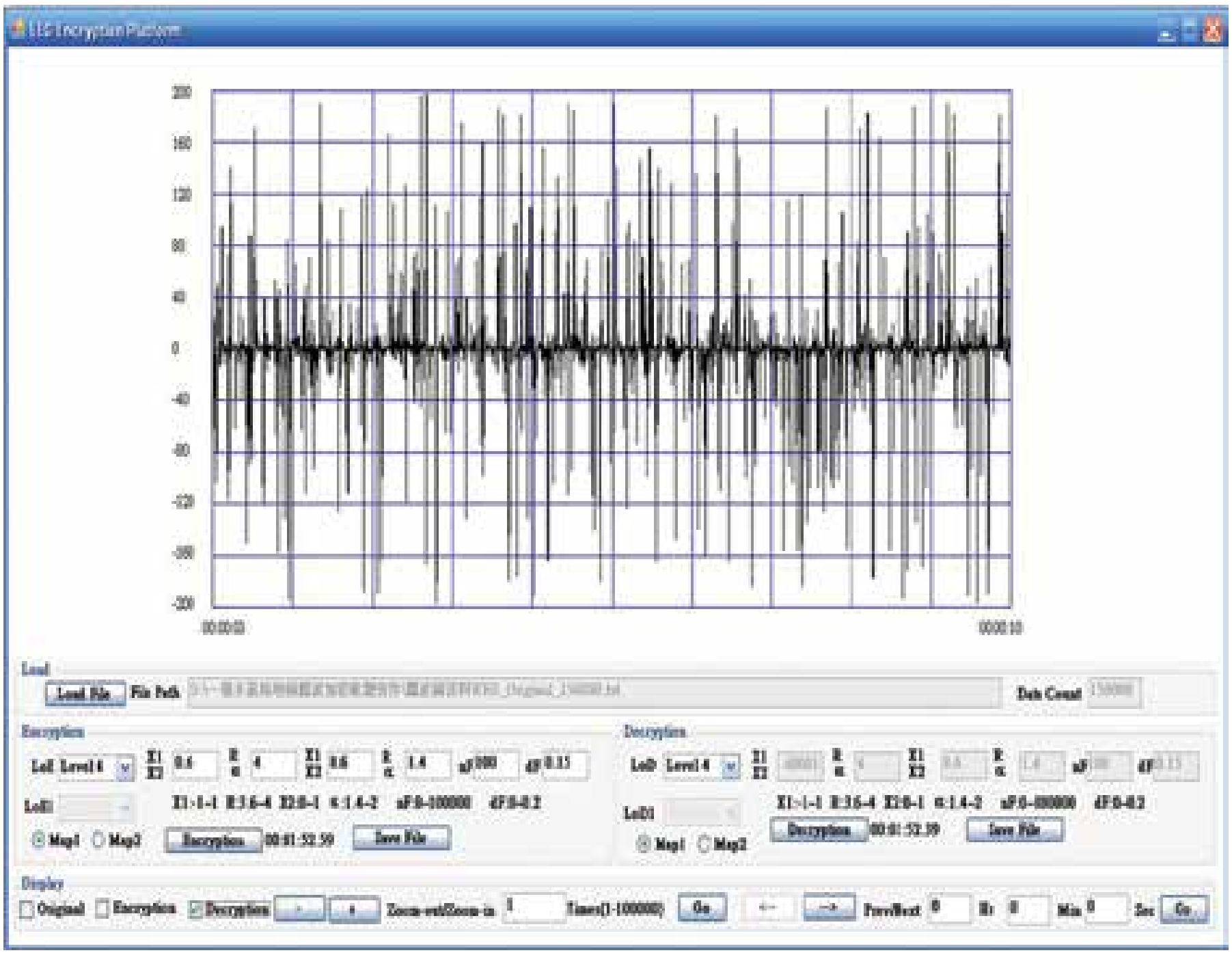

**Figure 7.** The decrypted one-channel clinical EEG signal with error decryption parameters.

| Encryption | | | | | | PCC | Error decryption time (ms) |
|---|---|---|---|---|---|---|---|
| $x$ | $r$ | $x_n$ | $\alpha$ | $n_F$ | $\delta_F$ | | |
| 0.60001 | 4 | 0.6 | 1.4 | 100 | 0.05 | 0.0020 | 112 |
| 0.6 | 3.9999 | 0.6 | 1.4 | 100 | 0.05 | 0.0061 | 113 |
| 0.6 | 4 | 0.601 | 1.4 | 100 | 0.05 | 0.0020 | 114 |
| 0.6 | 4 | 0.6 | 1.4 | 200 | 0.05 | 0.0015 | 114 |
| 0.6 | 4 | 0.6 | 1.4 | 100 | 0.15 | 0.0015 | 114 |

**Table 2.** The decryption parameters, PCC values of error decryption, and error encryption time of the proposed chaotic multimaps visual encryption mechanism for clinical EEG signal.

## 4. Conclusion

This chapter described the proposed chaotic multimaps visual encryption mechanism for one-channel clinical EEG signals. Chaotic logic and chaotic quadratic maps were employed in CPAIA and CCESGA, respectively. The proposed software was implemented using C# language and Microsoft Visual Studio IDE. The PRD and PCC values were used to evaluate the accuracy of the correctly decrypted clinical EEG signals and the robustness of error decryption clinical EEG signals, respectively. The testing results showed that the proposed chaotic multimaps visual encryption software is an excellent encryption software. In the future, the chaotic maps with 2-D, i.e., Henon map, can be adopted to enhance the encryption robustness.

## Acknowledgements

The authors acknowledge the support of the NTOU Center for Teaching and Learning, Maritime Telemedicine Teaching and Learning Project, the support of the Union Teaching of the Ministry of Education for Medical Electronics in Taiwan, and the valuable comments of the reviewers.

## Author details

Chin-Feng Lin* and Che-Wei Liu

*Address all correspondence to: lcf1024@mail.ntou.edu.tw

Department of Electrical Engineering, National Taiwan Ocean University, Taiwan, Republic of China

# ADVANCES IN SECURITY IN COMPUTING AND COMMUNICATIONS

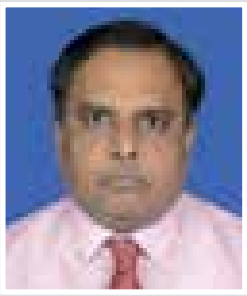

Edited by **Jaydip Sen**

Prof. Jaydip Sen has around 25 years of experience in the field of communication networks, protocol design, network analysis, cryptography, network security, and data analytics in reputed organizations like Oil and Natural Gas Corporation Ltd., India; Oracle India Pvt. Ltd., India; Akamai Technology Pvt. Ltd., India; Tata Consultancy Services Ltd., India; National Institute of Science and Technology, India; and Calcutta Business School, India. Currently, he is associated with Praxis Business School, Kolkata, India, as professor. His research areas include security and privacy issues in computing and communications, intrusion detection systems, secure routing protocols in wireless ad hoc and sensor networks, and privacy issues in the Internet of Things. Prof. Sen obtained his Bachelor of Engineering in Mechanical Engineering with honors from Jadavpur University, Kolkata, India, in the year 1988 and Master of Technology in Computer Science with honors from the Indian Statistical Institute, Kolkata, in 2001.

In the era of Internet of Things (IoT) and with the explosive worldwide growth of electronic data volume, and associated need of processing, analysis, and storage of such humongous volume of data, several new challenges are faced in protecting privacy of sensitive data and securing systems by designing novel schemes for secure authentication, integrity protection, encryption, and non-repudiation. Lightweight symmetric key cryptography and adaptive network security algorithms are in demand for mitigating these challenges. This book presents some of the state-of-the-art research work in the field of cryptography and security in computing and communications. It is a valuable source of knowledge for researchers, engineers, practitioners, graduates, and doctoral students who are working in the field of cryptography, network security, and security and privacy issues in the Internet of Things (IoT). It will also be useful for faculty members of graduate schools and universities.